%% file: arxiv-july-31.tex
\documentclass[mnsc,nonblindrev]{informs3}
\OneAndAHalfSpacedXI

\usepackage{natbib}
 \bibpunct[, ]{(}{)}{,}{a}{}{,}%
 \def\bibfont{\small}%

\usepackage{amsmath}
\usepackage[dvipsnames]{xcolor}
\usepackage{algorithm}
\usepackage[noend]{algorithmic}
\usepackage{enumitem}
\usepackage{subfiles}
\usepackage{subcaption}
\usepackage{pstricks}
\usepackage{bm}
\usepackage{booktabs}

\definecolor{royalazure}{rgb}{0.0, 0.22, 0.66}
\usepackage[hypertexnames=false]{hyperref}
\hypersetup{
	colorlinks = true,
	linkcolor=royalazure,
	citecolor=royalazure,
	urlcolor=black,
	linkbordercolor = {white}
}

\newtheorem{lem}{Lemma}
\newtheorem{thm}{Theorem}
\newtheorem{prop}{Proposition}

\newtheorem{clm}{Claim}
\newtheorem{defn}{Definition}

\usepackage{bbm}

\usepackage{mathtools}
\usepackage{amsmath}
\usepackage{tikz}
\usepackage{tikz-cd}
\usepackage{pgfplots}
\usepackage[utf8]{inputenc}

\usepackage{pgf}

\usepackage{mfirstuc}
\MFUnocap{are}
\MFUnocap{or}
\MFUnocap{etc}

\newcommand{\revcolor}[1]{{{{#1}}}}
\newcommand{\opps}{opportunities} 
\newcommand{\opp}{opportunity} 
\newcommand{\vol}{volunteer} 
\newcommand{\vols}{volunteers} 
\newcommand{\extfrac}{\beta}
\newcommand{\convrate}{\mu}

\newcommand{\capa}{c}
\newcommand{\exttraf}{external traffic}
\newcommand{\inttraf}{internal traffic}
\newcommand{\intorexttraf}{internal or external traffic}
\newcommand{\intandexttraf}{internal and external traffic}

\newcommand{\arrseq}{\vec{\mathbf{A}}}
\newcommand{\instance}{\mathcal{I}}

\newcommand{\instancedomain}{\mathcal{I}}
\newcommand{\MSVV}{\textnormal{\texttt{MSVV}}}
\newcommand{\adaptivecapfull}{Adaptive Capacity}
\newcommand{\OPT}{\textnormal{\texttt{OPT}}}
\newcommand{\adaptivecapmath}{\textnormal{\texttt{AC}}}
\newcommand{\adaptivecap}{\adaptivecapmath}
\newcommand{\AC}{\adaptivecapmath}
\newcommand{\ACrank}{\textnormal{\texttt{AC-R}}}
\newcommand{\CP}{\textnormal{\texttt{CP}}}
\newcommand{\SCP}{\textnormal{\texttt{SCP}}}
\newcommand{\GPG}{\textnormal{\texttt{GPG}}}
\newcommand{\RC}{\textnormal{\texttt{RC}}}

\newcommand{\exttrafmath}{\textsc{ext}}
\newcommand{\inttrafmath}{\textsc{int}}

\newcommand{\volset}{\mathcal{V}}
\newcommand{\exttimes}{\volset^{\textsc{ext}}}
\newcommand{\inttimes}{\volset^{\textsc{int}}}
\newcommand{\oppset}{\mathcal{S}}
\newcommand{\samplepath}{\bm{\omega}}
\newcommand{\samplepathcomponent}{\omega}

\newcommand{\balancefunc}{\psi}

\newcommand{\cascadeclick}{\nu}
\newcommand{\cascadequit}{q}
\newcommand{\convprob}{\mu}

\newcommand{\fillrate}{\textsc{FR}}
\newcommand{\weightcap}{\sigma}

\newcommand{\invbidtobudget}{\underline{c}}
\newcommand{\compratiofunc}{f}
\newcommand{\ALG}{\pi}
\newcommand{\algdomain}{\Pi}
\newcommand{\randomdraw}{\bm{\omega}}
\newcommand{\extrecommend}{i^*_t}
\newcommand{\opprecommendbase}{S}
\newcommand{\opprecommend}[1]{\opprecommendbase^{#1}_t}
\newcommand{\vecopprecommend}[1]{\vec{\opprecommendbase}^{#1}_t}

\newcommand{\numopps}{n}
\newcommand{\horizon}{T}
\newcommand{\choicefunc}{\xi}

\newcommand{\mathprog}{(\textsc{MP})}
\newcommand{\MP}{\mathprog}
\newcommand{\ballvar}{L}
\newcommand{\binvar}{K}
\newcommand{\ballvarwarmup}{\hat{L}}
\newcommand{\binvarwarmup}{\hat{K}}
\newcommand{\bonustimes}{\mathcal{V}^0}

\newcommand{\acbonus}{\AC^0}
\newcommand{\ACrankbonus}{\ACrank^0}
\newcommand{\counter}{k}
\newcommand{\compratio}{\text{CompRatio}}
\newcommand{\signup}{sign-up}
\newcommand{\signups}{sign-ups}

\newcommand{\volchoicet}[1]{\choicefunc_t(\opprecommend{#1})}
\newcommand{\vecvolchoicet}[1]{\choicefunc_t(\vecopprecommend{#1})}

\newcommand{\volchoicetsuccess}[1]{\tilde{\choicefunc}_t(\opprecommend{#1})}
\newcommand{\vecvolchoicetsuccess}[1]{\tilde{\choicefunc}_t(\vecopprecommend{#1})}

\newcommand{\msvvhelper}{\alpha}
\newcommand{\largecapacity}{C}
\newcommand{\largeopps}{N}
\newcommand{\fracextname}{{EFET}}
\newcommand{\fracextnamefull}{{Effective Fraction of External Traffic}}
\newcommand{\cascademodelname}{Opportunity-Agnostic Cascade Model}
\newcommand{\permutation}{\mathcal{P}}
\newcommand{\rankmath}{\mathcal{R}}

\newcommand{\ballvarranking}{L^{\rankmath}}
\newcommand{\binvarranking}{K^{\rankmath}}
\newcommand{\rankchoice}{\phi}
\newcommand{\MCPR}{MCPR}
\newcommand{\MCPRfull}{Maximum Conversion Probability Ratio}
\newcommand{\cascademath}{\mathcal{C}}
\newcommand{\cascadebonus}{\mathcal{V}^\cascademath}
\newcommand{\ballvarcascade}{\ballvar^\cascademath}
\newcommand{\binvarcascade}{\binvar^\cascademath}

\newcommand{\ACcascadebonus}{\ACrank^\cascademath}
\newcommand{\MPsub}[1]{(\textsc{MP}_{#1})}

\newcommand{\numbins}{k}
\newcommand{\indexa}{i}
\newcommand{\indexb}{j}
\newcommand{\dualindex}{\hat{\indexa}}
\newcommand{\oppindex}{u}
\newcommand{\oppindexb}{v}
\newcommand{\capn}[1]{n_{#1}}
\newcommand{\ext}[1]{b_{#1}}
\newcommand{\intern}[1]{a_{#1}}
\newcommand{\totalcap}{N}
\newcommand{\typeset}[1]{O_{#1}}
\newcommand{\cola}{black}
\newcommand{\colb}{black}
\newcommand{\colc}{black}

\newtheorem{exmp}{Example}
\newtheorem{remark}{Remark}

\begin{document}
\TITLE{Online Algorithms for Matching Platforms with Multi-Channel Traffic}
\MANUSCRIPTNO{MS-RMA-22-00910.R1}

\RUNAUTHOR{Manshadi et al.}

\RUNTITLE{Algorithms for Multi-Channel Traffic}

\ARTICLEAUTHORS{%
\AUTHOR{Vahideh Manshadi}
\AFF{Yale School of Management, New Haven, CT, \EMAIL{vahideh.manshadi@yale.edu}}
\AUTHOR{Scott Rodilitz}
\AFF{University of California - Los Angeles, Los Angeles, CA, \EMAIL{scott.rodilitz@anderson.ucla.edu}}
\AUTHOR{Daniela Saban}
\AFF{Stanford Graduate School of Business, Stanford, CA, \EMAIL{dsaban@stanford.edu}}
\AUTHOR{Akshaya Suresh}
\AFF{Yale School of Management, New Haven, CT, \EMAIL{akshaya.suresh@yale.edu}}
} 

\ABSTRACT{
Two-sided platforms rely on their recommendation algorithms to help visitors successfully find a match. However, on platforms such as VolunteerMatch -- which has facilitated millions of connections between volunteers and nonprofits -- a sizable fraction of website traffic arrives directly to a nonprofit's volunteering page via an external link, thus bypassing the platform's recommendation algorithm. We study how such platforms should account for this \emph{external traffic} in the design of their recommendation algorithms, given the goal of maximizing successful matches. We model the platform's problem as a special case of online matching, where (using VolunteerMatch terminology) volunteers arrive sequentially and probabilistically match with one opportunity, each of which has finite need for volunteers. In our framework, external traffic is interested only in their targeted opportunity; by  contrast, \emph{internal traffic} may be interested in many opportunities, and the platform's online algorithm selects which opportunity to recommend. In evaluating the performance of different algorithms, we refine the notion of competitive ratio by parameterizing it based on the amount of external traffic. After demonstrating the shortcomings of a commonly-used algorithm that is optimal in the absence of external traffic, we propose a new algorithm -- Adaptive Capacity (\AC) -- which accounts for  matches differently based on whether they originate from internal or external traffic. We provide a lower bound on \AC's competitive ratio that is increasing in the amount of external traffic and that is close to \revcolor{(and, in some regimes, exactly matches)} the parameterized upper bound we establish on the competitive ratio of any online algorithm. 
\revcolor{We complement our theoretical results with a numerical  study motivated by VolunteerMatch data where we demonstrate the strong performance of \AC\ relative to current practice and further our understanding of the difference between \AC\ and other commonly-used algorithms.}  
}


\maketitle

\vspace{-20pt}

\section{Introduction}
\label{sec:intro}

Online platforms have become increasingly prominent in facilitating social and economic connections  in both the private and nonprofit sectors. In the private sector, the e-commerce platform Etsy has empowered over 2 million small-scale sellers to showcase their products to over 40 million buyers and has facilitated transactions on the scale of \$4 billion.\footnote{https://www.sec.gov/Archives/edgar/data/1370637/000137063719000028/etsy1231201810k.htm}
In the nonprofit sector, {the crowdfunding platform DonorsChoose has} helped public school teachers to successfully solicit \$314 million in donations for 1.7 million classroom projects.\footnote{https://www.donorschoose.org/about/impact.html}  Similarly, VolunteerMatch has enabled over 18 million connections between organizations and {individuals} looking for volunteering opportunities.

These platforms attract traffic through multiple channels. Some users organically visit the platform and rely on its recommendation algorithm to find a desired  product or volunteering opportunity---we refer to these users as {\em \inttraf}. Other users, which we refer to as {\em \exttraf}, follow an external direct link to a particular page. This \exttraf\ is  generated through a variety of off-platform outreach mechanisms, such as posting on 
social media or sending customized notifications. 
For example, an artist who sells their handmade products on Etsy may tweet about them, or an NGO may publicize their volunteering/donation opportunities on their Facebook page. 
In this paper, we aim to understand how these matching platforms can 
efficiently leverage  traffic from {\em all} sources in order to maximize the number of successful transactions/connections.

This work is partly motivated by our collaboration with VolunteerMatch (VM), the largest nationwide  platform that connects nonprofits with volunteers.  More than {130,000} organizations---supporting a variety of social causes, ranging from human rights and literacy to helping seniors---have posted their volunteering \opps\ on the VM website. Most of these organizations rely on volunteers who sign-up after browsing the VM website. Some of these organizations also generate sign-ups by publicizing their opportunities on other websites, such as LinkedIn or Facebook. Our analysis of VM data reveals two key facts. First,  a significant portion of volunteer \signups\ come from \exttraf: for example, 30\% of all \signups\ made by NYC-based volunteers between August 1, 2020 and March 1, 2021 came from \exttraf. Second,  there is a significant disparity {across \opps} in terms of both the total number of \signups\ and the source of those sign-ups.

\begin{figure}[t]
    \begin{center}
        \resizebox {0.75\textwidth} {!} {\input{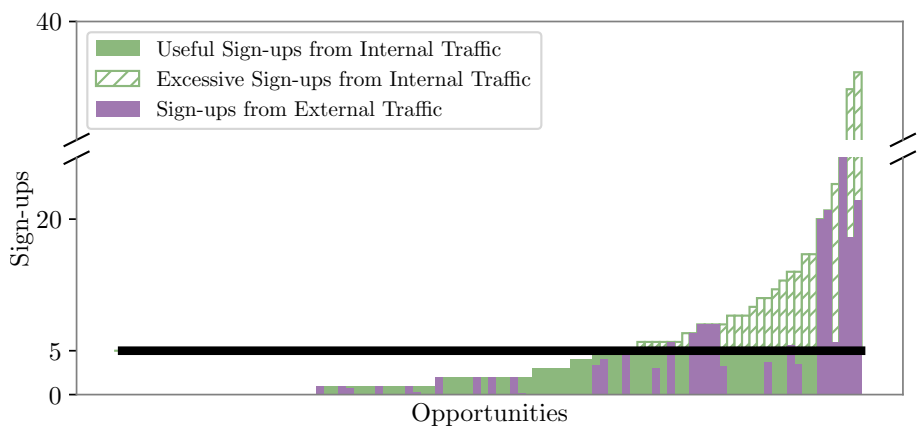}}
    \end{center}
    \caption{Distribution of sign-ups on VM across a subset of opportunities requesting 5 volunteer sign-ups. 
    }
    \label{fig:signup_disparity}
\end{figure}

To illustrate these two facts, in Figure~\ref{fig:signup_disparity} we plot the distribution of the number of \signups\ for a subset of \opps\ that all requested 5 \vol\ \signups.\footnote{This subset of 100 \opps\ is a random sample of all virtual \opps\ requesting 5 volunteer \signups\ between August 2020 and March 2021.} 
Partitioning the \signups\ into two groups based on their source, we observe that
the volume of \signups\ from external traffic (in purple) and from internal traffic (in green) varies substantially across opportunities.\footnote{We only observe the source for a subset of \signups, as described in Appendix \ref{app:data_availability}. We estimate the source of each \opp's \signups\ proportionally, based on this subset.}  From the platform's perspective, a key difference between external and internal traffic {comes from whether or not the user's choice can be influenced}: the platform cannot control the ``landing page'' for external traffic, but it can impact what internal traffic views (and thus the decisions made) via its recommendation algorithm. Through its search design, the platform can (potentially) re-distribute ``excessive'' \signups\ from \inttraf\ (i.e., sign-ups that exceed an \opp's need) to \opps\ with insufficient sign-ups, thereby helping VM achieve its strategic goal of maximizing the total number of ``useful'' \signups\ across \opps.\footnote{We note that the skewed \signup\ distribution not only hurts \opps\ with insufficient \signups, but it also harms other stakeholders. For instance, individuals that sign up for \opps\ with excessive \signups\ may be discouraged if their attempts to \vol\ are ignored or if they exert unnecessary effort. Additionally, organizations that receive excessive \signups\ may also incur/impose costs due to screening or training unnecessary \vols.} For instance, for the subset of \opps\ presented in Figure~\ref{fig:signup_disparity}, in hindsight, {$49\%$} of \signups\ from \inttraf\ (the dashed green portions of the bars) {could potentially have been re-directed} to opportunities with insufficient \signups.

The above observations motivate our main research question: {\em how can matching platforms, such as VM, integrate external and internal traffic to maximize the number of useful \signups?} As the traffic pattern is generally unknown a priori and there is heterogeneity in the level of external traffic, making better {\em real-time} recommendations to \inttraf\ may be challenging.

\subsection{Our Contributions}
To study the above question, we introduce a framework for online matching with multi-channel traffic. Taking a competitive analysis approach, we show that existing algorithms---that are optimal in the absence of external traffic---fail to integrate such traffic efficiently; thus, we develop a new algorithm that effectively incorporates external traffic, resulting in near-optimal guarantees in certain regimes. {Beyond worst-case guarantees,} we illustrate the effectiveness of our algorithm in a simulation study calibrated on VM data. We describe each contribution in more detail next.

{\bf A model for online matching with multi-channel traffic:}
For concreteness, we utilize terminology from the context of  VM 
and refer to  the two sides of the matching platform as ``opportunities'' and ``volunteers.''  
{In our setting, a fixed set of opportunities are posted on the platform, each requiring a certain number of volunteers which we refer to as its ``capacity.''
Volunteers arrive sequentially {(in an arbitrary order)}
and are either external or internal traffic. \revcolor{External traffic directly views a specific opportunity's page and signs up for it.  By contrast, internal traffic can be influenced by the platform's recommendation algorithm as follows.  When an \inttraf\ \vol\ arrives, the platform observes their {\em conversion probability} for each \opp\ (i.e., the probability that the volunteer signs up for that opportunity conditional on viewing it), and then must immediately and irrevocably  recommend one such opportunity.}\footnote{{In our base model (introduced in Section \ref{sec:model}), we assume that the platform recommends a single \opp. We consider a more general setting where the platform can present a ranking of \opps\ in Appendix \ref{app:rank:fix}.}} The  goal of the platform is to maximize the total number of ``useful'' \signups, i.e., the total number of sign-ups that don't exceed an \opp's capacity. 
}

In the absence of external traffic, the above problem can be viewed as  an instance of the online bipartite B-matching problem {with stochastic rewards and an adversarial arrival sequence}. In this general framework, it has been shown that a simple myopic algorithm commonly-referred to as $\MSVV$ achieves the best-possible competitive ratio of $1-1/e$ \citep{mehta2007adwords}.\footnote{Though \citet{mehta2007adwords} considers a setting with deterministic rewards, as noted in \citet{mehta2013online}, the guarantee and the optimality of \MSVV\ extend (asymptotically) to a B-matching setting with stochastic rewards when all capacities are sufficiently large. 
We will henceforth describe results only for the large-capacity setting; however, our technical results are all parameterized by the minimum capacity. } 
We augment this framework by modeling \exttraf\ as arrivals with only one possible edge (e.g., \vols\ that only consider one \opp). The presence of \exttraf\ reduces the complexity of making real-time decisions: the platform cannot change what \exttraf\ volunteers will view, as they are only interested in one opportunity. Thus, in the extreme case where all capacity can be filled by \exttraf, the platform trivially maximizes the number of useful \signups.

In light of the above observation, we parameterize problem instances based on the fraction of total capacity that can be filled by external traffic, which we call the \textit{\MakeLowercase{\fracextnamefull}} (\fracextname), as formalized in Definition \ref{def:beta}. For a given \fracextname, we define the competitive ratio of an algorithm to be the worst-case ratio between its outcome and that of a benchmark, among all instances with that \fracextname\ (see Definition \ref{def:compratio}). {Our benchmark (denoted $\OPT$) is a clairvoyant solution that a priori knows the sequence of arrivals, but only observes the \signup\ realizations of internal traffic after recommending an opportunity (see Definition \ref{def:opt}). }
We study how the addition of \exttraf\ improves the achievable competitive ratio.

{\bf Failure of channel-agnostic algorithms:} To gain intuition, we first focus on a thought experiment where all of the external traffic arrives before any of the internal traffic. In such a setting, after the \signups\ from \exttraf\ realize, the platform is faced with a standard instance of the  online matching problem. Thus, by making recommendations in the  \emph{remaining} problem according to an optimal algorithm like $\MSVV$, we would hope to achieve a competitive ratio that is a convex combination of $1$ and $1-1/e$. Indeed, in Proposition \ref{prop:warmupupperbound}, we prove that this convex combination is an upper bound on any online algorithm.
However, somewhat surprisingly, applying $\MSVV$ to the \emph{entire} problem instance does not achieve this intuitive bound (Proposition \ref{prop:warmupmsvv}). 
The suboptimality of this algorithm stems precisely from a lack of differentiation between external and internal traffic.

{\bf Adaptive Capacity (\AC) algorithm:} 
Building on the intuition developed in the thought experiment above, we introduce a new algorithm called {\em Adaptive Capacity} ($\adaptivecapmath$) which reduces an \opp's capacity by one whenever that \opp\ receives a \signup\ from \exttraf. If all external traffic arrives before any internal traffic, $\AC$ achieves the upper bound in Proposition~\ref{prop:warmupmsvv}. However, in a general setting where external traffic can arrive at arbitrary times, $\AC$ does not have the information needed to reduce capacities up-front; instead, it {\em adaptively}  reduces capacity after each \signup\ from \exttraf\ (see Algorithm \ref{alg:acpolicy}).

\revcolor{To shed light on the inherent difficulty of making real-time decisions under arbitrary arrival sequences, we first establish an upper-bound (as a function of the \fracextname) on the competitive ratio of any online algorithm (Theorem \ref{thm:hardness}). Our main theoretical results establish performance guarantees (also as a function of the \fracextname) on the competitive ratio of $\adaptivecapmath$ that depend on the conversion probabilities (see Figures \ref{fig:CR_weightcap} and \ref{fig:weight_caps} for an illustration). In the special case with deterministic conversion, i.e., where conversion probabilities are either $0$ or $1$, we use combinatorial arguments inspired by the ideas of \citet{mehta2007adwords} to establish a lower bound on the performance of \AC\ that converges to our upper bound as capacities increase (Theorem~\ref{thm:tightproof}).\footnote{\revcolor{In recent work, \citet{udwani2021adwords} proposes a randomized algorithm whose competitive ratio (for arbitrary capacities) matches our upper bound. We emphasize that \citet{udwani2021adwords} only considers the case with deterministic conversion.}}

For more general settings, Theorem \ref{thm:AClower} parameterizes \AC's achievable competitive ratios by the \emph{maximum conversion probability ratio} (\MCPR), which we formally introduce in Definition \ref{def:MCPR}.
Fixing any \MCPR\ and focusing on the large-capacity regime, our lower bound curve  
starts at $1-1/e$ (when there is no \exttraf) but weakly increases with the \fracextname\ and eventually breaks the barrier of $1 -1/e$.  
As the \MCPR\ approaches $1$, our lower bound nearly matches our upper bound on $\AC$ for any \fracextname. The lower bounds that we establish on the competitive ratio of \AC\ compare favorably to the competitive ratio of \MSVV, which we show is strictly worse in some regimes. Beyond worst-case guarantees, in Section \ref{subsubsec:thm2:discussion} we provide insight into instance characteristics that favor one algorithm over the other.}

\revcolor{Our theoretical results are particularly intriguing because our algorithm does not require \emph{a priori} knowledge of the volume of external traffic; yet by adaptively reducing capacities, \AC\ achieves a near-optimal (and in some settings, exactly optimal) competitive ratio. This is an appealing quality for practitioners, as it is often infeasible to know the volume of \exttraf\ in advance. Moreover, for high-information settings where practitioners have advance knowledge about the volume of \exttraf\ \emph{per opportunity}, we describe how a variant of \AC\ can attain stronger performance guarantees (see Remark \ref{remark:extknown}).}

\revcolor{To analyze the competitive ratio of $\AC$ in the general case, we build on the LP-free approach in  \citet{goyal2020asymptotically}, which establishes a system of inequalities involving path-based ``pseudo-rewards.'' To break the barrier of $1-1/e$ we leverage the observation that an algorithm cannot make a bad decision for external traffic, and thus we define pseudo-rewards based on the source of the traffic. Moreover, as the volume of external traffic varies across opportunities, we move beyond an opportunity-level analysis, and instead bound the ``global'' value of $\AC$ relative to $\OPT$.}  

{\bf Case study based on VM:} 
\revcolor{To explore the performance of our algorithm beyond worst-case settings, we evaluate it on problem instances constructed using data from the VM platform. {We show that our $\AC$ algorithm significantly outperforms its worst-case guarantee and performs similar to or better than several benchmarks (Table \ref{table:policyperf}). In particular, we show that our $\AC$ algorithm compares favorably against a proxy for current practice on VM} by reducing the number of excessive \signups, thereby utilizing \inttraf\ more efficiently 
(Figure \ref{fig:ac_vs_cp}). Furthermore, we explore how instance characteristics such as the EFET and the arrival sequence impact the relative performance of $\AC$ and $\MSVV$ (Figure~\ref{fig:ac_vs_msvv}).}

\section{Related Work}
\label{sec:literature}
Our work relates to and contributes to several streams of literature.

{\bf Generalized Online Matching:}
The rich literature on online matching started with the seminal work of \citet{karp1990optimal};  given the scope of this literature, we discuss only a few papers  and kindly refer
the reader to \citet{mehta2013online} for a comprehensive survey. We model the platform's problem as a generalized instance of online B-matching \citep{kalyanasundaram2000optimal}, which has been extensively studied in the context of online advertising \citep{mehta2007adwords, buchbinder2007online, balseiro2020best, udwani2021adwords}.\footnote{Our framework allows for stochastic rewards, which can introduce additional challenges \citep{mehta2012online, goyal2019online}. We sidestep this challenge by parameterizing our results based on the minimum capacity and by focusing on the large-capacity regime, following the approach of this literature.} 
Variants of online B-matching problems have been recently proposed to study a variety of problems arising in online platforms, including real-time assortment decisions \citep{golrezaei2014real, ma2020algorithms, aouad2020online, desir2021capacitated} and online allocation of reusable resources \citep{Rad2019, goyal2020asymptotically, rusmevichientong2020dynamic, gong2019online}. We contribute to this line of work by introducing a variant of online matching motivated by platforms with multi-channel traffic. 

In our model, each external traffic \vol\ corresponds to a degree-one arriving node. Our $\adaptivecapmath$ algorithm effectively incorporates these degree-one nodes, and not only breaks the barrier of $1-1/e$ given a sufficient amount of external traffic, but also achieves a near-optimal competitive ratio in certain parameter regimes. 
In a similar vein, the work of 
\citet{buchbinder2007online} and \citet{naor2018near} impose a bound on the degree of {\em all} nodes in one or both  sides and show that one can improve upon a competitive ratio of $1-1/e$ for such structured instances.  We emphasize that our work differs from these papers, as we make no assumption on the degree of \inttraf.  

\revcolor{Our algorithm builds on ideas in \citet{mehta2007adwords} and introduces a different notion for an opportunity's \emph{fill rate}, leading to an improved competitive ratio; in a different context (where arrivals are batched), \citet{feng2022batching} likewise adapts the notion of a fill rate to establish improved performance guarantees. Our proof technique builds on the flexible LP-free approach of \citet{goyal2019online} and \citet{goyal2020asymptotically}, which we use to distinguish between external and  \inttraf\ in our analysis. 

The flexibility of this approach is further displayed in \citet{udwani2021adwords}, which studies a ``capacity-oblivious'' variant of the AdWords problem (i.e., a setting in which the algorithm does not know the capacity of the offline side until its capacity has been filled); for this setting, it introduces and analyzes a randomized algorithm called Generalized Perturbed Greedy. In addition to establishing intriguing results in the AdWords setting, \citet{udwani2021adwords} extends the analysis of their algorithm to a special case of our setting where sign-ups are deterministic. Even for arbitrary opportunity capacities, the competitive ratio of Generalized Perturbed Greedy matches the upper bound we establish on any online algorithm (see Theorem \ref{thm:hardness}).}

{\bf Hybrid Traffic Models:}
The challenge of integrating different channels of traffic arises in other application domains as well, such as retail and e-commerce. 
\citet{dzyabura2018offline} study a retail setting where the firm offers products through  both offline and online channels. Consumers are a mixture of three types: those who visit only online or only offline, and those who visit the store before making a purchasing decision online (and thus their preference may be impacted by the products showcased in the offline store). They study  assortment problems for this mixture of consumers.  
In the context of e-commerce, \citet{esfandiari2015online}, \citet{kumar2018semi},  and \citet{hwang2021online} consider online allocation problems where the traffic is composed of a predictable component (i.e., deterministic or from a known distribution) as well as an unpredictable component (i.e., adversarial). 
We contribute to this line of work by introducing a new hybrid traffic model that consists of external and internal traffic.

{\bf Design of Matching Platforms:}
Motivated by the rapid growth of online matching platforms, recent work has shed light on how platform design can influence matching outcomes, e.g., in the context of labor markets  \citep{aouad2020online}, crowdsourcing \citep{manshadi2020online}, affordable housing \citep{arnosti2020design}, ridesharing \citep{besbes2021surge}, and dating markets \citep{rios2020improving}. 
Among other insights, this line of research analyzes the relative merits of different pricing/compensation policies \citep{alaei2022revenue, elmachtoub2022revenue}, demonstrates the value of limiting user choice \citep{immorlica2021designing, kanoria2021facilitating}, and provides guidance on which assortments to show users of two-sided platforms \citep{ashlagi2019assortment, aouad2020online, feldman2022multinomial}.
We add to the platform design literature by studying how online matching platforms should adjust their recommendations to account for \exttraf.

\section{Model}
\label{sec:model}

In this section, we formally introduce our model for the problem that a platform faces when providing recommendations in the presence of multi-channel traffic, which is a variant of online matching. (For ease of exposition, we will use terminology from the context of a volunteer matching platform to describe the model.)
We then describe the platform's objective and the metric of a competitive ratio, which we will use to evaluate any online algorithm. 

Each problem instance $\instance$ consists of a static set of \opps\ on the platform (denoted $\oppset$), a finite horizon of $\horizon$ periods,  and a sequence of $T$ \vols\ who arrive to the platform (denoted $\arrseq$). We index \opps\ with $i$ from $i = 1$ to $\numopps = |\oppset|$.  Each \opp\ $i$ has \emph{capacity} $\capa_i$, which represents the total number of \vol\ \signups\ needed by \opp\ $i$. 
In each period $t$, the $t^{th}$  \vol\ in sequence $\arrseq$ arrives to the platform. As each period corresponds a  \vol\ arrival, we index volunteers according to their arrival time, i.e., \vol\ $t$ arrives at time $t$ for $t \in [\horizon]$.\footnote{For any $n \in \mathbb{N}$, we use $[n]$ to denote the set $\{1, 2, \dots, n\}$.}


\smallskip

\textbf{Volunteer dynamics:} 
When volunteer $t$  arrives, the platform observes its type, which consists of two components. The first component of a \vol's type is its source, either $\exttrafmath$ or $\inttrafmath$, which indicates whether the volunteer arrives to the platform as external or internal traffic, respectively. This is our way of modeling the multi-channel nature of the platform's traffic. We use
$\exttimes$ (resp. $\inttimes$) to denote the set of \vols\ who arrive as \exttraf\ (resp. \inttraf).

The second component of a \vol's type is a vector $\bm{\convprob}_t=\{\convprob_{i,t} : i \in \oppset \}$, where $\convprob_{i,t} \in [0,1]$ is the pair-specific \textit{conversion probability} with which \vol\ $t$ will \signup\ for \opp\ $i$ if the \vol\ views \opp\ $i$.
\revcolor{As motivated in the introduction, we assume that whenever \exttraf\ arrives, they cannot be influenced by the platform and instead automatically view their targeted \opp, denoted $\extrecommend$. To simplify exposition, we assume that each \exttraf\ volunteer deterministically signs up, i.e., the conversion probability for their targeted opportunity is $1$. (In Appendix \ref{apx:exttrafstochastic}, we discuss how our model and results generalize to account for settings where \exttraf\ volunteers \textit{probabilistically} sign up for their targeted opportunity.)
} 
By contrast, the platform chooses the \opp\ that \inttraf\ views (as formalized below). After viewing an \opp\ and making a \signup\ decision, the internal \vol\ leaves the platform.

\smallskip
    
\textbf{Platform's Decisions and Objective:} Upon each arrival, the platform observes the \vol's type, i.e., their source as well as their pair-specific conversion probabilities.
The platform then must immediately and irrevocably recommend a single \opp\ to \vol\ $t$, denoted $\opprecommend{} \in \oppset \cup \{0\}$.\footnote{We introduce a ``dummy'' \opp\ with index $0$, which we use to indicate when the platform does not recommend an \opp\ and when a \vol\ does not \signup\ for an \opp.} (In Appendix \ref{sec:ranking}, we discuss how our model and results generalize to settings where the platform provides a ranked set of recommendations.)
For \exttraf, even though the platform plays no role in the \vol's decision, we adopt the convention that the platform recommends $\opprecommend{} = \extrecommend$.
The platform's recommendation for \inttraf\ can depend on the current \vol's type, \opp\ capacities, and the full history of \vol\ arrivals and decisions. The \vol\ then (deterministically) views the recommended \opp, and signs up according to their pair specific conversion probability. We use the random variable $\volchoicet{} \in \{\opprecommend{}, 0 \}$ to denote the \vol's \signup\ decision when presented with the recommendation $\opprecommend{}$.

The platform's objective is to maximize the amount of capacity filled by all \vols\ (either \intorexttraf). We assume that all the \signups\ for an \opp\ beyond its capacity \emph{provide no value}. In the context of volunteer matching, these ``excessive'' sign-ups represent an ineffective use of \vols, but can also have significant negative side effects, such as overwhelming the \vol-management staff for that \opp\ due to costly screening and reducing \vol\ engagement due to under-utilization \citep{sampson2006optimization}. (In other contexts such as e-commerce, the platform may be naturally constrained based on capacities.)

In pursuit of this objective, the platform follows an online recommendation algorithm $\ALG \in \algdomain$. For a volunteer arriving at time $t$, let \opp\ $\opprecommend{\ALG}$ denote the (possibly random) opportunity recommended by algorithm $\ALG$. 
Then, the expected amount of filled capacity generated by $\ALG$ (henceforth referred to as the expected \emph{value} of $\ALG$) on instance $\instance$ is given by  
$$\ALG(\instance) \quad = \quad \mathbb{E}\bigg[\sum_{i \in \oppset} \min\Big\lbrace \capa_i, \sum_{t \in [\horizon]} \mathbbm{1} [\volchoicet{\ALG} = i]\Big\rbrace\bigg],$$
where the expectation is taken with respect to the \vols' \signup\ realizations and, possibly, the randomized decisions by the algorithm.

\smallskip

\textbf{Performance metric:} 
\revcolor{To assess the quality of any proposed online algorithm $\ALG$, we compare its expected value to that of an optimal clairvoyant algorithm $\OPT$ on the same instance, denoted by $\OPT(\instance)$. Consistent with the literature, we assume that $\OPT$ operates with \emph{a priori} knowledge of the exact sequence of \vol\ arrivals $\arrseq$ but without \emph{a priori} knowledge of the realizations of their sign-up decisions. We formalize our notion of the benchmark $\OPT$ in the following definition.

\begin{defn}[Optimal Clairvoyant Algorithm]
\label{def:opt}
The optimal clairvoyant algorithm is the solution to a dynamic program (of exponential size) which takes as input the instance $\instance$. 
Upon the arrival of each \vol\ $t$, the optimal clairvoyant algorithm recommends an \opp\ $\opprecommend{\OPT} \in \oppset \cup \{0\}$ that maximizes the total amount of filled capacity, given the instance and the \signup\ history up to that point.\footnote{\revcolor{If there are multiple optimal solutions to this dynamic program, we use the convention that $\OPT$ is one such solution that never exceeds the capacity of an opportunity and maximizes the amount of capacity filled by \exttraf.} \label{footnote:OPT}}
\end{defn}
}

\revcolor{The performance of an algorithm relative to that of $\OPT$ can depend significantly on the amount of capacity that can be filled by \exttraf. For instance, if \exttraf\ can fill the entire capacity of each \opp, then we can easily design an algorithm that achieves the same value as $\OPT$. {In this case, it would not matter how \inttraf\ was allocated, since \exttraf\ alone will suffice to fill all capacity.} 
Based on this observation, our performance metric will be a function of both the online algorithm $\ALG$ as well as the fraction of capacity which can be filled by \exttraf, as formalized below.}
\revcolor{
\begin{defn}[\fracextnamefull]
\label{def:beta}
For a fixed instance $\instance$, the \textit{\MakeLowercase{\fracextnamefull}} (\fracextname) is the fraction of capacity which can be filled by \exttraf. We use $\extfrac$ to denote the \fracextname, where  \begin{equation}
\extfrac(\instance) = \frac{\sum_{i \in \oppset} \min\{\capa_i, \sum_{t \in \exttimes}\mathbbm{1}[\extrecommend = i] \}}{\sum_{i \in \oppset} \capa_i}. \label{eq:betadef}
\end{equation}
\end{defn}
}
For a given $\extfrac \in [0,1]$, we let $\instancedomain_\extfrac$ be the set of all possible instances where the \fracextname\ is $\extfrac$.
Having defined our benchmark $\OPT$ and the parameter $\extfrac$, we now define our performance metric. We will evaluate the performance of any online algorithm via the competitive ratio parameterized by $\extfrac$.
\begin{defn}[Competitive Ratio]
\label{def:compratio}
The competitive ratio of an algorithm $\ALG$ for any given \MakeLowercase{\fracextnamefull} $\extfrac \in [0,1]$ is defined as:
\begin{equation}\label{eq:compratiodef}
    \compratio(\ALG, \extfrac) \quad = \quad
\min_{\instance \in \instancedomain_{\extfrac}} \frac{\ALG(\instance)}{\OPT(\instance)} 
\end{equation}
\end{defn}

By taking the minimum value of this ratio over all instances in $\instancedomain_\extfrac$, the competitive ratio provides a guarantee against even an adversarially-chosen instance.
To conclude this section, we revisit the connection with the online matching problems discussed in Section~\ref{sec:literature}. The competitive ratio is a standard metric in this literature (see, e.g., \citealt{mehta2007adwords}), though the competitive ratio is  commonly taken with respect to \emph{all} possible instances. (In our setting, the domain of all possible instances is equivalent to the union over domains $\instancedomain_\extfrac$ for all $\extfrac \in [0,1]$.) 
In this work, motivated by the nature of \exttraf\ that constitutes a considerable portion of traffic on some matching platforms, we explore how imposing structure on the problem (in the form of the \fracextname\ $\extfrac$) impacts the achievable competitive ratio.


\section{Results}
\label{sec:results}
\revcolor{
We now investigate different settings which together paint a clear picture of the impact of \exttraf\ on the design of online algorithms. We start in Section \ref{subsec:results:extknown} by considering a setting where all \exttraf\ arrives before any \inttraf. This special case provides intuition behind the shortcomings of known algorithms and motivates the need for our \adaptivecapfull\ (\AC) algorithm. 
Building on this intuition, in Section~\ref{subsec:general_arrivals} we turn our focus to a setting with general arrivals. We establish an upper bound on the competitive ratio of any online algorithm, and for the case where sign-ups are deterministic,
we establish an asymptotically matching lower bound on the competitive ratio of \AC\ (i.e., as capacities increase).
We then characterize a family of lower bounds on the competitive ratio of \AC\ in a general setting with probabilistic sign-ups. Finally, we elaborate on implications and insights from these results in Section~\ref{subsubsec:thm2:discussion}.}
 
\subsection{Warm-up: External Traffic Arrives First}
\label{subsec:results:extknown}
Let us first consider a setting where the platform observes all the \exttraf\ before the arrival of any \inttraf.  
Any recommendation algorithm would use the same amount of \exttraf\ as $\OPT$, as we assume that the platform cannot influence \exttraf.
However, 
an online algorithm may make sub-optimal recommendations to \inttraf, as it does not know which \opps\ can be filled by future \vols\ and which \opps\ cannot. In settings without \exttraf, this leads to a ``barrier'' of $1-1/e$. Building on this intuition, the following proposition establishes an upper bound on the competitive ratio of any online algorithm.

\begin{prop}[Upper Bound when All External Traffic Arrives First]
\label{prop:warmupupperbound}
Suppose that all  \exttraf\ arrives before \inttraf. Then, for any \MakeLowercase{\fracextnamefull} $\extfrac$ and any minimum capacity, no online algorithm can achieve a competitive ratio greater than $\extfrac + (1-\extfrac)(1-1/e)$.
\end{prop}

The proof of Proposition \ref{prop:warmupupperbound} (which is presented in Appendix \ref{proof:prop:warmupupperbound}) adjusts the hard instance presented in \citet{mehta2007adwords} by appending \exttraf\ at the beginning of the arrival sequence, such that the \fracextname\ is equal to $\extfrac$.

Based on Proposition~\ref{prop:warmupupperbound}, one may ask: is it possible to design an online algorithm that achieves this {upper bound}, at least asymptotically as the minimum capacity $\invbidtobudget = \min_{i \in [\numopps]}\capa_i$ tends to infinity?\footnote{\revcolor{Henceforth, we use “asymptotically” to
refer to the regime where $\invbidtobudget \rightarrow \infty$. Notably, in the finite-capacity regime a competitive ratio of $1-1/e$ is \emph{not} attainable by a deterministic algorithm. To see this, suppose there are two opportunities ($i$ and $j$) with capacities $\capa_{i} = \capa_{j} = 1$ and two volunteers ($1$ and $2$). Consider two different arrival sequences. In both arrival sequences, volunteer $1$ is deterministically compatible with both opportunities. In the first (resp. second) arrival sequence, volunteer $2$ is only compatible with opportunity $i$ (resp. $j$). If the algorithm deterministically matches volunteer $1$ to opportunity $i$ (resp. $j$), then in the first (resp. second) arrival sequence, it cannot match volunteer $2$ and will only obtain a $1/2$ fraction of the value of the clairvoyant solution.}\label{footnote:finitecapacity}} 
Intuitively, the answer should be yes. In the absence of \exttraf, it is possible to design algorithms that asymptotically achieve a competitive ratio of $1-1/e$ \citep{mehta2007adwords}. Building on such results, we should be able to design an algorithm that first fills a $\extfrac$ fraction of capacity with \exttraf, and then --- based on the capacities that remain --- treats the \inttraf\ portion of the problem as a typical instance of online matching, for which we can achieve a  $1-1/e$  fraction of the offline solution $\OPT$. Overall, this would lead to an asymptotic competitive ratio of at least $\beta + (1-\beta)(1-\frac{1}{e})$, as desired.
However, a naive approach that only relies on existing algorithms does not achieve such a competitive ratio.

\subsubsection{The failure of \MSVV.} A prime candidate to achieve this level of performance is the well-known algorithm introduced in \citet{mehta2007adwords}, commonly referred to as $\MSVV$. This algorithm achieves, asymptotically, the best-possible competitive ratio of $1-1/e$ for our online matching problem in the absence of \exttraf, i.e., when $\extfrac = 0$.

The idea behind the \MSVV\ algorithm is very simple. To balance the trade-off between the upside of recommending the \opp\ with the highest conversion probability and the downside of reaching an \opp's capacity before the end of the horizon, $\MSVV$ weighs each \opp's conversion probability
with the following decreasing {trade-off function of the \opp's fill rate:}
\begin{equation}
    \label{eq:potentialfunc}
    \balancefunc(\fillrate) = 1-\text{exp}(\fillrate - 1).
\end{equation}
Opportunity $i$'s fill rate under $\MSVV$ after the arrival of \vol\ $t$ (denoted $\fillrate_{i,t}^{\MSVV}$) is the fraction of \opp\ $i$'s capacity ($c_i$) that is filled at that time. 
We formally present $\MSVV$ in Algorithm~\ref{alg:msvv}.\footnote{If there are multiple recommendations that satisfy $\MSVV$'s optimality criteria, we follow the convention of recommending the one with the lowest index.}

\begin{algorithm}[t]
        \caption{\MSVV\ Algorithm \citep{mehta2007adwords}}
        \begin{algorithmic}
        \label{alg:msvv}
        {\small
                \STATE Initialize $\MSVV_{i,0} = 0$, $\fillrate_{i,0}^{\MSVV} = 0$ for all $i \in [\numopps]$.
                \FOR{$t \in [\horizon]$} 
                \IF{\vol\ $t \in \exttimes$}
                \STATE Recommend $\opprecommend{\MSVV} = \extrecommend$ (i.e., recommend the unique targeted \opp). 
                \ELSE 
                \STATE Recommend $\opprecommend{\MSVV} \in \text{argmax}_{\opprecommendbase \in [\numopps] \cup \{0\}}\convrate_{\opprecommendbase,t}\cdot \balancefunc(\fillrate_{\opprecommendbase,t-1}^{\MSVV})$, where $\balancefunc$ is defined in \eqref{eq:potentialfunc}.
                \ENDIF
                \FOR{$i \in [\numopps]$}
                \STATE $\MSVV_{i,t} = \min\{\capa_i,  \MSVV_{i,t-1} + \mathbbm{1}[\volchoicet{\MSVV} = i]\}$; $\quad$  $\fillrate_{i,t}^{\MSVV} = \MSVV_{i,t}/\capa_i$ 
                \ENDFOR
                \ENDFOR
                }
            \end{algorithmic}
            \end{algorithm}

Surprisingly, $\MSVV$ does not achieve the desired competitive ratio of $\beta + (1-\beta)(1-\frac{1}{e})$ in the setting where all external traffic comes first, 
as established by the following proposition.

\begin{prop}[Upper Bound on \MSVV\ when All External Traffic Arrives First]
\label{prop:warmupmsvv}
Suppose \exttraf\ arrives before \inttraf. Then for any \MakeLowercase{\fracextnamefull} $\extfrac$ and any minimum capacity, the competitive ratio of $\MSVV$ is at most 
\begin{equation}
    1- \frac{1-\hat{\msvvhelper}_1}{\text{\emph{exp}}\left(\text{\emph{exp}}(-\hat{\msvvhelper}_1/(1-\hat{\msvvhelper}_1))\right)} \label{eq:msvvwarmupbound}
\end{equation}
\noindent where $\hat{\msvvhelper}_1$ is the unique solution in $[0,1]$ to $\extfrac = \hat{\msvvhelper}_1 + (1-\hat{\msvvhelper}_1)\Big(\text{\emph{exp}}\big(-\hat{\msvvhelper}_1/(1-\hat{\msvvhelper}_1)\big)-1\Big)$. 
\end{prop}

In Figure \ref{fig:CR_upper}, we illustrate the upper bound on the competitive ratio of $\MSVV$ given by \eqref{eq:msvvwarmupbound}. There is a significant gap between the upper bound on the competitive ratio of $\MSVV$ (dashed red curve) and the potentially-achievable frontier characterized in Proposition~\ref{prop:warmupupperbound} (solid blue line). 
The shortcomings of $\MSVV$ stem from its definition of an \opp's fill rate, i.e., $\fillrate^{\MSVV}_{i,t} = \MSVV_{i,t} / c_i$, which accounts for \intandexttraf\ in an identical fashion. 
Under $\MSVV$, the \opps\ that receive \signups\ from \exttraf\ will have strictly positive fill rates when \inttraf\ arrives, and thus will be de-prioritized. 
The proof of Proposition \ref{prop:warmupmsvv} (presented in Appendix \ref{proof:prop:warmupmsvv}) builds on this intuition: we design a family of instances in which $\MSVV$ (sub-optimally) withholds \inttraf\ from  opportunities that initially receive external traffic.
In these instances, for $\extfrac \in (0,1)$, the amount of capacity filled by \inttraf\ under $\MSVV$ is less than a $1-1/e$ factor of the amount of capacity filled by \inttraf\ under $\OPT$.
Consequently, it would appear that in order to achieve a competitive ratio of $\beta + (1-\beta)(1-\frac{1}{e})$, we must design an algorithm that incorporates the source of traffic into its decision-making. To that end, we next introduce our \emph{\adaptivecapfull} ($\adaptivecapmath$) algorithm, which accounts for the amount of filled capacity separately based on source.

\subsubsection{Accounting for the source of traffic: the Adaptive Capacity algorithm.}
Similar to \MSVV, the $\AC$ algorithm uses the exponential trade-off function $\balancefunc$, as defined in \eqref{eq:potentialfunc}, and it recommends the \opp\ with the greatest weighted conversion probability, i.e., the opportunity $i$ that maximizes $\convrate_{i,t}\cdot \balancefunc(\fillrate_{i,t-1})$.\footnote{If there are multiple recommendations that satisfy $\AC$'s optimality criteria, we follow the convention of recommending the one with the lowest index.} However, \textit{\AC\ crucially differs from \MSVV\ in its definition of an \opp's fill rate}. 
 The fill rate definition used by $\MSVV$ aggregates all \signups\ in the numerator; that is, it defines an \opp's fill rate as $\fillrate_{i, t}^{\MSVV} = \left(\MSVV_{i,t}\right)/\capa_i$. By contrast, $\adaptivecapmath$ aggregates \signups\ separately based on source, using counters $\adaptivecapmath_{i,t}^{\exttrafmath}$ and $\adaptivecapmath_{i,t}^{\inttrafmath}$. It then removes any \exttraf\ \signups\ from the total capacity (the denominator), i.e., $\fillrate_{i, t} =  \adaptivecapmath_{i,t}^{\inttrafmath}/\left(\capa_i - \adaptivecapmath_{i,t}^{\exttrafmath}\right)$. In other words, every time capacity is filled by \exttraf, we \emph{adaptively} reduce the capacity of that opportunity by one.
We formally describe $\AC$ in Algorithm~\ref{alg:acpolicy}.

\begin{algorithm}[ht]
        \caption{\adaptivecap\ Algorithm}
        \begin{algorithmic}
        \label{alg:acpolicy}
        {\small
        \STATE Initialize $\adaptivecapmath_{i,0}^{\exttrafmath} = 0$, $\adaptivecapmath_{i,0}^{\inttrafmath} = 0$, and $\fillrate_{i,0} = 0$ for all $i$ in $[\numopps]$.
                \FOR{$t \text{ in } [\horizon]$} 
                \IF{\vol\ $t \text{ in } \exttimes$}
                \STATE Recommend $\opprecommend{\AC} := j$, where $j = \extrecommend$ (i.e., recommend the unique targeted \opp). 
                \STATE $\AC_{j,t}^{\exttrafmath} = \min\{\capa_j - \AC_{j,t}^{\inttrafmath},  \AC_{j,t-1}^\exttrafmath + \mathbbm{1}[\choicefunc_t(j) = j]\}$; $\quad \AC_{j,t}^\inttrafmath = \AC_{j,t-1}^\inttrafmath$
                \ELSE
                \STATE Recommend $\opprecommend{\AC} := j$, where $j \in \text{argmax}_{\opprecommendbase \in [\numopps] \cup \{0\}}\convrate_{\opprecommendbase,t}\cdot \balancefunc(\fillrate_{\opprecommendbase,t-1})$, where $\balancefunc$ is defined in \eqref{eq:potentialfunc}.
                \STATE $\AC_{j,t}^{\inttrafmath} = \min\{\capa_j - \AC_{j,t}^{\exttrafmath},  \AC_{j,t-1}^\inttrafmath + \mathbbm{1}[\choicefunc_t(j) = j]\}$; $\quad \AC_{j,t}^\exttrafmath = \AC_{j,t-1}^\exttrafmath$
                \ENDIF
               \STATE $\fillrate_{j,t} = \AC_{j,t}^{\inttrafmath}/(\capa_{j} -\AC_{j,t}^{\exttrafmath})$
                \FOR{$i \text{ in } [\numopps] \setminus \{j\}$}
                \STATE $\AC_{i,t}^\exttrafmath = \AC_{i,t-1}^\exttrafmath$; $\quad \AC_{i,t}^\inttrafmath = \AC_{i,t-1}^\inttrafmath$; $\quad \fillrate_{i,t} = \fillrate_{i,t-1} $ 
                \ENDFOR
                \ENDFOR
                }
            \end{algorithmic}
            \end{algorithm}

In the following, we establish that the competitive ratio of $\adaptivecapmath$ is asymptotically optimal  when \exttraf\ arrives before \inttraf. Intuitively, in this warm-up setting,  $\adaptivecapmath$ 
implements the solution discussed in the beginning of this section: it reduces capacities based on the number of \signups\ from \exttraf\ and then, for \inttraf, it runs $\MSVV$ on the \textit{remaining} capacities. Building on this intuition, the following proposition  lower-bounds the competitive ratio of  $\AC$.

\begin{prop}[Lower Bound on \AC\ when All External Traffic Arrives First]
\label{prop:warmup}
Suppose all \exttraf\ arrives before \inttraf. Then for any \MakeLowercase{\fracextnamefull} $\extfrac$ and any minimum capacity $\invbidtobudget$, the competitive ratio of $\adaptivecapmath$ is at least $\extfrac + (1-\extfrac)(1-1/e) - \invbidtobudget^{-1}$.
\end{prop}

The lower bound given in Proposition~\ref{prop:warmup} (which we prove in Appendix \ref{proof:prop:warmup}) asymptotically achieves the upper bound established in Proposition~\ref{prop:warmupupperbound} (shown by Figure~\ref{fig:CR_upper}). \revcolor{To conclude this section, we note that even though this warm-up setting is unrealistic and studied solely to develop intuition, it is roughly equivalent to the more realistic setting described in the following remark. 
\begin{remark}[Attainable Performance when External Traffic is Predictable]
\label{remark:extknown}
    Suppose the amount of \exttraf\ for each opportunity can be predicted in advance with perfect accuracy, but the arrival order of the external and internal traffic can be arbitrarily mixed. In such a setting, a variant of $\AC$ that reduces capacities up front based on the \emph{predicted} number of \exttraf\ arrivals for each opportunity (and then for \inttraf, runs $\MSVV$ on the remaining capacities) will asymptotically obtain the guarantee in Proposition \ref{prop:warmup}.
\end{remark} 

Intuitively, this is akin to the $\AC$ algorithm in our warm-up setting, which reduces capacities by the \emph{observed} amount of \signups\ from \exttraf. We now move beyond the case where \exttraf\ arrives first and analyze \AC\ in more general settings.}

\subsection{General arrivals}
\label{subsec:general_arrivals}
When \exttraf\ arrives to the platform first, we observed that the competitive ratio of  $\AC$ is asymptotically optimal and significantly improves upon the fundamental barrier of $1-1/e$ (which we remind is the upper-bound in the absence of external traffic).
We now investigate the competitive ratio of $\AC$  when the arrival sequence of \vol\ types is completely unknown.
In contrast with the setting previously described, the $\AC$ algorithm cannot always observe the \signups\ from \exttraf\ before making recommendations for \inttraf.
As a consequence, when \inttraf\ arrives, the $\AC$ algorithm may inadvertently recommend an \opp\ which could be filled entirely by later-arriving \exttraf.

This is not only a limitation of the $\AC$ algorithm: no online algorithm has access to information about future \exttraf. However, the information available to $\OPT$ is unchanged: it still has \emph{a priori} knowledge of entire arrival sequence, including the capacity that can be filled by \exttraf. We should intuitively expect the achievable competitive ratio will decrease in this setting (compared to the previous setting), as one could construct hard examples where valuable information about \exttraf\ is not revealed until the end of the arrival sequence (e.g., if all \exttraf\ arrives after all \inttraf).\footnote{We remark that even though many hard instances involve all \exttraf\ arriving after all \inttraf, the two algorithms that we consider (i.e., $\AC$ and $\MSVV$) do not exhibit performance that is monotonic in the arrival order of \exttraf\ vis-\`{a}-vis \inttraf.}

Building on this intuition, we modify the hard instance of \citet{mehta2007adwords} by replacing the tail end of the arrival sequence with carefully-designed \exttraf. This modification allows us to establish the following family of upper bounds on the competitive ratio of any online algorithm.

\begin{thm}[Upper Bound on Competitive Ratio]
\label{thm:hardness}
For any \MakeLowercase{\fracextnamefull} $\extfrac$ and any minimum capacity, no online algorithm can achieve a competitive ratio better than $\big(1-1/e\big)\mathbbm{1}_{\extfrac \leq 1/e} + \big(1+\extfrac \log(\extfrac)\big) \mathbbm{1}_{\extfrac > 1/e}$.\footnote{\revcolor{For a condition $c$ that is either true or false, $\mathbbm{1}_c$ equals $1$ if $c$ is true and $0$ otherwise.}}
\end{thm}

\begin{figure}[t]
 \centering
 \begin{subfigure}[b]{0.31\textwidth}
    \resizebox{\textwidth}{!}{\input{images/rev_upper_bounds_warmup.pgf}}
    \caption{}
    \label{fig:CR_upper}
 \end{subfigure}
 \begin{subfigure}[b]{0.31\textwidth}
    \resizebox{\textwidth}{!}{\input{images/rev_general_upper_bounds.pgf}}
    \caption{}
    \label{fig:CR_weightcap}
 \end{subfigure}
 \begin{subfigure}[b]{0.31\textwidth}
    \resizebox{\textwidth}{!}{\input{images/rev_weight_caps.pgf}}
    \caption{}
    \label{fig:weight_caps}
 \end{subfigure}
 \caption{\revcolor{In the asymptotic regime, we present
 bounds for settings where (a) all \exttraf\ arrives first, (b) arrivals are general and sign-ups are deterministic, and (c) arrivals and conversion probabilities are general.}
 }
\end{figure}
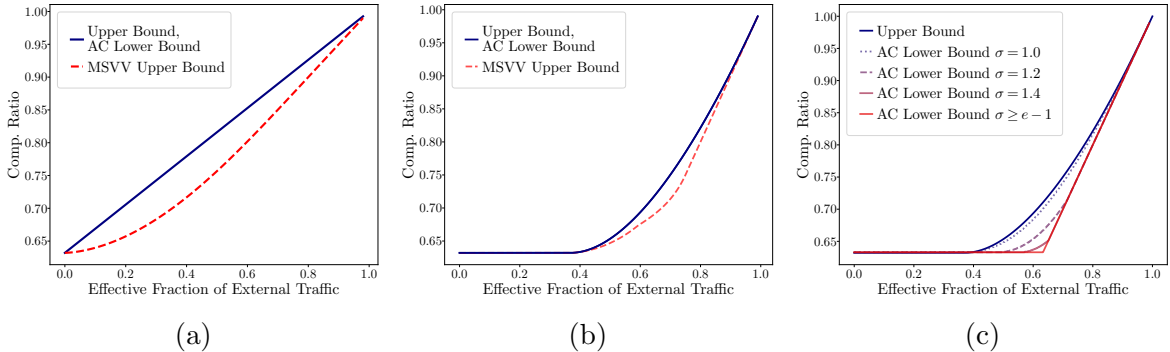

\revcolor{In contrast to the linear upper bound established in the warm-up setting (see Proposition \ref{prop:warmupupperbound}), the upper bound of Theorem \ref{thm:hardness} does not exceed $1-1/e$ until $\extfrac > 1/e$ (as shown in Figure \ref{fig:CR_weightcap}). We defer a discussion of this upper bound to Section \ref{subsubsec:thm2:discussion}, and we formally prove this result in Appendix~\ref{proof:thm:hardness}.}


\revcolor{Naturally, one wonders whether the $\AC$ algorithm can attain this upper bound. 
\textit{We find that the answer depends on the conversion probabilities.}} 
\revcolor{Specifically, we first show in Theorem \ref{thm:tightproof} that if conversion probabilities are either $0$ or $1$ (i.e., if sign-ups are deterministic), then \AC's competitive ratio asymptotically matches the upper bound.\footnote{\revcolor{It is worth noting that in the family of instances used to show the upper bound in Theorem~\ref{thm:hardness}, sign-ups are deterministic, i.e., $\bm{\convprob}_t \in \{0,1\}^{\numopps}$.}}
However, in Example \ref{ex:weightedAC}, a tight analysis of \AC\ shows that it cannot attain this upper bound for arbitrary conversion probabilities.
}

\begin{thm}[\revcolor{Lower Bound on \AC\ when Sign-ups are Deterministic}]
\label{thm:tightproof}
\revcolor{Let the smallest capacity be given by $\invbidtobudget$ and let $\bm{\convprob}_t \in \{0,1\}^{\numopps}$ for all $t \in [T]$. Then, for any \MakeLowercase{\fracextnamefull} $\extfrac$, the competitive ratio of the \adaptivecap\ algorithm defined in Algorithm \ref{alg:acpolicy} (with $\balancefunc$ as defined in Eq. \eqref{eq:potentialfunc}) is at least $\big(1-1/e\big)\mathbbm{1}_{\extfrac \leq 1/e} + \big(1+\extfrac \log(\extfrac)\big) \mathbbm{1}_{\extfrac > 1/e} - 2/\invbidtobudget$.}
\end{thm}

\revcolor{This setting corresponds to the online B-matching problem introduced in \citet{kalyanasundaram2000optimal} and commonly studied in the online matching literature.\footnote{\revcolor{In this special case of our model (i.e., when sign-ups are deterministic), a \emph{randomized} algorithm introduced and analyzed in \citet{udwani2021adwords} is shown to attain our upper bound from Theorem \ref{thm:hardness}, even for arbitrary capacities. We note that randomness is required to attain a matching bound for arbitrary capacities, as discussed in Footnote \ref{footnote:finitecapacity}.}}
The proof of Theorem \ref{thm:tightproof} builds on the approach from \citet{mehta2007adwords}, with several modifications and additional analysis needed to account for the presence of \exttraf. While this proof technique enables us to achieve asymptotically matching bounds for \AC, it does not easily generalize to other settings (e.g., with probabilistic sign-ups); hence, we defer the details of the proof to Appendix \ref{apx:proof:tightproof}.}

\revcolor{Though the \AC\ algorithm obtains the optimal asymptotic guarantee when sign-ups are deterministic, its optimal performance does not necessarily extend to the setting where conversion probabilities are general, as illustrated by the following example.}

\begin{exmp}[Limitation of $\AC$ when Conversion Probabilities are Arbitrary]
\label{ex:weightedAC}
Consider an instance with two \opps\ ($1$ and $2$) with capacities $\capa_1 = \largeopps$ and $\capa_2 = \frac{1}{e-1} \largeopps$ for sufficiently large $\largeopps$. There are $2\largeopps$ \vols, and the first \largeopps\ \vols\ are \inttraf\ with conversion probabilities given by 
$$\convprob_{1,t} = 1, \qquad \convprob_{2,t} = \frac{1-\text{\emph{exp}}\Big(\frac{t-1}{\largeopps} - 1\Big)}{1-\text{\emph{exp}}(-1)} - \frac{1}{2\largeopps}.$$ The remaining $\largeopps$ \vols\ are \exttraf\ with conversion probabilities of~$1$ for \opp\ $1$ and $0$ for \opp~$2$.
\end{exmp}
In Example \ref{ex:weightedAC}, the \fracextname\ $\extfrac = 1-1/e$, as the capacity of \opp\ $1$ can be entirely filled with \exttraf, and the minimum capacity $\invbidtobudget$ is arbitrarily large. In this instance, $\OPT$ will recommend \opp\ $2$ to all \inttraf, and in expectation \opp\ $2$ will receive $\frac{1}{e-1} \largeopps - o(\largeopps)$ \signups.\footnote{For two functions $d, l : \mathbb{N}\rightarrow \mathbb{R}$, $l(n) = o(d(n))$ if $\lim_{n \rightarrow \infty} \frac{l(n)}{d(n)} = 0 $.} Then, \exttraf\ arrives and fills \opp\ $1$, which means the amount of filled capacity under $\OPT$ is $\frac{e}{e-1}\largeopps - o(\largeopps)$.

In sharp contrast, $\AC$  will recommend \opp\ $1$ to all \inttraf\ \vols, because the conversion probabilities in Example \ref{ex:weightedAC} are constructed such that $\convprob_{1,t} \balancefunc(\fillrate_{1,t}) > \convprob_{2,t} \balancefunc(0)$ for all $t \in [\largeopps]$. These \inttraf\ \vols\ completely fill \opp\ $1$. Consequently, even though the \fracextname\ is $1-1/e$, \emph{no capacity is filled by \exttraf\ under $\AC$}. In total, the amount of filled capacity under $\AC$ is $\largeopps$. Thus, in this example, the ratio between the expected value of $\AC$ and the expected value of $\OPT$ approaches $1-1/e$, despite the fact that the \fracextname\ $\extfrac = 1 - 1/e$. 

\revcolor{In this example, note that a volunteer's conversion probabilities vary unboundedly across \opps: we have $\convprob_{1,\largeopps} = 1$ while $\convprob_{2,\largeopps} = o(1)$. 
As we discuss further in Section \ref{subsubsec:thm2:discussion}, any improvement in \AC's competitive ratio over $1-1/e$ stems from guaranteeing that \AC\ uses some amount of \exttraf\ to fill remaining capacity. However, in instances such as Example \ref{ex:weightedAC}, the differences in conversion probabilities lead \AC\ to waste \emph{all} \exttraf, thereby limiting its competitive ratio to $1-1/e$. Based on this intuition, we expect the performance of \AC\ to depend on the maximum conversion probability ratio, a quantity we formally define below.} 

\begin{defn}[\MCPRfull]
\label{def:MCPR}
For each \vol\ $t$, let $\oppset_{t}$ denote the subset of \opps\ $i$ for which $\convprob_{i,t} > 0$.\footnote{Without loss of generality, we assume that for all \vols, there is at least one \opp\ for which they have a strictly positive conversion probability. Otherwise, we can simply remove that \vol\ and re-index.} The \emph{conversion probability ratio} (CPR) for \vol\ $t$ is given by $\frac{\max_{i \in \oppset_t}\convprob_{i,t}}{\min_{i \in \oppset_t}\convprob_{i,t}}$. The \emph{maximum conversion probability ratio} (\MCPR), denoted by $\weightcap$, is the maximum CPR across all volunteers, i.e.
\begin{equation}
    \weightcap  = \max_{t \in [\horizon]} \left(\frac{\max_{i \in \oppset_t}\convprob_{i,t}}{\min_{i \in \oppset_t}\convprob_{i,t}}\right) \label{eq:weightcap}
\end{equation}
\end{defn}

We now present the main result of this section, which is a family of lower bounds on the competitive ratio of the $\AC$ algorithm. These bounds are parameterized by the \fracextname\ $\extfrac$, the minimum capacity $\invbidtobudget$, and the \MCPR\ $\weightcap$.

\begin{thm}[Lower Bound on \AC's Competitive Ratio]
\label{thm:AClower} 
Let the smallest capacity be given by $\invbidtobudget$ and let the \MakeLowercase{\MCPRfull} be at most $\weightcap$. Then, for any \MakeLowercase{\fracextnamefull} $\extfrac$, the competitive ratio of the \adaptivecap\ algorithm defined in Algorithm \ref{alg:acpolicy} (with $\balancefunc$ as defined in Eq. \eqref{eq:potentialfunc}) is at least $\compratiofunc(\beta, \invbidtobudget, \weightcap) = \max\{\extfrac, z^*\}$, where
\begin{align}
    z^* \quad = \quad \min_{z \in [0,1]} \ z& \label{eq:defzstar} \\ \text{\emph{subject to}}& \qquad z \quad \geq \quad e^{-1/\invbidtobudget}(1-1/e) \nonumber
    \\& \qquad z \quad \geq \quad  e^{-1/\invbidtobudget} \breve{g}\left(\max\{0,\extfrac - \weightcap + z\}, z - \max\{0,\extfrac - \weightcap + z\}\right), \nonumber 
\end{align}
and $\breve{g}(x_1,x_2)$ denotes the lower convex envelope of $g(x_1,x_2)$ over the domain $\mathcal{D} = \{(x_1, x_2)\in \mathbb{R}^2_{\geq 0}:  x_1+x_2 \leq 1\}$,\footnote{The lower convex envelope of a function $g$ over a domain $\mathcal{D}$ is the supremum of all convex functions that are less than or equal to $g$ on domain $\mathcal{D}$.} where
\begin{equation}
    g(x_1,x_2) = 1-\frac{1}{e} + x_1 + (1-x_1)\balancefunc \left(\frac{x_2}{ 1-x_1}\right) - \balancefunc \left(x_2\right).   \label{eq:convexenv}
\end{equation}
\end{thm}
\revcolor{We defer a proof sketch of Theorem \ref{thm:AClower} to Section \ref{subsubsec:proof}, with the remaining details provided in Appendix \ref{apx:thm1proofdetails}.
We highlight that the technique used to prove Theorem \ref{thm:AClower} is flexible enough to not only account for arbitrary conversion probabilities but also can incorporate other practical considerations such as probabilistic sign-ups from \exttraf\ (as discussed in Appendix \ref{apx:exttrafstochastic}) or settings where the platform presents a ranked set of recommendations (as discussed in Appendix~\ref{sec:ranking}).}

\subsection{Discussion of Results and Managerial Implications}
\label{subsubsec:thm2:discussion}
\revcolor{Throughout this section, we restrict our attention to the asymptotic regime where $\invbidtobudget$ approaches infinity.
In Figure~\ref{fig:CR_weightcap}, we plot the lower bound on the competitive ratio of the $\AC$ algorithm when sign-ups are deterministic, as well as the matching upper bound on the competitive ratio of any online algorithm (solid blue line).\footnote{We note that the upper bound holds for any minimum capacity $\invbidtobudget$.}
The existence of matching bounds is particularly intriguing because our $\AC$ algorithm does not need to know the value of $\extfrac$ in order to achieve the best-possible guarantee for that \fracextname. In other words, knowing the \emph{aggregate} amount of \exttraf\ in advance cannot improve the attainable guarantee (at least, for the deterministic sign-up setting). In contrast, knowing the amount of \exttraf\ for \emph{each} opportunity in advance leads to an improved competitive ratio, as stated in Remark \ref{remark:extknown}. These observations can guide practitioners in settings where it may be feasible to acquire \emph{a priori} information about \exttraf\ for each \opp.}

We next aim to better understand the relationship between the \fracextname\ $\extfrac$ and our tight bound on the competitive ratio of $\AC$ in the setting where conversion probabilities are either $0$ or $1$.
Similar to our bound in the setting where \exttraf\ arrives first  (as given by Propositions \ref{prop:warmupupperbound} and \ref{prop:warmup} in Section \ref{subsec:results:extknown}), the lower bound on the competitive ratio of the $\AC$ algorithm is non-decreasing in $\extfrac$. However, in the previous setting, the competitive ratio was linearly increasing in $\extfrac$. In contrast, in this setting, no online algorithm can break the barrier of $1-1/e$ unless $\extfrac$ exceeds $\extfrac^* = 1/e$.

As the dependence on $e$ might suggest, there is a nice relationship between the fundamental barrier of $1-1/e$ (which we remind is the upper-bound in the absence of \exttraf) and the threshold $\extfrac^*$ on the \fracextname. Whenever \AC\ generates a \signup\ from \exttraf, we know that \OPT\ could not have made a ``better'' decision because \exttraf\ (by definition) targets that particular \opp. By leveraging the value of \AC's ``correct'' decisions, we can demonstrate that \AC\ has a competitive ratio strictly above $1-1/e$ \emph{if} it fills a strictly positive amount of capacity with \exttraf, and the competitive ratio is increasing in that amount. Unfortunately, when the \fracextname\ is less than $\extfrac^*$, we cannot guarantee that \AC\ fills \emph{any} capacity with \exttraf.

To see why, consider the following informal argument: suppose $\OPT$ allocates all \vols\ and exactly fills all capacities. Even though \AC\ attains the best-possible competitive ratio of $1-1/e$ in the absence of \exttraf, there exists at least one instance where it ``wastes'' a $\extfrac^* = 1/e$ fraction of \vols. In other words, as we are currently assuming that all sign-ups are deterministic, in that instance it must be the case that \AC\ \emph{cannot} fill capacity with those \vols. For any \fracextname\ $\extfrac \leq \extfrac^*$, we can construct a nearly-identical instance where the set of ``wasted'' \vols\ includes \emph{all} the \exttraf. Indeed, under the \AC\ algorithm, all \exttraf\ is wasted on the instances which establish the upper bound of Theorem \ref{thm:hardness} for $\extfrac \in [0, \extfrac^*]$. 
 However, when the \fracextname\ $\extfrac$ strictly exceeds $\extfrac^*$, $\AC$ must fill a strictly positive amount of capacity with \exttraf, which enables us to prove that $\AC$'s competitive ratio breaks the $1-1/e$ barrier. 

The informal argument of the prior paragraph falls apart, however, when applied to settings where conversion probabilities can be general. In such settings, achieving a competitive ratio of $1-1/e$ is no longer a sufficient condition to ensure that the fraction of wasted \vols\ is at most $\extfrac^*$, as shown by Example \ref{ex:weightedAC} in Section \ref{subsec:general_arrivals}. As a consequence, we can no longer guarantee that the $\AC$ algorithm will break the $1-1/e$ barrier for every \fracextname\ greater than $\extfrac^*$. We now turn our attention to this more general setting.

\revcolor{In Figure \ref{fig:weight_caps}, we illustrate \AC's guarantee for various values of the \MCPR\ $\weightcap$, as given by Theorem \ref{thm:AClower}. We first note that the lower bound is decreasing in the \MCPR\ $\weightcap$. In one extreme where $\weightcap = 1$, \AC\ asymptotically provides a near-optimal guarantee (as shown in Figure \ref{fig:weight_caps}), which suggests that stochasticity by itself (i.e., in the absence of heterogeneous conversion probabilities) does not substantially impact the performance of \AC. However, when the \MCPR\ is sufficiently large (i.e., when $\weightcap \geq e-1$), the lower bound on the competitive ratio of \AC\ does not exceed $1-1/e$ until the \fracextname\ $\extfrac$ exceeds $1-1/e$, in which case at least that fraction of capacity must be filled by \exttraf. In the other extreme where $\weightcap \rightarrow \infty$, Example \ref{ex:weightedAC} demonstrates that our analysis of the \AC\ algorithm is tight for that set of parameters: it establishes an upper bound on the competitive ratio of the $\AC$ algorithm that matches the lower bound of Theorem \ref{thm:AClower} when $\extfrac = 1-1/e$.}

\revcolor{Having discussed the relationship between \fracextname\ $\extfrac$ and the competitive ratio of $\AC$, as well as the comparative statics of our main result with respect to the \MCPR\ $\weightcap$, we refer the interested reader to Section~\ref{subsubsec:proof} for an overview of our proof technique.}

\revcolor{
\subsubsection*{Comparison between \AC\ and \MSVV.}
Although we have demonstrated the superior competitive ratio of $\AC$ compared to $\MSVV$ when \exttraf\ comes first (Propositions \ref{prop:warmupmsvv} and \ref{prop:warmup}), it is natural to wonder if $\AC$ continues to outperform $\MSVV$ in more general settings. 
To shed light on this question, in Proposition \ref{prop:generalmsvv} (presented in Appendix \ref{proof:prop:generalmsvv}) we provide an upper bound on the competitive ratio of $\MSVV$ in the deterministic sign-ups setting (shown by the dashed red curve in Figure \ref{fig:CR_weightcap}). This upper bound on \MSVV\ is strictly below the corresponding lower bound on \AC\ for all $\extfrac \in [0.40, 1.00)$ (by a multiplicative factor up to $5.2\%$). 
Consequently, in terms of robustness to arbitrary arrival sequences (i.e., worst-case guarantees), the degree to which one should prefer \AC\ depends on the setting: if $\extfrac < 0.4$ or if conversion probabilities are stochastic and sufficiently heterogeneous, our results do not establish a separation between the worst-case guarantees of \AC\ and \MSVV.

Moving beyond a comparison of worst-case guarantees, it is illuminating to consider the settings in which \MSVV\ performs poorly relative to \AC. Each of the examples generating our upper bounds on \MSVV\ has a similar structure: some opportunities receive \exttraf\ at the beginning of the horizon, causing \MSVV\ to mistakenly withhold \inttraf\ from those same opportunities. These examples rely on the fact that \MSVV\ applies a harsher ``punishment'' than \AC\ for capacity filled by \exttraf\ (as determined by the respective fill rates under each algorithm). Said another way, \MSVV\ performs relatively poorly in settings where opportunities with early-arriving \exttraf\ should not be harshly penalized (e.g., if opportunities with early-arriving \exttraf\ have fewer compatible arrivals in the future).
Such non-stationary arrival patterns can naturally arise, for example, if attention from outside sources is fickle and fades quickly, perhaps due to a celebrity tweeting once about a particular opportunity before moving on to other topics. 
In contrast, \MSVV\ performs relatively well when past \exttraf\ is positively correlated with future compatible arrivals, which would be the case if, e.g., \exttraf\ for each opportunity is reasonably spread throughout the arrival sequence. 

{The above discussion focuses on furthering our theoretical understanding of the differences between \AC\ and \MSVV. In Section~\ref{sec:sim:acvsmsvv}, we complement these results by numerically studying the performance of these algorithms under different types of arrival patterns.}
}

\section{Proof Sketch of Theorem \ref{thm:AClower}}
\label{subsubsec:proof}
\label{sec:proof}
\revcolor{In this section, we present the proof sketch of Theorem \ref{thm:AClower}. We note that this section is self-contained and can be safely skipped.}

\revcolor{The lower bound on the competitive ratio of \AC\ given by Theorem \ref{thm:AClower} is the maximum of two terms, meaning that each term lower-bounds the competitive ratio. The first term, $\extfrac$, is clearly a lower bound: based on the definition of the EFET (Definition \ref{def:beta}), any algorithm will fill at least a $\extfrac$ fraction of capacity. In the following, we provide an overview of our proof that the second term, $z^*$, is also a lower bound on the competitive ratio. The formal proof can be found in Appendix~\ref{apx:thm1proofdetails}.}

Our analysis leverages the LP-free approach developed in \citet{goyal2019online} and \citet{goyal2020asymptotically}. This approach has proven useful in accounting for \emph{post-allocation} stochasticity, e.g., stochastic rewards (as in \citealt{goyal2019online}) or stochastic usage duration (as in \citealt{goyal2020asymptotically}); \revcolor{in our setting, the volunteers' conversion probabilities for \inttraf\ can be viewed as stochastic rewards. However, the novel part of our analysis is to crucially use the flexibility of this method to separately account for \signups\ based on their source, as the amount of \signups\ from \exttraf\ crucially impacts the guarantee that can be provided by the \AC\ algorithm.}


Central to this approach is the concept of \textit{path-based pseudo-rewards}, i.e., values that are defined so as to keep track of the rewards that accrue during a particular run of an online algorithm relative to \OPT. It is important to highlight that pseudo-rewards are defined purely for accounting purposes; in other words, they are not necessarily equivalent to the rewards of the algorithm on that particular run. (Nor are the pseudo-rewards equivalent to the dual solution of the underlying linear program, which is another commonly-used approach in the literature. See, e.g., \citealt{buchbinder2009design}.) These pseudo-rewards assist in the comparison between the online algorithm and \OPT\ and ultimately allow us to establish a lower bound of $z^*$ on the competitive ratio.

Implementing this approach in our setting requires three steps. In Step (1), we define appropriate pseudo-rewards for our setting. Our construction of pseudo-rewards departs from the approach of \citet{goyal2020asymptotically}, as we define pseudo-rewards that are source-dependent. 
In Step (2), we use these pseudo-rewards to establish a lower bound on the expected value of \AC\ that depends (in part) on the expected value of \OPT\ (Lemmas \ref{lem:lowerboundalg} and \ref{lem:lowerboundballvarbinvar}). 
In contrast to the approach taken in \citet{goyal2020asymptotically}, we cannot formulate a lower bound on the pseudo-rewards for each \opp, as the amount of \exttraf\ can be heterogeneous across \opps. Instead, our more complex lower bound (on the expected sum of \emph{all} pseudo-rewards) eventually enables us to break the competitive ratio barrier of $1-1/e$, but doing so requires an additional step.
In this final step, Step (3), we construct a factor-revealing mathematical program (see Table \ref{table:MP}) based, in part, on the lemmas of the previous step. Through analysis of this program, we place a lower bound of $z^*$ on the competitive ratio of the $\AC$ algorithm (Lemmas \ref{lem:MP} and \ref{lem:MPlower}).

\medskip

\noindent \textbf{Step 1: Defining Pseudo-Rewards} 

\noindent We begin by fixing a problem instance $\instance$. We then define a \emph{sample path} $\samplepath = \{\randomdraw_1, \dots, \randomdraw_\horizon\}$, as the realizations of random variables that govern \vol\ choices in this instance.\footnote{Fixing a set of realizations $\samplepath$, the path of \emph{any} deterministic algorithm (such as the $\AC$ algorithm) is uniquely determined. Hence, we refer to  $\samplepath$ as a sample path. That said, we emphasize that these realizations determine \emph{all} possible choices for \vols, not just the choices along the resulting sample path (i.e., the choices that result from the recommendations made by an algorithm).}
Formally, we interpret $\randomdraw_t$ as a vector of length $n$, where the $i$\textsuperscript{th} component of $\randomdraw_t$ (denoted $\samplepathcomponent_{i, t}$) indicates \vol\ $t$'s \signup\ decision if the platform were to recommend \opp\ $i$.\footnote{If the platform recommends \opp\ $0$, then the \vol\ deterministically does not view (or sign up for) any \opp.} For the fixed instance $\instance$ and for any fixed sample path $\samplepath$, we will define pseudo-rewards $\ballvar_t(\instance, \samplepath)$ for each \vol\ $t \in [\horizon]$, along with pseudo-rewards $\binvar_i(\instance, \samplepath)$ for each \opp\ $i \in [\numopps]$.  Henceforth, to ease exposition, we suppress the dependence on the instance and the sample path.

Our pseudo-rewards $\ballvar_t$ and $\binvar_i$ will depend on an \opp's fill rate under \adaptivecap\ along this fixed sample path, i.e., $\fillrate_{i,t} = \frac{\adaptivecapmath_{i,t}^{\inttrafmath}}{\capa_i - \adaptivecapmath_{i,t}^{\exttrafmath}}$, as well as on the realizations of \vols' \signup\ decisions under both $\AC$ (denoted $\volchoicet{\adaptivecapmath}$) and $\OPT$ (denoted $\volchoicet{\OPT}$).\footnote{As noted above, we are suppressing these variables' dependence on the instance and the sample path. We emphasize that for a fixed instance and sample path, these variables are all deterministic.} Recall our convention that any algorithm (including $\AC$) always recommends the targeted \opp\ to \exttraf. To ensure that we do not count \signups\ that exceed the capacity of an \opp, we define $\volchoicetsuccess{\adaptivecapmath}$ as the \opp\ that \vol\ $t$ \emph{fills capacity of} under $\AC$. To be precise, if \opp\ $\volchoicet{\adaptivecapmath}$ has remaining capacity at time $t$, then $\volchoicetsuccess{\AC} = \volchoicet{\AC}$; otherwise, $\volchoicetsuccess{\AC} = 0$. 

\revcolor{Although one would expect that in ``hard'' instances, \OPT\ will not waste any arriving volunteers, for full rigor we must account for this possibility. To that end, for this fixed instance $\instance$ and along this fixed sample path $\samplepath$, let $\bonustimes$ represent the subset of \inttraf\ for which $\OPT$ recommends \opp\ $0$, i.e., \OPT\ does not recommend any \opp.\footnote{The set $\bonustimes$ is a function of the instance and the sample path, but we remind that we are suppressing that dependence.} 
Based on our convention that \OPT\ is an optimal solution that maximizes the amount of capacity filled by \exttraf\ (as stated in Footnote \ref{footnote:OPT}), any capacity that will eventually be filled by \exttraf\ is effectively reserved. If an \inttraf\ arrival cannot fill any of the remaining capacity, \OPT\ will recommend \opp\ $0$ and this arrival will be in the set $\bonustimes$.}

With the above definitions, we are now ready to define the pseudo-rewards $\ballvar_t$ and $\binvar_i$.
\begin{align}
    \ballvar_t &= \begin{cases}
    \sum_{i \in [\numopps]} \balancefunc(\fillrate_{i,t-1})\mathbbm{1}[\volchoicetsuccess{\adaptivecapmath} = i], & t \in  \exttimes \cup \bonustimes \\
    \sum_{i \in [\numopps]} \balancefunc(\fillrate_{i,t-1})\mathbbm{1}[\volchoicet{\OPT} = i], & t \in \inttimes \setminus \bonustimes \\
    \end{cases} \label{eq:ballvar}
    \\
    \binvar_i &=  \sum_{t \in [\horizon]} \left(1-\balancefunc(\fillrate_{i,t-1})\right)\mathbbm{1}[\volchoicetsuccess{\adaptivecapmath} = i] \label{eq:binvar}
\end{align}

For intuition behind our design of the \vols' pseudo-rewards (i.e., the two cases in \eqref{eq:ballvar}), recall that our goal is to bound the difference between the values of $\adaptivecapmath$ and $\OPT$, which depends on the number of times $\OPT$ makes a ``better'' recommendation than $\AC$. Whenever \exttraf\ arrives, $\OPT$ will recommend the targeted \opp, which cannot be better than the recommendation made by $\AC$. Similarly, for \inttraf\ where $\OPT$ does not recommend an \opp\ (i.e., for $t \in \bonustimes$), then the recommendation made by $\OPT$ cannot be better, in the sense that the objective is (weakly) increasing in the total number of \signups. In contrast, when \inttraf\ arrives and $\OPT$ does make a recommendation, then this recommendation can be ``better'' than the recommendation made by $\AC$.
Hence, we define different pseudo-rewards for these arriving \vols. \revcolor{We note that if we defined pseudo-rewards identically for all volunteers (i.e., $\ballvar_t = \sum_{i \in [\numopps]} \balancefunc(\fillrate_{i,t-1})\mathbbm{1}[\volchoicet{\OPT} = i]$ for all $t$), a simpler proof would suffice to establish an asymptotic lower bound of $1-1/e$, e.g., building on Lemma 1 in \citet{udwani2021adwords}. However, to break the barrier of $1-1/e$, we crucially rely on differentiating the pseudo-rewards based on a volunteer's source.}

\medskip

\noindent \textbf{Step 2: Lower-bounding the Value of \AC}

\noindent This step of the proof involves two lemmas. First, in Lemma \ref{lem:lowerboundalg}, we use the optimality criteria for the recommendations provided by the $\AC$ algorithm to show that the expected sum of the $\ballvar_t$ and $\binvar_i$ pseudo-rewards is a lower bound on the expected value of $\AC$. (We use $\AC$ to denote the value of the $\AC$ algorithm along a fixed sample path for a fixed instance, again suppressing the dependence for ease of exposition.) Then, in Lemma \ref{lem:lowerboundballvarbinvar}, we use properties of the function $\balancefunc$ (as defined in \eqref{eq:potentialfunc}) to lower bound the expected sum of these pseudo-rewards with a function that depends on the quantity and the source of \signups\ under both $\OPT$ and $\adaptivecapmath$. By combining these lemmas, we establish a (non-linear) relationship between the expected value of \AC\ and that  of \OPT. 

\begin{lem}[Lower Bound on $\adaptivecapmath$ via Pseudo-rewards] 
\label{lem:lowerboundalg}
For any instance $\instance$, the expected sum of all of the pseudo-rewards is a lower bound on the expected value of $\adaptivecapmath$, i.e.,
\begin{equation}
      \mathbbm{E}_{\samplepath}[\adaptivecapmath] \quad \geq \quad \mathbbm{E}_{\samplepath}\Big[\sum_{t \in [\horizon]} \ballvar_t + \sum_{i \in [\numopps]} \binvar_i\Big],
\end{equation}
where $\ballvar_t$ and $\binvar_i$ are defined in \eqref{eq:ballvar} and \eqref{eq:binvar}, respectively.
\end{lem}
The proof of Lemma \ref{lem:lowerboundalg} crucially relies on the fact that whenever \inttraf\ arrives, $\adaptivecapmath$ recommends the \opp\ which maximizes $\convrate_{i,t}\balancefunc(\fillrate_{i,t-1})$. Due to stochasticity in \vols' realized \signup\ decisions, this inequality holds only in expectation over all sample paths. We present the full proof in Appendix \ref{proof:lem:lowerboundalg}. 

In the subsequent lemma, we establish a lower bound on the expected sum of the pseudo-rewards. Recall that, for a fixed instance and sample path, we use counters such as $\AC_{i,\horizon}^\inttrafmath$ to indicate the number of \signups\ for \opp\ $i$ made by \vols\ $t \in \inttimes$ under the $\AC$ algorithm. Similarly, we will use $\acbonus_{i,\horizon}$ 
to represent the amount of \opp\ $i$'s capacity filled by \vols\ $t \in \bonustimes$ under the $\AC$ algorithm.
Mathematically, we have $\acbonus_{i,\horizon} = \sum_{t \in \bonustimes}\mathbbm{1}[\volchoicetsuccess{\AC} = i]$. Furthermore, to mirror our notation for the $\AC$ algorithm, we define $\OPT_{i, \horizon}^\inttrafmath$ (resp. $\OPT_{i, \horizon}^\exttrafmath$) as the amount of \opp\ $i$'s capacity filled by \inttraf\ (resp. \exttraf) under $\OPT$ at the end of the horizon.

\begin{lem}[Lower Bound on Pseudo-Rewards]
\label{lem:lowerboundballvarbinvar}
For any instance $\instance$, we have the following lower bound on the expected sum of all of the pseudo-rewards: 
\begin{align}
    \mathbbm{E}_{\samplepath}\left[\sum_{t \in [\horizon]} \ballvar_t + \sum_{i \in [\numopps]} \binvar_i\right]  \quad \geq \quad  e^{-1/\invbidtobudget}\mathbbm{E}_{\samplepath}\left[\sum_{i \in [\numopps]}\right.& \adaptivecapmath_{i,\horizon}^\exttrafmath + \acbonus_{i,\horizon} +  \OPT_{i,\horizon}^{\inttrafmath} \cdot \balancefunc \left(\frac{\adaptivecapmath_{i, \horizon}^{\inttrafmath}}{\capa_i - \adaptivecapmath_{i, \horizon}^{\exttrafmath}}\right)  \nonumber \\ &\left.  +  \capa_i \left(1-\balancefunc \left(\frac{\adaptivecapmath_{i, \horizon}^{\inttrafmath} - \acbonus_{i,\horizon}}{\capa_i}\right) - 1/e\right)\right], \label{eq:lem:lowerboundballvarbinvar}
\end{align}
where $\ballvar_t$ and $\binvar_i$ are defined in \eqref{eq:ballvar} and \eqref{eq:binvar}, respectively.
\end{lem}
Though we present \eqref{eq:lem:lowerboundballvarbinvar} in expectation over all sample paths, in the proof of Lemma \ref{lem:lowerboundballvarbinvar} we show that the inequality holds along each sample path by separately bounding the sum of the $\ballvar_t$ pseudo-rewards and the sum of the $\binvar_i$ pseudo-rewards. 
The proof relies on properties of the function $\balancefunc$, and the full proof details can be found in Appendix \ref{proof:lem:lowerboundballvarbinvar}.

\medskip

\noindent \textbf{Step 3: Bounding the Competitive Ratio of \AC} 
 
\noindent \revcolor{The final step involves the creation of an instance-specific, factor-revealing mathematical program $\MP$ that serves as a lower bound on the ratio between $\mathbb{E}_{\samplepath}[\AC]$ and $\mathbb{E}_{\samplepath}[\OPT]$ on that instance. The program $\MP$ for instance $\instance$ is designed such that we can construct a feasible solution using the outputs of $\AC$ and $\OPT$ on that instance.\footnote{We emphasize that $\MP$ depends on the instance $\instance$, even though we suppress that dependence. The program $\MP$ partly consists of decision variables specific to each sample path $\samplepath$ that can occur in instance $\instance$. We use $\Omega$ to denote this set of sample paths, which has an associated probability measure (determined by a set of independent Bernoulli random variables) induced by the instance $\instance$.}}

\begin{table}[ht]
  \caption{Definition of the mathematical program \MP.} \label{table:MP}
{\small
$
\arraycolsep=1.4pt\def\arraystretch{1}
\begin{array}{|cllll|}
\hline
&&&&\\[-.5em]
\multicolumn{5}{|c|}{\text{Given an instance $\instance$, the inputs to $\MP$ are the set of \opps\ $\oppset$, the \fracextname\ $\extfrac$, the \MCPR\ $\weightcap$, }} \\
\multicolumn{5}{|c|}{\text{and the set of feasible sample paths $\Omega$, along with its associated probability measure.}} \\[.5em]
\multicolumn{5}{|c|}{\text{$\MP$ uses the set of variables } \  \vec{x} \in \mathbb{R}_{\geq 0}^{3 \times \numopps \times |\Omega|} \text{ and } \  \vec{y} \in \mathbb{R}_{\geq 0}^{2 \times n \times |\Omega|} \setminus \vec{\bf{0}} \text{, along with } z \in [0,1]} \\
&&&&\\[-.5em]
\hline
&&&& \\[-.5em]
\ \underset{\displaystyle \vec{x}, \vec{y}, z }{\textrm{min}}\quad\quad&
 \multicolumn{3}{l}{\displaystyle z}&\quad \quad \textbf{$\MP$ \ } \quad \\[2.5em]
\text{s.t.}  &\ \forall i, \samplepath, \ \quad \capa_i \geq y_{1, i, \samplepath} + y_{2, i, \samplepath} \quad \ \ \text{\bf (i)}&\qquad \capa_i \geq x_{1, i, \samplepath} + x_{2, i, \samplepath} \quad \ \ \text{\bf (ii)}\qquad x_{2,i,\samplepath} \geq x_{3,i,\samplepath} &&\quad \text{\bf (iii)} \\[.7em]
&\multicolumn{3}{l}{\qquad \qquad \ \ \capa_i = x_{1, i, \samplepath} + x_{2, i, \samplepath} \qquad \text{\bf OR} \qquad x_{1, i, \samplepath} = y_{1, i, \samplepath}} &\quad \text{\bf (iv)}\\[.7em]
&\multicolumn{3}{l}{\mathbb{E}_{\samplepath}\left[\sum_{i \in [\numopps]}x_{1, i, \samplepath} + x_{2, i, \samplepath}\right] \quad  \leq \quad z
\sum_{i \in [\numopps]}\capa_i
} &\quad \text{\bf (v)}
\\[.7em]
&\multicolumn{3}{l}{\mathbb{E}_{\samplepath}\left[\sum_{i \in [\numopps]} x_{1, i, \samplepath}+x_{3, i, \samplepath}\right] \quad  \geq \quad (\extfrac - \weightcap + z)
\sum_{i \in [\numopps]}\capa_i
}  &\quad \text{\bf (vi)}\\[.7em]
&\multicolumn{3}{l}{
\mathbb{E}_{\samplepath}\left[\sum_{i \in [\numopps]} x_{1,i,\samplepath} + x_{3,i,\samplepath} + y_{2, i, \samplepath} \cdot \balancefunc \left(\frac{x_{2,i,\samplepath}}{\capa_i - x_{1,i,\samplepath}}\right) + \capa_i \left(1 - \balancefunc \left(\frac{x_{2,i,\samplepath} - x_{3,i,\samplepath}}{\capa_i}\right) - 1/e\right)\right]}& \\[0.7em]
&\multicolumn{3}{l}{\qquad \qquad \qquad \qquad \qquad \qquad \qquad  \leq e^{1/\invbidtobudget} z \mathbb{E}_{\samplepath}\left[\sum_{i \in [\numopps]}y_{1, i, \samplepath} + y_{2, i, \samplepath}\right]} & \quad \text{\bf(vii)} \\[.7em]
\hline
\end{array}
$
}
\end{table}

\revcolor{The constraints are inspired by the results from Step 2 as well as the physical constraints of the problem. In particular, the seventh constraint should be thought of as a bound on the ratio between \AC\ and \OPT\ that builds on Lemmas \ref{lem:lowerboundalg} and \ref{lem:lowerboundballvarbinvar}. In addition, the sixth constraint should be thought of as a lower bound on the capacity filled by \vols\ in $\exttimes$ and $\bonustimes$, which we remind are the two sets of \vols\ for which $\OPT$ could not have made a better decision than $\AC$ (see Eq. \eqref{eq:ballvar} and the following discussion). As the \MCPR\ $\weightcap$ increases, this constraint becomes looser, meaning that (all else equal) the value of the program weakly decreases. Consequently, the lower bound on the competitive ratio of \AC\ is decreasing in $\weightcap$, as shown in Figure \ref{fig:weight_caps}.}

\revcolor{We analyze $\MP$ via two additional lemmas, whose statements and proofs we defer to Appendix \ref{apx:thm1proofdetails}. First, in Lemma \ref{lem:MP}, we show that the optimal value of $\MP$ is a lower bound on the ratio between the expected value of $\AC$ and the expected value of $\OPT$ in instance $\instance$.\footnote{\revcolor{We remark that this may be a \emph{strict} lower bound, and hence the source of some slack in our analysis (i.e., the small gap shown in Figure \ref{fig:weight_caps} between the upper bound and the lower bound on \AC's competitive ratio when $\weightcap = 1$). Specifically, fixing an instance $\instance$, the feasibility region of $\MP$ can be superset of the possible values of $\vec{x}, \vec{y}$, and $z$ that correspond to the outputs of $\AC$ and $\OPT$ on instance $\instance$. To rule out such ``unrealistic'' solutions to $\MP$, additional constraints would be necessary (e.g., on the feasible distribution of \inttraf\ across opportunities under \AC). Due to the inherent difficulty of operationalizing such constraints, we turned to a different (and less flexible) approach to establish a tight performance guarantee in the special case of deterministic sign-ups (see Theorem \ref{thm:tightproof}).}}}
\revcolor{Then, in Lemma \ref{lem:MPlower}, we place a lower bound of $z^*$ on the value of $\MP$, where we remind that $z^*$ only depends on three properties of the instance $\instance$: the \fracextname\ $\extfrac$, the minimum capacity $\invbidtobudget$, and the \MCPR\ $\weightcap$.   Together, these lemmas prove that $z^*$ is a lower bound on the competitive ratio of the \AC\ algorithm.}



    
                

\section{\revcolor{Evaluating Algorithm Performance on VM Data}}
\label{sec:case_study}
\revcolor{In this section, we go beyond a worst-case analysis and use data from VM to numerically evaluate the performance of the $\adaptivecapmath$ algorithm in more realistic problem instances. In Section \ref{subsec:background}, we 
describe how we use VM data to construct instances of our model. In Section~\ref{subsec:policy:eval} we compare the performance of $\adaptivecapmath$ on those instances to the performance of relevant benchmarks. We conclude in Section \ref{sec:sim:acvsmsvv} with a more in-depth comparison of $\AC$ and $\MSVV$ on modified instances to provide insight into the key features that influence the relative performance of those two algorithms.}

\subsection{Instance Construction} 
\label{subsec:background}
As introduced in Section \ref{sec:intro}, VM is the world’s largest online platform for connecting \vols\ and \opps.  To carry out our case study, we draw upon VM's database (which provides us with information on opportunity characteristics such as capacities) as well as VM's Google Analytics (GA) dataset (which consists of a sample of around 20\% of session-level activity on the platform). For the sessions included in the GA data, we know the number of views and sign-ups for each opportunity as well as the source of the volunteer (internal or external). We use data from August 1, 2020 through March 1, 2021 and provide more details about the available data in Appendix~\ref{app:data_availability}.

Constructing an instance of our model requires (i) a set of opportunities and their capacities; (ii) a set of external traffic volunteers, including their targeted opportunities and arrival times; and (iii) a set of internal traffic volunteers, including their conversion probabilities and arrival times. The overall arrival sequence is then determined by the arrival times of both external and internal traffic. We now describe how we draw upon the available data to specify each of these three components.

\subsubsection*{Opportunities.} 
We will only consider the 10,737 virtual opportunities appearing in GA data between August 2020 and March 2021 for which we have precise data on capacity, i.e., the number of volunteers needed. The results we present in this section are based on an instance constructed using a random sample of 100 \opps\ from this set of 10,737; among this subset of 100 opportunities, the average capacity is 4.49 and the minimum capacity is 1. 
We have performed several relevant robustness checks (sampling different opportunities, varying the number of opportunities sampled, etc.) which show qualitatively similar results and are omitted for the sake of brevity.

\subsubsection*{External Traffic.}
Consistent with our base model, we will assume each view from \exttraf\ deterministically results in a sign-up for its targeted opportunity.\footnote{\revcolor{For this subset of 100 opportunities, only 14.9\% of views from \exttraf\ result in sign-ups. As discussed in Appendix \ref{apx:exttrafstochastic}, our results in Theorem \ref{thm:AClower} continue to hold in settings where the sign-up decisions of \exttraf\ are not deterministic.}} For each \exttraf\ arrival, we sample a view uniformly at random from the pool of external views in the GA data for the aforementioned $100$ \opps. We preserve the time stamp and the targeted opportunity of this external view. To estimate the total volume of external traffic, we compute the number of sign-ups from external traffic for this subset of 100 opportunities in GA data. As this GA data represents roughly 20\% of all traffic, we scale this number by a factor of 5 to arrive at $T^{\textsc{ext}} = 225$ arrivals (equivalently, sign-ups) from external traffic

We note that only 49 of the 100 opportunities receive any views from external traffic in the GA data, and a majority of the views go to only 4 opportunities. Consequently, even though we have 225 external traffic arrivals (which deterministically lead to sign-ups) in this constructed instance, only 86 of them can be ``useful,'' leading to an EFET of $\extfrac = 0.19$. The remaining sign-ups from \exttraf\ exceed the capacity of the targeted opportunity. 

\subsubsection*{Internal Traffic.}
In order to keep the relative number of expected sign-ups from internal and external traffic approximately consistent with what we observe in GA data, we assume that there are $T^{\textsc{int}} = 3,539$ internal traffic volunteers for our subset of 100 opportunities. For each such volunteer, we sample a view uniformly at random from the GA data on internal views, and we preserve the time stamp of this view. The overall arrival sequence is obtained by interleaving the arrival times of \exttraf\ and \inttraf: each arrival is associated with a time stamp, and we order the arrivals based on those time stamps.

\begin{figure}[t]
    \begin{center}
        \resizebox{.8\textwidth}{!}{\input{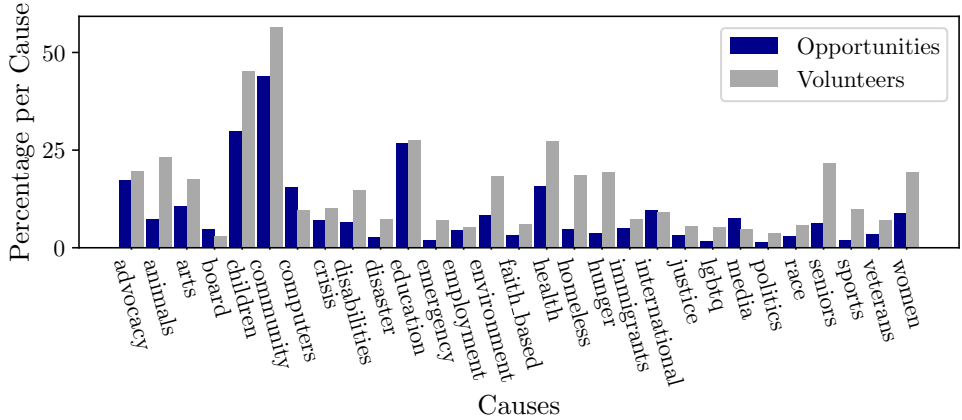}}
    \end{center}
    \caption{Percentage of \vols\ and \opps\ associated with each cause. These percentages need not sum to $100\%$, as \opps\ and \vols\ can be associated with multiple causes.}
    \label{fig:causes}
\end{figure}

Moving beyond the arrival time of an internal volunteer, specifying their sign-up behavior is more involved, as there are many potential \opps\ that they can sign up for, each with a possibly different conversion probability. Unfortunately, the available data does not allow for precise counterfactual estimates of conversion probabilities.
We only observe that, conditional on viewing an opportunity, an internal traffic volunteer signs up roughly 10\% of the time, which does not vary predictably across opportunities. Based on this limitation, to approximate volunteers' conversion probabilities, we leverage data on \emph{causes}.\footnote{\revcolor{We note that for in-person \opps, location likely also plays a role in conversion probabilities. However, here we focus on virtual \opps.}} When an opportunity is created, it selects up to three associated causes, out of a list of 29 (e.g., seniors, hunger, etc.). Volunteers also select an arbitrary-sized subset of the different causes when creating an account on VM. In Figure \ref{fig:causes}, we display the percentage of \opps\ and volunteers that are associated with each cause. 

We preserve the cause profile of the \opps; however, we cannot do the same for volunteers because our data does not connect each view to a particular volunteer profile. As a result, we resort to sampling from the aggregate data on volunteer preferences. In particular, whenever an internal traffic volunteer $t$ arrives, we determine their associated causes by independently drawing one Bernoulli random variable for each cause with probability equal to the proportion of volunteers associated with that cause. We use the data on causes to construct the following three instances:
\begin{itemize}
    \item \textbf{Base Instance.} We say that \vol\ $t$ and \opp\ $i$ are \emph{compatible} if they share at least one common cause, and we set their conversion probability to $\convprob_{i,t} = 0.1$. On the other hand, if they share no causes, then we say they are incompatible, i.e., $\convprob_{i,t} = 0$. Roughly speaking, this captures an improved version of the VM platform that has visibility into the cause profile of every searching \vol.
\end{itemize}
Though this base instance reflects aggregate volunteering preferences, it cannot capture time-varying aspects of preferences. For example, volunteers may be drawn to certain opportunities for short periods of time (e.g., due to real-world events), and some opportunities may not be available for the entire time horizon.\footnote{\revcolor{For one example of time-varying interest in particular causes, see \url{https://blogs.volunteermatch.org/two-data-points-should-give-you-hope}.}
} 
As an illustration of temporal variation in compatibility, we construct two additional instances. 
\begin{itemize}
    \item \textbf{Auxiliary Instance I (minimal time-varying compatibility).} For each opportunity $i$, we restrict its compatibility to a sub-interval of the arrival sequence of \inttraf, i.e., $[\underline{t}_i, \underline{t}_i + \tau_i] \subseteq [1, T^{\textsc{int}}]$. For the $t\textsuperscript{th}$ arriving \inttraf\ volunteer, if $t \in [\underline{t}_i, \underline{t}_i + \tau_i]$, we leave $\convprob_{i,t}$ the same as it was in the base instance, but if $t \notin [\underline{t}_i, \bar{t}_i]$, we set $\convprob_{i,t} = 0$, representing that \opp\ $i$ is unavailable or unappealing to \vols\ arriving outside of this compatibility interval. (For simplicity and to keep the EFET the same across instances, we assume that \exttraf\ always signs-up for its targeted \opp, regardless of its arrival time.)
    To specify each opportunity's sub-interval, we first randomly draw the length of the interval (i.e., $\tau_i$) in proportion to the number of volunteers opportunity $i$ needs (and truncated such that $\tau_i < T^{\textsc{int}}$). 
    In this instance, opportunities' average interval length is approximately 75\% of the arrival sequence of \inttraf, i.e., $\frac{1}{100} \sum_{i=1}^{100} \tau_i \approx 0.75 \cdot T^{\textsc{int}}$.
    The start point of the interval (i.e., $\underline{t}_i$) is drawn uniformly at random from the set $\{1, \dots, T^{\textsc{int}} - \tau_i\}$. 
   This implies that the interval $[\underline{t}_i, \underline{t}_i + \tau_i]$ occurs uniformly at random throughout the arrival sequence of \inttraf.

    \item \textbf{Auxiliary Instance II (significant time-varying compatibility).} The construction of this instance is identical to the previous instance except that we further restrict opportunities' compatibility to a shorter sub-interval. In this instance, opportunities' average interval length is approximately 25\% of the arrival sequence of \inttraf, i.e., $\frac{1}{100} \sum_{i=1}^{100} \tau_i \approx 0.25 \cdot T^{\textsc{int}}$.
\end{itemize}
 These three instances all correspond to a setting with stochastic rewards where the MCPR $\sigma = 1$.

\subsection{Policy Evaluation}
\label{subsec:policy:eval}
We now aim to understand how well $\AC$ performs --- both in comparison to its theoretical guarantee and also in comparison to other reasonable alternatives --- in the instances that we constructed based on VM data. To that end, we consider the following benchmarks: 

\begin{itemize}
\item[] {\bf{Upper bound on \OPT}} ($\mathbf{\overline{\OPT}}$): Our benchmark $\OPT$ (see Definition \ref{def:opt}) is a dynamic program, which can be of exponential size given the stochastic nature of volunteer sign-ups. Therefore, instead of using \OPT, we follow the standard approach of using an LP relaxation {(see, e.g., \citealt{Alaei2013TheOS})}, denoted $\overline{\OPT}$ as our normalization factor when evaluating the performance of $\AC$ and the other benchmarks. 
We formally define $\overline{\OPT}$ in Appendix \ref{app:OPT_UB}, and we show that it is an upper bound on $\OPT$.

 \item[] {\bf Current Practice (\CP):} 
When a volunteer arrives, CP recommends the compatible opportunity that has been most recently updated (based on data for the 100 opportunities in our sample), regardless of its remaining capacity. This benchmark serves as a stylized proxy for VM's actual recommendation algorithm for virtual \opps, which provides a ranked list of opportunities based on the ``recency'' of an opportunity's actions. Importantly, this benchmark (similar to the actual algorithm on VM) does not account for opportunities' current fill rates or the traffic source (i.e., internal or external). 
 
  \item[] {\bf Smart Current Practice (\SCP): } 
We next consider a more sophisticated version of \CP\ that only considers compatible opportunities \emph{with remaining capacity}, and among those recommends the most recently updated one.
 
 \item[] {\bf{Remaining Capacity Heuristic (\RC)}:} We also consider a simple heuristic that recommends the compatible opportunity with the largest remaining capacity. 

\item[] {\bf Generalized Perturbed Greedy (\GPG): } 
 Here we consider a randomized algorithm from \citet{udwani2021adwords} which is shown to obtain the best-possible competitive ratio in the special case where sign-ups are deterministic, even if capacities are small. This algorithm can also be applied in our constructed instances where sign-ups are stochastic. Specifically, for each opportunity $i$, \GPG\ independently generates a random value $y_i$ uniformly from $[0,1]$. When a volunteer arrives, it considers the opportunities with remaining capacity and recommends the one with the largest value of $\mu_{i,t} \Big(1 - e^{(y_i-1)}\Big)$. 

\item[] {\bf MSVV:} Our final benchmark is the algorithm introduced in \citet{mehta2007adwords}, which we previously described in Algorithm \ref{alg:msvv}.
\end{itemize}

\begin{table}[t]
\centering
\caption{\revcolor{Performance of \AC\ and benchmarks on the three instances constructed in Section \ref{subsec:background}, averaged over 10,000 simulations and normalized by $\overline{\OPT}$. The standard errors are negligible.}}
 \begin{tabular}{|l c c c c c c c|} 
 \hline
 &&&&&&&\\
 \cline{2-7}
 & \AC & \CP & \SCP & \RC & \GPG & \MSVV &\\ [-0.4ex] 
 \cline{2-7} &&&&&&& \\[-1.2ex]
 \textbf{Base Instance} & $ 0.945 $&$ 0.302 $&$ 0.898 $&$ 0.984 $&$ 0.933 $&$ 0.952 $ &\\ [0.5ex]
\textbf{Auxiliary Instance I} &  $ 0.946 $&$ 0.316 $&$ 0.862 $&$ 0.942 $&$ 0.929 $&$ 0.952 $ & \\ [0.5ex]
\textbf{Auxiliary Instance II} & $ 0.876 $&$ 0.421 $&$ 0.802 $&$ 0.834 $&$ 0.845 $&$ 0.877 $ &\\[1.0ex]
\hline
 \end{tabular}
 \label{table:policyperf}
\end{table}

\subsubsection*{Performance of \AC\ and Benchmarks:}

For each of the three constructed instances, in Table \ref{table:policyperf} we present the value of \AC\ and the benchmarks introduced above (\CP, \SCP, \RC, \GPG, and \MSVV), normalized by $\overline{\OPT}$ and averaged over $10,000$ simulations. (In each simulation, the randomness only comes from the sign-up decisions of \inttraf, as well as any randomness within the algorithm itself.)
In the first column of Table \ref{table:policyperf}, we observe that in all three instances \AC\ performs quite close to $\overline{\OPT}$ and far above its competitive ratio in this finite-capacity regime.\footnote{\revcolor{For $\invbidtobudget = 1$, $\extfrac = 0.19$, and $\weightcap = 1$ --- as is the case in these instances --- the competitive ratio of \AC\ is $e^{-1}(1-1/e) \approx 0.23$ according to Theorem \ref{thm:AClower}.}} 
We note the relative performance of \AC\ is worst in the third instance, when compatibility varies significantly over time, which is aligned with our intuition that foreknowledge of the arrival sequence is more valuable when the sequence is non-stationary (i.e., when compatibility varies over time).

\begin{figure}[t]
    \begin{center}
        \includegraphics[width=0.98\textwidth]{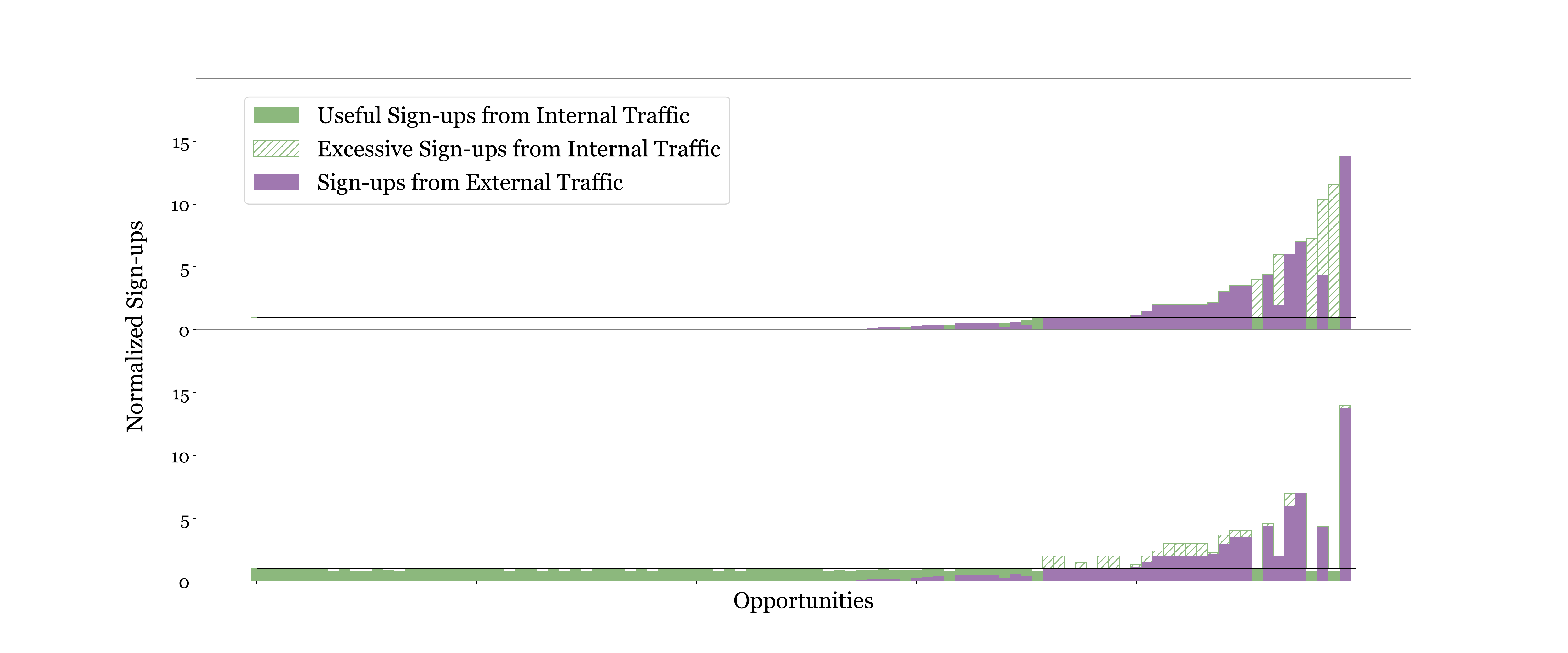}
    \end{center}
    \caption{\revcolor{Distribution of sign-ups for each opportunity (normalized by its capacity) as well as the source of sign-ups in one simulation of \CP\ (top) and \AC\ (bottom) on the base instance constructed in Section \ref{subsec:background}.}}
    \label{fig:ac_vs_cp}
\end{figure}

We next observe that \AC\ dramatically outperforms \CP\ across all instances. To further explore the differences between \AC\ and \CP, in Figure \ref{fig:ac_vs_cp} we focus on a single simulation of the base instance and plot the quantity and source of sign-ups for each opportunity, normalized by that opportunity's capacity. We emphasize that the total number of sign-ups is identical under the two algorithms. (This may not be obvious from the figure, given that we normalize each vertical bar based on that opportunity's capacity.) However, many of the sign-ups under \CP\ are not ``useful'' because it may continue to recommend an opportunity even after it has filled its capacity. This observation partially explains why the performance of \CP\ actually improves in the third instance: the time-varying compatibility leads to additional variation in \CP's recommendations.

A simple adjustment of \CP\ that accounts for capacity (i.e., \SCP) bridges a large portion of the gap, though \AC\ continues to outperform \SCP\ by 5-10\%, depending on the instance. Our other simple heuristic, \RC, actually performs the best out of all the considered algorithms in the base instance ($0.984$). However, its performance suffers when compatibility varies over time: in the third instance, \AC\ outperforms \RC\ by 5\%.

We conclude this discussion by comparing \AC\ to two other algorithms with strong theoretical guarantees in certain special cases: \GPG\ (which obtains the optimal guarantee when sign-ups are deterministic) and \MSVV\ (which is asymptotically optimal in the absence of \exttraf). Potentially because it does not adapt based on the current fill rate of opportunities, \GPG\ performs worse than \AC\ in all three instances, by up to $3.7\%$. In contrast, the performance of \AC\ and \MSVV\ is quite similar, as one might expect given the relatively low EFET. Their relative performance is always within $0.7\%$ across the three instances, with \MSVV\ performing slightly better in all three instances. Due to the similarities in performance between these two algorithms in the constructed instances, we next dig a bit deeper into instance features that drive performance gaps between \AC\ and \MSVV.

\begin{figure}[t]
    \begin{center}
        \includegraphics[width=0.7\textwidth]{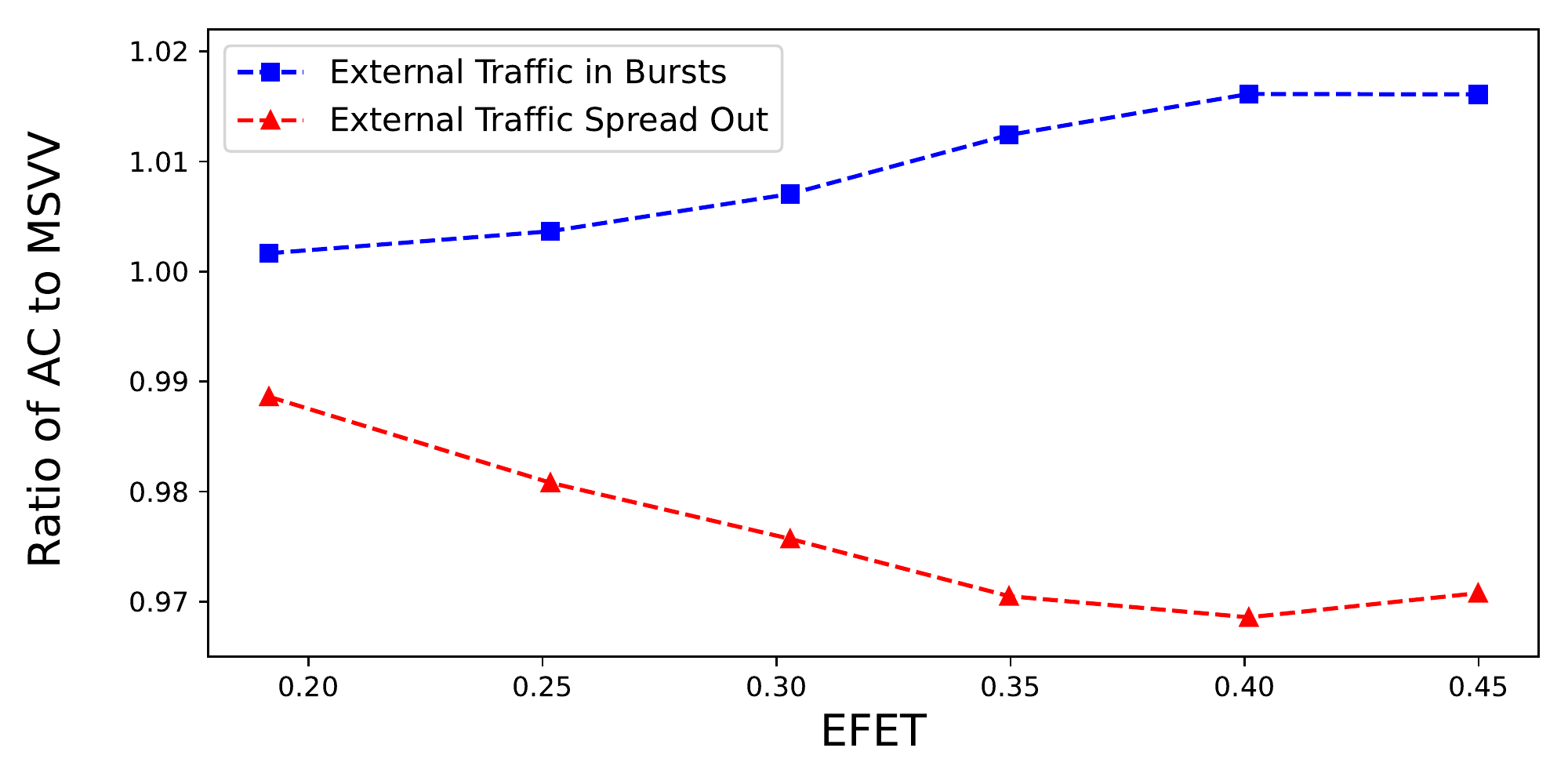}
    \end{center}
    \caption{\revcolor{Relative performance of \AC\ and \MSVV\ (averaged over 10,000 simulations) for two different arrival patterns as a function of the EFET $\extfrac$. The standard errors are negligible.}}
    \label{fig:ac_vs_msvv}
\end{figure}

\subsection{Comparison Between \AC\ and \MSVV:}
\label{sec:sim:acvsmsvv}
As discussed in Section \ref{subsubsec:thm2:discussion} and illustrated in Figure \ref{fig:CR_weightcap}, the competitive ratio of \AC\ is only strictly better than the competitive ratio of \MSVV\ when the EFET $\extfrac$ is sufficiently large. In the instances we constructed in Section \ref{subsec:background}, we have $\extfrac = 0.19$. 
To obtain instances with a larger EFET, we scale up the number of \exttraf\ arrivals while continuing to sample from the view distribution of \exttraf\ to determine their targeted opportunity. This enables us to construct instances where the EFET varies from 0.19 to 0.45. (Even large numbers of \exttraf\ arrivals cannot meaningfully exceed this EFET given our sampling approach, as only 49 of the 100 opportunities in our subset receive any views from \exttraf.) 
To keep the achievable performance as similar as possible in each of these instances, we correspondingly reduce the number of \inttraf\ arrivals such that our benchmark $\overline{\OPT}$ remains the same. We determine compatibility for these \inttraf\ arrivals using the same procedure described in our base instance.

For a sufficient amount of \exttraf, the key factor that separates the performance of \AC\ and \MSVV\ lies in how they account for \exttraf\ in their definition of an opportunity's fill rate. In particular, \AC\ imposes a weaker penalty than \MSVV\ when capacity is filled by \exttraf. Based on this observation, as discussed at the end of Section \ref{subsubsec:thm2:discussion}, we would expect \AC\ to perform better in instances where opportunities with \exttraf\ should not be harshly penalized (e.g., if opportunities with early-arriving \exttraf\ have fewer compatible arrivals in the future), whereas \MSVV\ should perform better when opportunities with \exttraf\ can safely be penalized (e.g., if some external traffic is predictive of more \exttraf\ in the future).

Fixing an EFET $\extfrac$, we numerically assess this hypothesis by considering two different modifications of the arrival sequence.  First, we consider an arrival sequence where external traffic arrives in \emph{bursts}. Specifically, building on the intuition developed in our hardness examples for \MSVV\ (see Propositions \ref{prop:warmupmsvv} and \ref{prop:generalmsvv}), we partition the \exttraf\ into two groups. If an \exttraf\ volunteer $t$ targets an \opp\ that can be entirely filled with \exttraf, then $t$ arrives at the end of the arrival sequence (i.e., after all \inttraf). Otherwise, if $t$ targets an \opp\ that can only be partially filled with \exttraf, then $t$ arrives at the beginning of the arrival sequence (i.e., before all \inttraf).
Roughly speaking, in this arrival sequence, it is better to minimally penalize opportunities for \exttraf\ arrivals; hence, \AC\ should outperform \MSVV.

To complement this arrival sequence, we considered an arrival sequence where \exttraf\ is \emph{spread out}. For this arrival sequence, we assume \exttraf\ arrives in a random order, which means each \exttraf\ arrival is predictive of future \exttraf\ arrivals. Roughly speaking, larger penalties for \exttraf\ should be better in this setting, which means \MSVV\ should perform better than \AC.\footnote{\revcolor{Of course, if we knew that \exttraf\ was spread out, we could consider a variant of \AC\ similar to what we describe in Remark \ref{remark:extknown} that continuously adjusts capacities based on the \emph{predicted} future amount of \exttraf\ for each \opp, potentially improving performance.}}

For each of these modified instances, we simulate both algorithms 10,000 times and we report the results in Figure \ref{fig:ac_vs_msvv} as a function of the EFET $\beta$. Here, the ratio of \AC\ to \MSVV\ for the arrival sequence where external traffic arrives in bursts (resp. where \exttraf\ is spread out) is shown by the blue line with squares (resp. red line with triangles). Consistent with our intuition, \AC\ does better than \MSVV\ when \exttraf\ arrives in bursts because it appropriately limits the penalty for sign-ups from \exttraf, and its relative improvement grows as a function of $\beta$. In contrast, \MSVV\ performs better when \exttraf\ is spread out over time. 

This can guide practitioners in determining which algorithm is most suitable for their setting: if external traffic represents a significant portion of traffic and arrives in unpredictable bursts, \AC\ may be preferable due to its superior worst-case performance. However, if \exttraf\ arrives consistently and spread over time, then the platform should either implement \MSVV\ or account for this in the design of \AC\ (i.e., by continuously predicting the amount of future \exttraf\ for each \opp\ and reducing capacities accordingly).

\section{Conclusion}
\label{sec:conclusion}
In this paper, we introduce a framework for making online recommendations to maximize matches in the presence of external traffic, motivated by platforms such as VolunteerMatch (the largest online volunteer engagement network in the US, and our industry partner). Our recommendation algorithm, Adaptive Capacity (\AC), does not know the amount of \exttraf\ \emph{a priori}, yet it nevertheless provides strong parameterized guarantees (relative to both the commonly-used $\MSVV$ algorithm and the upper bound we establish on any online algorithm). 

Beyond theoretical guarantees, we demonstrate \AC's practical effectiveness in simulations based on VM data. We are currently collaborating with VM  to implement a version of our algorithm.

More generally, our work shows the importance of accounting for the source of traffic in decision-making on platforms with multi-channel traffic, which opens up opportunities for further research. For instance, while we have focused on settings where the platform cannot influence external traffic, some platforms may have some degree of control over the timing or the destination of this traffic (e.g., via marketing campaigns or curated email recommendations). Also, platforms with external traffic may have objectives beyond maximizing the number of matches (e.g., platforms such as DonorsChoose may aim to maximize the number of donation campaigns that reach a certain threshold). Studying the platform design in such settings is an interesting direction for future work.

\bibliographystyle{plainnat}
\bibliography{online_matching_ranking}
\newpage

\begin{APPENDICES}
\mathtoolsset{showonlyrefs=true}

\section{Omitted Proofs of Section \ref{sec:results}}
We first present the proofs of the three theorems from Section \ref{sec:results}, in order of their appearance in the text. We then proceed to the proofs of the propositions, also in order of appearance.


\subsection{Proof of Theorem \ref{thm:hardness} (Section \ref{subsec:general_arrivals})}
\label{proof:thm:hardness}
This proof is an adaptation of the proof of Theorem 7.1 in \cite{mehta2007adwords}, which we have generalized to apply in our setting. We aim to prove that no online algorithm (deterministic or randomized) can provide a competitive ratio greater than $\big(1-1/e\big)\mathbbm{1}_{\extfrac \leq 1/e} + \big(1+\extfrac \log(\extfrac)\big) \mathbbm{1}_{\extfrac > 1/e}$. By Yao's Lemma \citep{yao1977probabilistic}, it is sufficient to show that there exists a distribution over a set of instances for which no deterministic algorithm can provide an expected value greater than $\Big(\big(1-1/e\big)\mathbbm{1}_{\extfrac \leq 1/e} + \big(1+\extfrac \log(\extfrac)\big) \mathbbm{1}_{\extfrac > 1/e}\Big) \OPT$.

We begin by fixing an \fracextname\ $\extfrac$ and describing an instance $\instance_1(\extfrac)$. In this instance, the set of \opps\ is of size $\largeopps$, each with identical large capacity $\largecapacity$. The arrival sequence consists of $\largeopps \largecapacity$ \vols. The first $(1-\extfrac)\largeopps\largecapacity$ of these \vols\ are \inttraf, and the remaining $\extfrac\largeopps\largecapacity$ are \exttraf.\footnote{We assume that $\extfrac\largeopps$ is an integer. This assumption does not impact the upper bound in the statement of Theorem \ref{thm:hardness}, as the expression comes from taking the limit as $\largeopps$ and $\largecapacity$ approach $\infty$.} All \vols\ have conversion probabilities of $1$ or $0$, and if $\mu_{i,t}=1$ (resp. 0), we will refer to \opp\ $i$ and \vol\ $t$ as \emph{compatible} (resp. incompatible). 

The arrival sequence of $\instance_1(\extfrac)$ can be broken down into $\largeopps$ batches of $\largecapacity$ sequentially-arriving identical \vols. For each $j \in \{1, \ldots, (1-\extfrac)\largeopps\}$, the $j$\textsuperscript{th} batch of \vols\ consists of \inttraf\ that is compatible with all opportunities $i \geq j$. For each $j \in \{(1-\extfrac)\largeopps +1, \ldots, \largeopps\}$, the $j$\textsuperscript{th} batch of \vols\ consists of \exttraf\ which views (and is compatible with) \opp\ $i^*_j = j$. This \exttraf\ can fill the entire capacity of each of these $\extfrac \largeopps$ \opps, which implies that the \fracextname\ is equal to $\extfrac$ in such an instance.

We first establish the value of $\OPT$ on instance $\instance_1(\extfrac)$. 

\begin{clm}
\label{clm:acopt}
For any \fracextname\ $\extfrac$, \OPT\ achieves a value of $\largeopps\largecapacity$ on $\instance_1(\extfrac)$.
\end{clm}
\begin{proof}{Proof of Claim \ref{clm:acopt}:}
Consider a solution which matches each of the $\largecapacity$ \vols\ in the $j$\textsuperscript{th} batch to \opp\ $j$, for all $j \in [\largeopps]$. These \vol-\opp\ pairs are all compatible based on the compatibility structure previously described, and (since the conversion probabilities are exactly equal to $1$) each \opp\ will exactly reach its capacity of $\largecapacity$. As this solution fills all capacity, $\OPT$ must also fill all capacity, thereby achieving a total value of $\largeopps\largecapacity$, regardless of the \fracextname\ $\extfrac$. \halmos 
\end{proof}

We now consider the set of instances which can be obtained from $\instance_1(\extfrac)$ by permuting the indices of the \opps. Specifically, we apply a permutation $\permutation$ to the set of \opps\ such that any algorithm sees \opps\ with indices $\{\permutation(1), \ldots, \permutation(\largeopps)\}$. We highlight that \emph{a priori} the \opps\ appear identical to an online algorithm, aside from their indices. We augment our previous notation and describe such an instance as $\instance_1(\extfrac, \permutation)$.

Suppose that the permutation $\permutation$ is drawn uniformly at random from the set of all permutations of $\largeopps$ indices. For the set of instances generated by this distribution over permutations, in the following lemma, we place an upper-bound on the expected value of any deterministic online algorithm. 
\begin{clm}
\label{clm:universal_ub}
Consider any deterministic online algorithm $\ALG$.
For any \fracextname\ $\extfrac$,
\begin{equation}\mathbb{E}_\permutation[\ALG(\instance_1(\extfrac, \permutation))] \leq \sum_{i \in [(1-\extfrac)\largeopps]}\min\left\{\largecapacity,\sum_{j=1}^{i} \frac{\largecapacity}{\largeopps-j+1}\right\} + \sum_{i \in [\largeopps]\setminus[(1-\extfrac)\largeopps]} \largecapacity,
\end{equation}
where the expectation is taken with respect to the uniform distribution over permutations $\permutation$.
\end{clm}
\begin{proof}{Proof of Claim \ref{clm:universal_ub}:}
To aid in this proof, let us define $d_{i,j}$ as the amount of \vols\ allocated to \opp\ $i$ from the $j$\textsuperscript{th} batch of arriving \vols.
Recall that for  $j \in \{1, \ldots, (1-\extfrac)\largeopps\}$, \vols\ in the $j$\textsuperscript{th} batch of arrivals are compatible with all \opps\ $i \geq j$.  Thus, we have:
\[
    E_{\permutation}[d_{i,j}] \leq 
\begin{dcases}
    \frac{\largecapacity}{\largeopps-j+1},& \text{if } i \geq j\\
    0              & \text{if } i < j
\end{dcases}
\]

To see why this must be the case, note that for each \vol\ in the $j$\textsuperscript{th} batch of \vols, there are a total of $\largeopps-j+1$ compatible \opps. The online algorithm cannot distinguish between these compatible \opps, as it only observes the indices $\{\permutation(j), \ldots, \permutation(\largeopps)\}$. Specifically, if $i$ is one such compatible \opp, the online algorithm does not know which index in the set $\{\permutation(j), \ldots, \permutation(\largeopps)\}$ is equal to $\permutation(i)$. Hence, the expected amount of \vols\ allocated to \opp\ $i$ cannot exceed $\frac{\largecapacity}{\largeopps-j+1}$, when taking expectation with respect to the uniform distribution over permutations $\permutation$.

More simply, for batches $j \in \{(1-\extfrac)\largeopps +1, \largeopps \}$, the \vols\ are only compatible with \opp\ $i$ if $i = j$. Hence,
\[
    E_{\permutation}[d_{i,j}] \leq 
\begin{dcases}
    \largecapacity & \text{if } i= j\\
    0              & \text{if } i \neq j
\end{dcases}
\]
After the arrival of all \vols, the expected fill of \opp\ $i$ is upper-bounded by $\sum_{j \in [\largeopps]} E_{\permutation}[d_{i,j}]$. This quantity is either $\largecapacity$ (if $i > (1-\extfrac)\largeopps$) or $\min\{\largecapacity,\sum_{j=1}^{i} \frac{\largecapacity}{\largeopps-j+1}\}$ (if $i \leq (1-\extfrac)\largeopps$). Summing over all \opps, we have the following upper bound on the value of any online algorithm:
\begin{equation}
    \sum_{i \in [(1-\extfrac)\largeopps]}\min\left\{\largecapacity,\sum_{j=1}^{i} \frac{\largecapacity}{\largeopps-j+1}\right\} + \sum_{i \in [\largeopps]\setminus[(1-\extfrac)\largeopps]} \largecapacity
\end{equation}
This completes the proof of Claim \ref{clm:universal_ub}.\halmos
\end{proof}
Together, and in combination with Yao's lemma, these claims establish an upper-bound on the achievable competitive ratio of any online algorithm of
\begin{align}
    \frac{1}{\largeopps\largecapacity}\left(\sum_{i \in [(1-\extfrac)\largeopps]}\min\left\{\largecapacity,\sum_{j=1}^{i} \frac{\largecapacity}{\largeopps-j+1}\right\} + \sum_{i \in [\largeopps]\setminus[(1-\extfrac)\largeopps]} \largecapacity\right) \quad \rightarrow \quad \big(1-1/e\big)\mathbbm{1}_{\extfrac \leq 1/e} + \big(1+\extfrac \log(\extfrac)\big) \mathbbm{1}_{\extfrac > 1/e},
\end{align}
where the limit holds as $\largecapacity$ and $\largeopps$ approach infinity.\footnote{To show that this upper bound holds for any minimum capacity $\invbidtobudget$, it suffices to add an additional \opp\ with capacity $\invbidtobudget$ for which \vols\ have conversion probability of $0$. The value of $\OPT$ and the upper bound on the performance of any algorithm do not change, and the \fracextname\ also remains the same in the limit as $\largeopps$ approaches infinity.}
\halmos

\subsection{Proof of Theorem \ref{thm:tightproof} (Section \ref{subsec:general_arrivals})}
\label{apx:proof:tightproof}
The proof builds on proof ideas in \citet{mehta2007adwords}, with additional intricacies to account for external traffic. As in \citet{mehta2007adwords}, our analysis relies on a partition of opportunities into a sufficiently large number of \emph{types}, where an opportunity's type is determined by its fill rate after a run of our algorithm. We begin by presenting the outline of our proof; we then present the omitted proofs of intermediate results in Appendices \ref{proof:lem:tightproofACbound} through \ref{proof:lem:tightproofLPbound}.

Recall that the fill rate of opportunity $\oppindex$ under \AC\ at the end of the arrival sequence is given by $\fillrate_{\oppindex,T} = \AC_{\oppindex,T}^{\inttrafmath}/(\capa_\oppindex -\AC_{\oppindex,T}^{\exttrafmath})$. This fill rate definition represents a crucial difference between \AC\ and \MSVV, and our analysis relies on \AC's fill rate definition to establish an improved competitive ratio. Fixing $\numbins \in \mathbb{N}$ (we will think of $\numbins$ as a large number), an opportunity $\oppindex$ belongs to \textit{type} $\indexa$ if its fill rate under \AC\ at the end of the arrival sequence is in the interval $[(\indexa-1)/\numbins, \indexa/\numbins)$.\footnote{\revcolor{By convention, if an opportunity is full at the end of the arrival sequence, it is of type $\numbins$.}}

\paragraph{Notation:} Throughout the proof, we will fix an instance and we fix the number of opportunity types $\numbins$, where we assume $\numbins \geq 4\invbidtobudget/\extfrac$.\footnote{\revcolor{Here we assume $\extfrac > 0$. In the special case of $\extfrac = 0$, Theorem \ref{thm:tightproof} follows directly from \citet{mehta2007adwords}.}} Let  $\typeset{\indexa}$ denote the set of opportunities of type $\indexa$. We will use $\capn{\indexa}$ to denote the \textit{aggregate capacity} of type $\indexa$ opportunities (i.e., $\capn{\indexa} = \sum_{\oppindex \in \typeset{\indexa}} \capa_\oppindex$), and the total capacity is given by $\totalcap = \sum_{\indexa =1}^{\numbins} \capn{\indexa}$. We will use $\ext{\indexa}$ to denote the total capacity of type $\indexa$ opportunities that can be filled by external traffic, which implies that $\sum_{\indexa =1}^{\numbins} \ext{\indexa} =  \extfrac \totalcap$ based on our definition of the \fracextname. Furthermore, let $\intern{\indexb}$ denote the total capacity of type $\indexb$ opportunities that $\OPT$ fills with internal traffic.

Using the partition of opportunities into types, we prove Theorem \ref{thm:tightproof} in four steps. (i) First, in Lemma \ref{lem:tightproofACbound}, we lower-bound the value of \adaptivecap\ as a function of $\capn{\indexa}$ and $\ext{\indexa}$. (ii) Then, in Lemma \ref{lem:tightproofconstraints}, we introduce an important constraint that relates the performance of \OPT\ to the value of each $\capn{\indexa}$. (iii) We combine these two lemmas with other natural constraints to formulate an LP that lower-bounds the value of \AC, as formally established in Lemma \ref{lem:tightproofLPboundsAC}. (iv) We conclude in Lemma \ref{lem:tightproofLPbound} by establishing that the value of this LP is at least $\OPT\bigg(\big(1-1/e\big)\mathbbm{1}_{\extfrac \leq 1/e} + \big(1+\extfrac \log(\extfrac)\big) \mathbbm{1}_{\extfrac > 1/e} - 2/\invbidtobudget\bigg)$.

\begin{lem}
\label{lem:tightproofACbound}
The value of \AC\ is at least
\begin{equation}
    \sum_{\indexa = 1}^\numbins \Big(\frac{\indexa}{\numbins} \capn{\indexa} + \frac{\numbins - \indexa}{\numbins} \ext{\indexa}\Big) - \frac{1}{\numbins} \totalcap
\end{equation}
\end{lem}
The proof of Lemma \ref{lem:tightproofACbound} is algebraic and follows from the fact that $\AC_{\oppindex,T}^{\inttrafmath}/(\capa_\oppindex -\AC_{\oppindex,T}^{\exttrafmath}) \geq (\indexa-1)/\numbins$ for any opportunity of type $\indexa$, based on \AC's fill rate definition. We formally prove Lemma \ref{lem:tightproofACbound} in Appendix \ref{proof:lem:tightproofACbound}. 
\begin{figure}[t]
\centering
\includegraphics[trim={5cm 16.8cm 5cm 3.9cm},clip,width=.9\textwidth]{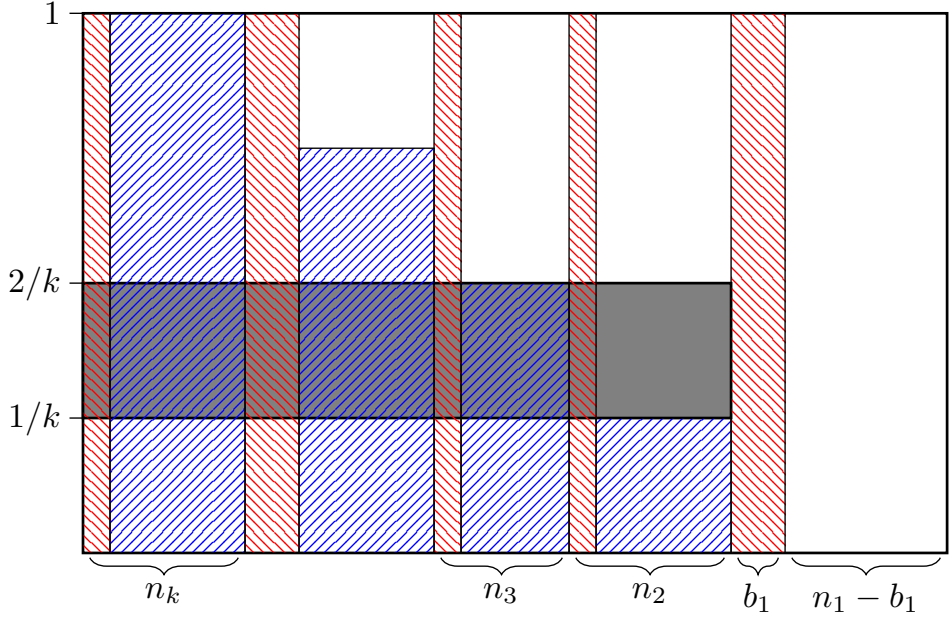}
    \caption{\revcolor{Visualization of the outcome of \AC\ where opportunities are grouped by type. The area of red down-sloping rectangles represent the capacity that can be filled by external traffic, and the area of the blue up-sloping rectangles is a lower bound on the capacity filled by internal traffic under \AC. The shaded region corresponds to an upper bound on the total capacity in slab 2 filled under \AC.}
    }
    \label{fig:tightproof}
\end{figure}

We can also establish this result via a proof-by-picture, using Figure \ref{fig:tightproof} as our visual aid. 
The overall length of Figure \ref{fig:tightproof} is equal to the total capacity $\totalcap$, and we organize the opportunities from right to left in increasing order of their type.  Each type $\indexa \in \{1, \dots, \numbins\}$ corresponds to a segment of length $\capn{\indexa}$ (i.e., the total capacity of opportunities of type $\indexa$). Within each segment, there are two rectangles, one filled with red down-sloping diagonals and another filled with blue up-sloping diagonals. The former rectangles have width $\ext{\indexa}$ and height $1$; their area represents the total amount of capacity of type $\indexa$ opportunities that can filled by external traffic. The latter rectangles have width $\capn{\indexa}-\ext{\indexa}$ and height $(\indexa-1)/\numbins$; their area represents a lower bound on the total amount of capacity of type $\indexa$ opportunities that \AC\ fills with internal traffic. Due to the fact that \AC\ uses all external traffic for an opportunity unless the opportunity is full, the total area of the red and blue rectangles represents a lower bound on the value of \AC. This total area is equivalent to the bound in Lemma \ref{lem:tightproofACbound}. We next relate this area to the value of $\OPT$.

\begin{lem}
\label{lem:tightproofconstraints}
Let $\intern{\indexb}$ denote the total capacity of type $\indexb$ opportunities that \OPT\ fills with internal traffic. Then for all $\indexa \in \{1, \dots, \numbins-1\}$,
\begin{equation}
    \sum_{\indexb=1}^{\indexa}\intern{\indexb} \quad \leq \quad \frac{\indexa}{\numbins} \totalcap - \sum_{\indexb=1}^\indexa \frac{\indexa - \indexb}{\numbins} \capn{\indexb} + \sum_{\indexb = \indexa+1}^\numbins \frac{1}{\invbidtobudget}\capn{\indexb} \label{eq:lemmatightproofconstraints}
\end{equation}
\end{lem}

We formally prove Lemma \ref{lem:tightproofconstraints} in Appendix \ref{proof:lem:tightproofconstraints}, but the proof idea can likewise be visualized in Figure \ref{fig:tightproof}. To see this, suppose we discretize each opportunity's capacity into $\numbins$ equally sized \emph{slabs}, numbered from $1$ to $\numbins$, where slab $\indexb$ consists of the $(\indexb-1)/\numbins$ to $\indexb/\numbins$ percentile of each opportunity's capacity. Based on \AC's optimality condition, in a setting with unweighted deterministic rewards, every internal traffic arrival $t$ assigned by \OPT\ to opportunities of type (at most) $\indexa$ must fill capacity in slab $\indexb \leq \indexa$ under \AC. (Such an assignment can span multiple slabs, which we rigorously account for in the formal proof of Lemma \ref{eq:lemmatightproofconstraints}.)\footnote{\revcolor{By including capacity that is eventually filled by \exttraf\ in each slab, we are accounting for the fact that $\AC$ may need to assign internal traffic \emph{before} observing the quantity and destination of external traffic.}}  We emphasize that this crucial observation no longer holds when rewards are weighted: in that case, an assignment by \OPT\ to an opportunity of type (at most) $\indexa$ may fill (less) capacity in slab $\indexb > \indexa$ under \AC. This highlights the difficulty of extending such a result to settings with weighted rewards.

Note that under \AC, opportunities with type at most $\indexa$ have a fill rate of at most $\indexa/\numbins$. Hence, for these opportunities, we know that capacity in slab $\indexb > \indexa$ is not filled by internal traffic (e.g., capacity in slab 2 is not filled by internal traffic under \AC\ for opportunities of type 1). In Figure \ref{fig:tightproof}, we use a shaded rectangle to represent an upper bound on the total capacity in slab 2 filled under \AC.  The total capacity of slabs $\indexb \leq \indexa$ filled by \AC\ is upper-bounded by the first two terms in Equation \eqref{eq:lemmatightproofconstraints}. The final term in the upper bound incorporates the minimum capacity to account for the fact that a sign-up can span multiple slabs when capacities are small. 

Based on Lemmas \ref{lem:tightproofACbound} and \ref{lem:tightproofconstraints} as well as other natural constraints, we introduce an instance-specific LP whose solution serves as a lower bound on the value of \AC, as established in Lemma \ref{lem:tightproofLPboundsAC}.

{\footnotesize
\begin{equation*}
\arraycolsep=1.4pt\def\arraystretch{1}
\begin{array}{|llll|}
\hline &&&\\
\min_{\mathbf{\intern,\ext,\capn}}\quad\quad&\displaystyle 
\sum_{\indexa = 1}^\numbins \Big(\frac{\indexa}{\numbins} \capn{\indexa} + \frac{\numbins - \indexa}{\numbins} \ext{\indexa}\Big) - \frac{1}{\numbins} \totalcap
&\qquad \qquad \qquad \qquad &(\textsc{LP}) \\[1.4em]
\text{s.t.} \qquad & \displaystyle  \sum_{\indexb=1}^{\indexa}\intern{\indexb} \quad \leq \quad \frac{\indexa}{\numbins} \totalcap - \sum_{\indexb=1}^\indexa \frac{\indexa - \indexb}{\numbins} \capn{\indexb} + \sum_{\indexb = \indexa+1}^\numbins \frac{1}{\invbidtobudget}\capn{\indexb}  \qquad\qquad & \indexa \leq \numbins -1 &(\alpha_\indexa)  \\[1.8em]
& \displaystyle  \ext{\indexa} + \intern{\indexa} \quad \leq \quad \capn{\indexa}  \qquad\qquad & \indexa \leq \numbins &(\gamma_\indexa)  \\[1em]
& \displaystyle  \extfrac \totalcap \quad \leq \quad \sum_{\indexa = 1}^\numbins \capn{\indexa}   &  &(\textcolor{\colc}{\lambda})  \\[1.4em]
& \displaystyle  \OPT \quad \leq \quad \extfrac \totalcap + \sum_{\indexa = 1}^\numbins \intern{\indexa}   &  &(\textcolor{\colb}{\theta})  \\[1.4em]
& \displaystyle  \totalcap \quad \leq \quad \sum_{\indexa = 1}^\numbins \capn{\indexa}   &  &(\textcolor{\cola}{\mu})  \\[1.4em]
\hline
\end{array}
\end{equation*}
}

\begin{lem}
\label{lem:tightproofLPboundsAC}
For any problem instance, the value of LP is a lower bound on the value of \AC\ on that instance.
\end{lem}

We prove Lemma \ref{lem:tightproofLPboundsAC} in Appendix \ref{proof:lem:tightproofLPboundsAC}. Based on this lemma, to establish a bound on the competitive ratio of \AC\ it suffices to bound the value of LP, which we do in the following lemma.

\begin{lem}
\label{lem:tightproofLPbound}
The value of LP is at least $\OPT\Big(\big(1-1/e\big)\mathbbm{1}_{\extfrac \leq 1/e} + \big(1+\extfrac \log(\extfrac)\big) \mathbbm{1}_{\extfrac > 1/e} - 2/\invbidtobudget\Big)$.
\end{lem}

To prove Lemma \ref{lem:tightproofLPbound}, we find a feasible solution to the dual of LP which obtains a value of $\OPT\Big(\big(1-1/e\big)\mathbbm{1}_{\extfrac \leq 1/e} + \big(1+\extfrac \log(\extfrac)\big) \mathbbm{1}_{\extfrac > 1/e} - 2/\invbidtobudget\Big)$. We provide the full details in Appendix \ref{proof:lem:tightproofACbound}.
Because the value of LP provides a lower bound on the value of \AC\ for \emph{any} problem instance (by Lemma \ref{lem:tightproofLPboundsAC}), it immediately follows from Lemma \ref{lem:tightproofLPbound} that the competitive ratio of \AC\ is lower-bounded by $\big(1-1/e\big)\mathbbm{1}_{\extfrac \leq 1/e} + \big(1+\extfrac \log(\extfrac)\big) \mathbbm{1}_{\extfrac > 1/e} - 2/\invbidtobudget$. This completes the proof of Theorem \ref{thm:tightproof}. \halmos

\subsubsection{Proof of Lemma \ref{lem:tightproofACbound}}
\label{proof:lem:tightproofACbound}
    Let $\typeset{\indexa}$ denote the set of opportunities of type $\indexa$. By definition, for all $\oppindex \in \typeset{\indexa}$ we have $\AC_{\oppindex,T}^{\inttrafmath}/(\capa_\oppindex -\AC_{\oppindex,T}^{\exttrafmath}) \geq (\indexa - 1)/\numbins$. This implies $\AC_{\oppindex,T}^{\inttrafmath} \geq (\capa_\oppindex -\AC_{\oppindex,T}^{\exttrafmath}) \cdot (\indexa - 1)/\numbins$.

    We now aggregate the total filled capacity for each opportunity under \AC:
    \begin{align}
        \AC &= \sum_{\indexa = 1}^{\numbins} \sum_{\oppindex \in \typeset{\indexa}} \AC_{\oppindex,T}^{\inttrafmath} + \AC_{\oppindex,T}^{\exttrafmath} \nonumber \\ 
        &\geq \sum_{\indexa = 1}^{\numbins} \sum_{\oppindex \in \typeset{\indexa}} \frac{\indexa-1}{\numbins}(\capa_\oppindex -\AC_{\oppindex,T}^{\exttrafmath}) + \AC_{\oppindex,T}^{\exttrafmath} \nonumber \\ 
        &= \sum_{\indexa = 1}^{\numbins} \sum_{\oppindex \in \typeset{\indexa}} \frac{\indexa-1}{\numbins}\capa_\oppindex + \frac{\numbins - \indexa +1}{\numbins}\AC_{\oppindex,T}^{\exttrafmath} \nonumber \\ 
        &\geq \sum_{\indexa = 1}^{\numbins} \sum_{\oppindex \in \typeset{\indexa}} \frac{\indexa-1}{\numbins}\capa_\oppindex + \frac{\numbins - \indexa}{\numbins}\AC_{\oppindex,T}^{\exttrafmath} \nonumber \\ 
        & = \sum_{\indexa = 1}^{\numbins} \frac{\indexa-1}{\numbins}\capn{\indexa} + \frac{\numbins - \indexa}{\numbins}\ext{\indexa} \nonumber
    \end{align}
    To see why the final step holds with equality, note that the first terms are identical by definition of $\capn{\indexa}$. For the second term, if $\oppindex$ is not of type $\numbins$ (i.e., if the opportunity has remaining capacity after a run of \AC), then \AC\ must use all external traffic for $\oppindex$. In that case, $\sum_{\oppindex \in \typeset{\indexa}} \AC_{\oppindex,T}^{\exttrafmath} = \ext{\indexa}$ by the definition of $\ext{\indexa}$. If $\oppindex$ is of type $\numbins$, then \AC\ may not use all external traffic for $\oppindex$; however, the coefficient is $0$.

\subsubsection{Proof of Lemma \ref{lem:tightproofconstraints}}
\label{proof:lem:tightproofconstraints}
Suppose \OPT\ uses internal traffic arrival $t$ to fill one unit of capacity of $\oppindexb \in \typeset{\indexa}$. We want to understand what \AC\ did with that arrival. If $\indexa < \numbins$, then there must have been at least one compatible opportunity for $t$ under \AC\ that had capacity left, and hence it must have been matched. Suppose it was matched to $\oppindex$. The definition of \AC\ ensures that $\oppindex \in \text{argmin}_{\oppindex'} \AC_{\oppindex',t-1}^{\inttrafmath}/(\capa_{\oppindex'} -\AC_{\oppindex',t-1}^{\exttrafmath})$. Based on this optimality condition and the fact that this fill rate is non-decreasing in $t$ and $\oppindexb \in \typeset{\indexa}$,
\begin{equation}
    \frac{\AC_{\oppindex,t-1}^{\inttrafmath}}{\capa_\oppindex -\AC_{\oppindex,t-1}^{\exttrafmath}} \quad \leq \quad \frac{\AC_{\oppindexb,t-1}^{\inttrafmath}}{\capa_\oppindexb -\AC_{\oppindexb,t-1}^{\exttrafmath}} \quad \leq \quad \frac{\AC_{\oppindexb,T}^{\inttrafmath}}{\capa_\oppindexb-\AC_{\oppindexb,T}^{\exttrafmath}} \quad \leq \quad \frac{\indexa}{\numbins} \label{eq:tightproof:assign1}
\end{equation}

In other words, every time \OPT\ assigns an internal arrival to an opportunity of type $\indexa$ (where $\indexa \leq \numbins$), \AC\ must assign that arrival to an opportunity with a bounded amount of internal traffic. How many of these assignments can \AC\ make? 

Suppose the opportunity $\oppindex$ that this \inttraf\ is matched with under \AC\ is of type $j$, i.s., $\oppindex \in \typeset{\indexb}$. Then, 
\begin{equation}
    \frac{\AC_{\oppindex,t-1}^{\inttrafmath}}{\capa_\oppindex -\AC_{\oppindex,t-1}^{\exttrafmath}}  \quad \leq \quad \frac{\AC_{\oppindex,T}^{\inttrafmath}}{\capa_\oppindex -\AC_{\oppindex,T}^{\exttrafmath}} \quad \leq \quad \frac{\indexb}{\numbins} \label{eq:tightproof:assign2}
\end{equation} If $\indexb \leq \indexa$, then $\oppindex$ receives at most $(\indexb/\numbins)\capa_\oppindex$ sign-ups from internal traffic under \AC\ (based on equation \eqref{eq:tightproof:assign2}). Otherwise, \AC\ may only assign this arrival to $\oppindex$ if  $\AC_{\oppindex,t-1}^{\inttrafmath} \leq \frac{\indexa}{\numbins} \capa_\oppindex$ (based on Equation \eqref{eq:tightproof:assign1}), so there can be at most $\frac{\indexa}{\numbins} \capa_\oppindex+1$ such assignments.

Putting these facts together, we can bound the total amount of internal traffic that \OPT\ assigns to opportunities of type $1$:
$$\sum_{\oppindexb \in \typeset{1}} \OPT_{\oppindexb,T}^{\inttrafmath} \quad \leq \quad \sum_{\oppindex \in \typeset{1}} \frac{1}{\numbins}\capa_{\oppindex} + \sum_{\indexb = 2}^{\numbins} \sum_{\oppindex \in \typeset{\indexb}} \left( \frac{1}{\numbins} \capa_{\oppindex} + 1 \right)$$
By the same logic, we can bound the total amount of internal traffic that \OPT\ assigns to opportunities with type at most $\indexa$:
$$\sum_{\indexb =1}^{\indexa} \sum_{\oppindexb \in \typeset{\indexb}} \OPT_{\oppindexb,T}^{\inttrafmath} \quad \leq \quad \sum_{\indexb =1}^{\indexa}\sum_{\oppindex \in \typeset{\indexb}} \frac{\indexb}{\numbins}\capa_{\oppindex} + \sum_{\indexb = \indexa+1}^{\numbins} \sum_{\oppindex \in \typeset{\indexb}} \left( \frac{\indexa}{\numbins}\capa_{\oppindex} + 1 \right)$$
We now re-write this bound, recalling that $\intern{\indexb}$ denotes the total capacity of type $\indexb$ opportunities that \OPT\ fills with internal traffic, and that $\capn{\indexb}$ denotes the total capacity of type $\indexb$ opportunities.
\begin{align}
        \sum_{\indexb=1}^{\indexa}\intern{\indexb} \quad &\leq \quad \sum_{\indexb=1}^\indexa \frac{\indexb}{\numbins} \capn{\indexb} + \sum_{\indexb = \indexa+1}^{\numbins} \frac{\indexa}{\numbins}\capn{\indexb} + \sum_{\indexb = \indexa+1}^\numbins \sum_{\oppindex \in \typeset{\indexb}} \frac{\capa_{\oppindex}}{\capa_{\oppindex}} \nonumber \\
         &\leq \quad \sum_{\indexb=1}^\indexa \Big(\frac{\indexa}{\numbins} - \frac{\indexa - \indexb}{\numbins}\Big)\capn{\indexb} + \sum_{\indexb = \indexa+1}^{\numbins} \frac{\indexa}{\numbins}\capn{\indexb} + \sum_{\indexb = \indexa+1}^\numbins \sum_{\oppindex \in \typeset{\indexb}} \frac{\capa_{\oppindex}}{\invbidtobudget} \nonumber \\
        &\leq \quad \frac{\indexa}{\numbins} \totalcap - \sum_{\indexb=1}^\indexa \frac{\indexa - \indexb}{\numbins} \capn{\indexb} + \sum_{\indexb = \indexa+1}^\numbins \frac{1}{\invbidtobudget}\capn{\indexb} \nonumber
\end{align}
The first inequality follows by the definition of $\capn{\indexb}$, i.e., $\capn{\indexb} = \sum_{\oppindex \in \typeset{\indexb}} \capa_{\oppindex}$, and the last inequality follows by the definition of $\totalcap$, i.e., $\totalcap = \sum_{\indexb=1}^\numbins \capn{\indexb}$.
This completes the proof of Lemma \ref{lem:tightproofconstraints}. \halmos

\subsubsection{Proof of Lemma \ref{lem:tightproofLPboundsAC}}
\label{proof:lem:tightproofLPboundsAC}
We reproduce LP here for ease of reference. By Lemma \ref{lem:tightproofconstraints}, the first set of constraints must hold. The second set of constraints holds because \OPT\ cannot fill an opportunity beyond its capacity. The third constraint holds by definition of the \fracextname\ $\extfrac$ (see Definition \ref{def:beta}). The fourth constraint holds because \OPT\ cannot use more external traffic than what exists. The final constraint holds because the sum of capacities across all types is equal to the total capacity. \halmos

{\footnotesize
\begin{equation*}
\arraycolsep=1.4pt\def\arraystretch{1}
\begin{array}{|llll|}
\hline &&&\\
\min_{\mathbf{\intern,\ext,\capn}}\quad\quad&\displaystyle 
\sum_{\indexa = 1}^\numbins \Big(\frac{\indexa}{\numbins} \capn{\indexa} + \frac{\numbins - \indexa}{\numbins} \ext{\indexa}\Big) - \frac{1}{\numbins} \totalcap
&\qquad \qquad \qquad \qquad &(\textsc{LP}) \\[1.4em]
\text{s.t.} \qquad & \displaystyle  \sum_{\indexb=1}^{\indexa}\intern{\indexb} \quad \leq \quad \frac{\indexa}{\numbins} \totalcap - \sum_{\indexb=1}^\indexa \frac{\indexa - \indexb}{\numbins} \capn{\indexb} + \sum_{\indexb = \indexa+1}^\numbins \frac{1}{\invbidtobudget}\capn{\indexb}  \qquad\qquad & \indexa \leq \numbins -1 &(\alpha_\indexa)  \\[1.8em]
& \displaystyle  \ext{\indexa} + \intern{\indexa} \quad \leq \quad \capn{\indexa}  \qquad\qquad & \indexa \leq \numbins &(\gamma_\indexa)  \\[1em]
& \displaystyle  \extfrac \totalcap \quad \leq \quad \sum_{\indexa = 1}^\numbins \capn{\indexa}   &  &(\textcolor{\colc}{\lambda})  \\[1.4em]
& \displaystyle  \OPT \quad \leq \quad \extfrac \totalcap + \sum_{\indexa = 1}^\numbins \intern{\indexa}   &  &(\textcolor{\colb}{\theta})  \\[1.4em]
& \displaystyle  \totalcap \quad \leq \quad \sum_{\indexa = 1}^\numbins \capn{\indexa}   &  &(\textcolor{\cola}{\mu})  \\[1.4em]
\hline
\end{array}
\end{equation*}
}

\subsubsection{Proof of Lemma \ref{lem:tightproofLPbound}}
\label{proof:lem:tightproofLPbound}
To prove Lemma \ref{lem:tightproofLPbound}, it is sufficient to find a feasible solution to the dual of LP (reproduced above) which obtains a value of at least $\OPT\Big(\big(1-1/e\big)\mathbbm{1}_{\extfrac \leq 1/e} + \big(1+\extfrac \log(\extfrac)\big) \mathbbm{1}_{\extfrac > 1/e} - 2/\invbidtobudget\Big)$. For now, we will assume that $\invbidtobudget \geq 3$. At the end of the proof, we will briefly revisit the case where $\invbidtobudget < 3$ and show that the lemma holds. We present the dual of LP below.

\noindent \paragraph{\bf The dual of LP.}
\begin{equation*}
\arraycolsep=1.4pt\def\arraystretch{1}
\begin{array}{llll}
\max_{\mathbf{\alpha,\gamma},\textcolor{\colc}{\lambda}, \textcolor{\colb}{\theta}, \textcolor{\cola}{\mu}}\quad\quad&\displaystyle 
\textcolor{\colb}{\theta} \OPT + (\textcolor{\colc}{\lambda}-\textcolor{\colb}{\theta})\extfrac\totalcap + \textcolor{\cola}{\mu} \totalcap - \sum_{\indexb = 1}^{\numbins-1}\Big(\frac{\indexb}{\numbins}\alpha_{\indexb}\totalcap\Big) - \frac{1}{\numbins} \totalcap 
&\qquad \qquad \qquad \qquad &(\textsc{DUAL}) \\[1.4em]
\text{s.t.} \qquad & \displaystyle  0 \quad \leq \quad \gamma_\indexa - \textcolor{\colc}{\lambda} + \frac{\numbins - \indexa}{\numbins} & \indexa \leq \numbins &(\ext{\indexa})  \\[1.4em]
& \displaystyle  0 \quad \leq \quad \frac{\indexa}{\numbins} - \gamma_\indexa - \textcolor{\cola}{\mu} +\sum_{\indexb=\indexa}^{\numbins-1} \Big(\frac{\indexb - \indexa}{\numbins}\Big)\alpha_{\indexb} - \sum_{\indexb=\numbins - \indexa + 1}^{\numbins-1} \Big(\frac{1}{\invbidtobudget}\Big)\alpha_{\indexb} \qquad \qquad & \indexa \leq \numbins &(\capn{\indexa})  \\[1.4em]
& \displaystyle 0 \quad \leq \quad \gamma_\indexa - \textcolor{\colb}{\theta} +\sum_{\indexb=\indexa}^{\numbins-1} \alpha_{\indexb}& \indexa \leq \numbins &(\intern{\indexa})  \\[1.4em]
\end{array}
\end{equation*}

\noindent \paragraph{\bf Candidate solution.}
Our candidate solution, presented below, will depend on the \fracextname\ $\extfrac$ via an index $\dualindex : = \min\{\numbins - 1, \lfloor -\numbins \log(\extfrac)\rfloor\}$. 
\begin{align}
\alpha_{\indexa} \quad &= \quad \begin{cases} \frac{\Big(1+\frac{1}{\numbins}\Big)^{\indexa}}{\numbins\Big(1+\frac{1}{\numbins}\Big)^{\dualindex+1}}, & \indexa \leq \dualindex \\
0, & \indexa > \dualindex
\end{cases} \\
\gamma_{\indexa} \quad &= \quad  \frac{\indexa}{\numbins}\Big(1 - \sum_{\indexb = \indexa}^{\numbins-1} \alpha_\indexb \Big) - \sum_{\indexb=1}^{\indexa-1} \Big(\frac{\indexb}{\numbins}\Big)\alpha_\indexb - \sum_{\indexb = \numbins - \indexa +1}^{\numbins-1}\Big(\frac{1}{\invbidtobudget}\Big) \alpha_{\indexb}\\
\textcolor{\colc}{\lambda} \quad &= \quad 1 - \sum_{\indexb=1}^{\numbins-1} \Big(\frac{\indexb}{\numbins} + \frac{1}{\invbidtobudget}\Big)\alpha_{\indexb} \\
\textcolor{\cola}{\mu} \quad &= \quad \sum_{\indexb=1}^{\numbins-1} \Big(\frac{\indexb}{\numbins}\Big)\alpha_{\indexb} \\
\textcolor{\colb}{\theta} \quad &= \quad \sum_{\indexb=1}^{\numbins-1} \Big(1-\frac{1}{\invbidtobudget}\Big)\alpha_{\indexb} 
\end{align}

We prove Lemma \ref{lem:tightproofLPbound} in three steps: (i) we show that this candidate solution satisfies the three dual constraints, (ii) we show that the dual variables are non-negative, and (iii) we bound the objective value of this candidate solution.

In these three steps, we will leverage the following two properties of the $\alpha_{\indexa}$ terms. 
\begin{align}
    \sum_{\indexb=\indexa}^{\numbins-1} \alpha_{\indexb} \quad &= \quad  \begin{cases} 1 - \frac{\Big(1+\frac{1}{\numbins}\Big)^{\indexa}}{\Big(1+\frac{1}{\numbins}\Big)^{\dualindex+1}}, & \indexa \leq \dualindex \\
0, & \indexa > \dualindex
\end{cases} \label{eq:tightproofprop1}\\
 \sum_{\indexb=1}^{\indexa-1} \Big(\frac{\numbins + \indexb}{\numbins}\Big)\alpha_{\indexb} \quad &= \quad  \begin{cases} \Big(\frac{\indexa-1}{\numbins}\Big)\frac{\Big(1+\frac{1}{\numbins}\Big)^{\indexa}}{\Big(1+\frac{1}{\numbins}\Big)^{\dualindex+1}}, & \indexa \leq \dualindex \\
\dualindex / \numbins, & \indexa > \dualindex
\end{cases} \label{eq:tightproofprop2}
\end{align}
To establish the first property, we use the fact that for any $m \in \mathbb{N}$, $\sum_{j=i}^m r^j = \frac{r^{m+1} - r^{\min\{i,m+1\}}}{r-1}$:
\begin{align}
    \sum_{\indexb=\indexa}^{\numbins-1} \alpha_{\indexb} \quad &= \quad \left(\frac{1}{\numbins\Big(1+\frac{1}{\numbins}\Big)^{\dualindex+1}}\right)\sum_{\indexb=\indexa}^{\dualindex}\Big(1+\frac{1}{\numbins}\Big)^{\indexb} \nonumber \\
    &= \quad \left(\frac{1}{\numbins\Big(1+\frac{1}{\numbins}\Big)^{\dualindex+1}}\right)\frac{\Big(1+\frac{1}{\numbins}\Big)^{\dualindex+1} - \Big(1+\frac{1}{\numbins}\Big)^{\min\{\indexa, \dualindex+1\}}}{\Big(1+\frac{1}{\numbins}\Big)-1} \nonumber \\ &= \quad \begin{cases} 1 - \frac{\Big(1+\frac{1}{\numbins}\Big)^{\indexa}}{\Big(1+\frac{1}{\numbins}\Big)^{\dualindex+1}}, & \indexa \leq \dualindex \\
0, & \indexa > \dualindex
\end{cases} \nonumber
\end{align}
The second property also uses the fact that for any $m \in \mathbb{N} \cup \{0\}$, $\sum_{j=1}^m j \cdot r^j = \frac{m \cdot r^{m+2}-(m+1)r^{m+1}+r}{(r-1)^2}$. For now, let us assume that $\indexa \leq \dualindex + 1$, thereby ensuring that each term in the summation is strictly positive.
\begin{align}
 \sum_{\indexb=1}^{\indexa-1} \Big(\frac{\numbins + \indexb}{\numbins}\Big)\alpha_{\indexb} \quad &= \quad  \left(\frac{1}{\numbins\Big(1+\frac{1}{\numbins}\Big)^{\dualindex+1}}\right)\sum_{\indexb=1}^{\indexa-1}\Big(1+\frac{\indexb}{\numbins}\Big)\Big(1+\frac{1}{\numbins}\Big)^{\indexb} \nonumber \\
 &= \quad \left(\frac{1}{\numbins\Big(1+\frac{1}{\numbins}\Big)^{\dualindex+1}}\right)\left(\frac{\Big(1+\frac{1}{\numbins}\Big)^{\indexa} - \Big(1+\frac{1}{\numbins}\Big)}{\Big(1+\frac{1}{\numbins}\Big)-1} \right. \nonumber \\
 & \qquad \qquad \left. + \frac{1}{\numbins}\Bigg(\frac{(\indexa-1) \big(1+\frac{1}{\numbins}\big)^{\indexa+1}-\indexa \cdot \big(1+\frac{1}{\numbins}\big)^{\indexa}+\big(1+\frac{1}{\numbins}\big)}{\Big(\big(1+\frac{1}{\numbins}\big)-1\Big)^2}\Bigg) \right) \nonumber \\
 &= \quad \left(\frac{1}{\numbins\Big(1+\frac{1}{\numbins}\Big)^{\dualindex+1}}\right)\left( \frac{(i-1)\Big(1+\frac{1}{\numbins}\Big)^{\indexa+1} - (i-1)\Big(1+\frac{1}{\numbins}\Big)^{\indexa}}{1/\numbins} \right) \nonumber \\
&= \quad  \Big(\frac{\indexa-1}{\numbins}\Big)\frac{\Big(1+\frac{1}{\numbins}\Big)^{\indexa}}{\Big(1+\frac{1}{\numbins}\Big)^{\dualindex+1}} \nonumber
\end{align}
This corresponds to the first case of Equation \eqref{eq:tightproofprop2}. If $\indexa > \dualindex + 1$, all additional terms in the summation are $0$; hence the value of the summation remains constant and equal to the case where $\indexa = \dualindex + 1$, which is captured by the second case of Equation \eqref{eq:tightproofprop2}

\vspace{10pt} 

\noindent {\bf Step (i): The candidate solution satisfies constraints $(\ext{\indexa})$, $(\capn{\indexa})$, and $(\intern{\indexa})$}

\noindent We begin with the first set of constraints, corresponding to $\ext{\indexa}$:
\begin{align}
\gamma_\indexa - \textcolor{\colc}{\lambda} + \frac{\numbins - \indexa}{\numbins} \quad &= \quad \frac{\indexa}{\numbins}\Big(1 - \sum_{\indexb = \indexa}^{\numbins-1} \alpha_\indexb \Big) - \sum_{\indexb=1}^{\indexa-1} \Big(\frac{\indexb}{\numbins}\Big)\alpha_\indexb - \sum_{\indexb = \numbins - \indexa +1}^{\numbins-1}\frac{1}{\invbidtobudget} \alpha_{\indexb} - \textcolor{\colc}{\Big(1 - \sum_{\indexb=1}^{\numbins-1} \Big(\frac{\indexb}{\numbins} + \frac{1}{\invbidtobudget}\Big)\alpha_{\indexa}\Big)} + \frac{\numbins - \indexa}{\numbins} \nonumber \\
& \geq \quad \textcolor{\colc}{\sum_{\indexb=1}^{\numbins-1}\Big(\frac{\indexb}{\numbins}\Big)\alpha_{\indexb}} - \sum_{\indexb = \indexa}^{\numbins-1} \Big(\frac{\indexa}{\numbins}\Big)\alpha_\indexb  - \sum_{\indexb=1}^{\indexa-1} \Big(\frac{\indexb}{\numbins}\Big)\alpha_\indexb \nonumber \\
&= \quad \sum_{\indexb=\indexa}^{\numbins-1} \Big(\frac{\indexb - \indexa}{\numbins}\Big)\alpha_\indexb \quad \geq \quad 0
\end{align}
For the second set of constraints, corresponding to $\capn{\indexa}$:
\begin{align}
    \frac{\indexa}{\numbins} - \gamma_\indexa - \textcolor{\cola}{\mu} +\sum_{\indexb=\indexa}^{\numbins-1} \Big(\frac{\indexb - \indexa}{\numbins}\Big)\alpha_{\indexb} & - \sum_{\indexb=\numbins - \indexa + 1}^{\numbins-1} \Big(\frac{1}{\invbidtobudget}\Big)\alpha_{\indexb}  \nonumber \\
     & = \quad \frac{\indexa}{\numbins} - \Big(\frac{\indexa}{\numbins}\Big(1 - \sum_{\indexb = \indexa}^{\numbins-1} \alpha_\indexb \Big) - \sum_{\indexb=1}^{\indexa-1} \Big(\frac{\indexb}{\numbins}\Big)\alpha_\indexb - \sum_{\indexb = \numbins - \indexa +1}^{\numbins-1}\Big(\frac{1}{\invbidtobudget}\Big) \alpha_{\indexb}\Big) \nonumber \\ & \qquad - \textcolor{\cola}{\sum_{\indexb=1}^{\numbins-1} \Big(\frac{\indexb}{\numbins}\Big)\alpha_{\indexb}} +\sum_{\indexb=\indexa}^{\numbins-1} \Big(\frac{\indexb - \indexa}{\numbins}\Big)\alpha_{\indexb} - \sum_{\indexb=\numbins - \indexa + 1}^{\numbins-1} \Big(\frac{1}{\invbidtobudget}\Big)\alpha_{\indexb} \nonumber \\
     & = \quad  \sum_{\indexb = \indexa}^{\numbins-1} \Big(\frac{\indexa}{\numbins}\Big)\alpha_\indexb +  \sum_{\indexb=1}^{\indexa-1} \Big(\frac{\indexb}{\numbins}\Big)\alpha_\indexb - \textcolor{\cola}{\sum_{\indexb=1}^{\numbins-1} \Big(\frac{\indexb}{\numbins}\Big)\alpha_{\indexb}} +\sum_{\indexb=\indexa}^{\numbins-1} \Big(\frac{\indexb - \indexa}{\numbins}\Big)\alpha_{\indexb} \nonumber \\
     & = \quad 0
\end{align}
For the third set of constraints, corresponding to $\intern{\indexa}$:
\begin{align}
\gamma_\indexa - \textcolor{\colb}{\theta} +\sum_{\indexb=\indexa}^{\numbins-1} \alpha_{\indexb} \quad &= \quad \frac{\indexa}{\numbins}\Big(1 - \sum_{\indexb = \indexa}^{\numbins-1} \alpha_\indexb \Big) - \sum_{\indexb=1}^{\indexa-1} \Big(\frac{\indexb}{\numbins}\Big)\alpha_\indexb - \sum_{\indexb = \numbins - \indexa +1}^{\numbins-1}\frac{1}{\invbidtobudget} \alpha_{\indexb} - \textcolor{\colb}{\sum_{\indexb=1}^{\numbins-1} \Big(1-\frac{1}{\invbidtobudget}\Big)\alpha_{\indexb} } +\sum_{\indexb=\indexa}^{\numbins-1} \alpha_{\indexb} \nonumber \\
&\geq \quad \frac{\indexa}{\numbins}\Big(1 - \sum_{\indexb = \indexa}^{\numbins-1} \alpha_\indexb \Big) - \sum_{\indexb=1}^{\indexa-1} \Big(\textcolor{\colb}{1} + \frac{\indexb}{\numbins}\Big)\alpha_\indexb \nonumber \\
& = \quad \begin{cases} 
\Big(\frac{1}{\numbins}\Big)\frac{\Big(1+\frac{1}{\numbins}\Big)^{\indexa}}{\Big(1+\frac{1}{\numbins}\Big)^{\dualindex+1}}, & \indexa \leq \dualindex \\
\frac{\indexa - \dualindex}{\numbins}, & \indexa > \dualindex
\end{cases}
\end{align}
The final step follows by applying the two properties of $\alpha$ given in Equations \eqref{eq:tightproofprop1} and \eqref{eq:tightproofprop2}. In either case, the value is positive, hence the constraint holds.

\medskip

\noindent {\bf Step (ii): Verifying the non-negativity of the Dual Variables}

\noindent By construction, each $\alpha_\indexa$ is weakly positive. It immediately follows that $\textcolor{\cola}{\mu}$ 
and $\textcolor{\colb}{\theta}$ are also weakly positive. For $\textcolor{\colc}{\lambda}$, we have 
$$\textcolor{\colc}{\lambda} \quad = \quad 1 - \sum_{\indexb=1}^{\numbins-1} \Big(\frac{\indexb}{\numbins} + \frac{1}{\invbidtobudget}\Big)\alpha_{\indexb} \quad \geq \quad 1 - \sum_{\indexb=1}^{\numbins-1} \Big(\frac{\indexb}{\numbins} + 1\Big)\alpha_{\indexb} \quad \geq \quad 1 - \frac{\numbins-1}{\numbins} \quad \geq \quad 0$$
Note that we bound the summation by applying Equation \eqref{eq:tightproofprop2}. All that remains is to show that $\gamma_\indexa$ is non-negative for all $\indexa \in \{1, \dots, \numbins\}$. For $\invbidtobudget =1$ or $\invbidtobudget = 2$, the $\gamma$ variables may in fact be negative. Therefore, we prove non-negativity for $\invbidtobudget \geq 3$, and we revisit the case of $\invbidtobudget < 3$ at the end of the proof. We partition these variables into two groups based on their indices, and we will consider the two groups separately. We will once again leverage the properties of the $\alpha$ variables established in Equations \eqref{eq:tightproofprop1} and \eqref{eq:tightproofprop2}.

\smallskip

\noindent {\it Case (i): $\indexa > \dualindex$}
\begin{align}
\gamma_{\indexa} \quad &= \quad  \frac{\indexa}{\numbins}\Big(1 - \sum_{\indexb = \indexa}^{\numbins-1} \alpha_\indexb \Big) - \sum_{\indexb=1}^{\indexa-1} \Big(\frac{\indexb}{\numbins}\Big)\alpha_\indexb - \sum_{\indexb = \numbins - \indexa +1}^{\numbins-1}\Big(\frac{1}{\invbidtobudget}\Big) \alpha_{\indexb} \nonumber \\
&\geq \quad \frac{\indexa}{\numbins}\Big(1 - \sum_{\indexb = \indexa}^{\numbins-1} \alpha_\indexb \Big) - \sum_{\indexb=1}^{\indexa-1} \Big(1+\frac{\indexb}{\numbins}\Big)\alpha_\indexb + \sum_{\indexb=1}^{\indexa-1} \alpha_{\indexb} - \sum_{\indexb = \numbins - \indexa +1}^{\numbins-1} \alpha_{\indexb} \nonumber \\
&=\quad \frac{\indexa}{\numbins}\Big(1 -0\Big) - \frac{\dualindex}{\numbins} + \sum_{\indexb=1}^{\numbins-1} \alpha_{\indexb} - \sum_{\indexb = \numbins - \indexa +1}^{\numbins-1} \alpha_{\indexb} \nonumber \\
&=\quad \frac{\indexa - \dualindex}{\numbins} + \sum_{\indexb=1}^{ \numbins - \indexa } \alpha_{\indexb} \nonumber \\
& \geq \quad 0 \nonumber
\end{align}

\smallskip

\noindent {\it Case (ii): $\indexa \leq \dualindex$}
\begin{align}
\gamma_{\indexa} \quad &= \quad  \frac{\indexa}{\numbins}\Big(1 - \sum_{\indexb = \indexa}^{\numbins-1} \alpha_\indexb \Big) - \sum_{\indexb=1}^{\indexa-1} \Big(\frac{\indexb}{\numbins}\Big)\alpha_\indexb - \sum_{\indexb = \numbins - \indexa +1}^{\numbins-1}\Big(\frac{1}{\invbidtobudget}\Big) \alpha_{\indexb} \nonumber \\
&= \quad \frac{\indexa}{\numbins}\Big( 1 - \sum_{\indexb = \indexa}^{\numbins-1} \alpha_\indexb \Big) - \sum_{\indexb=1}^{\indexa-1} \Big(1+\frac{\indexb}{\numbins}\Big)\alpha_\indexb + \sum_{\indexb=1}^{\numbins-1} \alpha_{\indexb} - \sum_{\indexb=\indexa}^{\numbins-1} \alpha_{\indexb} - \sum_{\indexb = \numbins - \indexa +1}^{\numbins-1} \Big(\frac{1}{\invbidtobudget}\Big)\alpha_{\indexb} \nonumber \\
&= \quad \Big(\frac{\indexa}{\numbins}\Big) \frac{\Big(1+\frac{1}{\numbins}\Big)^{\indexa}}{\Big(1+\frac{1}{\numbins}\Big)^{\dualindex+1}} - \Big(\frac{\indexa-1}{\numbins}\Big)\frac{\Big(1+\frac{1}{\numbins}\Big)^{\indexa}}{\Big(1+\frac{1}{\numbins}\Big)^{\dualindex+1}} + \sum_{\indexb=1}^{\numbins-1} \alpha_{\indexb} - \sum_{\indexb=\indexa}^{\numbins-1} \alpha_{\indexb} - \sum_{\indexb = \numbins - \indexa +1}^{\numbins-1} \Big(\frac{1}{\invbidtobudget}\Big)\alpha_{\indexb} \nonumber \\
& \geq \quad \sum_{\indexb=1}^{\numbins-1} \alpha_{\indexb} - \sum_{\indexb=\indexa}^{\numbins-1} \alpha_{\indexb} - \sum_{\indexb = \numbins - \indexa +1}^{\numbins-1} \Big(\frac{1}{\invbidtobudget}\Big)\alpha_{\indexb} \nonumber \\
& = \quad \left(1 - \frac{\Big(1+\frac{1}{\numbins}\Big)}{\Big(1+\frac{1}{\numbins}\Big)^{\dualindex+1}}\right) - \left(1 - \frac{\Big(1+\frac{1}{\numbins}\Big)^{\indexa}}{\Big(1+\frac{1}{\numbins}\Big)^{\dualindex+1}}\right) - \frac{1}{\invbidtobudget}\left(1 - \frac{\Big(1+\frac{1}{\numbins}\Big)^{\min\{\dualindex+1,\numbins-\indexa+1\}}}{\Big(1+\frac{1}{\numbins}\Big)^{\dualindex+1}}\right) \nonumber \\
& = \quad \Big(1+\frac{1}{\numbins}\Big)^{-\dualindex}\Bigg(-1 + \Big(1+\frac{1}{\numbins}\Big)^{\indexa-1}  - \frac{1}{\invbidtobudget}\Big(1+\frac{1}{\numbins}\Big)^{\dualindex} + \frac{1}{\invbidtobudget}\Big(1+\frac{1}{\numbins}\Big)^{\min\{\dualindex,\numbins-\indexa\}} \Bigg) \nonumber
\end{align}
It is straightforward to verify that this expression is non-negative when $\dualindex$ is the minimum, i.e., when $\dualindex \leq \numbins - \indexa$. Therefore, it is sufficient to show that this expression is also non-negative when $\dualindex > \numbins - \indexa$. In the following, we establish that this holds as long as $\invbidtobudget \geq 3$.
\begin{align}
    \Big(1+\frac{1}{\numbins}\Big)^{\indexa-1} - 1 - \frac{1}{\invbidtobudget}\Big(1+\frac{1}{\numbins}\Big)^{\dualindex} + \frac{1}{\invbidtobudget}\Big(1+\frac{1}{\numbins}\Big)^{\numbins-\indexa} \ & \geq \quad \Big(1+\frac{1}{\numbins}\Big)^{\indexa-1} -1 - \frac{1}{\invbidtobudget}\Big(1+\frac{1}{\numbins}\Big)^{\numbins - 1} + \frac{1}{\invbidtobudget}\Big(1+\frac{1}{\numbins}\Big)^{\numbins-\indexa} \nonumber \\
    &=  \Bigg(\Big(1+\frac{1}{\numbins}\Big)^{\indexa-1}-1\Bigg)  - \frac{1}{\invbidtobudget}\Big(1+\frac{1}{\numbins}\Big)^{\numbins - \indexa}\Bigg(\Big(1+\frac{1}{\numbins}\Big)^{\indexa-1}-1\Bigg) \nonumber \\
    &= \Bigg(1-\frac{1}{\invbidtobudget}\Big(1+\frac{1}{\numbins}\Big)^{\numbins - \indexa}\Bigg)\Bigg(\Big(1+\frac{1}{\numbins}\Big)^{\indexa-1}-1\Bigg) \nonumber \\
    &\geq \Bigg(1-\frac{e}{\invbidtobudget}\Bigg)\Bigg(\Big(1+\frac{1}{\numbins}\Big)^{\indexa-1}-1\Bigg) \nonumber \\
    &\geq 0 \nonumber
\end{align}

\medskip

\noindent {\bf Step (iii): Bounding the Objective Value}

We now plug this candidate solution into the objective of the dual:
\begin{align}
  &\textcolor{\colb}{\theta} \OPT + (\textcolor{\colc}{\lambda}-\textcolor{\colb}{\theta})\extfrac\totalcap + \textcolor{\cola}{\mu} \totalcap - \sum_{\indexa = 1}^{\numbins-1}\Big(\frac{\indexb}{\numbins}\alpha_{\indexb}\totalcap\Big) - \frac{1}{\numbins} \totalcap \nonumber \\ 
  & \qquad = \quad \Bigg(\textcolor{\colb}{\sum_{\indexb=1}^{\numbins-1} \Big(1-\frac{1}{\invbidtobudget}\Big)\alpha_{\indexb}} \Bigg) \OPT + \Bigg(\textcolor{\colc}{1 - \sum_{\indexb=1}^{\numbins-1} \Big(\frac{\indexb}{\numbins} + \frac{1}{\invbidtobudget}\Big)\alpha_{\indexb}}-\textcolor{\colb}{\sum_{\indexb=1}^{\numbins-1} \Big(1-\frac{1}{\invbidtobudget}\Big)\alpha_{\indexb}}\Bigg)\extfrac\totalcap  \nonumber \\ 
  & \qquad \qquad + \textcolor{\cola}{\sum_{\indexb=1}^{\numbins-1} \Big(\frac{\indexb}{\numbins}\Big)\alpha_{\indexb}} \totalcap - \sum_{\indexa = 1}^{\numbins-1}\Big(\frac{\indexb}{\numbins}\alpha_{\indexb}\totalcap\Big)  - \frac{1}{\numbins} \totalcap \nonumber \\
  & \qquad = \quad \textcolor{\colb}{\Bigg(\sum_{\indexb=1}^{\numbins-1} \Big(1-\frac{1}{\invbidtobudget}\Big)\alpha_{\indexb} \Bigg)} \OPT + \Bigg(\textcolor{\colc}{1} - \sum_{\indexb=1}^{\numbins-1} \Big(\textcolor{\colc}{\frac{\indexb}{\numbins}} + \textcolor{\colb}{1}\Big)\alpha_{\indexb}\Bigg)\extfrac\totalcap - \frac{1}{\numbins} \totalcap \nonumber \\
  & \qquad = \quad \textcolor{\colb}{\Big(1-\frac{1}{\invbidtobudget}\Big) \Bigg(1 - \frac{1}{\Big(1+\frac{1}{\numbins}\Big)^{\dualindex}}\Bigg) } \OPT + \Bigg(\textcolor{\colc}{1} - \frac{\dualindex}{\numbins}\Bigg)\extfrac\totalcap - \frac{1}{\numbins} \totalcap \nonumber
\end{align}
There are now two cases to consider, depending on the value of $\extfrac$. Recall our assumption that $\numbins \geq 4\invbidtobudget/\extfrac$, which implies that $\numbins \geq 4\invbidtobudget\totalcap/\OPT$ since $\OPT$ must fill at least a $\extfrac$ fraction of capacity (with external traffic). In addition, we will make use of the fact that $(1+1/x)^{yx-1} \geq e^y(1+1/x)^{-2}$ for all $x \geq 1$ and $y \in [0,1]$.

\smallskip

\noindent {\it Case (i): $\extfrac \leq e^{-1} \ \implies \ \dualindex = \numbins-1$}
\begin{align}
\Bigg(1-\frac{1}{\invbidtobudget}\Big) \Bigg(1 - \frac{1}{\Big(1+\frac{1}{\numbins}\Big)^{\numbins-1}}\Bigg) \OPT + \Bigg(1 - \frac{\numbins-1}{\numbins}\Bigg)\extfrac\totalcap - \frac{1}{\numbins} \totalcap \quad &\geq \quad \Bigg(1-\frac{1}{\invbidtobudget} - \Big(1+\frac{1}{\numbins}\Big)^{2}e^{-1}\Bigg) \OPT - \frac{1}{\numbins} \totalcap \nonumber \\
&\geq \quad \Bigg(1-e^{-1} - \frac{2}{\invbidtobudget}\Bigg) \OPT \nonumber
\end{align}
We note that as long as $\extfrac \leq e^{-1}$, this expression is equal to $\OPT\Big(\big(1-1/e\big)\mathbbm{1}_{\extfrac \leq 1/e} + \big(1+\extfrac \log(\extfrac)\big) \mathbbm{1}_{\extfrac > 1/e} - 2/\invbidtobudget\Big)$. Hence, we have proved Lemma \ref{lem:tightproofLPbound} for $\extfrac \leq e^{-1}$ and $\invbidtobudget \geq 3$.

\smallskip

\noindent {\it Case (ii): $\extfrac > e^{-1} \ \implies \ \dualindex  \in [- \numbins \log(\extfrac) - 1, -\numbins\log(\extfrac)] $}
\begin{align}
&\Big(1-\frac{1}{\invbidtobudget}\Big) \Bigg(1 - \frac{1}{\Big(1+\frac{1}{\numbins}\Big)^{\dualindex}}\Bigg) \OPT + \Bigg(1 - \frac{\dualindex}{\numbins}\Bigg)\extfrac\totalcap - \frac{1}{\numbins} \totalcap 
\nonumber \\
& \qquad \qquad \geq \quad \Big(1-\frac{1}{\invbidtobudget}\Big) \Bigg(1 - \frac{1}{\Big(1+\frac{1}{\numbins}\Big)^{- \numbins \log(\extfrac) - 1}}\Bigg) \OPT + \Bigg(1+ \log(\extfrac)\Bigg)\extfrac\totalcap - \frac{1}{\numbins} \totalcap \nonumber   \\
& \qquad \qquad \geq \quad \Bigg(1 - \frac{1}{\invbidtobudget} - \Big(1+\frac{1}{\numbins}\Big)^2e^{\log(\extfrac)} \Bigg) \OPT + \Bigg(1+ \log(\extfrac)\Bigg)\extfrac\OPT - \frac{1}{\numbins} \totalcap \nonumber   \\
& \qquad \qquad \geq \quad \Bigg(1 + \extfrac \log(\extfrac) - \frac{2}{\invbidtobudget} \Bigg) \OPT \nonumber
\end{align}
We note that when $\extfrac > e^{-1}$, this expression is equal to $\OPT\Big(\big(1-1/e\big)\mathbbm{1}_{\extfrac \leq 1/e} + \big(1+\extfrac \log(\extfrac)\big) \mathbbm{1}_{\extfrac > 1/e} - 2/\invbidtobudget\Big)$. This completes the proof of Lemma \ref{lem:tightproofLPbound} for $\invbidtobudget \geq 3$.

If $\invbidtobudget < 3$, then the bound in Lemma \ref{lem:tightproofLPbound} is weakly negative for all $\extfrac \in [0,1]$. Hence, it is sufficient to show that the value of LP is weakly positive. To see this, consider the objective of LP:
$$\sum_{\indexa = 1}^\numbins \Big(\frac{\indexa}{\numbins} \capn{\indexa} + \frac{\numbins - \indexa}{\numbins} \ext{\indexa}\Big) - \frac{1}{\numbins} \totalcap$$
The first term of the objective, $\sum_{\indexa = 1}^\numbins (\indexa/\numbins) \capn{\indexa}$, must be at least $(1/\numbins)\totalcap$ due to the fact that $\sum_{\indexa = 1}^\numbins \capn{\indexa} = \totalcap$ (i.e., the final constraint of LP). Furthermore, the second term of the objective must be weakly positive because $\numbins \geq \indexa$ and $\ext{\indexa} \geq 0$. Therefore, the value of LP must be weakly positive, which completes the proof of Lemma~\ref{lem:tightproofLPbound} for $\invbidtobudget < 3$. \halmos 

\subsection{Proof of Theorem \ref{thm:AClower}}
\label{apx:thm1proofdetails}
We first formalize the proof sketch provided in Section \ref{sec:proof}. We first establish that the first term in the maximum of $\compratiofunc(\extfrac, \invbidtobudget, \weightcap)$ (i.e., $\extfrac$) is a lower bound on the competitive ratio (Lemma \ref{lem:aclowerboundpart1}); then, we prove that the second term, $z^*$, is also a lower bound on the competitive ratio (Lemma \ref{lem:aclowerboundpart2}). After providing the full proof outline, we then present the omitted proofs of intermediate results in Appendices \ref{proof:lem:lowerboundalg} through \ref{proof:lem:MPlower}. We conclude in Appendix \ref{apx:exttrafstochastic} by describing how this proof can extend to settings where sign-ups from \exttraf\ are stochastic.

\begin{lem}[Lower Bound of $\extfrac$ on $\compratiofunc(\extfrac, \invbidtobudget, \weightcap)$]
\label{lem:aclowerboundpart1}
Let the smallest capacity be given by $\invbidtobudget$ and let the \MakeLowercase{\MCPRfull} be at most $\weightcap$. Then, for any \MakeLowercase{\fracextnamefull} $\extfrac$, the competitive ratio of the \adaptivecap\ algorithm defined in Algorithm \ref{alg:acpolicy} (with $\balancefunc$ as defined in \eqref{eq:potentialfunc}) is at least~$\extfrac$.
\end{lem}
\begin{proof}{Proof:}
The proof of Lemma \ref{lem:aclowerboundpart1} is immediate: we simply note that $\adaptivecapmath$ always recommends the targeted \opp\ to \exttraf. Applying the definition of the \fracextname\ (Definition \ref{def:beta}) then ensures that at least a $\extfrac$ fraction of capacity is filled in expectation. \halmos
\end{proof}

\begin{lem}[Lower Bound of $z^*$ on $\compratiofunc(\beta, \invbidtobudget, \weightcap)$]
\label{lem:aclowerboundpart2}
Let the smallest capacity be given by $\invbidtobudget$ and let the \MakeLowercase{\MCPRfull} be at most $\weightcap$. Then, for any \MakeLowercase{\fracextnamefull} $\extfrac$, the competitive ratio of the \adaptivecap\ algorithm defined in Algorithm \ref{alg:acpolicy} (with $\balancefunc$ as defined in \eqref{eq:potentialfunc}) is at least $z^*$ (with $z^*$ as defined in \eqref{eq:defzstar}).
\end{lem}

\begin{proof}{Proof of Lemma \ref{lem:aclowerboundpart2}:}
This proof follows the three steps described in Section \ref{sec:proof} and relies on the construction of an instance-specific mathematical program $\MP$.

We use the following two lemmas to go from the formulation of $\MP$ to a bound on the competitive ratio of \AC.

 \begin{lem}[Lower-Bound on Ratio of Expected Values via $\MP$] 
 \label{lem:MP}
 For any instance $\instance$, the ratio between the expected value of $\AC$ (i.e., $\mathbb{E}_{\samplepath}[\AC]$) and the expected value of $\OPT$ (i.e., $\mathbb{E}_{\samplepath}[\OPT]$) on instance $\instance$ is at least the optimal value of $\MP$. 
 \end{lem}


The proof of Lemma \ref{lem:MP} uses the values of $\AC$ and $\OPT$ along each sample path to construct a feasible solution for the instance-specific $\MP$. We defer the details to Appendix \ref{proof:lem:MPlower}.

In general, $\MP$ is non-convex. Despite this, we are able to bound the optimal value of $\MP$ for any instance with $z^*$, which we remind is only a function of the instance's \fracextname\ $\extfrac$, its minimum capacity $\invbidtobudget$, and its \MCPR\ $\weightcap$.

\begin{lem}[Lower Bound on the Optimal Value of $\MP$]
\label{lem:MPlower}
For any instance $\instance$, the optimal value of $\MP$ is at least $z^*$, where $z^*$ is defined in \eqref{eq:defzstar} in the statement of Theorem \ref{thm:AClower}.
\end{lem}

The proof of Lemma \ref{lem:MPlower} is mainly algebraic and relies on repeatedly relaxing the program's constraints and restricting its domain (without loss of optimality) until we can ultimately establish a lower bound of $z^*$. We defer the details to Appendix \ref{proof:lem:MPlower}. Together, Lemmas \ref{lem:MP} and \ref{lem:MPlower} prove Lemma \ref{lem:aclowerboundpart2}, namely, that $z^*$ is a lower bound on the competitive ratio of the \AC\ algorithm.
\end{proof}

In combination with Lemma \ref{lem:aclowerboundpart1}, we have shown that the competitive ratio of the $\adaptivecapmath$ algorithm is at least $\compratiofunc(\beta, \invbidtobudget, \weightcap)$, as defined in the statement of Theorem~\ref{thm:AClower}.~\halmos

\subsubsection{Proof of Lemma \ref{lem:lowerboundalg}}
\label{proof:lem:lowerboundalg}
The proof of Lemma \ref{lem:lowerboundalg} follows from the definition of the pseudo-rewards $\ballvar_t$ and $\binvar_i$ (which we replicate below for ease of reference) as well as the definition of the \adaptivecap\ algorithm. 
\begin{align*}
    \ballvar_t &= \begin{cases}
    \sum_{i \in [\numopps]} \balancefunc(\fillrate_{i,t-1})\mathbbm{1}[\volchoicetsuccess{\adaptivecapmath} = i], & t \in  \exttimes \cup \bonustimes \\
     \sum_{i \in [\numopps]} \balancefunc(\fillrate_{i,t-1})\mathbbm{1}[\volchoicet{\OPT} = i], & t \in \inttimes \setminus \bonustimes \\
    \end{cases}
    \\
    \binvar_i &=  \sum_{t \in [\horizon]} \left(1-\balancefunc(\fillrate_{i,t-1})\right)\mathbbm{1}[\volchoicetsuccess{\adaptivecapmath} = i]
\end{align*}
Recall that $\volchoicetsuccess{\AC}$ represents the \opp\ that \vol\ $t$ \emph{contributes to} under \AC. To be precise, if \opp\ $\volchoicet{\AC}$ has remaining capacity at time $t$, then $\volchoicetsuccess{\AC} = \volchoicet{\AC}$. Otherwise, $\volchoicetsuccess{\AC} = 0$. In addition, recall that $\bonustimes$ represents the set of arriving \inttraf\ for which \OPT\ recommends \opp~$0$. 

Based on these definitions, 
\begin{align}
    \mathbbm{E}_{\samplepath}[\adaptivecapmath] &= \mathbb{E}_{\samplepath} \left[ \sum_{t\in \inttimes \setminus \bonustimes}\sum_{i \in [\numopps]} \mathbbm{1}[\volchoicetsuccess{\adaptivecapmath} = i] + \sum_{t\in \exttimes \cup \bonustimes}\sum_{i \in [\numopps]}\mathbbm{1}[\volchoicetsuccess{\adaptivecapmath} = i]\right] \\
    &=\mathbb{E}_{\samplepath} \left[ \sum_{i \in [\numopps]} \left(\sum_{t\in \inttimes \setminus \bonustimes}\balancefunc(\fillrate_{i,t-1})\mathbbm{1}[\volchoicetsuccess{\adaptivecapmath} = i] +  \sum_{t\in \inttimes \setminus \bonustimes}(1-\balancefunc(\fillrate_{i,t-1}))\mathbbm{1}[\volchoicetsuccess{\adaptivecapmath} = i]\right.\right. \nonumber \\
    &\qquad \qquad \left.\left.+  \sum_{t\in \exttimes \cup \bonustimes}\balancefunc(\fillrate_{i,t-1})\mathbbm{1}[\volchoicetsuccess{\adaptivecapmath} = i] +   \sum_{t\in \exttimes \cup \bonustimes}(1-\balancefunc(\fillrate_{i,t-1}))\mathbbm{1}[\volchoicetsuccess{\adaptivecapmath} = i]\right) \right] \\
    & = \mathbb{E}_{\samplepath} \left[ \sum_{i \in [\numopps]} \sum_{t\in \inttimes \setminus \bonustimes}\balancefunc(\fillrate_{i,t-1})\mathbbm{1}[\volchoicetsuccess{\adaptivecapmath} = i]\right] +\mathbb{E}_{\samplepath} \left[\sum_{t \in \exttimes \cup \bonustimes} \ballvar_t + \sum_{i \in [\numopps]}\binvar_i\right] \\
    & = \mathbb{E}_{\samplepath} \left[ \sum_{i \in [\numopps]} \sum_{t\in \inttimes \setminus \bonustimes}\balancefunc(\fillrate_{i,t-1})\mathbbm{1}[\volchoicet{\adaptivecapmath} = i]\right] +\mathbb{E}_{\samplepath} \left[\sum_{t \in \exttimes \cup \bonustimes} \ballvar_t + \sum_{i \in [\numopps]}\binvar_i\right] \label{eq:successsignups} \\
    &\geq \mathbb{E}_{\samplepath} \left[ \sum_{i \in [\numopps]} \sum_{t\in \inttimes \setminus \bonustimes}\balancefunc(\fillrate_{i,t-1})\mathbbm{1}[\volchoicet{\OPT} = i]\right] +\mathbb{E}_{\samplepath} \left[\sum_{t \in \exttimes \cup \bonustimes} \ballvar_t + \sum_{i \in [\numopps]}\binvar_i\right] \label{eq:acoptcondition} \\
    &= \mathbbm{E}_{\samplepath}\left[\sum_{t \in [\horizon]} \ballvar_t + \sum_{i \in [\numopps]} \binvar_i\right]
\end{align}
All steps are algebraic except for \eqref{eq:successsignups} and \eqref{eq:acoptcondition}. To establish the former, we will show that $\sum_{i \in [\numopps]}\balancefunc(\fillrate_{i,t-1})\mathbbm{1}[\volchoicet{\adaptivecapmath} = i] = \sum_{i \in [\numopps]}\balancefunc(\fillrate_{i,t-1})\mathbbm{1}[\volchoicetsuccess{\adaptivecapmath} = i]$. We consider two cases. First, if $\fillrate_{\volchoicet{\adaptivecapmath},t-1} < 1$, then $\volchoicet{\adaptivecapmath} = \volchoicetsuccess{\adaptivecapmath}$ and the equality holds. Alternatively, if $\fillrate_{\volchoicet{\adaptivecapmath},t-1} = 1$, then $\volchoicetsuccess{\adaptivecapmath} = 0$ and $\balancefunc(\fillrate_{\volchoicet{\adaptivecapmath},t-1}) = 0$. Thus, both summations equal $0$, and the equality holds. 

Inequality \eqref{eq:acoptcondition} follows from the $\AC$ algorithm's optimality condition (see Algorithm \ref{alg:acpolicy}), which ensures that it recommends the \opp\ that maximizes the weighted probability of generating a \signup\ (where the weight for \opp\ $i$ at time $t$ is given by $\balancefunc(\fillrate_{i,t-1})$). Since the recommendation provided by \OPT\ to any \vol\ must be independent of their \signup\ realization, the inequality holds.  Applying the definition of the pseudo-rewards $\ballvar_t$ for $t \in \inttimes \setminus \bonustimes$ completes the proof of Lemma \ref{lem:lowerboundalg}.

\subsubsection{Proof of Lemma \ref{lem:lowerboundballvarbinvar}} 
\label{proof:lem:lowerboundballvarbinvar}
We will prove a stronger version of this lemma by establishing the following inequality along any fixed sample path $\samplepath$: 

\begin{align*}
    \sum_{t \in [\horizon]} \ballvar_t + \sum_{i \in [\numopps]} \binvar_i  \quad \geq \quad  e^{-1/\invbidtobudget}\sum_{i \in [\numopps]}& \left(\adaptivecapmath_{i,\horizon}^\exttrafmath + \acbonus_{i,\horizon} +  \OPT_{i,\horizon}^{\inttrafmath} \cdot \balancefunc \left(\frac{\adaptivecapmath_{i, \horizon}^{\inttrafmath}}{\capa_i - \adaptivecapmath_{i, \horizon}^{\exttrafmath}}\right) \right.  \nonumber \\ &\left. +  \capa_i \left(1-\balancefunc \left(\frac{\adaptivecapmath_{i, \horizon}^{\inttrafmath} - \acbonus_{i,\horizon}}{\capa_i}\right) - 1/e\right)\right),
\end{align*}

 We proceed by separately deriving lower bounds on the $\ballvar_t$ pseudo-rewards and the $\binvar_i$ pseudo-rewards. For the former, 
\begin{align}
    \sum_{t \in [\horizon]} \ballvar_t \quad &= \quad \sum_{t \in \exttimes \cup \bonustimes}\ballvar_t + \sum_{t \in \inttimes \setminus \bonustimes} \ballvar_t \\
    &= \quad \sum_{t \in \exttimes \cup \bonustimes}\ballvar_t + \sum_{t \in \inttimes \setminus \bonustimes} \sum_{i\in[\numopps]} \balancefunc(\fillrate_{i,t-1})\mathbbm{1}[\volchoicet{\OPT} = i] \label{eq:bonustimesextra} \\
    &\geq \quad \sum_{t \in \exttimes \cup \bonustimes}\ballvar_t + \sum_{t \in \inttimes \setminus \bonustimes} \sum_{i\in[\numopps]} \balancefunc(\fillrate_{i,\horizon})\mathbbm{1}[\volchoicet{\OPT} = i]  \label{eq:ballvarlowerbound} \\
     &= \quad \sum_{t \in \exttimes \cup \bonustimes}\ballvar_t +\sum_{i \in [\numopps]} \balancefunc \left(\frac{\adaptivecapmath_{i, \horizon}^{\inttrafmath}}{\capa_i - \adaptivecapmath_{i, \horizon}^{\exttrafmath}}\right) \OPT_{i,\horizon}^{\inttrafmath}  \label{eq:ballvarfinalbound}\end{align}
Equality in \eqref{eq:bonustimesextra} follows from the definition of $\ballvar_t$. Inequality in \eqref{eq:ballvarlowerbound} holds because $\balancefunc$ is a decreasing function in its argument, and $\fillrate_{i,\horizon} \geq \fillrate_{i,t-1}$ for all $t \in [\horizon]$. Equality in \eqref{eq:ballvarfinalbound} comes from applying the definition of the fill rate as well as the fact that $\OPT_{i,\horizon}^{\inttrafmath} = \sum_{t \in \inttimes \setminus \bonustimes} \mathbbm{1}[\volchoicet{\OPT} = i] $.

We next turn our attention to the $\binvar_i$ pseudo-rewards, which we further separate into two summations: 
\begin{align}
    \sum_{i \in [\numopps]}\binvar_i \quad =& \quad \sum_{i \in [\numopps]}\sum_{t \in \exttimes \cup \bonustimes} \left(1-\balancefunc(\fillrate_{i,t-1})\right)\mathbbm{1}[\volchoicetsuccess{\adaptivecapmath} = i] + \sum_{i \in [\numopps]}\sum_{t \in \inttimes \setminus \bonustimes} \left(1-\balancefunc(\fillrate_{i,t-1})\right)\mathbbm{1}[\volchoicetsuccess{\adaptivecapmath} = i]  
    \end{align}
We note that the first summation has a nice relationship with the first term in \eqref{eq:ballvarfinalbound}. To see this, recall that we define $\acbonus_{i,\horizon} = \sum_{t \in \bonustimes}\mathbbm{1}[\volchoicetsuccess{\AC} = i]$ as the sum of \signups\ under $\AC$ for \opp\ $i$ by \vols\ who did not receive a recommendation under $\OPT$. Then,
\begin{align}
   \sum_{i \in [\numopps]} \sum_{t \in \exttimes \cup \bonustimes} \left(1-\balancefunc(\fillrate_{i,t-1})\right)\mathbbm{1}[\volchoicetsuccess{\adaptivecapmath} = i] \quad & = \sum_{i \in [\numopps]}\left(\sum_{t \in \exttimes \cup \bonustimes}\mathbbm{1}[\volchoicetsuccess{\adaptivecapmath} = i] - \balancefunc(\fillrate_{i,t-1})\mathbbm{1}[\volchoicetsuccess{\adaptivecapmath} = i]\right) \\
    &=\quad \sum_{i \in [\numopps]} \adaptivecapmath_{i,\horizon}^\exttrafmath + \acbonus_{i,\horizon} - \sum_{t \in \exttimes \cup \bonustimes} \ballvar_t \label{eq:binvarintermediate}
\end{align}
Now focusing on the second summation, which deals with \inttraf\ for which \OPT\ provides a recommendation:
\begin{align}    \sum_{i \in [\numopps]}\sum_{t \in \inttimes \setminus \bonustimes} \left(1-\balancefunc(\fillrate_{i,t-1})\right)\mathbbm{1}[\volchoicetsuccess{\adaptivecapmath} = i]
    \quad \geq& \quad \sum_{i \in [\numopps]}\sum_{t \in \inttimes \setminus \bonustimes} \left(1-\balancefunc\left(\frac{\adaptivecapmath_{i, t-1}^{\inttrafmath}}{\capa_i}\right)\right) \mathbbm{1}[\volchoicetsuccess{\adaptivecapmath} = i] \label{eq:binvarlowerbound} \\
    \quad \geq& \quad
    \sum_{i \in [\numopps]}\sum_{\counter \in [\adaptivecapmath_{i, \horizon}^{\inttrafmath} - \acbonus_{i,\horizon}]} \left(1-\balancefunc\left(\frac{\counter-1}{\capa_i}\right)\right)
    \label{eq:reimannsum1} \\
    \quad \geq& \quad \sum_{i \in [\numopps]}e^{-1/\capa_i}\sum_{\counter \in [\adaptivecapmath_{i, \horizon}^{\inttrafmath}- \acbonus_{i,\horizon}]} \left(1-\balancefunc\left(\frac{\counter}{\capa_i}\right)\right)  \label{eq:reimannsum2} \\
    \quad \geq& \quad  e^{-1/\invbidtobudget}\sum_{i \in [\numopps]}\int_{0}^{\adaptivecapmath_{i, \horizon}^{\inttrafmath}- \acbonus_{i,\horizon}} 1 - \balancefunc(x/\capa_i) \ \partial x \label{eq:reimannbound} \\
    \quad =& \quad  e^{-1/\invbidtobudget}\sum_{i \in [\numopps]}\capa_i\left(1 - \balancefunc \left(\frac{\adaptivecapmath_{i, \horizon}^{\inttrafmath}- \acbonus_{i,\horizon}}{\capa_i}\right) - 1/e\right) \label{eq:binvarfinalbound}
\end{align}
In \eqref{eq:binvarlowerbound}, we use the fact that $\balancefunc$ is decreasing and $\frac{\adaptivecapmath_{i, t-1}^{\inttrafmath}}{\capa_i} \leq \frac{\adaptivecapmath_{i, t-1}^{\inttrafmath}}{\capa_i - \adaptivecapmath_{i, t-1}^{\exttrafmath}} = \fillrate_{i,t-1}$. We then further reduce the argument in $\balancefunc$ in \eqref{eq:reimannsum1} by noting that the lowest possible values of $\adaptivecapmath_{i, t}^{\inttrafmath}$ are $\{1, \dots,\adaptivecapmath_{i, \horizon}^{\inttrafmath}- \acbonus_{i,\horizon}\}$, since $\adaptivecapmath_{i, t}^{\inttrafmath}$ increases by $1$ for any $t \in \inttimes$ where $\volchoicetsuccess{\adaptivecapmath} = i$. 

The summation in \eqref{eq:reimannsum1} represents a left Riemann sum of an increasing function. In \eqref{eq:reimannsum2}, we utilize the fact that for any $\counter$, $1-\balancefunc((\counter-1)/\capa_i) \geq e^{1/\invbidtobudget}(1-\balancefunc(\counter/\capa_i))$. As the summation in \eqref{eq:reimannsum2} is now a right Riemann sum of an increasing function, we bound the sum with an appropriate integral in \eqref{eq:reimannbound}. We evaluate the integral to arrive at \eqref{eq:binvarfinalbound}.

Combining \eqref{eq:ballvarfinalbound}, \eqref{eq:binvarintermediate}, and \eqref{eq:binvarfinalbound} along with the observation that $e^{-1/\invbidtobudget} < 1$, we see that for any sample path $\samplepath$,
\begin{align*} \sum_{t \in [\horizon]} \ballvar_t + \sum_{i \in [\numopps]} \binvar_i  \quad \geq \quad e^{-1/\invbidtobudget}\sum_{i \in [\numopps]}&\left( \adaptivecapmath_{i,\horizon}^\exttrafmath + \acbonus_{i,\horizon} +  \OPT_{i,\horizon}^{\inttrafmath} \balancefunc \left(\frac{\adaptivecapmath_{i, \horizon}^{\inttrafmath}}{\capa_i - \adaptivecapmath_{i, \horizon}^{\exttrafmath}}\right)\right. \nonumber \\ &\left.  +  \capa_i \left(1-\balancefunc \left(\frac{\adaptivecapmath_{i, \horizon}^{\inttrafmath} - \acbonus_{i,\horizon}}{\capa_i}\right) - 1/e\right)\right)
\end{align*}
Taking expectations over all sample paths completes the proof of Lemma \ref{lem:lowerboundballvarbinvar} \halmos

\subsubsection{Proof of Lemma \ref{lem:MP}:}
\label{proof:lem:MP}
To prove Lemma \ref{lem:MP}, it is sufficient to show that for any instance $\instance$, we can construct a feasible solution to $\MP$ which has a value of $\frac{\mathbb{E}_{\samplepath}[\adaptivecapmath]}{\mathbb{E}_{\samplepath}[\OPT]}$. (We remind that $\AC$ and $\OPT$ depend on both the instance $\instance$ and the sample path $\samplepath$, but we suppress that dependence to ease exposition). 

To construct such a feasible solution, we define the values of $\vec{x}$ based on the value of $\AC$ along a particular sample path.\footnote{{We emphasize that fixing a sample path $\samplepath$, the entire sequence of \opp\ recommendations and \vol\ \signups\ are entirely deterministic under both \AC\ and \OPT. To see this, note that for any fixed history, the \AC\ algorithm makes a deterministic recommendation, and the \vol's decision in response to that recommendation is deterministic, conditional on $\samplepath$. Similarly, \OPT\ makes a deterministic recommendation for any fixed history and fixed inputs. The history as well as inputs (i.e., the instance $\instance$ as well as the \signup\ decisions of all \exttraf) are deterministic for any fixed $\samplepath$.}} Specifically, $x_{1,i,\samplepath}$ (resp. $x_{2,i,\samplepath}$) represents the amount of \exttraf\ (resp. \inttraf) that contributes to \opp\ $i$ under $\AC$, given by $\adaptivecapmath_{i,\horizon}^{\exttrafmath}$ (resp. $\adaptivecapmath_{i,\horizon}^{\inttrafmath}$). The third component, $x_{3,i,\samplepath}$, accounts for the value of $\AC$ on the \vols\ for which \OPT\ recommends \opp\ $0$, which we denote as $\acbonus_{i, \horizon} := \sum_{t \in \bonustimes}\mathbbm{1}[\volchoicetsuccess{\AC} = i]$.
In a similar fashion, we define the values of $\vec{y}$ based on the value of $\OPT$ along a particular sample path. Specifically, $y_{1,i,\samplepath}$ (resp. $y_{2,i,\samplepath}$) represents the amount of \exttraf\ (resp. \inttraf) that contributes to \opp\ $i$ under $\OPT$, given by $\OPT_{i,\horizon}^{\exttrafmath}$ (resp. $\OPT_{i,\horizon}^{\inttrafmath}$).
Finally, we define $z$ as the ratio between the expected value of $\AC$ and the expected value of $\OPT$ on this instance. 

To summarize, we consider the following feasible solution:

\begin{align*}
    &x_{1,i,\samplepath} = \adaptivecapmath_{i,\horizon}^{\exttrafmath}, \qquad x_{2,i,\samplepath} = \adaptivecapmath_{i,\horizon}^{\inttrafmath}, \qquad x_{3,i,\samplepath} = \acbonus_{i,\horizon}, \\
    &y_{1,i,\samplepath} = \OPT_{i,\horizon}^{\exttrafmath}, \qquad y_{2,i,\samplepath} = \OPT_{i,\horizon}^{\inttrafmath}, \qquad
    z = \frac{\mathbb{E}_{\samplepath}[\adaptivecapmath]}{\mathbb{E}_{\samplepath}[\OPT]}
\end{align*}
If such a solution is feasible, then the optimal value of $\MP$ is at most $\frac{\mathbb{E}_{\samplepath}[\adaptivecapmath]}{\mathbb{E}_{\samplepath}[\OPT]}$, since the optimal value of $\MP$ is less than or equal to the value of any feasible solution. We proceed by sequentially showing that each constraint is met under this candidate solution.\footnote{We remark that we restrict our attention to instances where $\mathbb{E}_{\samplepath}[\OPT] > 0$; thus, $\vec{y}$ can be constrained to have at least one strictly positive element.} 

First, observe that neither $\adaptivecapmath$ nor $\OPT$ can exceed the capacity of the \opp\ along any sample path $\samplepath$. Hence, constraints (i) and (ii) are never violated. Similarly, $\adaptivecapmath_{i,\horizon}^{\inttrafmath}$ is the sum of \signups\ from \inttraf\ under $\AC$, while $\acbonus_{i,\horizon}$ is the sum of \signups\ from a subset of \inttraf\ under $\AC$. Thus, constraint (iii) must hold.

For constraint (iv), we first fix an \opp\ $i$. Based on Definition \ref{def:opt}, $\OPT$ will never use \inttraf\ to fill capacity that would otherwise be filled by \exttraf. As a consequence, \OPT\ uses all \exttraf\ for $i$ (or fills \opp\ $i$ with \exttraf) along each sample path. In contrast, $\AC$ may use \inttraf\ to fill capacity that could otherwise have been filled by \exttraf. In other words, if an \opp\ reaches full capacity under \AC, then some \exttraf\ may be excessive. Thus, along a fixed sample path, either $\AC$ uses the same amount of \exttraf\ as $\OPT$ for \opp\ $i$, or \opp\ $i$ reaches capacity under $\AC$.\footnote{By our convention for \exttraf, $\AC$ will always \emph{recommend} the \vol's targeted \opp\ $\extrecommend$. However, if this \opp\ has already reached capacity, the \signup\ does not \emph{fill} any capacity.} These two possibilities give rise to constraint (iv).

Constraint (v) holds based on the definitions of $\vec{x}, \vec{y},$ and $z$: 
$$z \quad = \quad  \frac{\mathbb{E}_{\samplepath}[\adaptivecapmath]}{\mathbb{E}_{\samplepath}[\OPT]} \quad = \quad \frac{\mathbb{E}_{\samplepath}[\sum_{i \in [\numopps]}x_{1, i, \samplepath} + x_{2, i, \samplepath}]}{\mathbb{E}_{\samplepath}[\sum_{i \in [\numopps]}y_{1, i, \samplepath} + y_{2, i, \samplepath}]} \quad \geq \quad \frac{\mathbb{E}_{\samplepath}[\sum_{i \in [\numopps]}x_{1, i, \samplepath} + x_{2, i, \samplepath}]}{\sum_{i \in [\numopps]} \capa_i}.$$

We now consider constraint (vi), which crucially provides a lower bound on the number of \signups\ generated by $\adaptivecapmath$ where $\OPT$ either generates a \signup\ to the same \opp\ or does not generate a \signup\ at all.  Fixing a sample path and an \opp, note that the total amount of \opp\ $i$'s capacity filled by $\adaptivecapmath$ in periods $t \in \inttimes \setminus \bonustimes$ is given by $x_{2, i, \samplepath} - x_{3, i, \samplepath}$, while the total amount of \opp\ $i$'s capacity filled by $\OPT$ in periods $t \in \inttimes \setminus \bonustimes$ is given by $y_{2, i, \samplepath}$. \revcolor{Furthermore, for all $t \in \inttimes \setminus \bonustimes$, $\OPT$ provides a recommendation, which means it fills a unit of capacity with probability at least $\min_{i \in \oppset_t}\convprob_{i,t}$,
while  $\adaptivecapmath$ will fill a unit of capacity with probability at most $\max_{i \in \oppset_t}\convprob_{i,t}$. (We remind that $\oppset_t$ represents the subset of \opps\ $i$ for which $\convprob_{i,t} > 0$.)} As a consequence, we can apply the definition of the \MCPR\ (Definition \ref{def:MCPR}) to show that $x_{2, i, \samplepath} - x_{3, i, \samplepath} \leq \weightcap y_{2, i, \samplepath} $, or equivalently, $x_{2, i, \samplepath} \leq \weightcap y_{2, i, \samplepath} + x_{3, i, \samplepath}$

Based on the constructed values of $\vec{x}, \vec{y},$ and $z$, as well as the upper bound on $x_{2, i, \samplepath}$ identified above,
\begin{align}\mathbb{E}_{\samplepath}\left[\sum_{i \in [\numopps]}x_{1,i,\samplepath}\right] & = z\cdot \mathbb{E}_{\samplepath}\left[\sum_{i \in [\numopps]}y_{1,i,\samplepath} + y_{2,i,\samplepath}\right] - \mathbb{E}_{\samplepath}\left[\sum_{i \in [\numopps]}x_{2,i,\samplepath}\right]  \\
&\geq z\cdot \mathbb{E}_{\samplepath}\left[\sum_{i \in [\numopps]}y_{1,i,\samplepath} + y_{2,i,\samplepath}\right] - \mathbb{E}_{\samplepath}\left[\sum_{i \in [\numopps]}\weightcap \cdot y_{2, i, \samplepath} + x_{3, i, \samplepath}\right] \\
&= \mathbb{E}_{\samplepath}\left[\sum_{i \in [\numopps]}y_{1,i,\samplepath} \right] - \mathbb{E}_{\samplepath}\left[\sum_{i \in [\numopps]} (1-z)\cdot y_{1,i,\samplepath} + (\weightcap - z)\cdot y_{2,i,\samplepath} \right] - \mathbb{E}_{\samplepath}\left[\sum_{i \in [\numopps]}x_{3, i, \samplepath}\right] \\
&\geq \mathbb{E}_{\samplepath}\left[\sum_{i \in [\numopps]}y_{1,i,\samplepath} \right] - (\weightcap - z) \cdot \mathbb{E}_{\samplepath}\left[\sum_{i \in [\numopps]} y_{1,i,\samplepath} + y_{2,i,\samplepath} \right] - \mathbb{E}_{\samplepath}\left[\sum_{i \in [\numopps]}x_{3, i, \samplepath}\right] \label{eq:weightcapbiggerthan1} \\
&\geq \beta \sum_{i \in [\numopps]} c_i - (\weightcap-z)\sum_{i \in [\numopps]} c_i  - \mathbb{E}_{\samplepath}\left[\sum_{i \in [\numopps]}x_{3, i, \samplepath}\right]
.\end{align}
Inequality \eqref{eq:weightcapbiggerthan1} uses the fact that $\weightcap \geq 1$. The final inequality uses the fact that $\mathbb{E}_{\samplepath}\left[\sum_{i \in [\numopps]}y_{1,i,\samplepath} \right] = \beta \sum_{i \in [\numopps]} c_i$ based on the definitions of the optimal clairvoyant algorithm $\OPT$ and the \fracextname\ $\extfrac$ (see Definitions \ref{def:opt} and \ref{def:beta}). This final inequality establishes that our proposed solution respects constraint (vi).

Finally, we turn our attention to constraint (vii). Given the constructed values of $\vec{x}, \vec{y},$ and $z$,
\begin{align}
    e^{1/\invbidtobudget} z \mathbb{E}_{\samplepath}\left[\sum_{i \in [\numopps]}y_{1,i,\samplepath} + y_{2,i,\samplepath}\right] &=  e^{1/\invbidtobudget} \mathbb{E}_{\samplepath}\left[\sum_{i \in [\numopps]}x_{1,i,\samplepath} + x_{2,i,\samplepath}\right] \\
    &\geq e^{1/\invbidtobudget}\mathbb{E}_{\samplepath} \left[\sum_{t \in [\horizon]} \ballvar_t + \sum_{i \in [\numopps]} \binvar_i \right] \label{eq:consvi1} \\
    &\geq \mathbb{E}_{\samplepath}\left[\sum_{i \in [\numopps]} x_{1,i,\samplepath} + x_{3,i,\samplepath} + y_{2, i, \samplepath} \balancefunc \left(\frac{x_{2,i,\samplepath}}{\capa_i - x_{1,i,\samplepath}}\right) \right.\nonumber
    \\&\qquad \qquad \left.+ \capa_i \left(1 - \balancefunc \left(\frac{x_{2,i,\samplepath} - x_{3,i,\samplepath}}{\capa_i}\right) - 1/e\right)\right] \label{eq:consvi2}
\end{align}
Inequality \eqref{eq:consvi1} comes from applying Lemma \ref{lem:lowerboundalg}, while \eqref{eq:consvi2} comes from applying Lemma \ref{lem:lowerboundballvarbinvar}. This establishes that constraint (vii) is met under our proposed solution.

In aggregate, we have shown that the proposed solution of $\vec{x}, \vec{y},$ and $z$ are feasible in $\MP$. This solution attains a value of $z = \frac{\mathbb{E}_{\samplepath}[\adaptivecapmath(\instance, \samplepath)]}{\mathbb{E}_{\samplepath}[\OPT(\instance, \samplepath)]}$, which completes the proof of Lemma \ref{lem:MP}. \halmos

\subsubsection{Proof of Lemma \ref{lem:MPlower}}
\label{proof:lem:MPlower}
To prove Lemma~\ref{lem:MPlower}, we will derive a valid lower bound on the value of $\MP$ (for a fixed instance $\instance$) that is parameterized by the \fracextname\ $\extfrac$, the minimum capacity $\invbidtobudget$, and the \MCPR\ $\weightcap$. We then argue that our lower bound is uniform given $\extfrac$, $\invbidtobudget$, and $\weightcap$, in that it is valid for any given instance $\instance$ with those parameters.

To derive the lower bound on the value of $\MP$, we propose a series of transformations to the optimization problem that will ultimately result in a solvable program. The solution to that transformed program serves as a lower bound on $\MP$, and its value can be characterized as a function that depends only on the \fracextname\ $\extfrac$, the minimum capacity $\invbidtobudget$, and the \MCPR\ $\weightcap$. We divide this process into five (algebraic) steps, each of which results in a new formulation for the optimization problem.  

First, in {\bf Step (a)} we show there is no feasible solution for $z < e^{-1/\invbidtobudget}(1 - 1/e)$. Thus, we create a new program ($\MPsub{a}$) where we restrict the feasible domain. The value of this new program serves as a lower bound on the value of $\MP$. In {\bf Step (b)}, we show that in $\MPsub{a}$, it is without loss of generality to consider only feasible solutions where constraint (i) binds for all $i$ and $\samplepath$ pairs. Based on this, we constuct a new program $\MPsub{b}$ which replaces the inequality in constraint (i) with an equality. In {\bf Step (c)}, we relax $\MPsub{b}$ by replacing constraints (i), (iv), and (vii)  with a unified constraint (viii). We define this new program as $\MPsub{c}$. In {\bf Step (d)}, we transform $\MPsub{c}$ by replacing the inequalities in constraints (iii), (v), and (vi) with three equalities, thereby creating the program $\MPsub{d}$. Finally, in {\bf Step (e)}, we convexify the simplified program from the previous step, to arrive at the (solvable) $\MPsub{e}$. We highlight that the value of each new program serves as a lower bound on the value of the previous program; i.e., the value of $\MPsub{b}$ is a lower bound on the value of $\MPsub{a}$, which is a lower bound on the value of $\MP$.

\vspace{0.2cm}

{\bf Step (a):} Suppose for a moment that there is a feasible solution where $z < e^{-1/\invbidtobudget} (1-1/e)$. We will show a contradiction by demonstrating that if such a solution satisfies constraints (ii) and (iv), it cannot satisfy constraint (vii). 
We begin by fixing a particular \opp\ $i$ and a particular sample path $\samplepath$. If constraint (iv) holds, there are two cases to consider: either $x_{1,i,\samplepath} + x_{2,i,\samplepath} = \capa_i$ or $x_{1,i,\samplepath} = y_{1,i,\samplepath}$. In the first case, we have that $\balancefunc \left(\frac{x_{2,i,\samplepath}}{\capa_i - x_{1,i,\samplepath}}\right) = 0$, as  $\balancefunc(1) = 0$ by definition. Note that the left hand side of constraint (vii) is a weighted summation over \opps\ and sample paths, where the weights depend on the probability of the sample path. Let us consider the term in that summation which corresponds to the fixed \opp\ $i$ and the fixed sample path $\samplepath$. This term is bounded by
\begin{align}
    x_{1,i,\samplepath} + x_{3,i,\samplepath}+ \capa_i(1 -  \balancefunc \left(\frac{x_{2,i,\samplepath} - x_{3,i,\samplepath}}{\capa_i}\right) - 1/e) &= x_{1,i,\samplepath} + x_{3,i,\samplepath}+ \capa_i \text{exp}\left( \frac{x_{2,i,\samplepath}-x_{3,i,\samplepath}}{\capa_i} - 1\right) - \frac{\capa_i}{e} \\
    &= x_{1,i,\samplepath} + x_{3,i,\samplepath}+ \capa_i \text{exp}\left( \frac{-x_{1,i,\samplepath} - x_{3,i,\samplepath}}{\capa_i}\right) - \frac{\capa_i}{e} \\
    &\geq x_{1,i,\samplepath} +x_{3,i,\samplepath}+ \capa_i\left(1-\frac{x_{1,i,\samplepath}}{\capa_i} - \frac{x_{3,i,\samplepath}}{\capa_i}\right) - \frac{\capa_i}{e} \label{eq:infeasiblecase1} \\
    &\geq (1-1/e)\capa_i  \\
    &\geq (1-1/e)(y_{1,i,\samplepath}+y_{2,i,\samplepath})\label{eq:infeasiblecase2}
\end{align}
Inequality \eqref{eq:infeasiblecase1} comes from the fact that $\text{exp}(-x) \geq 1-x$ for all $x$, and the remaining steps are algebraic. 

We now address the second case, where $x_{1,i\samplepath} = y_{1,i,\samplepath}$ for this particular $i$ and $\samplepath$. Let us again consider the term in the summation on the left hand side of constraint (vii) which corresponds to the fixed \opp\ $i$ and the fixed sample path $\samplepath$. This term is bounded by
\begin{align}
    &x_{1,i,\samplepath} + x_{3,i,\samplepath} + y_{2, i, \samplepath}\balancefunc \left(\frac{x_{2,i,\samplepath}}{\capa_i - x_{1,i,\samplepath}}\right) + \capa_i \left(1 - \balancefunc \left(\frac{x_{2,i,\samplepath} - x_{3,i,\samplepath}}{\capa_i}\right) - 1/e\right) \nonumber \\& \qquad \qquad = \quad y_{1,i,\samplepath} + x_{3,i,\samplepath} + y_{2, i, \samplepath} - y_{2, i, \samplepath}\text{exp}\left(\frac{x_{2,i,\samplepath}}{\capa_i - y_{1,i,\samplepath}}-1\right) + \capa_i \text{exp} \left(\frac{x_{2,i,\samplepath} - x_{3,i,\samplepath}}{\capa_i} -1\right) - \capa_i/e 
    \\
    &\qquad \qquad \geq \quad y_{1,i,\samplepath} + y_{2, i, \samplepath} - y_{2, i, \samplepath}\text{exp}\left(\frac{x_{2,i,\samplepath}}{\capa_i - y_{1,i,\samplepath}}-1\right) + \capa_i \text{exp} \left(\frac{x_{2,i,\samplepath}}{\capa_i} -1\right) - \capa_i/e 
    \label{eq:quasiconvexity}
\end{align}
The second inequality holds because the expression is increasing in $x_{3,i,\samplepath}$. Note that the right hand side of \eqref{eq:quasiconvexity} is quasi-concave in $x_{2,i,\samplepath}$. We demonstrate quasi-concavity by first noting that the expression is a continuously differentiable function of $x_{2,i,\samplepath}$, and then by establishing that this function cannot have a local minimum. To prove the latter, we begin by calculating the derivative of the right hand side (RHS) with respect to $x_{2,i,\samplepath}$.
$$\frac{\partial}{\partial x_{2,i,\samplepath}} \text{RHS} = \frac{-y_{2,i,\samplepath}}{\capa_i - y_{1,i,\samplepath}}\text{exp}\left(\frac{x_{2,i,\samplepath}}{\capa_i - y_{1,i,\samplepath}} - 1\right) + \text{exp}\left(\frac{x_{2,i,\samplepath}}{\capa_i}-1\right),$$
which is equal to $0$ only when $\frac{y_{2,i,\samplepath}}{\capa_i-y_{1,i,\samplepath}}\text{exp}\left(x_{2,i,\samplepath}/(\capa_i - y_{1,i,\samplepath})-1\right) = \text{exp}(x_{2,i,\samplepath}/\capa_i - 1)$. When this first-order condition holds, we see that the second derivative of the right hand side with respect to $x_{2,i,\samplepath}$ must be strictly negative: 
$$\frac{\partial^2}{\partial x_{2,i,\samplepath}^2} \text{RHS} = \frac{-y_{2,i,\samplepath}}{(\capa_i - y_{1,i,\samplepath})^2}\text{exp}\left(\frac{x_{2,i,\samplepath}}{\capa_i - y_{1,i,\samplepath}} - 1\right) + \frac{1}{\capa_i}\text{exp}\left(\frac{x_{2,i,\samplepath}}{\capa_i}-1\right) = \frac{-y_{1,i,\samplepath}}{\capa_i(\capa_i - y_{1,i,\samplepath})}\text{exp}\left(\frac{x_{2,i,\samplepath}}{\capa_i}\right)$$
Hence, this expression is quasi-concave in $x_{2,i,\samplepath}$, and as a consequence is minimized at one of the extreme points of $x_{2,i,\samplepath}$.

The two extreme points for $x_{2,i,\samplepath}$ are $0$ and $\capa_i - x_{1,i,\samplepath}$ (based on constraint (ii)). If $x_{2,i,\samplepath} = 0$, the RHS of \eqref{eq:quasiconvexity} is equal to $y_{1,i,\samplepath}+ (1-1/e)y_{2,i,\samplepath}$. If $x_{2,i,\samplepath} = \capa_i -x_{1,i,\samplepath}$, we have returned to the first case for constraint (iv), where we established a lower bound of $(1-1/e)(y_{1,i,\samplepath}+y_{2,i,\samplepath})$ in \eqref{eq:infeasiblecase2}.

Therefore, we have shown that for any particular $i$ and $\samplepath$, if constraints (ii) and (iv) are satisfied, 
$$x_{1,i,\samplepath} + (y_{2, i, \samplepath}+x_{3,i,\samplepath}) \balancefunc \left(\frac{x_{2,i,\samplepath}}{\capa_i - x_{1,i,\samplepath}}\right) + \capa_i \left(1 - \balancefunc \left(\frac{x_{2,i,\samplepath}}{\capa_i}\right) - 1/e\right) \ \geq \ (1-1/e)(y_{1,i,\samplepath}+y_{2,i,\samplepath}) $$
Summing this up over all \opps\ and taking expectations over all sample paths,\footnote{Because we restrict our attention to arrival sequences where $\mathbb{E}[\OPT]$ is non-zero, this includes at least one \opp\ and sample path for which $y_{1,i,\samplepath} + y_{2,i,\samplepath} > 0$.} we see that constraint (vii) must be violated for any $z < e^{-1/\invbidtobudget} (1-1/e)$. This completes Step (a).

In the subsequent step, we will work with a modified version of $\MP$, which we refer to as $\MPsub{a}$ (shown below), that restricts the domain by imposing that $ z \geq e^{-1/\invbidtobudget}(1-1/e)$. Any feasible solution to $\MP$ remains feasible in $\MPsub{a}$, and thus the value of $\MPsub{a}$ is a valid lower bound on the value of $\MP$.

{\small
\begin{equation*}
\arraycolsep=1.4pt\def\arraystretch{1}
\begin{array}{|cllll|}
\hline
&&&&\\[-.5em]
\multicolumn{5}{|c|}{\text{Given an instance $\instance$, the inputs to $\MPsub{a}$ are the set of \opps\ $\oppset$, the \fracextname\ $\extfrac$, the \MCPR\ $\weightcap$, }} \\
\multicolumn{5}{|c|}{\text{and the set of feasible sample paths $\Omega$, along with its associated probability measure.}} \\[.5em]
\multicolumn{5}{|c|}{\text{$\MPsub{a}$ uses the set of variables } \  \vec{x} \in \mathbb{R}_{\geq 0}^{3 \times \numopps \times |\Omega|} \text{ and } \  \vec{y} \in \mathbb{R}_{\geq 0}^{2 \times n \times |\Omega|} \setminus \vec{\bf{0}} \text{, along with } z \in [e^{-1/\invbidtobudget} (1-1/e), 1]} \\
&&&&\\[-.5em]
\hline
&&&& \\[-.5em]
\ \underset{\displaystyle \vec{x}, \vec{y}, z }{\textrm{min}}\quad\quad&
 \multicolumn{3}{l}{\displaystyle z}&\quad \textbf{$\MPsub{a}$ \ } \quad \\[2.5em]
\text{s.t.}  &\ \forall i, \samplepath, \ \quad \capa_i \geq y_{1, i, \samplepath} + y_{2, i, \samplepath} \quad \ \ \text{\bf (i)}&\qquad \capa_i \geq x_{1, i, \samplepath} + x_{2, i, \samplepath} \quad \ \ \text{\bf (ii)}\qquad x_{2,i,\samplepath} \geq x_{3,i,\samplepath} &&\quad \text{\bf (iii)} \\[.7em]
&\multicolumn{3}{l}{\qquad \qquad \ \ \capa_i = x_{1, i, \samplepath} + x_{2, i, \samplepath} \qquad \text{\bf OR} \qquad x_{1, i, \samplepath} = y_{1, i, \samplepath}} &\quad \text{\bf (iv)}\\[.7em]
&\multicolumn{3}{l}{\mathbb{E}_{\samplepath}\left[\sum_{i \in [\numopps]}x_{1, i, \samplepath} + x_{2, i, \samplepath}\right] \quad  \leq \quad z
\sum_{i \in [\numopps]}\capa_i
} &\quad \text{\bf (v)}
\\[.7em]
&\multicolumn{3}{l}{\mathbb{E}_{\samplepath}\left[\sum_{i \in [\numopps]} x_{1, i, \samplepath}+x_{3, i, \samplepath}\right] \quad  \geq \quad (\extfrac - \weightcap + z)
\sum_{i \in [\numopps]}\capa_i
}  &\quad \text{\bf (vi)}\\[.7em]
&\multicolumn{3}{l}{
\mathbb{E}_{\samplepath}\left[\sum_{i \in [\numopps]} x_{1,i,\samplepath} + x_{3,i,\samplepath} + y_{2, i, \samplepath} \cdot \balancefunc \left(\frac{x_{2,i,\samplepath}}{\capa_i - x_{1,i,\samplepath}}\right) + \capa_i \left(1 - \balancefunc \left(\frac{x_{2,i,\samplepath} - x_{3,i,\samplepath}}{\capa_i}\right) - 1/e\right)\right]}& \\[0.7em]
&\multicolumn{3}{l}{\qquad \qquad \qquad \qquad \qquad \qquad \qquad  \leq e^{1/\invbidtobudget} z \mathbb{E}_{\samplepath}\left[\sum_{i \in [\numopps]}y_{1, i, \samplepath} + y_{2, i, \samplepath}\right]} & \quad \text{\bf(vii)} \\[.7em]
\hline
\end{array}
\end{equation*}
}

{\bf Step (b):} In this step, we will show that we can restrict our attention to feasible solutions of $\MPsub{a}$ where constraint (i) is tight for all $i$ and $\samplepath$ without loss of optimality. Consider any feasible solution $\{\vec{x}, \vec{y}, z\}$ where constraint (i) is loose for some $i, \samplepath$ pair. We will construct a new solution  $\{\vec{x}', \vec{y}', z'\}$ which is feasible and has the same objective value. Let $y_{2,i, \samplepath}' = \capa_i - y_{1,i, \samplepath}$. The other decision variables are unchanged: $y_{1,i,\samplepath}' = y_{1,i,\samplepath}$,  $\vec{x}' = \vec{x}$, and $z' = z$. 

The objective value is identical in both solutions, and only constraints (i) and (vii) are impacted by (weakly) increasing $y_{2,i, \samplepath}$ to $y_{2,i, \samplepath}'$. Constraint (i) is satisfied by construction, and constraint (vii) remains satisfied because for all $x \in [0,1]$, $\balancefunc(x) \leq 1-1/e \leq e^{1/\invbidtobudget} z$, where the second inequality holds as a result of the restricted domain on $z$ imposed in $\MPsub{a}$. This completes Step (b).

In the subsequent step, we will work with a modified version of $\MPsub{a}$, which we refer to as $\MPsub{b}$ (shown below), that replaces the inequality in constraint (i) with equality. As demonstrated in this step, the tightening of constraint (i) is without loss of optimality; thus, the value of $\MPsub{b}$ is a valid lower bound on the value of $\MPsub{a}$.

{\small
\begin{equation*}
\arraycolsep=1.4pt\def\arraystretch{1}
\begin{array}{|cllll|}
\hline
&&&&\\[-.5em]
\multicolumn{5}{|c|}{\text{Given an instance $\instance$, the inputs to $\MPsub{b}$ are the set of \opps\ $\oppset$, the \fracextname\ $\extfrac$, the \MCPR\ $\weightcap$, }} \\
\multicolumn{5}{|c|}{\text{and the set of feasible sample paths $\Omega$, along with its associated probability measure.}} \\[.5em]
\multicolumn{5}{|c|}{\text{$\MPsub{b}$ uses the set of variables } \  \vec{x} \in \mathbb{R}_{\geq 0}^{3 \times \numopps \times |\Omega|} \text{ and } \  \vec{y} \in \mathbb{R}_{\geq 0}^{2 \times n \times |\Omega|} \setminus \vec{\bf{0}} \text{, along with } z \in [e^{-1/\invbidtobudget} (1-1/e), 1]} \\
&&&&\\[-.5em]
\hline
&&&& \\[-.5em]
\ \underset{\displaystyle \vec{x}, \vec{y}, z }{\textrm{min}}\quad\quad&
 \multicolumn{3}{l}{\displaystyle z}&\quad \textbf{$\MPsub{b}$ \ } \quad \\[2.5em]
\text{s.t.}  &\ \forall i, \samplepath, \ \quad \capa_i = y_{1, i, \samplepath} + y_{2, i, \samplepath} \quad \ \ \text{\bf (i)}&\qquad \capa_i \geq x_{1, i, \samplepath} + x_{2, i, \samplepath} \quad \ \ \text{\bf (ii)}\qquad x_{2,i,\samplepath} \geq x_{3,i,\samplepath} &&\quad \text{\bf (iii)} \\[.7em]
&\multicolumn{3}{l}{\qquad \qquad \ \ \capa_i = x_{1, i, \samplepath} + x_{2, i, \samplepath} \qquad \text{\bf OR} \qquad x_{1, i, \samplepath} = y_{1, i, \samplepath}} &\quad \text{\bf (iv)}\\[.7em]
&\multicolumn{3}{l}{\mathbb{E}_{\samplepath}\left[\sum_{i \in [\numopps]}x_{1, i, \samplepath} + x_{2, i, \samplepath}\right] \quad  \leq \quad z
\sum_{i \in [\numopps]}\capa_i
} &\quad \text{\bf (v)}
\\[.7em]
&\multicolumn{3}{l}{\mathbb{E}_{\samplepath}\left[\sum_{i \in [\numopps]} x_{1, i, \samplepath}+x_{3, i, \samplepath}\right] \quad  \geq \quad (\extfrac - \weightcap + z)
\sum_{i \in [\numopps]}\capa_i
}  &\quad \text{\bf (vi)}\\[.7em]
&\multicolumn{3}{l}{
\mathbb{E}_{\samplepath}\left[\sum_{i \in [\numopps]} x_{1,i,\samplepath} + x_{3,i,\samplepath} + y_{2, i, \samplepath} \cdot \balancefunc \left(\frac{x_{2,i,\samplepath}}{\capa_i - x_{1,i,\samplepath}}\right) + \capa_i \left(1 - \balancefunc \left(\frac{x_{2,i,\samplepath} - x_{3,i,\samplepath}}{\capa_i}\right) - 1/e\right)\right]}& \\[0.7em]
&\multicolumn{3}{l}{\qquad \qquad \qquad \qquad \qquad \qquad \qquad  \leq e^{1/\invbidtobudget} z \mathbb{E}_{\samplepath}\left[\sum_{i \in [\numopps]}y_{1, i, \samplepath} + y_{2, i, \samplepath}\right]} & \quad \text{\bf(vii)} \\[.7em]
\hline
\end{array}
\end{equation*}
}

{\bf Step (c):} We will show that we can relax $\MPsub{b}$ by replacing constraints (i), (iv), and (vii) with the following constraint: 
\begin{equation}
    \mathbb{E}_{\samplepath}\left[\sum_{i \in [\numopps]} \capa_i \hat{g}\left(\frac{x_{1,i,\samplepath}}{\capa_i}, \frac{x_{2, i, \samplepath}}{\capa_i}, \frac{x_{3, i, \samplepath}}{\capa_i}\right) \right]  \leq e^{1/\invbidtobudget} z \sum_{i \in [\numopps]}\capa_i, \nonumber \tag{\bf viii}
\end{equation}
where 
\begin{align}\hat{g}(x_1, x_2, x_3) = x_1 + x_3 + (1-x_1) \cdot \balancefunc \left(\frac{x_2}{ 1-x_1}\right) + 1-  \balancefunc \left(x_2 - x_3\right) - 1/e. \label{eq:hatg}
\end{align}

This relaxation results in a new program, which we refer to as $\MPsub{c}$.
We now prove that the value of $\MPsub{c}$ provides a lower bound on the value of $\MPsub{b}$ by showing that any solution which satisfies constraints (i), (iv), and (vii) must necessarily satisfy constraint (viii). In $\MPsub{b}$, constraint (i) binds, which means that the right hand sides of constraints (vii) and (viii) are identical. The difference between the left hand sides of constraints (vii) and (viii) is simply the expected sum of $(y_{2,i,\samplepath} - \capa_i + x_{1,i,\samplepath}) \cdot \balancefunc \left(\frac{x_{2,i,\samplepath}}{\capa_i - x_{1,i,\samplepath}}\right)$. 
Given a solution where constraint (iv) is satisfied for every $i, \samplepath$ pair, we must have
either $\balancefunc \left(\frac{x_{2,i,\samplepath}}{\capa_i - x_{1,i,\samplepath}}\right) = 0$, or $\capa_i-x_{1, i, \samplepath} = \capa_i-y_{1, i, \samplepath} = y_{2,i, \samplepath}$.
(The second equality comes from the fact that constraint (i) binds.) As a consequence, the difference between the left hand sides of constraints (vii) and (viii) is $0$.
Thus, any solution satisfying constraints (i), (iv), and (vii) must also satisfy constraint (viii). This completes step (c), and in the subsequent step, we will work with $\MPsub{c}$ (shown below). We note that the variables $\vec{y}$ do not appear in either the objective or the constraints of $\MPsub{c}$. As a result, we remove these variables from the program.

{\small
\begin{equation*}
\arraycolsep=1.4pt\def\arraystretch{1}
\begin{array}{|cllll|}
\hline
&&&&\\[-.5em]
\multicolumn{5}{|c|}{\text{Given an instance $\instance$, the inputs to $\MPsub{c}$ are the set of \opps\ $\oppset$, the \fracextname\ $\extfrac$, the \MCPR\ $\weightcap$, }} \\
\multicolumn{5}{|c|}{\text{and the set of feasible sample paths $\Omega$, along with its associated probability measure.}} \\[.5em]
\multicolumn{5}{|c|}{\text{$\MPsub{c}$ uses the set of variables } \  \vec{x} \in \mathbb{R}_{\geq 0}^{3 \times \numopps \times |\Omega|} \text{ and } \ z \in [e^{-1/\invbidtobudget} (1-1/e), 1]} \\
&&&&\\[-.5em]
\hline
&&&& \\[-.5em]
\ \underset{\displaystyle \vec{x}, z }{\textrm{min}}\quad\quad&
 \multicolumn{3}{l}{\displaystyle z}&\quad \textbf{$\MPsub{c}$ \ } \quad \\[2.5em]
\text{s.t.}  &\ \forall i, \samplepath, \ \qquad \capa_i \geq x_{1, i, \samplepath} + x_{2, i, \samplepath} \quad \quad \ \ \text{\bf (ii)} \qquad \qquad \qquad x_{2,i,\samplepath} \geq x_{3,i,\samplepath} &&&\quad \quad \text{\bf (iii)} \\[.7em]
&\multicolumn{3}{l}{\mathbb{E}_{\samplepath}\left[\sum_{i \in [\numopps]}x_{1, i, \samplepath} + x_{2, i, \samplepath}\right] \quad  \leq \quad z
\sum_{i \in [\numopps]}\capa_i
} &\quad \quad \text{\bf (v)}
\\[.7em]
&\multicolumn{3}{l}{\mathbb{E}_{\samplepath}\left[\sum_{i \in [\numopps]} x_{1, i, \samplepath}+x_{3, i, \samplepath}\right] \quad  \geq \quad (\extfrac - \weightcap + z)
\sum_{i \in [\numopps]}\capa_i
}  &\quad \quad \text{\bf (vi)}\\[.7em]
&\multicolumn{3}{l}{
\mathbb{E}_{\samplepath}\left[\sum_{i \in [\numopps]} \capa_i \hat{g}\left(\frac{x_{1,i,\samplepath}}{\capa_i}, \frac{x_{2, i, \samplepath}}{\capa_i}, \frac{x_{3, i, \samplepath}}{\capa_i}\right) \right]  \leq e^{1/\invbidtobudget} z \sum_{i \in [\numopps]}\capa_i} & \quad \quad \text{\bf(viii)} \\[.7em]
\hline
\end{array}
\end{equation*}
}

{\bf Step (d):} In this step, we transform $\MPsub{c}$ by replacing constraints (iii), (v), and (vi) with equalities for $\mathbb{E}_{\samplepath}\left[\sum_{i \in [\numopps]} x_{1, i, \samplepath}\right]$, $\mathbb{E}_{\samplepath}\left[\sum_{i \in [\numopps]} x_{2, i, \samplepath}\right]$, and $\mathbb{E}_{\samplepath}\left[\sum_{i \in [\numopps]} x_{3, i, \samplepath}\right]$. We will show that such a transformation is without loss of optimality, and we will refer to the resulting program as $\MPsub{d}$.
To aid in this step, below we compute the derivatives of $\hat{g}(x_1, x_2, x_3)$, as defined in \eqref{eq:hatg}.
\begin{align}
    \frac{\partial  \hat{g}}{\partial x_{1}} \quad &= 
    \quad \text{exp}\left(\frac{x_2}{1-x_1} - 1\right)\left(1 - \frac{x_2}{1-x_1} \right) \\
    \frac{\partial  \hat{g}}{\partial x_{2}} \quad &= 
    \quad -\text{exp}\left(\frac{x_2}{1-x_1} - 1\right) + \text{exp}\left(x_2 - x_3 - 1\right)\\
    \frac{\partial  \hat{g}}{\partial x_{3}} \quad &= 
    \quad 1 - \text{exp}\left(x_2 - x_3 - 1\right)
\end{align}
Based on these derivatives, we can replace constraints (iii), (v), and (vi) with the following constraints:
\begin{align}
\mathbb{E}_{\samplepath}\left[\sum_{i \in [\numopps]} x_{1, i, \samplepath}\right] \quad &= \quad \max\{0, \extfrac - \weightcap + z\}\sum_{i \in [\numopps]}\capa_i \tag{ix} \\ \mathbb{E}_{\samplepath}\left[\sum_{i \in [\numopps]} x_{2, i, \samplepath}\right] \quad &= \quad \left(z - \max\{0, \extfrac - \weightcap + z\}\right)\sum_{i \in [\numopps]}\capa_i \tag{x} \\ \mathbb{E}_{\samplepath}\left[\sum_{i \in [\numopps]} x_{3, i, \samplepath}\right] \quad &= \quad 0 \tag{xi}
\end{align}
To see why, first consider any feasible solution for $\MPsub{c}$, $\{\vec{x}, z\}$, such that $x_{3,i,\samplepath} > 0$ for some $i, \samplepath$ pair. We construct a new solution $\{\vec{x}', z'\}$, where $x_{1,i,\samplepath}' = x_{1,i,\samplepath} + x_{3,i,\samplepath}, \ x_{2,i,\samplepath}' = x_{2,i,\samplepath} - x_{3,i,\samplepath}, \ x_{3,i,\samplepath}' = 0 $ and all other variables remain the same, including $z' = z$. Clearly, this solution has an equivalent objective value, and we can show that such a solution remains feasible.

For constraint (ii) and constraint (v), note that $x_{1,i,\samplepath}' + x_{2,i,\samplepath}' = x_{1,i,\samplepath} + x_{2,i,\samplepath}$. Similarly, for constraint (iii), note that $x_{2,i,\samplepath}' - x_{3,i,\samplepath}' = x_{2,i,\samplepath} - x_{3,i,\samplepath}$, and for constraint (vi), we have $x_{1,i,\samplepath}' + x_{3,i,\samplepath}' = x_{1,i,\samplepath} + x_{3,i,\samplepath}$. 
Finally, note that based on the derivatives calculated above, any increase in $x_1$ and proportional decrease in $x_2$ and $x_3$ must (weakly) decrease the left hand side of constraint (viii):
\begin{align}
   \frac{\partial  \hat{g}}{\partial x_{1}} - \frac{\partial  \hat{g}}{\partial x_{2}} - \frac{\partial  \hat{g}}{\partial x_{3}} &= \text{exp}\left(\frac{x_2}{1-x_1} - 1\right)\left(2 - \frac{x_2}{1-x_1} \right) - 1 \\
   &= \text{exp}\left(\frac{x_2}{1-x_1} - 1\right)\left(2 - \frac{x_2}{1-x_1} - \text{exp}\left(1-\frac{x_2}{1-x_1}\right)\right) \\
   &\leq \text{exp}\left(\frac{x_2}{1-x_1} - 1\right)\left(2 - \frac{x_2}{1-x_1} -2+\frac{x_2}{1-x_1}\right) \\
   &\leq 0
\end{align}
Note that the second-to-last inequality uses the fact that $e^x \geq 1+x$ for any $x$. This proves that the constructed solution remains feasible, and thus it is without loss of optimality to impose the constraint that $\mathbb{E}_{\samplepath}\left[\sum_{i \in [\numopps]} x_{3, i, \samplepath}\right] = 0$. 

Using a similar approach that relies on the fact that $\frac{\partial  \hat{g}}{\partial x_{1}} \geq 0$, we can show that it is without loss of generality to assume that either constraint (vi) binds or (if the right hand side of constraint (vi) is negative) every $x_{1,i,\samplepath} = 0$. Otherwise, we can simply reduce any non-zero $x_{1,i,\samplepath}$ and remain feasible. Coupled with the constraint $\mathbb{E}_{\samplepath}\left[\sum_{i \in [\numopps]} x_{3, i, \samplepath}\right] = 0$, this establishes the equality for $\mathbb{E}_{\samplepath}\left[\sum_{i \in [\numopps]} x_{1, i, \samplepath}\right]$.

Again using a similar approach, this time relying on the fact that $\frac{\partial  \hat{g}}{\partial x_{2}} \leq 0$, we can show that it is without loss of generality to assume that constraint (v) binds. Otherwise, we can simply increase any $x_{2,i,\samplepath}$ where constraint (ii) is loose (such an $i,\samplepath$ pair must exist if constraint (v) is loose). Coupled with the constraint on $\mathbb{E}_{\samplepath}\left[\sum_{i \in [\numopps]} x_{1, i, \samplepath}\right]$, this establishes the equality for $\mathbb{E}_{\samplepath}\left[\sum_{i \in [\numopps]} x_{2, i, \samplepath}\right]$.

Therefore, we can impose the three equality constraints without loss of optimality, and we can then relax the program by dropping constraints (iii), (v), and (vi). This transforms $\MPsub{c}$ into a new program $\MPsub{d}$ (shown below), where the value of $\MPsub{d}$ is a lower bound on the value of $\MPsub{c}$. This completes step (d), and for the next and final step, we will use $\MPsub{d}$ as the starting point.

{\small
\begin{equation*}
\arraycolsep=1.4pt\def\arraystretch{1}
\begin{array}{|cllll|}
\hline
&&&&\\[-.5em]
\multicolumn{5}{|c|}{\text{Given an instance $\instance$, the inputs to $\MPsub{d}$ are the set of \opps\ $\oppset$, the \fracextname\ $\extfrac$, the \MCPR\ $\weightcap$, }} \\
\multicolumn{5}{|c|}{\text{and the set of feasible sample paths $\Omega$, along with its associated probability measure.}} \\[.5em]
\multicolumn{5}{|c|}{\text{$\MPsub{d}$ uses the set of variables } \  \vec{x} \in \mathbb{R}_{\geq 0}^{3 \times \numopps \times |\Omega|} \text{ and } \ z \in [e^{-1/\invbidtobudget} (1-1/e), 1]} \\
&&&&\\[-.5em]
\hline
&&&& \\[-.5em]
\ \underset{\displaystyle \vec{x}, z }{\textrm{min}}\quad\quad&
 \multicolumn{3}{l}{\displaystyle z}&\quad \textbf{$\MPsub{d}$ \ } \quad \\[2.5em]
\text{s.t.}  &\ \forall i, \samplepath, \qquad x_{2,i,\samplepath}\quad  \geq \quad x_{3,i,\samplepath} &&&\quad  \text{\bf (iii)} \\[.7em]
&\multicolumn{3}{l}{
\mathbb{E}_{\samplepath}\left[\sum_{i \in [\numopps]} \capa_i \hat{g}\left(\frac{x_{1,i,\samplepath}}{\capa_i}, \frac{x_{2, i, \samplepath}}{\capa_i}, \frac{x_{3, i, \samplepath}}{\capa_i}\right) \right]  \quad \leq \quad e^{1/\invbidtobudget} z \sum_{i \in [\numopps]}\capa_i} & \quad \text{\bf(viii)} \\[.7em]
&\multicolumn{3}{l}{
\mathbb{E}_{\samplepath}\left[\sum_{i \in [\numopps]} x_{1, i, \samplepath}\right] \quad = \quad \max\{0, \extfrac - \weightcap + z\}\sum_{i \in [\numopps]}\capa_i} & \quad \text{\bf(ix)} \\[.7em]
&\multicolumn{3}{l}{
\mathbb{E}_{\samplepath}\left[\sum_{i \in [\numopps]} x_{2, i, \samplepath}\right] \quad = \quad \left(z - \max\{0, \extfrac - \weightcap + z\}\right)\sum_{i \in [\numopps]}\capa_i} & \quad \text{\bf(x)} \\[.7em]
&\multicolumn{3}{l}{
\mathbb{E}_{\samplepath}\left[\sum_{i \in [\numopps]} x_{3, i, \samplepath}\right] \quad = \quad 0} & \quad \text{\bf(xi)} \\[.7em]
\hline
\end{array}
\end{equation*}
}

{\bf Step (e):} In the final step, we relax $\MPsub{d}$ by replacing $\hat{g}(x_1, x_2, 0) := g(x_1, x_2)$ with its lower convex envelope over the domain $\mathcal{D} = \{(x_1, x_2)\in \mathbb{R}^2_{\geq 0}:  x_1+x_2 \leq 1\}$. We denote this lower convex envelope by $\breve{g}(x_1, x_2)$. Any solution which satisfies constraint (viii) in $\MPsub{d}$ will continue to satisfy constraint (viii) after this change, due to the lower convex envelope being a lower bound (by definition) on the original function $g$.

Furthermore, as the function $\breve{g}$ is convex, we can require constraints (ix), (x), and (xi) to hold pointwise (i.e., for any $i, \samplepath$ pair) without loss of optimality. To see why, note that any feasible solution in $\MPsub{d}$ will remain feasible when averaging over \opps\ and sample paths such that $\frac{x_{1,i,\samplepath}}{c_i}$ is the same for all $i, \samplepath$ pairs. (This averaging would not impact the value of the solution, $z$). Similarly, any feasible solution in $\MPsub{d}$ will remain feasible when averaging over \opps\ and sample paths such that $\frac{x_{2,i,\samplepath}}{c_i}$ is the same for all $i, \samplepath$ pairs. Additionally, constraint (xi) ensures that $x_{3,i,\samplepath} = 0$ for all  $i, \samplepath$ pairs, which eliminates the need for constraint (iii).

Based on these observations, we can construct a new program, which we denote by $\MPsub{e}$, where $x_{1,i,\samplepath} = \max\{0, \extfrac - \weightcap + z\}$, $x_{2,i,\samplepath} = z - \max\{0, \extfrac - \weightcap + z\}$, and $x_{3,i,\samplepath} = 0$ for all $i, \samplepath$ pairs. We then plug these values into constraint (viii), the only remaining constraint, to arrive at $\MPsub{e}$ (shown below). As this transformation was without loss of optimality, we note that $\MPsub{e}$ (shown below) represents a lower bound on $\MPsub{d}$.  

{\small
\begin{equation*}
\arraycolsep=1.4pt\def\arraystretch{1}
\begin{array}{|cllll|}
\hline
&&&&\\[-.5em]
\multicolumn{5}{|c|}{\text{Given an instance $\instance$, the inputs to $\MPsub{e}$ are the \fracextname\ $\extfrac$, the minimum capacity $\invbidtobudget$, }} \\
\multicolumn{5}{|c|}{\text{and the \MCPR\ $\weightcap$.}} \\[.5em]
\multicolumn{5}{|c|}{\text{$\MPsub{e}$ uses the variable } \ z \in [e^{-1/\invbidtobudget} (1-1/e), 1]} \\
&&&&\\[-.5em]
\hline
&&&& \\[-.5em]
\ \underset{\displaystyle z }{\textrm{min}}\quad\quad&
 \multicolumn{3}{l}{\displaystyle z}&\quad \textbf{$\MPsub{e}$ \ } \quad \\[2.5em]
\text{s.t.}  &\multicolumn{3}{l}{
\breve{g}\left( \max\{0, \extfrac - \weightcap + z\}, z - \max\{0, \extfrac - \weightcap + z\}\right)   \quad \leq \quad e^{1/\invbidtobudget} z } & \quad \text{\bf(viii)} \\[.7em]
\hline
\end{array}
\end{equation*}
}

We note that the value of $\MPsub{e}$ is equivalent to $z^*$, as defined in \eqref{eq:defzstar} (see Theorem \ref{thm:AClower}). Furthermore, by steps (a) through (e) and the transitivity property, we have shown that the value  of $\MPsub{e}$ represents a lower bound on the value of $\MP$ for any instance $\instance$. We emphasize that this lower bound depends only on the \fracextname\ $\extfrac$ of the instance, the minimum capacity $\invbidtobudget$ of the instance, and the \MCPR\ $\weightcap$ of the instance. This completes the proof of Lemma \ref{lem:MPlower}. \halmos

\subsubsection{Extension to Settings where External Sign-ups are Stochastic}
\label{apx:exttrafstochastic}
\revcolor{We now discuss how to extend the proof of Theorem \ref{thm:AClower} to settings where sign-ups from \exttraf\ are stochastic, which is often the case in practical settings. In fact, the proof of Theorem \ref{thm:AClower} never relies on the assumption that sign-ups from \exttraf\ are deterministic --- instead, the proof only assumes that $\OPT$ will never use \inttraf\ to fill capacity that would otherwise be filled by \exttraf. (This assumption is used to show that Constraint (iv) is satisfied by our feasible solution in the proof of Lemma \ref{lem:MP}.) This assumption about \OPT\ is without loss of optimality when sign-ups from \exttraf\ are deterministic, as \OPT\ knows the arrival sequence in advance and can ``reserve'' the precise amount of capacity for each \opp\ that can be filled by \exttraf; however, this is not necessarily the case when sign-ups from \exttraf\ are stochastic. To sidestep this issue, we can strengthen our definition of \OPT\ by assuming that it not only knows that arrival sequence in advance but also the realized sign-up decisions of all \exttraf. Based on this information, it always knows the precise amount of capacity for each \opp\ that can be filled by \exttraf. We denote this stronger benchmark $\widehat{\OPT}$ and formalize it in the following definition:

\begin{defn}[Benchmark when Sign-ups from External Traffic are Stochastic ($\widehat{\OPT}$)]
\label{def:opthat}
This benchmark is the solution to a dynamic program (of exponential size) which takes as input the instance $\instance$ as well as the \signup\ decisions of all \exttraf\ throughout the time horizon. 
Upon the arrival of each \inttraf\ \vol, $\widehat{\OPT}$ recommends an \opp\ $\opprecommend{\widehat{\OPT}} \in \oppset \cup \{0\}$ that maximizes the total expected amount of filled capacity, given the \signup\ history up to that point and the inputs to the program.
Whenever there is more than one \opp\ in this set of optimal \opps, we use the convention (without loss of optimality) that $\widehat{\OPT}$ deterministically recommends the \opp\ in this set with lowest index.  
\end{defn}

We highlight that this definition of $\widehat{\OPT}$ ensures that it fills as much capacity as possible with \exttraf. To see this, first note that $\widehat{\OPT}$ knows in advance how much capacity \emph{can} be filled by \exttraf. Furthermore, if capacity can be filled by \exttraf, then $\widehat{\OPT}$ will never fill it with \inttraf\ instead: our convention for breaking ties in favor of \opps\ with the lowest index implies that $\widehat{\OPT}$ will recommend \opp\ $0$ (i.e., no \opp) rather than wasting the \signup\ from \exttraf\ that will realize later. 

We note that $\widehat{\OPT}$ is a stronger benchmark than one that does not have foreknowledge of the \signup\ decisions of any arrivals (i.e., $\OPT$), as it can always choose to ignore this knowledge. Nevertheless, one can prove that Theorem \ref{thm:AClower} holds even against this stronger benchmark by simply replacing $\OPT$ with $\widehat{\OPT}$ in each step of the proof.
}

\subsection{Proof of Proposition \ref{prop:warmupupperbound} (Section \ref{subsec:results:extknown})}
\label{proof:prop:warmupupperbound}
The proof of Proposition \ref{prop:warmupupperbound} follows from the more general hardness result of Theorem \ref{thm:hardness}, which establishes an upper bound of $1-1/e$ on the competitive ratio of any online algorithm in the special case where there is no \exttraf\ (i.e., when $\extfrac = 0$). 

We start from the instance that establishes this result ($\instance_1(0)$, described in Appendix \ref{proof:thm:hardness}), which consists of a total capacity of $\largeopps \largecapacity$ and an equal number of \inttraf\ \vols. Fixing a particular $\extfrac \in [0,1)$,\footnote{For $\extfrac = 1$, we have the trivial result that the upper bound on the competitive ratio is $1$.} we add one \opp\ to that instance with capacity  $\frac{\extfrac}{1-\extfrac}\largeopps \largecapacity$. To exactly fill this \opp, we append $\frac{\extfrac}{1-\extfrac}\largeopps \largecapacity$ \exttraf\ \vols\ to the start of the arrival sequence, where each of these arriving \vols\ has a conversion probability of $1$ for the newly-added \opp.

By design, (i) all \exttraf\ arrives first, (ii) the \fracextname\ is exactly equal to $\extfrac$,  (iii) the new \opp\ will be entirely filled with \exttraf\ under any algorithm, as this traffic directly views the \opp, but (iv) by Theorem \ref{thm:hardness}, no online algorithm can achieve a competitive ratio better than $1-1/e$ on the remaining \opps\ (none of the added \vols\ are compatible with the remaining \opps). Putting these four observations together, we have established an upper bound of $\extfrac + (1-\extfrac)(1-1/e)$ on the competitive ratio of any online algorithm when the \exttraf\ arrives first.\footnote{To show that this upper bound holds for any minimum capacity $\invbidtobudget$, it suffices to add an additional \opp\ with capacity $\invbidtobudget$ for which \vols\ have conversion probability of $0$. The value of $\OPT$ and the upper bound on the performance of any algorithm do not change, and the \fracextname\ also remains the same in the limit as $\largeopps$ approaches infinity.}

\subsection{Proof of Proposition \ref{prop:warmupmsvv} (Section \ref{subsec:results:extknown})}
\label{proof:prop:warmupmsvv}
Consider a family of instances $\instance_2(\extfrac)$ parameterized by the \fracextname\ $\extfrac$. In each instance, there are a large number of \opps\ $\largeopps$, each with identical large capacity $\largecapacity$.  The arrival sequence consists of $\largeopps \largecapacity$ \vols, and for a given \MakeLowercase{\fracextnamefull} $\extfrac$, the first $\extfrac \largeopps \largecapacity$ of these \vols\ are \exttraf.\footnote{We assume that $(1-\extfrac)\largeopps\largecapacity$ is an integer. This assumption does not impact the upper bound in the statement of Proposition \ref{prop:warmupmsvv}, as the expression comes from taking the limit as $\largeopps$ approaches $\infty$.} All \vols\ have conversion probabilities of $1$ or $0$, and if $\mu_{i,t}=1$ (resp. 0), we will refer to \opp\ $i$ and \vol\ $t$ as \emph{compatible} (resp. incompatible). 

To help describe the compatibility structure of the arriving \vols, we first define constants $\hat{\msvvhelper}_1$ and $\hat{\msvvhelper}_2$, where the former is the unique solution in $[0,1]$\footnote{We note that for any $\extfrac \in [0,1]$, it is easy to verify algebraically that there is a unique solution in the interval $[0,1]$ for $\hat{\msvvhelper}_1$.} to $$\extfrac = \hat{\msvvhelper}_1 + (1-\hat{\msvvhelper}_1)\Big(\text{{exp}}\big(-\hat{\msvvhelper}_1/(1-\hat{\msvvhelper}_1)\big)-1\Big),$$ and the latter is defined as $$\hat{\msvvhelper}_2 = 1- \frac{1-\hat{\msvvhelper}_1}{\text{{exp}}\left(\text{{exp}}(-\hat{\msvvhelper}_1/(1-\hat{\msvvhelper}_1))\right)}.$$

We illustrate the arrival sequence (and its associated compatibility structure) for this family of instances in Figure \ref{fig:hardnessexample_upper}. To be precise, the $\extfrac \largeopps \largecapacity$ \exttraf\ \vols\ arrive first, and for each \opp\ $i \in \{1, \dots, \hat{\msvvhelper}_1\largeopps\}$, there are $\largecapacity\left(1-\left(\frac{(1-\hat{\msvvhelper}_1)\largeopps}{(1-\hat{\msvvhelper}_1)\largeopps + 1}\right)^i\right)$ compatible \exttraf\ arrivals for that \opp. 
After the arrival of the last \exttraf, the \inttraf\ arrives, according to the following compatibility structure: for each \opp\ $i \in [\largeopps]$, there is a batch of $\Delta_i$ sequentially-arriving homogeneous \vols. For each $i \in \{1, \dots, \hat{\msvvhelper}_1\largeopps\}$, there are $\Delta_i = \largecapacity\left(\frac{(1-\hat{\msvvhelper}_1)\largeopps}{(1-\hat{\msvvhelper}_1)\largeopps + 1}\right)^i$ \vols\ who are compatible with all \opps\ $j \geq i$. In addition, for each $i \in \{\hat{\msvvhelper}_1\largeopps + 1, \dots, \largeopps\}$, there are $\Delta_i = \largecapacity$ \vols\ who are again compatible with all \opps\ $j \geq i$. 

\begin{figure}[t]
 \centering
    \includegraphics[trim={5cm 10cm 3.5cm 12.5cm},clip,width=.55\textwidth]{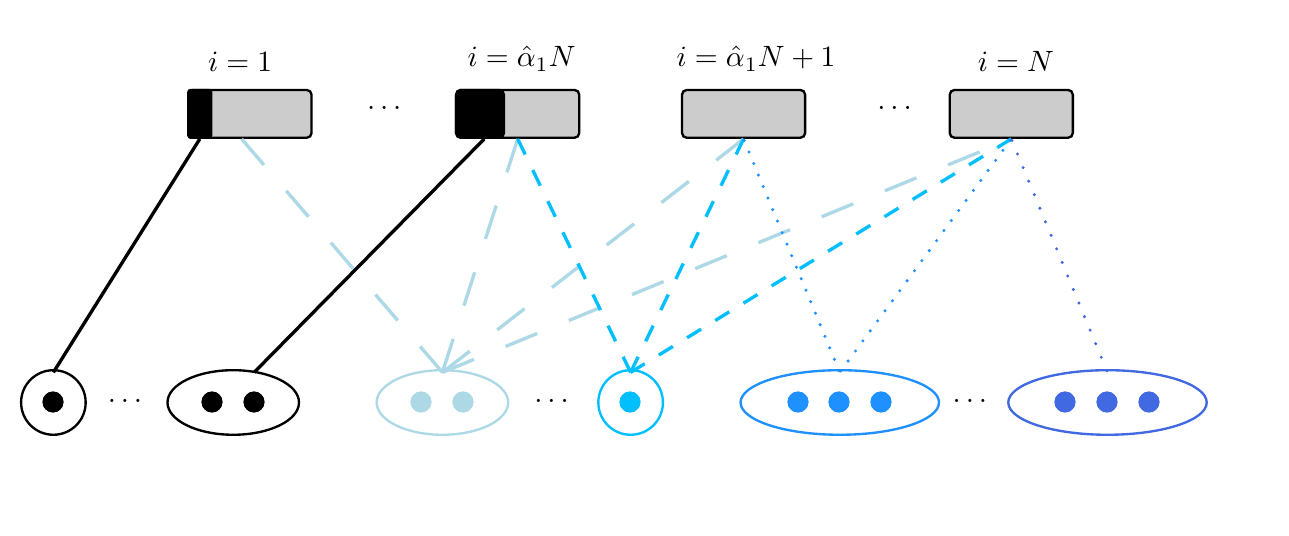}
 \caption{The family of instances generating the upper bound on $\MSVV$ when all \exttraf\ arrives first.}
     \label{fig:hardnessexample_upper}
\end{figure}

First, we verify that the \fracextname\ is equal to $\extfrac$ in the limit as $\largeopps$ gets large.

\begin{align}
    \frac{1}{\largeopps\largecapacity}\sum_{i = 1}^{\hat{\msvvhelper}_1\largeopps} \largecapacity\left(1-\left(\frac{(1-\hat{\msvvhelper}_1)\largeopps}{(1-\hat{\msvvhelper}_1)\largeopps + 1}\right)^i\right) & = \frac{1}{\largeopps\largecapacity}\sum_{i = 1}^{\hat{\msvvhelper}_1\largeopps} \left[\largecapacity\left(1-\left(1- \frac{1}{(1-\hat{\msvvhelper}_1)\largeopps + 1}\right)^i\right)\right]
    \nonumber \\ 
    & \xrightarrow{\largeopps \rightarrow \infty}  \int_0^{\hat{\msvvhelper}_1} \left[1 - \text{exp}\left(\frac{-x}{1-\hat{\msvvhelper}_1}\right) \ \partial x \right] \label{eq:prop2:beta1} \\
    & =  \left(\hat{\msvvhelper}_1 + (1-\hat{\msvvhelper}_1)\left(\text{exp}\left(\frac{-\hat{\msvvhelper}_1}{1-\hat{\msvvhelper}_1}\right) -1 \right)\right) \nonumber \\
    & =  \extfrac \label{eq:prop2:beta2}
\end{align}
In \eqref{eq:prop2:beta1}, we use the fact that $(1-1/n)^{nx}$ approaches $\text{exp}^{-x}$ as $n$ approaches infinity. Furthermore, \eqref{eq:prop2:beta2} follows by applying the definition of $\hat{\msvvhelper}_1$. Next, we analyze the value of \MSVV\ and \OPT\ on the above family of instances via the following two claims.

\begin{clm}
\label{clm:msvv}
For any $\fracextname$ $\extfrac$, the fraction of total capacity filled under \MSVV\ on $\instance_2(\extfrac)$ is at most 
$\hat{\msvvhelper}_2$. 
\end{clm}
\begin{proof}{Proof of Claim \ref{clm:msvv}}
To prove this claim, we will bound the amount of filled capacity for each \opp\ under \MSVV. First, we will show that the $\hat{\msvvhelper}_1\largeopps$ \opps\ that receive \exttraf\ do not receive any matches from \inttraf; i.e., for each $i \in [\hat{\msvvhelper}_1\largeopps]$, we will show that  $\MSVV_{i,\horizon} = \largecapacity\left(1-\left(\frac{(1-\hat{\msvvhelper}_1)\largeopps}{(1-\hat{\msvvhelper}_1)\largeopps + 1}\right)^i\right)$. Suppose towards a contradiction that there exists some \opp\ $j \in [\hat{\msvvhelper}_1\largeopps]$ which receives a match from \inttraf\ under \MSVV. Due to restrictions on compatibility, this match must have come from one of the first $j$ batches of \inttraf, which in  total represents
\begin{align}
\sum_{i=1}^j \Delta_i = \sum_{i=1}^j\largecapacity\left(\frac{(1-\hat{\msvvhelper}_1)\largeopps}{(1-\hat{\msvvhelper}_1)\largeopps + 1}\right)^i = \largecapacity \left((1-\hat{\msvvhelper}_1)\largeopps\right) \left(1-\left(\frac{(1-\hat{\msvvhelper}_1)\largeopps}{(1-\hat{\msvvhelper}_1)\largeopps + 1}\right)^j\right) \label{eq:fillmsvvprop}
\end{align}
\inttraf\ \vols. We are supposing that one of these \vols\ was allocated to \opp\ $j$. In that case, due to the pigeonhole principle, there must be at least one \opp\ $j'$ -- from among the $(1-\hat{\msvvhelper}_1)\largeopps$ \opps\ that \emph{did not} receive \exttraf\ -- with a filled capacity strictly less than $\largecapacity\left(1-\left(\frac{(1-\hat{\msvvhelper}_1)\largeopps}{(1-\hat{\msvvhelper}_1)\largeopps + 1}\right)^j\right)$ upon the arrival of the last \vol\ in batch $j$. By definition, $\MSVV$ should never have recommended $j$ ahead of $j'$, giving us a contradiction.

Next, we show that each \opp\ $i \in \{\hat{\msvvhelper}_1\largeopps+1, \dots, \largeopps\}$ has a filled capacity of
$$\MSVV_{i,\horizon} = \min\left\{\largecapacity, \largecapacity\left(1-\left(\frac{(1-\hat{\msvvhelper}_1)\largeopps}{(1-\hat{\msvvhelper}_1)\largeopps + 1}\right)^{\hat{\msvvhelper}_1\largeopps} + \sum_{j=\hat{\msvvhelper}_1\largeopps+1}^{i}\frac{1}{\largeopps - j + 1}\right)\right\}.$$

Note that in this matching setting, $\MSVV$ recommends \opps\ to equalize their fill rate. Thus, after the arrival of the ${\hat{\msvvhelper}_1\largeopps}$\textsuperscript{th} batch of \vols, all \opps\ $j \in \{\hat{\msvvhelper}_1\largeopps+1, \dots, \largeopps\}$ have an equal amount of filled capacity of $\largecapacity\left(1-\left(\frac{(1-\hat{\msvvhelper}_1)\largeopps}{(1-\hat{\msvvhelper}_1)\largeopps + 1}\right)^{\hat{\msvvhelper}_1\largeopps}\right)$, based on the analysis in the above paragraph (i.e., Equation \eqref{eq:fillmsvvprop}).\footnote{We allow $\largecapacity$ to be sufficiently large such that there is vanishing integrality gap.} For the subsequent batches of \vols, i.e., for $j \in \{\hat{\msvvhelper}_1\largeopps + 1, \dots, \hat{\msvvhelper}_2\largeopps\}$,   \MSVV\ will maintain an equal fill rate among all compatible \opps\ by evenly distributing the $\Delta_j = \largecapacity$ arriving \vols\ in batch $j$ among the $\largeopps - j + 1$ compatible \opps. \revcolor{After the final arrival in batch $\hat{\msvvhelper}_2\largeopps$, all remaining compatible \opps\ will have reached their capacity.} Consequently, after the final arrival in batch $j$ (which is the last \vol\ compatible with \opp\ $j$), \opp\ $j$ will either have reached capacity or will have a filled capacity of 
$$\largecapacity\left(1-\left(\frac{(1-\hat{\msvvhelper}_1)\largeopps}{(1-\hat{\msvvhelper}_1)\largeopps + 1}\right)^{\hat{\msvvhelper}_1\largeopps}\right) + \sum_{j=\hat{\msvvhelper}_1\largeopps+1}^{i}\frac{\largecapacity}{\largeopps - j + 1}.$$


To compute the fraction of total capacity filled under $\MSVV$ on $\instance_2(\extfrac)$, we then take an average over the fill rate of all \opps. To that end, we first compute the fill rate for each \opp\ in the limit as the number of \opps\ approaches infinity. 

For $i \in [\hat{\msvvhelper}_1\largeopps]$,
\begin{align*}
    \fillrate_{i, \horizon} \quad = \quad 1-\left(\frac{(1-\hat{\msvvhelper}_1)\largeopps}{(1-\hat{\msvvhelper}_1)\largeopps + 1}\right)^i
\end{align*}
Each \opp\ $i \in \{\hat{\msvvhelper}_1\largeopps+1, \dots, \hat{\msvvhelper}_2\largeopps\}$ will not reach capacity, and thus its fill rate approaches:
\begin{align*}
    \fillrate_{i, \horizon} \quad & = \quad  1-\left(\frac{(1-\hat{\msvvhelper}_1)\largeopps}{(1-\hat{\msvvhelper}_1)\largeopps + 1}\right)^{\hat{\msvvhelper}_1\largeopps} + \sum_{j=\hat{\msvvhelper}_1\largeopps+1}^{i}\frac{1}{\largeopps - j + 1} \\
    & = \quad 1-\left(\frac{(1-\hat{\msvvhelper}_1)\largeopps}{(1-\hat{\msvvhelper}_1)\largeopps + 1}\right)^{\hat{\msvvhelper}_1\largeopps} + \sum_{k = \largeopps - i +1}^{(1-\hat{\msvvhelper}_1)\largeopps} \frac{1}{k}
\end{align*}
It is easy to verify algebraically that for $i = \hat{\msvvhelper}_2 \largeopps$, the fill rate of \opp\ $i$, $\fillrate_{i,\horizon}$, asymptotically approaches $1$. The remaining \opps\ reach capacity.

With this in mind, the fraction of filled capacity under $\MSVV$ can be computed as follows:
\begin{align}
\frac{1}{\largeopps}\sum_{i \in [\largeopps]}\fillrate_{i,\horizon} =& \frac{1}{\largeopps}\left(\sum_{i=1}^{\hat{\msvvhelper}_1\largeopps}\left(1-\left(\frac{(1-\hat{\msvvhelper}_1)\largeopps}{(1-\hat{\msvvhelper}_1)\largeopps + 1}\right)^i\right) \right. \nonumber \\  &\left.\qquad \quad \quad +\sum_{i=\hat{\msvvhelper}_1\largeopps+1}^{\hat{\msvvhelper}_2 \largeopps} \left(1-\left(\frac{(1-\hat{\msvvhelper}_1)\largeopps}{(1-\hat{\msvvhelper}_1)\largeopps + 1}\right)^{\hat{\msvvhelper}_1\largeopps} + \sum_{k = \largeopps - i +1}^{(1-\hat{\msvvhelper}_1)\largeopps} \frac{1}{k}\right) +\sum_{i=\hat{\msvvhelper}_2\largeopps+1}^{\largeopps} 1 \right) \nonumber \\
\xrightarrow{\largeopps \rightarrow \infty}& \int_0^{\hat{\msvvhelper}_1} 1-\text{exp}\left(\frac{-x}{1-\hat{\msvvhelper}_1}\right) \ \partial x + 
    \int_{\hat{\msvvhelper}_1}^{\hat{\msvvhelper}_2} 1-\text{exp}\left(\frac{-\hat{\msvvhelper}_1}{1-\hat{\msvvhelper}_1}\right) + \log\left(\frac{1-\hat{\msvvhelper}_1}{1-x} \right) \ \partial x  + (1- \hat{\msvvhelper}_2) \label{eq:prop2limitline} \\
    =&  
    \hat{\msvvhelper}_1 - (1-\hat{\msvvhelper}_1)\left(1-\text{exp}\left(\frac{-\hat{\msvvhelper}_1}{1-\hat{\msvvhelper}_1}\right)\right)
    + (\hat{\msvvhelper}_2-\hat{\msvvhelper}_1)\left[ 1-\text{exp}\left(\frac{-\hat{\msvvhelper}_1}{1-\hat{\msvvhelper}_1}\right)\right] \nonumber \\ & \qquad \quad \quad + \int_{\hat{\msvvhelper}_1}^{\hat{\msvvhelper}_2}\log\left(\frac{1-\hat{\msvvhelper}_1}{1-x} \right) \ \partial x \nonumber + (1- \hat{\msvvhelper}_2) \\
    =& (1-\hat{\msvvhelper}_2)\Big(\text{{exp}}\big(-\hat{\msvvhelper}_1/(1-\hat{\msvvhelper}_1)\big)+\log\big((1-\hat{\msvvhelper}_2)/(1-\hat{\msvvhelper}_1)\big)\Big) + \hat{\msvvhelper}_2 \nonumber \\
    =& \hat{\msvvhelper}_2 \nonumber
\end{align}
In \eqref{eq:prop2limitline}, we again use the fact that $(1-1/n)^{nx}$ approaches $e^{-x}$ as $n$ approaches infinity. Furthermore, we use the fact that $\sum_{k = yn}^{xn}1/k$ approaches $\log(x/y)$ as $n$ approaches infinity. The last equality comes from applying the definition of $\hat{\msvvhelper}_2$ to see that $\log\big((1-\hat{\msvvhelper}_2)/(1-\hat{\msvvhelper}_1)\big) = -\text{{exp}}\big(-\hat{\msvvhelper}_1/(1-\hat{\msvvhelper}_1)\big)$. This completes the proof of Claim \ref{clm:msvv}. \halmos
\end{proof}

\begin{clm}
\label{clm:msvvopt}
For any \fracextname\ $\extfrac$, \OPT\ fills all capacity on $\instance_2(\extfrac)$.
\end{clm}
\begin{proof}{Proof of Claim \ref{clm:msvvopt}}
Consider a solution which matches all \exttraf\ and then matches each of the $\Delta_i$ \inttraf\ \vols\ in batch $i$ to \opp\ $i$. To see why such a solution gives a perfect matching, note that each \opp\ $i \in \{1, \dots, \hat{\msvvhelper}_1\largeopps\}$ will receive $\largecapacity\left(1-\left(\frac{(1-\hat{\msvvhelper}_1)\largeopps}{(1-\hat{\msvvhelper}_1)\largeopps + 1}\right)^i\right)$ matches from \exttraf\ and $\Delta_i = \largecapacity\left(\frac{(1-\hat{\msvvhelper}_1)\largeopps}{(1-\hat{\msvvhelper}_1)\largeopps + 1}\right)^i$ matches from \inttraf, leading to a total of $\largecapacity$ matches. Each \opp\ $i \in \{\hat{\msvvhelper}_1\largeopps + 1, \dots, \largeopps\}$ will receive $\Delta_i = \largecapacity$ matches (all from \inttraf). Thus, each \opp\ is filled to capacity under this solution, which implies that the optimal solution must also fill all capacity. \halmos 
\end{proof}

Combining Claims \ref{clm:msvv} and \ref{clm:msvvopt}, we see that \MSVV\ only fills a fraction $\hat{\msvvhelper}_2$ of the capacity filled by \OPT\ on this family of instances, which provides a parameterized upper bound on the competitive ratio of \MSVV\ in this setting.\footnote{To show that this upper bound holds for any minimum capacity $\invbidtobudget$, it suffices to add an additional \opp\ with capacity $\invbidtobudget$ for which \vols\ have conversion probability of $0$. The performance of both $\OPT$ and $\MSVV$ are unchanged, and the \fracextname\ remains the same in the limit as $\largeopps$ approaches infinity.}


\subsection{Proof of Proposition \ref{prop:warmup} (Section \ref{subsec:results:extknown})}
\label{proof:prop:warmup}
We prove Proposition \ref{prop:warmup} in two steps. In {\bf Step (a)}, fixing an instance $\instance \in \instancedomain_\extfrac$, we show that the expected fraction of capacity filled by \exttraf\ is $\extfrac$ under both $\AC$ and $\OPT$. Then, in {\bf Step (b)} we establish a lower bound on the amount of capacity filled by \inttraf\ under $\AC$, which depends on the amount of capacity filled by \inttraf\ under $\OPT$. Together, these steps enable us to place a lower bound on the competitive ratio of $\AC$, where the bound is parameterized by the \fracextname\ $\extfrac$.

\textbf{Step (a):} Both $\OPT$ and $\adaptivecapmath$ always recommend the targeted \opp\ $\extrecommend$ to \exttraf. Since \exttraf\ is assumed to arrive before all \inttraf, this \exttraf\ will fill a fraction of capacity given by
\begin{equation}
\extfrac(\instance) = \frac{\sum_{i \in [\numopps]} \mathbb{E}\left[\min\{\capa_i, \sum_{t \in \exttimes}\mathbbm{1}[\volchoicet{\extrecommend} = i] \}\right] }{\sum_{i \in [\numopps]} \capa_i}. \nonumber
\end{equation}
We note that this fraction of capacity is exactly equivalent to the definition of the \fracextname\ (see Definition \ref{def:beta}), which is equal to $\extfrac$ for any instance $\instance \in \instancedomain_\extfrac$.

\textbf{Step (b):}  Fixing an instance $\instance$, we now turn our attention to lower-bounding the expected amount of capacity filled by \inttraf\ under the \AC\ algorithm, where the expectation is taken over \emph{sample paths} $\samplepath = \{\randomdraw_1, \dots, \randomdraw_\horizon\}$, i.e., realizations of random variables that govern \vol\ choices in this instance.
Formally, we interpret $\randomdraw_t$ as a vector of length $n+1$, where the $i$\textsuperscript{th} component of $\randomdraw_t$ (denoted $\samplepathcomponent_{i, t}$) indicates \vol\ $t$'s \signup\ decision if the platform were to recommend \opp\ $i$. 
For a fixed instance $\instance$ and a fixed sample path $\samplepath$, we use $\widehat{\AC}$ to denote the amount of capacity filled by \inttraf\ under the $\AC$ algorithm.\footnote{Even though $\widehat{\AC}$ depends on the instance and the sample path, we hereafter suppress this dependence to ease exposition (for $\widehat{\AC}$ as well as for all other quantities that depend on the instance and the sample path).}

Our lower bound on $\mathbb{E}_{\samplepath}\big[\widehat{\AC}\big]$ will depend on the expected amount of capacity filled by \inttraf\ under $\OPT$, 
which we likewise denote with $\mathbb{E}_{\samplepath}\big[\widehat{\OPT}\big]$ \revcolor{(We note that the expectations are taken with respect to all possible realizations of volunteer sign-up decisions, i.e., all sample paths $\samplepath$.)} Note that in this step of the proof, we are concerned only with the remaining capacities for each \opp\ $i$ after the arrival of \exttraf\ 
(denoted $\hat{c}_i$), 
which depends on the realizations of sign-ups made by \exttraf. 
As such, $\hat{c}_i$ depends not only on the instance $\instance$, but also on the sample path $\samplepath$.

To provide such a lower bound, we leverage the LP-free approach developed in \citet{goyal2019online} and \citet{goyal2020asymptotically}, which involves the creation of path-based pseudo-rewards. (For a more complete discussion of the intuition behind this approach, we kindly refer to the proof sketch of Theorem \ref{thm:AClower} in Section \ref{subsubsec:proof}.)
For a fixed instance $\instance$ and a fixed sample path $\samplepath$, we define the pseudo-rewards $\ballvarwarmup_t$ for all $t \in \inttimes$ and $\binvarwarmup_i$ for all $i \in [\numopps]$ according to the following:

\begin{align}
    \ballvarwarmup_t &=
     \sum_{i \in [\numopps]} \balancefunc(\fillrate_{i,t-1})\mathbbm{1}[\volchoicet{\OPT} = i] \label{eq:ballvarwarmup}
    \\
    \binvarwarmup_i &=  \sum_{t \in \inttimes} \left(1-\balancefunc(\fillrate_{i,t-1})\right)\mathbbm{1}[\volchoicet{\adaptivecapmath} = i], \label{eq:binvarwarmup}
\end{align}
where we remind that under the $\AC$ algorithm, $\fillrate_{i,t-1} = \AC_{i,t}^{\inttrafmath}/(\capa_i -\AC_{i,t}^{\exttrafmath})$. This is equivalent to $\AC_{i,t}^{\inttrafmath}/\hat{c}_i$ in our warm-up setting where \exttraf\ arrives first and the remaining capacity for \opp\ $i$ is given by $\hat{c}_i$.
We now prove that the expected sum of these pseudo-rewards serves as a lower bound on the expected value of $\widehat{\AC}$. 
\begin{lem}
\label{lem:acboundwarmup}
For any instance $\instance$,
\begin{equation}
      \mathbbm{E}_{\samplepath}\big[\widehat{\AC}\big] \quad \geq \quad \mathbbm{E}_{\samplepath}\left[\sum_{t \in \inttimes} \ballvarwarmup_t + \sum_{i \in [\numopps]} \binvarwarmup_i\right],
\end{equation}
where $\ballvarwarmup_t$ and $\binvarwarmup_i$ are defined in \eqref{eq:ballvarwarmup} and \eqref{eq:binvarwarmup}, respectively.
\end{lem}
\begin{proof}{Proof of Lemma \ref{lem:acboundwarmup}:}
The proof follows from the definition of $\ballvarwarmup_t$ and $\binvarwarmup_i$ as well as the design of the \adaptivecap\ algorithm:
\begin{align}
    \mathbbm{E}_{\samplepath}\big[\widehat{\AC}\big] &= \mathbb{E}_{\samplepath} \left[ \sum_{t\in \inttimes}\sum_{i \in [\numopps]} \mathbbm{1}[\volchoicet{\adaptivecapmath} = i]\right] \label{eq:lem1warmup1} \\
    &= \mathbb{E}_{\samplepath} \left[ \sum_{t\in \inttimes}\sum_{i \in [\numopps]} \balancefunc(\fillrate_{i,t-1})\mathbbm{1}[\volchoicet{\adaptivecapmath} = i] + \sum_{t\in \inttimes}\sum_{i \in [\numopps]}\Big(1- \balancefunc(\fillrate_{i,t-1})\Big)\mathbbm{1}[\volchoicet{\adaptivecapmath} = i]\right] \\
    &\geq \mathbb{E}_{\samplepath} \left[ \sum_{t\in \inttimes}\sum_{i \in [\numopps]} \balancefunc(\fillrate_{i,t-1})\mathbbm{1}[\volchoicet{\OPT} = i] + \sum_{i \in [\numopps]}\binvarwarmup_i\right] \label{eq:acoptconditionwarmup} \\
    &= \mathbbm{E}_{\samplepath}\left[\sum_{t \in \inttimes} \ballvarwarmup_t + \sum_{i \in [\numopps]} \binvarwarmup_i\right]
\end{align}
Equality \eqref{eq:lem1warmup1} holds because the $\AC$ algorithm will never recommend an \opp\ that has already reached capacity to \inttraf.\footnote{To see this, note that if \opp\ $i$ has reached capacity before time $t$, then $\convprob_{i,t} \cdot \balancefunc(\fillrate_{i,t-1}) = 0$. Based on its convention for breaking ties in favor of the \opp\ with the lowest index, the $\AC$ algorithm would always recommend \opp\ $0$ instead of an at-capacity \opp\ $i$.} Consequently, the amount of capacity filled by \inttraf\ under the $\AC$ algorithm is exactly equal to the numbers of \signups\ from \inttraf.

Inequality \eqref{eq:acoptconditionwarmup} follows from the $\AC$ algorithm's optimality condition (see Algorithm \ref{alg:acpolicy}), which ensures that it recommends the \opp\ that maximizes the weighted probability of generating a \signup\ (where the weight for \opp\ $i$ at time $t$ is given by $\balancefunc(\fillrate_{i,t-1})$). \revcolor{As \OPT\ does not have foreknowledge of the realization of the sign-up decisions of \inttraf, the recommendation provided by \OPT\ to any \vol\ must be independent of their \signup\ realization. Hence, the inequality holds.}  Applying the definition of the pseudo-rewards $\ballvarwarmup_t$ for $t \in \inttimes \setminus \bonustimes$ completes the proof of Lemma \ref{lem:acboundwarmup}. \halmos

\end{proof}

Next, we place a lower bound on the expected sum of the pseudo-rewards, which depends on the amount of capacity of each \opp\ $i$ filled by \inttraf\ under $\OPT$ along a fixed sample path, which we denote by $\widehat{\OPT}_i$.
\begin{lem}
\label{lem:optboundwarmup}
For any instance $\instance$, 
\begin{equation}\mathbbm{E}_{\samplepath}\left[\sum_{t \in \inttimes} \ballvarwarmup_t + \sum_{i \in [\numopps]} \binvarwarmup_i\right] \quad \geq \quad (1-1/e)\mathbbm{E}_{\samplepath}\left[\widehat{\OPT}\right] - \sum_{i \in [\numopps]}\mathbbm{E}_{\samplepath}\left[\mathbbm{1}\big[\widehat{\OPT}_i = \hat{c}_i\big]\right] \end{equation}
\end{lem}
\begin{proof}{Proof of Lemma \ref{lem:optboundwarmup}:}
We will prove this claim along each sample path $\samplepath$ by separately placing lower bounds on the $\ballvarwarmup_t$ pseudo-rewards and the $\binvarwarmup_i$ pseudo-rewards. 
For the former,
\begin{align}
    \sum_{t \in \inttimes} \ballvarwarmup_t \quad &=  \quad \sum_{t \in \inttimes} \sum_{i \in [\numopps]} \balancefunc(\fillrate_{i,t-1})\mathbbm{1}[\volchoicet{\OPT} = i]
    \label{eq:warmuplem2eq1}\\
    &\geq \quad  \sum_{t \in \inttimes} \sum_{i \in [\numopps]} \balancefunc(\fillrate_{i,\horizon})\mathbbm{1}[\volchoicet{\OPT} = i] \label{eq:ballvarlowerboundwarmup} \\
     & = \quad \sum_{i \in [\numopps]} \widehat{\OPT}_i \balancefunc\left(\fillrate_{i,\horizon}\right) \label{eq:ballvarfinalboundwarmup}\end{align}
Equality in \eqref{eq:warmuplem2eq1} follows from the definition of $\ballvarwarmup_t$. Inequality in \eqref{eq:ballvarlowerboundwarmup} holds because $\balancefunc$ is a decreasing function in its argument, and $\fillrate_{i,\horizon} \geq \fillrate_{i,t-1}$ for all $t \in [\horizon]$. All other steps are algebraic.

We now turn our attention to the $\binvarwarmup_i$ pseudo-rewards:
\begin{align}
    \binvarwarmup_i \quad =& \quad \sum_{t \in \inttimes} \left(1-\balancefunc(\fillrate_{i,t-1})\right)\mathbbm{1}[\volchoicet{\adaptivecapmath} = i]  \\
    =& \quad \sum_{t \in \inttimes} \left(1-\balancefunc\left(\frac{\adaptivecapmath_{i, t-1}^{\inttrafmath}}{\hat{c}_i}\right)\right) \mathbbm{1}[\volchoicet{\adaptivecapmath} = i] \\
    \quad =& \quad
    \sum_{\counter \in [\adaptivecapmath_{i, \horizon}^{\inttrafmath}]} \left(1-\balancefunc\left(\frac{\counter-1}{\hat{c}_i}\right)\right)
    \label{eq:reimannsum1warmup} \\
    \quad =& \quad e^{\frac{-1}{\hat{c}_i}}\sum_{\counter \in [\adaptivecapmath_{i, \horizon}^{\inttrafmath}]} \left(1-\balancefunc\left(\frac{\counter}{\hat{c}_i}\right)\right)  \label{eq:reimannsum2warmup} \\
    \quad \geq& \quad  e^{\frac{-1}{\hat{c}_i}}\int_{0}^{\adaptivecapmath_{i, \horizon}^{\inttrafmath}} 1 - \balancefunc(x/\hat{c}_i) \ \partial x \label{eq:reimannboundwarmup} \\
    \quad =& \quad  e^{\frac{-1}{\hat{c}_i}}\hat{c}_i\left(1 - \balancefunc \left(\fillrate_{i,\horizon}\right) - 1/e\right) \\
    \quad \geq& \quad  (\hat{c}_i-1)\left(1 - \balancefunc \left(\fillrate_{i,\horizon}\right) - 1/e\right)\label{eq:binvarfinalboundwarmup}
\end{align}
Equality in \eqref{eq:reimannsum1warmup} holds because the counter $\adaptivecapmath_{i, t}^{\inttrafmath}$ will increase by $1$ for any $t \in \inttimes$ where $\volchoicet{\adaptivecapmath} = i$. The summation in \eqref{eq:reimannsum1warmup} represents a left Riemann sum of an increasing function. In \eqref{eq:reimannsum2warmup}, we utilize the fact that for any $\counter$, $1-\balancefunc((\counter-1)/\hat{c}_i) = e^{-1/\hat{c}_i}(1-\balancefunc(\counter/\hat{c}_i))$. As the summation in \eqref{eq:reimannsum2warmup} is now a right Riemann sum of an increasing function, we bound the sum with an appropriate integral in \eqref{eq:reimannboundwarmup}. Finally, \eqref{eq:binvarfinalboundwarmup} holds because $e^{-x} \geq 1-x$ for any $x$.

Combining \eqref{eq:ballvarfinalboundwarmup} and \eqref{eq:binvarfinalboundwarmup}, we see that for each sample path $\samplepath$, 
\begin{align}
    \sum_{t \in \inttimes} \ballvarwarmup_t + \sum_{i \in [\numopps]} \binvarwarmup_i &\geq \sum_{i \in [\numopps]} \widehat{\OPT}_i \balancefunc\left(\fillrate_{i,\horizon}\right) + (\hat{c}_i-1)\left(1 - \balancefunc \left(\fillrate_{i,\horizon}\right) - 1/e\right) \\
    &\geq \sum_{i \in [\numopps]} \left(\widehat{\OPT}_i - \mathbbm{1}[\widehat{\OPT}_{i} = \hat{c}_i]\right) \balancefunc\left(\fillrate_{i,\horizon}\right) + \left(\widehat{\OPT}_i - \mathbbm{1}[\widehat{\OPT}_{i} = \hat{c}_i]\right)\left(1 - \balancefunc \left(\fillrate_{i,\horizon}\right) - 1/e\right) \label{eq:warmuplemfinal}\\
    &= (1-1/e) \sum_{i \in [\numopps]} \left(\widehat{\OPT}_i - \mathbbm{1}\big[\widehat{\OPT}_i = \hat{c}_i\big]\right) \\
    &\geq (1-1/e)\cdot \widehat{\OPT} - \sum_{i \in [\numopps]}\mathbbm{1}\big[\widehat{\OPT}_i = \hat{c}_i\big]
\end{align}
Inequality in \eqref{eq:warmuplemfinal} comes from noting that $\left(\widehat{\OPT}_i - \mathbbm{1}\big[\widehat{\OPT}_i = \hat{c}_i\big]\right)$ cannot exceed either $\widehat{\OPT}_i$ or $\hat{c}_i - 1$. (We note that the binary indicator $\mathbbm{1}\big[\widehat{\OPT}_i = \hat{c}_i\big]$ is equal to $1$ if and only if \opp\ $i$ reaches capacity under $\OPT$ along the fixed sample path $\samplepath$.)
 Taking expectation across all sample paths completes the proof of Lemma \ref{lem:optboundwarmup}.
\halmos
\end{proof}

Combining Lemmas \ref{lem:acboundwarmup} and \ref{lem:optboundwarmup}, we see that we can bound the expected amount of capacity filled by \inttraf\ under $\AC$ via the following inequality:

\begin{equation}
      \mathbbm{E}_{\samplepath}\big[\widehat{\AC}\big] \quad \geq \quad  (1-1/e)\mathbbm{E}_{\samplepath}\big[\widehat{\OPT}\big] - \sum_{i \in [\numopps]}\mathbbm{E}_{\samplepath}\left[\mathbbm{1}\big[\widehat{\OPT}_i = \hat{c}_i\big]\right]
\end{equation}

Together with Step (a), we have shown that for any instance $\instance \in \instancedomain_\extfrac$,
\begin{align}
    \frac{\mathbbm{E}_{\samplepath}\left[\AC\right]}{\mathbbm{E}_{\samplepath}\left[\OPT\right]} \quad &= \quad \frac{\extfrac \cdot \sum_{i \in [\numopps]} \capa_i + \mathbbm{E}_{\samplepath}\left[\widehat{\adaptivecapmath}\right]}{\extfrac \cdot \sum_{i \in [\numopps]} \capa_i+\mathbbm{E}_{\samplepath}\left[\widehat{\OPT}\right]} \label{eq:prop3last1}\\
    &\geq \quad \frac{\extfrac \cdot \sum_{i \in [\numopps]} \capa_i + (1-1/e)\mathbbm{E}_{\samplepath}\left[\widehat{\OPT}\right] - \sum_{i \in [\numopps]}\mathbbm{E}_{\samplepath}\left[\mathbbm{1}\big[\widehat{\OPT}_i = \hat{c}_i\big]\right]}{\extfrac \cdot \sum_{i \in [\numopps]} \capa_i+\mathbbm{E}_{\samplepath}\left[\widehat{\OPT}\right]} \label{eq:prop3last2} \\
    & = \quad \frac{\extfrac \cdot \sum_{i \in [\numopps]} \capa_i + (1-1/e)\mathbbm{E}_{\samplepath}\left[ \widehat{\OPT}\right]}{\extfrac \cdot \sum_{i \in [\numopps]} \capa_i+\mathbbm{E}_{\samplepath}\left[\widehat{\OPT}\right]} - \frac{\sum_{i \in [\numopps]}\mathbbm{E}_{\samplepath}\left[\mathbbm{1}\big[\widehat{\OPT}_i = \hat{c}_i\big]\right]}{\mathbbm{E}_{\samplepath}\left[\OPT\right]} \label{eq:prop3last3}
    \\& \geq \quad \frac{\extfrac \cdot \sum_{i \in [\numopps]} \capa_i + (1-1/e)\mathbbm{E}_{\samplepath}\left[ \widehat{\OPT}\right]}{\extfrac \cdot \sum_{i \in [\numopps]} \capa_i+\mathbbm{E}_{\samplepath}\left[\widehat{\OPT}\right]} - \invbidtobudget^{-1}
     \label{eq:warmupcmintrick}
    \\ &\geq \quad \extfrac + (1-\extfrac)(1-1/e) - \invbidtobudget^{-1} \label{eq:prop3lastlast}
\end{align}
Equality in \eqref{eq:prop3last1} comes from applying the result of Step (a), while inequality in \eqref{eq:prop3last2} comes from applying the result of Step (b). Equality in \eqref{eq:prop3last2} follows from the definition of $\OPT$.
To see that \eqref{eq:warmupcmintrick} holds, we first fix a sample path. Along that sample path, if $\widehat{\OPT}_i = \hat{c}_i$, then \opp\ $i$ must have reached capacity under $\OPT$. The capacity of \opp\ $i$ is at least $\invbidtobudget$. Thus, along every sample path, $\OPT \geq \invbidtobudget \sum_{i \in [\numopps]}\mathbbm{E}_{\samplepath}\left[\mathbbm{1}\big[\widehat{\OPT}_i = \hat{c}_i\big]\right]$. This is a sufficient condition to establish \eqref{eq:warmupcmintrick}.  

Finally, \eqref{eq:prop3lastlast} comes from noting that the expression in \eqref{eq:warmupcmintrick} is decreasing in $\mathbbm{E}_{\samplepath}\left[\widehat{\OPT}\right]$, which can be at most $ \mathbbm{E}_{\samplepath}\left[\sum_{i \in [\numopps]}\hat{c}_i\right]$. Furthermore, $\mathbbm{E}_{\samplepath}\left[\sum_{i \in [\numopps]}  \hat{c}_i\right] = (1-\extfrac)\sum_{i \in [\numopps]} \capa_i$. We then plug in this upper bound for $\mathbbm{E}_{\samplepath}\left[\widehat{\OPT}\right]$. This final inequality establishes a lower bound for any instance $\instance \in \instancedomain_\extfrac$. Thus, it represents a lower bound on the competitive ratio parameterized by the \fracextname\ $\extfrac$, as desired. This completes the proof of Proposition \ref{prop:warmup}.
\halmos

\subsection{Upper Bound on \MSVV\ in General Settings (Section \ref{subsubsec:thm2:discussion})}
\label{proof:prop:generalmsvv}
In the following proposition, we provide an upper bound on the competitive ratio of \MSVV\ as a function of the \fracextname\ $\extfrac$.
\begin{prop}[Upper Bound on \MSVV]
\label{prop:generalmsvv}
For any \MakeLowercase{\fracextnamefull} $\extfrac$ and any minimum capacity, 
$\MSVV$ cannot achieve a competitive ratio better than
\begin{equation}
\begin{cases}
    1-1/e, & \extfrac \leq 1/e \\
    \min\left\{\msvvhelper_2, \msvvhelper_3\right\}, & \extfrac > 1/e
\end{cases}
\nonumber 
\end{equation}
where, for $\extfrac > 1/e$, $\msvvhelper_2$ is given by
\begin{equation}
\msvvhelper_2 =
    1- \frac{1-\msvvhelper_1}{\text{\emph{exp}}\left(\text{\emph{exp}}(-\msvvhelper_1/(1-\msvvhelper_1))\right)} \nonumber
\end{equation}
and $\msvvhelper_1$ and is the unique solution in $[0,1]$ to $\extfrac = \msvvhelper_1 + (1-\msvvhelper_1)\Big(\text{\emph{exp}}\big(-\msvvhelper_1/(1-\msvvhelper_1)\big)-1\Big) + \frac{1-\msvvhelper_1}{\text{\emph{exp}}\left(\text{\emph{exp}}(-\msvvhelper_1/(1-\msvvhelper_1))\right)}$. 
In addition, $$\msvvhelper_3 = 
    \min_{\msvvhelper_4 \in [0, \extfrac]} 1 - \frac{1-\extfrac}{1-\msvvhelper_4}\left(\msvvhelper_5 + (1-\msvvhelper_6)\log\left(\frac{1-\msvvhelper_5}{1-\msvvhelper_6}\right) \right),$$
where $\msvvhelper_5 = \min\{1-\msvvhelper_4, \frac{\msvvhelper_4(\extfrac-\msvvhelper_4)}{1-\extfrac}\}$ and $\msvvhelper_6 = \min\{1 - \msvvhelper_4, 1-(1-\msvvhelper_5)/e\}$.
\end{prop}

\begin{proof}{Proof of Proposition \ref{prop:generalmsvv}}
The first part of Proposition \ref{prop:generalmsvv} -- which establishes an upper bound of $1-1/e$ when $\extfrac \leq 1/e$ -- follows immediately from Theorem \ref{thm:hardness}, in which we prove such an upper bound on the competitive ratio of \emph{any} online algorithm. 

We prove the remainder of this proposition in two claims by showing two different upper bounds ($\msvvhelper_2$ and $\msvvhelper_3$) on the competitive ratio of \MSVV\ parameterized by the \MakeLowercase{\fracextnamefull} $\extfrac$. Proposition \ref{prop:generalmsvv} follows by taking the minimum of the two upper bounds for a given $\extfrac$. 

To prove each claim, we construct a family of instances parameterized by $\extfrac$. We then evaluate the value of \MSVV\ on that family of instances relative to the value of $\OPT$. Both of the instances that we design leverage the fact that the notion of a fill rate under \MSVV\ does not distinguish between \intandexttraf. As a result, \MSVV\ may mistakenly withhold \inttraf\ from \opps\ that have previously received \exttraf. Furthermore, all instances leverage the triangular structure of our general hardness result (see Appendix \ref{proof:thm:hardness}). 

\begin{clm}
\label{clm:generalmsvv_small}
For any \MakeLowercase{\fracextnamefull} $\extfrac \in (1/e, 1]$, the competitive ratio of $\MSVV$ is at most 
\begin{equation}
    1- \frac{1-\msvvhelper_1}{\text{\emph{exp}}\left(\text{\emph{exp}}(-\msvvhelper_1/(1-\msvvhelper_1))\right)}\label{eq:msvvsmallbound}
\end{equation}
where $\msvvhelper_1$ is the unique solution in $[0,1]$ to $\extfrac = \msvvhelper_1 + (1-\msvvhelper_1)\Big(\text{\emph{exp}}\big(-\msvvhelper_1/(1-\msvvhelper_1)\big)-1\Big) + \frac{1-\msvvhelper_1}{\text{\emph{exp}}\left(\text{\emph{exp}}(-\msvvhelper_1/(1-\msvvhelper_1))\right)}$. 
\end{clm}
\begin{proof}{Proof of Claim \ref{clm:generalmsvv_small}}
To prove this claim, we construct a family of instances $\instance_3(\extfrac)$ parameterized by the \fracextname\ $\extfrac$. (As we will highlight below, this family of instances will have a close relationship to the family of instances $\instance_2(\extfrac)$, introduced in the proof of Proposition \ref{prop:warmupmsvv}.) In each instance, there are a large number of \opps\ $\largeopps$, each with identical large capacity $\largecapacity$.  The arrival sequence consists of $\largeopps \largecapacity$ \vols, and for a given \MakeLowercase{\fracextnamefull} $\extfrac$, the first $\extfrac \largeopps \largecapacity$ of these \vols\ are \exttraf.\footnote{We assume that $(1-\extfrac)\largeopps\largecapacity$ is an integer. This assumption does not impact the upper bound in the statement of Claim \ref{clm:generalmsvv_small}, as the expression comes from taking the limit as $\largeopps$ approaches $\infty$.} All \vols\ have conversion probabilities of $1$ or $0$, and if $\mu_{i,t}=1$ (resp. 0), we will refer to \opp\ $i$ and \vol\ $t$ as \emph{compatible} (resp. incompatible). 

To help describe the compatibility structure of the arriving \vols, we first define constants ${\msvvhelper}_1$ and ${\msvvhelper}_2$. For $\extfrac \leq 1/e$, we define $\msvvhelper_1 = 0$, while for $\extfrac > 1/e$, we define $\msvvhelper_1$ as the unique solution in $[0,1]$\footnote{We note that for any $\extfrac \in (1/e,1]$, it is easy to verify numerically that there is a unique solution in the interval $[0,1]$ for ${\msvvhelper}_1$.} to $$\extfrac = {\msvvhelper}_1 + (1-{\msvvhelper}_1)\Big(\text{{exp}}\big(-{\msvvhelper}_1/(1-{\msvvhelper}_1)\big)-1\Big) + \frac{1-\msvvhelper_1}{\text{{exp}}\left(\text{{exp}}(-\msvvhelper_1/(1-\msvvhelper_1))\right)} ,$$ and $\msvvhelper_2$ is defined as $${\msvvhelper}_2 = 1- \frac{1-{\msvvhelper}_1}{\text{{exp}}\left(\text{{exp}}(-{\msvvhelper}_1/(1-{\msvvhelper}_1))\right)}.$$

The arrival sequence begins with \exttraf\ \vols\ for the first $\msvvhelper_1\largeopps$ \opps. Specifically, for each \opp\ $i \in \{1, \dots, {\msvvhelper}_1\largeopps\}$, there are $\largecapacity\left(1-\left(\frac{(1-{\msvvhelper}_1)\largeopps}{(1-{\msvvhelper}_1)\largeopps + 1}\right)^i\right)$ compatible \exttraf\ arrivals for that \opp. 
After the arrival of these \vols, the \inttraf\ arrives, according to the following compatibility structure: for each \opp\ $i \in \{1, \dots, {\msvvhelper}_2\largeopps\}$, there is a batch of $\Delta_i$ sequentially-arriving homogeneous \vols. The batches consist of $\Delta_i = \largecapacity\left(\frac{(1-{\msvvhelper}_1)\largeopps}{(1-{\msvvhelper}_1)\largeopps + 1}\right)^i$ \vols\ for each $i \in \{1, \dots, {\msvvhelper}_1\largeopps\}$, and they consist of $\Delta_i = \largecapacity$ \vols\  for each $i \in \{\msvvhelper_1 \largeopps + 1, \dots, {\msvvhelper}_2\largeopps\}$. Volunteers in batch $i$ are compatible with all \opps\ $j \geq i$. 
Finally, the arrival sequence concludes with $(1-\msvvhelper_2)\largeopps$ batches of $\largecapacity$ \exttraf\ \vols, where each batch views (and is compatible with) one \opp\ $i \in \{{\msvvhelper}_2\largeopps + 1, \dots, \largeopps\}$. 

Before analyzing this family of instances, we make two observations. First, this  arrival sequence is quite similar to the arrival sequence in the family of instances $\instance_2(\extfrac)$, which are visualized in Figure \ref{fig:hardnessexample_upper} and which provide our upper bound on \MSVV\ in the setting where all \exttraf\ arrives first (see Proposition \ref{prop:warmupmsvv}). The only difference comes from the last batches of arrivals, which are \exttraf\ in this family of instances (as opposed to \inttraf\ with broader compatibility, as in $\instance_2(\extfrac)$). In both cases, these \vols\ are unable to be allocated under \MSVV\ as their compatible \opps\ have already reached capacity, whereas these \vols\ are allocated under \OPT. Hence, the value of \MSVV\ and the value of \OPT\ are both unchanged. Crucially, though, the \fracextname\ is different in these two instances, due to the change in source of the last-arriving \vols. As a result, for a fixed $\extfrac$, the instance $\instance_2(\extfrac)$ and $\instance_3(\extfrac)$ differ significantly. Instead, $\instance_2(\extfrac)$ and $\instance_3(\extfrac + \hat{\msvvhelper}_2)$ are nearly identical (where $\hat{\msvvhelper}_2$ is a function of $\extfrac$, as defined in the proof of Proposition \ref{prop:warmupmsvv}). This relationship means the upper bound provided by the family of instances $\instance_3(\extfrac)$ is a non-linear transformation of the upper bound provided by the family of instances $\instance_2(\extfrac)$.
Furthermore, we remark that in the limit as $\extfrac$ approaches $1/e$, $\instance_3(\extfrac)$ approaches the instance $\instance_1(1/e)$, which provides our general hardness result presented in Theorem \ref{thm:hardness}. 

We now verify that the \fracextname\ is equal to $\extfrac$ in the limit as $\largeopps$ gets large.

\begin{align}
    \frac{1}{\largeopps\largecapacity}\left(\sum_{i = 1}^{{\msvvhelper}_1\largeopps} \largecapacity\left(1-\left(\frac{(1-{\msvvhelper}_1)\largeopps}{(1-{\msvvhelper}_1)\largeopps + 1}\right)^i\right) + (1-\msvvhelper_2)\largeopps\largecapacity\right)  = & \frac{1}{\largeopps}\left[\sum_{i = 1}^{{\msvvhelper}_1\largeopps} \left(1-\left(1- \frac{1}{(1-{\msvvhelper}_1)\largeopps + 1}\right)^i\right) + (1-\msvvhelper_2)\largeopps\right]
    \nonumber \\ 
     \xrightarrow{\largeopps \rightarrow \infty}&  \int_0^{{\msvvhelper}_1} \left[1 - \text{exp}\left(\frac{-x}{1-{\msvvhelper}_1}\right) \ \partial x \right] + (1-\msvvhelper_2) \label{eq:prop5:beta1} \\
 =&  \left({\msvvhelper}_1 + (1-{\msvvhelper}_1)\left(\text{exp}\left(\frac{-{\msvvhelper}_1}{1-{\msvvhelper}_1}\right) -1 \right)\right) + (1-\msvvhelper_2) \nonumber \\
     =&  \extfrac \label{eq:prop5:beta2}
\end{align}
In \eqref{eq:prop5:beta1}, we use the fact that $(1-1/n)^{nx}$ approaches $e^{-x}$ as $n$ approaches infinity. Furthermore, \eqref{eq:prop5:beta2} follows by applying the definitions of ${\msvvhelper}_2$ and ${\msvvhelper}_1$. Next, we analyze the value of \MSVV\ and \OPT\ on the above family of instances.

\medskip

\noindent {\bf Value of \MSVV\ on Instance $\instance_3(\extfrac)$:}
We will show that for any $\fracextname$ $\extfrac$, the fraction of total capacity filled under \MSVV\ on $\instance_3(\extfrac)$ is at most 
${\msvvhelper}_2$. 
To that end, we will first bound the amount of filled capacity for each \opp\ under \MSVV. First, we will show that the ${\msvvhelper}_1\largeopps$ \opps\ that initially receive \exttraf\ do not receive any matches from \inttraf; i.e., for each $i \in [{\msvvhelper}_1\largeopps]$, we will show that  $\MSVV_{i,\horizon} = \largecapacity\left(1-\left(\frac{(1-{\msvvhelper}_1)\largeopps}{(1-{\msvvhelper}_1)\largeopps + 1}\right)^i\right)$. Suppose towards a contradiction that there exists some \opp\ $j \in [{\msvvhelper}_1\largeopps]$ which receives a match from \inttraf\ under \MSVV. Due to restrictions on compatibility, this match must have come from one of the first $j$ batches of \inttraf, which in total represents $$\sum_{i=1}^j \Delta_i = \sum_{i=1}^j\largecapacity\left(\frac{(1-{\msvvhelper}_1)\largeopps}{(1-{\msvvhelper}_1)\largeopps + 1}\right)^i = \largecapacity \left((1-{\msvvhelper}_1)\largeopps\right) \left(1-\left(\frac{(1-{\msvvhelper}_1)\largeopps}{(1-{\msvvhelper}_1)\largeopps + 1}\right)^j\right)$$ 
\inttraf\ \vols. We are supposing that one of these \vols\ was allocated to \opp\ $j$. In that case, due to the pigeonhole principle, there must be at least one \opp\ $j'$ -- from among the $(1-{\msvvhelper}_1)\largeopps$ \opps\ that \emph{did not} initially receive \exttraf\ -- with a filled capacity strictly less than $\largecapacity\left(1-\left(\frac{(1-{\msvvhelper}_1)\largeopps}{(1-{\msvvhelper}_1)\largeopps + 1}\right)^j\right)$ upon the arrival of the last \vol\ in batch $j$. By definition, $\MSVV$ should never have recommended $j$ ahead of $j'$, giving us a contradiction.

Next, we show that each \opp\ $i \in \{{\msvvhelper}_1\largeopps+1, \dots, {\msvvhelper}_2\largeopps\}$ has a filled capacity of
\begin{align}
\MSVV_{i,\horizon} \leq  \largecapacity\left(1-\left(\frac{(1-{\msvvhelper}_1)\largeopps}{(1-{\msvvhelper}_1)\largeopps + 1}\right)^{{\msvvhelper}_1\largeopps} + \sum_{j={\msvvhelper}_1\largeopps+1}^{i}\frac{1}{\largeopps - j + 1}\right). \label{eq:prop7explainer}
\end{align}

Note that in this matching setting, $\MSVV$ recommends \opps\ to equalize their fill rate. Thus, after the arrival of the ${{\msvvhelper}_1\largeopps}$\textsuperscript{th} batch of \vols, all \opps\ $j \in \{{\msvvhelper}_1\largeopps+1, \dots, \msvvhelper_2 \largeopps\}$ have an equal amount of filled capacity of $\largecapacity\left(1-\left(\frac{(1-{\msvvhelper}_1)\largeopps}{(1-{\msvvhelper}_1)\largeopps + 1}\right)^{{\msvvhelper}_1\largeopps}\right)$, based on the analysis in the above paragraph (i.e., \eqref{eq:prop7explainer}).\footnote{We allow $\largecapacity$ to be sufficiently large such that there is vanishing integrality gap.} For the subsequent batches of \inttraf\ \vols, i.e., for $j \in \{{\msvvhelper}_1\largeopps + 1, \dots, {\msvvhelper}_2\largeopps\}$,   \MSVV\ will maintain an equal fill rate among all compatible \opps\ by evenly distributing the $\Delta_j = \largecapacity$ arriving \vols\ in batch $j$ among the $\largeopps - j + 1$ compatible \opps. Thus, after the final arrival in batch $j$ (which is the last \vol\ compatible with \opp\ $j$), \opp\ $j$ will have a filled capacity of at most
$$\largecapacity\left(1-\left(\frac{(1-{\msvvhelper}_1)\largeopps}{(1-{\msvvhelper}_1)\largeopps + 1}\right)^{{\msvvhelper}_1\largeopps}\right) + \sum_{j={\msvvhelper}_1\largeopps+1}^{i}\frac{\largecapacity}{\largeopps - j + 1}.$$
\revcolor{The remaining \opps\ (i.e., \opps\ $i$ for $i > \msvvhelper_2 \largeopps$) will all have reached capacity following the last arrival in batch $\msvvhelper_2 \largeopps$.}


To compute the fraction of total capacity filled under $\MSVV$ on $\instance_3(\extfrac)$, we then take an average over the fill rate of all \opps. To that end, we first compute the fill rate for each \opp\ in the limit as the number of \opps\ approaches infinity. 

For $i \in [{\msvvhelper}_1\largeopps]$,
\begin{align*}
    \fillrate_{i, \horizon} \quad = \quad 1-\left(\frac{(1-{\msvvhelper}_1)\largeopps}{(1-{\msvvhelper}_1)\largeopps + 1}\right)^i
\end{align*}
Each \opp\ $i \in \{{\msvvhelper}_1\largeopps+1, \dots, {\msvvhelper}_2\largeopps\}$ has a fill rate which is bounded by:
\begin{align*}
    \fillrate_{i, \horizon} \quad & = \quad  1-\left(\frac{(1-{\msvvhelper}_1)\largeopps}{(1-{\msvvhelper}_1)\largeopps + 1}\right)^{{\msvvhelper}_1\largeopps} + \sum_{j={\msvvhelper}_1\largeopps+1}^{i}\frac{1}{\largeopps - j + 1} \\
    & = \quad 1-\left(\frac{(1-{\msvvhelper}_1)\largeopps}{(1-{\msvvhelper}_1)\largeopps + 1}\right)^{{\msvvhelper}_1\largeopps} + \sum_{k = \largeopps - i +1}^{(1-{\msvvhelper}_1)\largeopps} \frac{1}{k} 
\end{align*}
It is easy to verify algebraically that for $i = {\msvvhelper}_2 \largeopps$, the fill rate of \opp\ $i$, $\fillrate_{i,\horizon}$, asymptotically approaches $1$. The remaining \opps\ reach capacity.

With this in mind, the fraction of filled capacity under \MSVV\ can be computed as follows:
\begin{align}
\frac{1}{\largeopps}\sum_{i \in [\largeopps]}\fillrate_{i,\horizon} =& \  \frac{1}{\largeopps}\left(\sum_{i=1}^{{\msvvhelper}_1\largeopps}
\left(1-\left(\frac{(1-{\msvvhelper}_1)\largeopps}{(1-{\msvvhelper}_1)\largeopps + 1}\right)^i\right)
\right. \nonumber \\  &\left.\qquad \quad \quad +\sum_{i={\msvvhelper}_1\largeopps+1}^{{\msvvhelper}_2 \largeopps}
 \left(1-\left(\frac{(1-{\msvvhelper}_1)\largeopps}{(1-{\msvvhelper}_1)\largeopps + 1}\right)^{{\msvvhelper}_1\largeopps} + \sum_{k = \largeopps - i +1}^{(1-{\msvvhelper}_1)\largeopps} \frac{1}{k} \right)
+\sum_{i={\msvvhelper}_2\largeopps+1}^{\largeopps} 1 \right) \nonumber \\
\xrightarrow{\largeopps \rightarrow \infty}& \int_0^{{\msvvhelper}_1} 1-\text{exp}\left(\frac{-x}{1-{\msvvhelper}_1}\right) \ \partial x + 
    \int_{{\msvvhelper}_1}^{{\msvvhelper}_2} 1-\text{exp}\left(\frac{-{\msvvhelper}_1}{1-{\msvvhelper}_1}\right) + \log\left(\frac{1-{\msvvhelper}_1}{1-x} \right) \ \partial x  + (1- {\msvvhelper}_2) \label{eq:prop7limitline} \\
    =&  \ 
    {\msvvhelper}_1 - (1-{\msvvhelper}_1)\left(1-\text{exp}\left(\frac{-{\msvvhelper}_1}{1-{\msvvhelper}_1}\right)\right)
    + ({\msvvhelper}_2-{\msvvhelper}_1)\left[ 1-\text{exp}\left(\frac{-{\msvvhelper}_1}{1-{\msvvhelper}_1}\right)\right] \nonumber \\ & \qquad \quad \quad + \int_{{\msvvhelper}_1}^{{\msvvhelper}_2}\log\left(\frac{1-{\msvvhelper}_1}{1-x} \right) \ \partial x \nonumber + (1- {\msvvhelper}_2) \\
    =& \ (1-{\msvvhelper}_2)\Big(\text{{exp}}\big(-{\msvvhelper}_1/(1-{\msvvhelper}_1)\big)+\log\big((1-{\msvvhelper}_2)/(1-{\msvvhelper}_1)\big)\Big) + {\msvvhelper}_2 \nonumber \\
    =& \ {\msvvhelper}_2 \nonumber
\end{align}
In \eqref{eq:prop7limitline}, we again use the fact that $(1-1/n)^{nx}$ approaches $e^{-x}$ as $n$ approaches infinity. Furthermore, we use the fact that $\sum_{k = yn}^{xn}1/k$ approaches $\log(x/y)$ as $n$ approaches infinity. The last equality comes from applying the definition of ${\msvvhelper}_2$ to see that $\log\big((1-{\msvvhelper}_2)/(1-{\msvvhelper}_1)\big) = -\text{{exp}}\big(-{\msvvhelper}_1/(1-{\msvvhelper}_1)\big)$. 
\medskip

\noindent {\bf Value of \OPT\ on Instance $\instance_3(\extfrac)$:}
We next show that for any \fracextname\ $\extfrac \in [1/e, 1]$, \OPT\ fills all capacity on $\instance_3(\extfrac)$.
To see this, consider a solution which matches all \exttraf\ and matches each of the $\Delta_i$ \inttraf\ \vols\ in batch $i$ to \opp\ $i$. To see why such a solution gives a perfect matching, note that each \opp\ $i \in \{1, \dots, {\msvvhelper}_1\largeopps\}$ will receive $\largecapacity\left(1-\left(\frac{(1-{\msvvhelper}_1)\largeopps}{(1-{\msvvhelper}_1)\largeopps + 1}\right)^i\right)$ matches from \exttraf\ and $\Delta_i = \largecapacity\left(\frac{(1-{\msvvhelper}_1)\largeopps}{(1-{\msvvhelper}_1)\largeopps + 1}\right)^i$ matches from \inttraf, leading to a total of $\largecapacity$ matches. Each \opp\ $i \in \{{\msvvhelper}_1\largeopps + 1, \dots, \largeopps\}$ will receive $\Delta_i = \largecapacity$ matches (either all from \inttraf\ or all from \exttraf). Thus, each \opp\ is filled to capacity under this solution, which implies that the optimal algorithm must also fill all capacity. 

Combining the upper bound on the fraction of capacity filled by \MSVV\ with the fact that \OPT\ fills all capacity, we see that \MSVV\ only fills a fraction ${\msvvhelper}_2$ of the capacity filled by \OPT\ on this family of instances. This provides a parameterized upper bound on the competitive ratio of \MSVV, and thereby completes the proof of Claim \ref{clm:generalmsvv_small}. \halmos
\end{proof}

\begin{clm}
\label{clm:generalmsvv_large}
For any \MakeLowercase{\fracextnamefull} $\extfrac \in [0,1]$, the competitive ratio of $\MSVV$ is at most 
\begin{equation}
    \min_{\msvvhelper_4 \in [0, \extfrac]} 1 - \frac{1-\extfrac}{1-\msvvhelper_4}\left(\msvvhelper_5 + (1-\msvvhelper_6)\log\left(\frac{1-\msvvhelper_5}{1-\msvvhelper_6}\right) \right) \label{eq:msvvlargebound}
\end{equation}
where $\msvvhelper_5 = \min\{1-\msvvhelper_4, \frac{\msvvhelper_4(\extfrac-\msvvhelper_4)}{1-\extfrac}\}$ and $\msvvhelper_6 = \min\{1 - \msvvhelper_4, 1-(1-\msvvhelper_5)/e\}$.
\end{clm}
\begin{proof}{Proof of Claim \ref{clm:generalmsvv_large}:}
Consider a family of instances $\instance_4(\extfrac)$, parameterized by the \fracextname\ $\extfrac$. Each instance has $\largeopps$ \opps, each with identical large capacity $\largecapacity$. Each instance in this family is also parameterized by $\msvvhelper_4 \in [0, \extfrac]$, which separates the $\largeopps$ into subsets of size $(1-\msvvhelper_4)\largeopps$ and $\msvvhelper_4\largeopps$.\footnote{We assume that $\msvvhelper_4\largeopps\largecapacity$ is an integer. This assumption does not impact the upper bound in the statement of Claim \ref{clm:generalmsvv_large}, as the expression comes from taking the limit as $\largeopps$ approaches $\infty$.} These two subsets will receive \exttraf\ at different times: the former subset will receive \exttraf\ at the beginning of the arrival sequence, while the latter will receive \exttraf\ at the end of the arrival sequence. The full arrival sequence consists of $\largeopps \largecapacity$ \vols, all of whom have conversion probabilities of $1$ or $0$. If $\mu_{i,t}=1$ (resp. 0), we will refer to opportunity $i$ and volunteer $t$ as compatible (resp. incompatible).  

Fixing an \fracextname\ $\extfrac$ and a parameter $\msvvhelper_4$, the arrival sequence begins with $\frac{\extfrac - \msvvhelper_4}{1-\msvvhelper_4}\cdot \largecapacity$ \exttraf\ \vols\ for each \opp\ $i \in \{1, \dots, (1-\msvvhelper_4)\largeopps\}$ (who are compatible with their targeted \opp). In total, this comprises $(\extfrac - \msvvhelper_4)\largeopps\largecapacity$ \exttraf\ \vols. Next, the \inttraf\ arrives, which consists of $(1-\extfrac)\largeopps\largecapacity$ \vols. These \vols\ can be separated into $(1-\msvvhelper_4)\largeopps$ batches of size $\frac{1-\extfrac}{1-\msvvhelper_4}\cdot\largecapacity$ sequentially-arriving homogeneous \vols, such that the \vols\ in the $i$\textsuperscript{th} batch are compatible with all opportunities $j \geq i$. Finally, additional \exttraf\ arrives, with $\largecapacity$ compatible \vols\ for each \opp\ $i \in \{(1-\msvvhelper_4)\largeopps + 1, \dots, \largeopps\}$.

We first note that the \fracextname\ in such instances is equal to $\extfrac$. To see that this is indeed the case, note that the \opps\ in the first subset (those that initially receive \exttraf) receive a total filled capacity of $(\extfrac - \msvvhelper_4)\largeopps\largecapacity$, while the \opps\ in the other subset (those that receive \exttraf\ at the end of the arrival sequence) receive a total filled capacity of $\msvvhelper_4 \largeopps \largecapacity$. In sum, this represents a fraction $\extfrac$ of total capacity. We now proceed to assessing the value of $\MSVV$ and $\OPT$ on this family of instances.

\medskip

\noindent {\bf Value of \MSVV\ on Instance $\instance_4(\extfrac)$:}
We now analyze the value of $\MSVV$ on this instance. All the initial \exttraf\ will be allocated to the appropriate \opp. At the conclusion of this process, each \opp\ $i \in \{1, \dots, (1-\msvvhelper_4) \largeopps\}$ will have a filled capacity of $\frac{\extfrac - \msvvhelper_4}{1-\msvvhelper_4}\cdot \largecapacity$.  Based on the allocation rule of \MSVV, at first all the \inttraf\ will be exclusively allocated (evenly) across the other $\msvvhelper_4$ compatible \opps, i.e., \opps\ $i \in \{(1-\msvvhelper_4)\largeopps + 1, \dots, \largeopps\}$, since those \opps\ will have less filled capacity. (Recall that \MSVV\ defines an \opp's fill rate as the ratio of filled capacity to total capacity, regardless of the source of the \vols.)  If there is enough \inttraf\ to fill all \opps\ to an equal fill rate of $\frac{\extfrac - \msvvhelper_4}{1-\msvvhelper_4}$, then the remaining \inttraf\ will be evenly split among compatible \opps, until the \inttraf\ runs out or the remaining compatible \opps\ have all reached capacity. Finally, the \exttraf\ fills \opps\ $i \in \{(1-\msvvhelper_4)\largeopps + 1, \dots, \largeopps\}$ to capacity.

This allocation corresponds to two different cases, based on the amount of \inttraf\ relative to the parameter $\msvvhelper_3$. To help define these cases, we introduce  $\msvvhelper_5 := \min\{1-\msvvhelper_4, \frac{\msvvhelper_4(\extfrac-\msvvhelper_4)}{1-\extfrac}\}$. 
\revcolor{As we will later show, $\msvvhelper_5 \largeopps$ represents the highest-indexed \opp\ that does not receive \inttraf\ under \MSVV, and as such, it will appear in the limits of the summations below that we use to calculate the total amount of filled capacity.}
In each case, we will demonstrate that the total amount of filled capacity is given by 
$$\sum_{i=1}^{\msvvhelper_5\largeopps} \frac{\extfrac - \msvvhelper_4}{1-\msvvhelper_4}\cdot \largecapacity + \sum_{i = \msvvhelper_5\largeopps+1}^{(1 - \msvvhelper_4)\largeopps} \min\left\{\left(\frac{\extfrac - \msvvhelper_4}{1-\msvvhelper_4}\cdot \largecapacity + \sum_{j = \msvvhelper_5\largeopps+1}^i \frac{1-\extfrac}{1-\msvvhelper_4}\cdot\frac{\largecapacity}{\largeopps - j + 1} \right), \largecapacity\right\} + \sum_{i = (1 - \msvvhelper_4)\largeopps+1}^{\largeopps} \largecapacity.$$

In case (i), the amount of \inttraf\ is insufficient to equalize the fill rate of all \opps, i.e., $(1-\extfrac)\largecapacity \leq \msvvhelper_4\left(\frac{\extfrac - \msvvhelper_4}{1-\msvvhelper_4}\cdot \largecapacity\right)$. Consequently, $\MSVV$ will simply divide all \inttraf\ equally among \opps\ $i \in \{(1-\msvvhelper_4)\largeopps + 1, \dots, \largeopps\}$. These \opps\ will then be filled to capacity by \exttraf. We note that $\msvvhelper_5 = 1-\msvvhelper_4$, since in this case, $1-\msvvhelper_4$ cannot exceed $\frac{\msvvhelper_4(\extfrac-\msvvhelper_4)}{1-\extfrac}$.  
Therefore, the total filled capacity in this case is given by 
$$\sum_{i=1}^{\msvvhelper_5\largeopps} \frac{\extfrac - \msvvhelper_4}{1-\msvvhelper_4}\cdot \largecapacity + \sum_{i = \msvvhelper_5\largeopps+1}^{(1 - \msvvhelper_4)\largeopps} \min\left\{\left(\frac{\extfrac - \msvvhelper_4}{1-\msvvhelper_4}\cdot \largecapacity + \sum_{j = \msvvhelper_5\largeopps+1}^i \frac{1-\extfrac}{1-\msvvhelper_4}\cdot\frac{\largecapacity}{\largeopps - j + 1} \right), \largecapacity\right\} + \sum_{i = (1 - \msvvhelper_4)\largeopps+1}^{\largeopps} \largecapacity,$$
as desired. We note that in this case, the middle sum is empty, as $\msvvhelper_5 = 1 - \msvvhelper_4$.

In case (ii), the amount of \inttraf\ is sufficient to equalize all fill rates, i.e., if $(1-\extfrac)\largecapacity > \msvvhelper_4\left(\frac{\extfrac - \msvvhelper_4}{1-\msvvhelper_4}\cdot \largecapacity\right)$ (which implies $\msvvhelper_5 =  \frac{\msvvhelper_4(\extfrac-\msvvhelper_4)}{1-\extfrac}$). In this case, the \opps\ will all reach an equal fill rate after the arrival of the $\msvvhelper_5 \largeopps$\textsuperscript{th} batch of \inttraf. From this point forward, \inttraf\ will be split among the compatible \opps, but none of the first $\msvvhelper_5 \largeopps$ \opps\ are compatible with remaining arrivals. As such, the total filled capacity is given by  
$$\sum_{i=1}^{\msvvhelper_5\largeopps} \frac{\extfrac - \msvvhelper_4}{1-\msvvhelper_4}\cdot \largecapacity + \sum_{i = \msvvhelper_5\largeopps+1}^{(1 - \msvvhelper_4)\largeopps} \min\left\{\left(\frac{\extfrac - \msvvhelper_4}{1-\msvvhelper_4}\cdot \largecapacity + \sum_{j = \msvvhelper_5\largeopps+1}^i \frac{1-\extfrac}{1-\msvvhelper_4}\cdot\frac{\largecapacity}{\largeopps - j + 1} \right), \largecapacity\right\} + \sum_{i = (1 - \msvvhelper_4)\largeopps+1}^{\largeopps} \largecapacity.$$

We now compute the fraction of total capacity that is filled under \MSVV. To help in this calculation, we define $\msvvhelper_6 := \min\{1 - \msvvhelper_4, 1-(1-\msvvhelper_5)/e\}$, which (asymptotically) represents the fraction of \opps\ that are not filled to capacity under \MSVV.
\begin{align}
    \frac{\MSVV(\instance_4(\extfrac))}{\largeopps\largecapacity} =& \frac{1}{\largeopps \largecapacity}\left[\sum_{i=1}^{\msvvhelper_5\largeopps} \frac{\extfrac - \msvvhelper_4}{1-\msvvhelper_4}\cdot \largecapacity + \sum_{i = \msvvhelper_5\largeopps+1}^{(1 - \msvvhelper_4)\largeopps} \min\left\{\left(\frac{\extfrac - \msvvhelper_4}{1-\msvvhelper_4}\cdot \largecapacity + \sum_{j = \msvvhelper_5\largeopps+1}^i \frac{1-\extfrac}{1-\msvvhelper_4}\cdot\frac{\largecapacity}{\largeopps - j + 1} \right), \largecapacity\right\} \right. \nonumber \\& \left.\qquad \qquad + \sum_{i = (1 - \msvvhelper_4)\largeopps+1}^{\largeopps} \largecapacity\right] \\
     \xrightarrow{\largeopps \rightarrow \infty}& \left(\frac{\extfrac - \msvvhelper_4}{1-\msvvhelper_4}\right)\msvvhelper_5 + \int_{\msvvhelper_5}^{1-\msvvhelper_4} \min\left\{\frac{\extfrac - \msvvhelper_4}{1-\msvvhelper_4} + \frac{1-\extfrac}{1-\msvvhelper_4} \log\left(\frac{1-\msvvhelper_5}{1-x}\right), 1\right\} \ \partial x + \msvvhelper_4 \label{eq:prop7update1} \\
    =& \left(\frac{\extfrac - \msvvhelper_4}{1-\msvvhelper_4}\right)\msvvhelper_5 + \int_{\msvvhelper_5}^{\msvvhelper_6} \frac{\extfrac - \msvvhelper_4}{1-\msvvhelper_4} + \frac{1-\extfrac}{1-\msvvhelper_4} \log\left(\frac{1-\msvvhelper_5}{1-x}\right) \ \partial x + \int_{\msvvhelper_6}^{1-\msvvhelper_4} 1 \ \partial x + \msvvhelper_4 \label{eq:prop7update2} \\
    =& \left(\frac{\extfrac - \msvvhelper_4}{1-\msvvhelper_4}\right)\msvvhelper_6 + \frac{1-\extfrac}{1-\msvvhelper_4}\int_{\msvvhelper_5}^{\msvvhelper_6}  \log\left(\frac{1-\msvvhelper_5}{1-x}\right) \ \partial x + (1-\msvvhelper_6) \\
     =& \frac{\extfrac - \msvvhelper_4}{1-\msvvhelper_4}\msvvhelper_6 + \frac{1-\extfrac}{1-\msvvhelper_4}\left(\msvvhelper_6 - \msvvhelper_5 -(1-\msvvhelper_6)\log\left(\frac{1-\msvvhelper_5}{1-\msvvhelper_6}\right)\right) + (1-\msvvhelper_6)  \\
    =& 1 - \frac{1-\extfrac}{1-\msvvhelper_4}\left(\msvvhelper_5 + (1-\msvvhelper_6)\log\left(\frac{1-\msvvhelper_5}{1-\msvvhelper_6}\right) \right) 
\end{align}
Equality \eqref{eq:prop7update1} uses the fact that $\sum_{k = yn}^{xn}1/k$ approaches $\log(x/y)$ as $n$ approaches infinity. Equality \eqref{eq:prop7update2} comes from applying the definition of $\msvvhelper_6$ and noting that for $x \geq \msvvhelper_6$, $1 \leq \frac{\extfrac - \msvvhelper_4}{1-\msvvhelper_4} + \frac{1-\extfrac}{1-\msvvhelper_4} \log\left(\frac{1-\msvvhelper_5}{1-x}\right)$. Taking the integrals and simplifying, we arrive at the final expression, which represents the fraction of total capacity that is filled under \MSVV.
\medskip

\noindent {\bf Value of \OPT\ on Instance $\instance_4(\extfrac)$:}
We now show that \OPT\ fills all capacity on this instance. Consider a solution that allocates all \exttraf\ to its targeted \opp, and allocates the $i$\textsuperscript{th} batch of \inttraf\ to \opp\ $i$. Under this solution, each \opp\ $i \in \{1, \dots, (1-\msvvhelper_4)\largeopps\}$ receives $\frac{\extfrac - \msvvhelper_4}{1-\msvvhelper_4}\cdot \largecapacity$ matches from \exttraf\ and $\frac{1-\extfrac}{1-\msvvhelper_4} \cdot \largecapacity$ matches from \inttraf, thereby reaching its capacity of $\largecapacity$. Furthermore, each \opp\ $i \in \{(1-\msvvhelper_4)\largeopps + 1, \dots, \largeopps\}$ receives $\largecapacity$ matches from \exttraf. Thus, under this solution, all capacity is filled, which means that $\OPT$ must also fill all capacity on this instance.

This establishes a competitive ratio of $1 - \frac{1-\extfrac}{1-\msvvhelper_4}\left(\msvvhelper_5 + (1-\msvvhelper_6)\log\left(\frac{1-\msvvhelper_5}{1-\msvvhelper_6}\right) \right)$, as desired. Taking the minimum over all $\msvvhelper_4 \in [0, \extfrac]$ completes the proof of the claim. \halmos
\end{proof}

Both claims establish an upper bound on the competitive ratio of $\MSVV$.\footnote{To show that these upper bounds hold for any minimum capacity $\invbidtobudget$, it suffices to add an additional \opp\ with capacity $\invbidtobudget$ for which \vols\ have conversion probability of $0$. The performance of both $\OPT$ and $\MSVV$ are unchanged, and the \fracextname\ remains the same in the limit as $\largeopps$ approaches infinity.} In Figure \ref{fig:CR_weightcap} of Section \ref{sec:results}, we illustrate the piecewise-defined upper bound on \MSVV\ that results from taking the minimum for any particular \fracextname\ $\extfrac > 1/e$, along with the universal upper bound of $1-1/e$ for $\extfrac \leq 1/e$. \halmos

\end{proof}


\if false

\subsection{Upper Bound on \MSVV\ in General Settings}
In the following proposition, we provide an upper bound on the competitive ratio of \MSVV\ as a function of the \fracextname\ $\extfrac$.
\begin{prop}[Upper Bound on \MSVV]
\label{prop:generalmsvv}
For any \MakeLowercase{\fracextnamefull} $\extfrac$, as well as for any minimum capacity $\invbidtobudget$ and any maximum heterogeneity across \vol's preferences $\weightcap$, $\MSVV$ cannot achieve a competitive ratio better than
\begin{equation}
    \min\left\{\msvvhelper_2, \msvvhelper_3\right\} \nonumber
\end{equation}
where $\msvvhelper_2 = 1- \frac{1-\msvvhelper_1}{\text{\emph{exp}}\left(\text{\emph{exp}}(-\msvvhelper_1/(1-\msvvhelper_1))\right)}$ and $\msvvhelper_1$ is the unique solution in $[0,1]$ to $$\extfrac = \msvvhelper_1 + (1-\msvvhelper_1)\Big(\text{\emph{exp}}\big(-\msvvhelper_1/(1-\msvvhelper_1)\big)-1\Big) + \frac{1-\msvvhelper_1}{\text{\emph{exp}}\left(\text{\emph{exp}}(-\msvvhelper_1/(1-\msvvhelper_1))\right)}.$$
In addition, $$\msvvhelper_3 = 
    \min_{\msvvhelper_4 \in [0, \extfrac]} 1 - \frac{1-\extfrac}{1-\msvvhelper_4}\left(\msvvhelper_5 + (1-\msvvhelper_6)\log\left(\frac{1-\msvvhelper_5}{1-\msvvhelper_6}\right) \right),$$
where $\msvvhelper_5 = \min\{1-\msvvhelper_4, \frac{\msvvhelper_4(\extfrac-\msvvhelper_4)}{1-\extfrac}\}$ and $\msvvhelper_6 = \min\{1 - \msvvhelper_4, 1-(1-\msvvhelper_5)/e\}$.
\end{prop}

\begin{proof}{Proof of Proposition \ref{prop:generalmsvv}}
\label{proof:prop:generalmsvv}
We prove this proposition in two claims by showing two different upper bounds on the competitive ratio of \MSVV\ parameterized by the \MakeLowercase{\fracextnamefull} $\extfrac$. Proposition \ref{prop:generalmsvv} follows by taking the minimum of the two upper bounds for a given $\extfrac$. 

To prove each claim, we construct a family of instances parameterized by $\extfrac$. We then evaluate the value of \MSVV\ on that family of instances relative to the value of $\OPT$. Both of the instances that we design leverage the fact that the notion of a fill rate under \MSVV\ does not distinguish between \intandexttraf. As a result, \MSVV\ may mistakenly withhold \inttraf\ from \opps\ that have previously received \exttraf. \revcolor{Furthermore, all instances leverage a similar structure to our general hardness result (see Appendix \ref{proof:thm:hardness}) where compatibility diminishes over time: each \inttraf\ \vol\ is only compatible with a subset of the \opps\ that the previous \inttraf\ \vol\ was compatible with.} 

\begin{clm}
\label{clm:generalmsvv_small}
For any \MakeLowercase{\fracextnamefull} $\extfrac$, the competitive ratio of $\MSVV$ is at most 
\begin{equation}
    1- \frac{1-\msvvhelper_1}{\text{\emph{exp}}\left(\text{\emph{exp}}(-\msvvhelper_1/(1-\msvvhelper_1))\right)} \label{eq:msvvsmallbound}
\end{equation}
where $\msvvhelper_1$ is the unique solution in $[0,1]$ to $\extfrac = \msvvhelper_1 + (1-\msvvhelper_1)\Big(\text{\emph{exp}}\big(-\msvvhelper_1/(1-\msvvhelper_1)\big)-1\Big) + \frac{1-\msvvhelper_1}{\text{\emph{exp}}\left(\text{\emph{exp}}(-\msvvhelper_1/(1-\msvvhelper_1))\right)}$. 
\end{clm}
\begin{proof}{Proof of Claim \ref{clm:generalmsvv_small}}
To prove this claim, we define a family of instances $\instance_3(\extfrac)$, which has a close relationship to the family of instances $\instance_2(\extfrac)$ that was introduced in the proof of Proposition \ref{prop:warmupmsvv} in Appendix \ref{proof:prop:warmupmsvv}.
Fixing an \fracextname\ ${\extfrac}$ and its associated instance $\instance_2({\extfrac})$, we define a nearly identical instance which differs only in the source of the final $(1-{\alpha}_{2})\largeopps\largecapacity$ \vols\ to arrive. (We note that ${\alpha}_{2}$, defined in Appendix \ref{proof:prop:warmupmsvv}, is a function of ${\extfrac}$.) In $\instance_2({\extfrac})$, these \vols\ are compatible with only one \opp, but they are counted as \inttraf, and hence the \fracextname\ is equal to ${\extfrac}$. However, by re-defining these last-arriving \vols\ as \exttraf, we change the \fracextname\ in this new instance. We denote the instance ``corresponding'' to $\instance_2({\extfrac})$ as $\instance_3({\extfrac} + 1 - {\alpha}_2)$, since the \fracextname\ in this adjusted instance has been increased by $1-{\alpha}_{2}$.

We emphasize that the change in the source of these \vols\ does not affect the value of $\OPT$ or $\MSVV$. Thus, for any ${\extfrac} \in [0,1]$, the competitive ratio of \MSVV\ on instance $\instance_2({\extfrac})$ is equivalent to the competitive ratio of \MSVV\ on instance $\instance_3({\extfrac} + 1-{\alpha}_{2})$.

With this in mind, given any $\extfrac$, we construct the instance $\instance_3(\extfrac)$ by defining $\msvvhelper_1$ as the unique solution in $[0,1]$ to the equation:
\begin{align*}
    \extfrac \quad &= \quad {\msvvhelper}_1 + (1-\msvvhelper_1)\Big(\text{{exp}}\big(-{\msvvhelper}_1/(1-{\msvvhelper}_1)\big)-1\Big) + \frac{1-{\msvvhelper}_1}{\text{{exp}}\left(\text{{exp}}(-{\msvvhelper}_1/(1-{\msvvhelper}_1))\right)}
\end{align*}
This ensures an \fracextname\ of $\extfrac$, as the initial amount of \exttraf\ is given by $\largeopps\largecapacity\left({\msvvhelper}_1 + (1-\msvvhelper_1)\Big(\text{{exp}}\big(-{\msvvhelper}_1/(1-{\msvvhelper}_1)\big)-1\Big)\right)$, while the last-arriving $\largeopps\largecapacity\frac{1-{\msvvhelper}_1}{\text{{exp}}\left(\text{{exp}}(-{\msvvhelper}_1/(1-{\msvvhelper}_1))\right)}$ \vols\ are also \exttraf\ in this adjusted instance. As shown in the proof of Proposition \ref{prop:warmupmsvv} (see Appendix \ref{proof:prop:warmupmsvv}), the competitive ratio of \MSVV\ in such an instance is $1- \frac{1-{\msvvhelper}_1}{\text{{exp}}\left(\text{{exp}}(-{\msvvhelper}_1/(1-{\msvvhelper}_1))\right)}$. This completes the proof of Claim \ref{clm:generalmsvv_small}. \halmos
\end{proof}

\begin{clm}
\label{clm:generalmsvv_large}
For any \MakeLowercase{\fracextnamefull} $\extfrac$, the competitive ratio of $\MSVV$ is at most 
\begin{equation}
    \min_{\msvvhelper_4 \in [0, \extfrac]} 1 - \frac{1-\extfrac}{1-\msvvhelper_4}\left(\msvvhelper_5 + (1-\msvvhelper_6)\log\left(\frac{1-\msvvhelper_5}{1-\msvvhelper_6}\right) \right) \label{eq:msvvlargebound}
\end{equation}
where $\msvvhelper_5 = \min\{1-\msvvhelper_4, \frac{\msvvhelper_4(\extfrac-\msvvhelper_4)}{1-\extfrac}\}$ and $\msvvhelper_6 = \min\{1 - \msvvhelper_4, 1-(1-\msvvhelper_5)/e\}$.
\end{clm}

Consider a family of instances $\instance_4(\extfrac)$, parameterized by the \fracextname\ $\extfrac$. Each instance has \largeopps\ \opps, each with identical large capacity \largecapacity. (Our results will hold in the limit as $\largeopps$ and $\largecapacity$ approach infinity.) Each instance in this family is also parameterized by $\msvvhelper_4 \in [0, \extfrac]$, which separates the $\largeopps$ into subsets of size $(1-\msvvhelper_4)\largeopps$ and $\msvvhelper_4\largeopps$. These two subsets will receive \exttraf\ at different times: the former subset will receive \exttraf\ at the beginning of the arrival sequence, while the latter will receive \exttraf\ at the end of the arrival sequence. The full arrival sequence consists of $\largeopps \largecapacity$ \vols, all of whom have binary conversion probabilities. If $\mu_{i,t}=1$ (resp. 0), we will refer to opportunity $i$ and volunteer $t$ as compatible (resp. incompatible).  

Fixing an \fracextname\ $\extfrac$ and a parameter $\msvvhelper_4$, the arrival sequence begins with $\frac{\extfrac - \msvvhelper_4}{1-\msvvhelper_4}\cdot \largecapacity$ \exttraf\ \vols\ for each \opp\ $i \in \{1, \dots, (1-\msvvhelper_4)\largeopps\}$. In total, this comprises $(\extfrac - \msvvhelper_4)\largeopps\largecapacity$ \exttraf\ \vols. Next, the \inttraf\ arrives, which consists of $(1-\extfrac)\largeopps\largecapacity$ \vols. These \vols\ can be separated into $(1-\msvvhelper_4)\largeopps$ subsets of size $\frac{1-\extfrac}{1-\msvvhelper_4}\cdot\largecapacity$ sequentially-arriving homogeneous \vols, such that the \vols\ in the $i$\textsuperscript{th} subset are compatible with all opportunities $j \geq i$. Finally, additional \exttraf\ arrives, with $\largecapacity$ \vols\ for each \opp\ $i \in \{(1-\msvvhelper_4)\largeopps + 1, \dots, \largeopps\}$.

To show that \OPT\ fills all capacity on this instance, consider an algorithm that allocates \exttraf\ to its targeted \opp, and allocates the $i$\textsuperscript{th} subset of \inttraf\ to \opp\ $i$. Under this algorithm, each \opp\ $i \in \{1, \dots, (1-\msvvhelper_4)\largeopps\}$ receives $\frac{\extfrac - \msvvhelper_4}{1-\msvvhelper_4}\cdot \largecapacity$ matches from \exttraf\ and $\frac{1-\extfrac}{1-\msvvhelper_4}\cdot\largecapacity$ matches from \inttraf, thereby reaching its capacity of $\largecapacity$. Furthermore, each \opp\ $i \in \{(1-\msvvhelper_4)\largeopps + 1, \dots, \largeopps\}$ receives $\largecapacity$ matches from \exttraf. Thus, under this algorithm, all capacity is filled, and by optimality, we must have that $\OPT$ also achieves a value of $\largeopps \largecapacity$ on this instance.

We now analyze the value of $\MSVV$ on this instance. All the initial \exttraf\ will be allocated to the appropriate \opp. At the conclusion of this process, each \opp\ $i \in \{1, \dots, \msvvhelper_4 \largeopps\}$ will have a filled capacity of $\frac{\extfrac - \msvvhelper_4}{1-\msvvhelper_4}\cdot \largecapacity$.  Based on the allocation rule of \MSVV, at first all the \inttraf\ will be exclusively allocated (evenly) across the other \opps, i.e., \opps\ $i \in \{(1-\msvvhelper_4)\largeopps + 1, \dots, \largeopps\}$, since those \opps\ will have less filled capacity. (Recall that \MSVV\ defines an \opp's fill rate as the ratio of filled capacity to total capacity, regardless of the source of the \vols.) If there is a sufficient amount of \inttraf, then eventually the filled capacity of all \opps\ will be equal. In particular, after the arrival of subset $\msvvhelper_5\largeopps$, where $\msvvhelper_5 := \min\{1-\msvvhelper_4, \frac{\msvvhelper_4(\extfrac-\msvvhelper_4)}{1-\extfrac}\}$, if any subsets of \inttraf\ remain (i.e., if the latter term is the minimum), then all \opps\ have an equal amount of filled capacity. From that point forward, remaining \inttraf\ will be evenly distributed across its compatible \opps, until capacity is reached or until the \inttraf\ is used up. Finally, the arriving \exttraf\ fills any remaining capacity of \opps\ $i \in \{(1-\msvvhelper_4)\largeopps + 1, \dots, \largeopps\}$. 

Based on this allocation, the filled capacity of each \opp\ $i \in \{1, \dots, \msvvhelper_5 \largeopps\}$ is equal to $\frac{\extfrac - \msvvhelper_4}{1-\msvvhelper_4}\cdot \largecapacity$, the filled capacity of each \opp\ $i \in \{\msvvhelper_5 \largeopps + 1, (1-\msvvhelper_4)\largeopps\}$ is equal to $$\min\left\{\frac{\extfrac - \msvvhelper_4}{1-\msvvhelper_4}\cdot \largecapacity + \sum_{j = \msvvhelper_5\largeopps+1}^i \frac{\frac{1-\extfrac}{1-\msvvhelper_4}\cdot\largecapacity}{\largeopps - j + 1}, 1\right\},$$ and the filled capacity of each \opp\ $i \in \{(1-\msvvhelper_4)\largeopps + 1, \dots, \largeopps\}$ is equal to $\largecapacity$. Adding up these filled capacities across all \opps\ and letting $\msvvhelper_6 := \min\{1 - \msvvhelper_4, 1-(1-\msvvhelper_5)/e\}$ denote the fraction of \opps\ which do not reach full capacity, we see that the value of \MSVV\ is given by:
\begin{align}
    \MSVV(\instance_4(\extfrac)) & = \sum_{i=1}^{\msvvhelper_5\largeopps} \frac{\extfrac - \msvvhelper_4}{1-\msvvhelper_4}\cdot \largecapacity + \sum_{i = \msvvhelper_5\largeopps+1}^{\msvvhelper_6\largeopps} \left(\frac{\extfrac - \msvvhelper_4}{1-\msvvhelper_4}\cdot \largecapacity + \sum_{j = \msvvhelper_5\largeopps+1}^i \frac{\frac{1-\extfrac}{1-\msvvhelper_4}\cdot\largecapacity}{\largeopps - j + 1} \right) + \sum_{i = \msvvhelper_6\largeopps +1}^\largeopps \largecapacity \\
    &\rightarrow \frac{\extfrac - \msvvhelper_4}{1-\msvvhelper_4}\msvvhelper_5\largeopps\largecapacity + \sum_{i = \msvvhelper_5\largeopps+1}^{\msvvhelper_6\largeopps}\left( \frac{\extfrac - \msvvhelper_4}{1-\msvvhelper_4}\cdot \largecapacity +  \frac{1-\extfrac}{1-\msvvhelper_4}\log\left(\frac{1-\msvvhelper_5}{1-i/\largeopps}\right)\largecapacity\right) + (1-\msvvhelper_6)\largeopps\largecapacity \\
    &\rightarrow \left(\frac{\extfrac - \msvvhelper_4}{1-\msvvhelper_4}\msvvhelper_6 + \frac{1-\extfrac}{1-\msvvhelper_4}\int_{\msvvhelper_5}^{\msvvhelper_6}\log\left(\frac{1-\msvvhelper_5}{1-x}\right) \ \partial x + (1-\msvvhelper_6)\right)\largeopps\largecapacity \\
    & = \left(\frac{\extfrac - \msvvhelper_4}{1-\msvvhelper_4}\msvvhelper_6 + \frac{1-\extfrac}{1-\msvvhelper_4}\left(\msvvhelper_6 - \msvvhelper_5 -(1-\msvvhelper_6)\log\left(\frac{1-\msvvhelper_5}{1-\msvvhelper_6}\right)\right) + (1-\msvvhelper_6)\right)\largeopps\largecapacity \\
    & = \left(1 - \frac{1-\extfrac}{1-\msvvhelper_4}\left(\msvvhelper_5 + (1-\msvvhelper_6)\log\left(\frac{1-\msvvhelper_5}{1-\msvvhelper_6}\right) \right) \right)\largeopps\largecapacity
\end{align}

This establishes a competitive ratio of $1 - \frac{1-\extfrac}{1-\msvvhelper_4}\left(\msvvhelper_5 + (1-\msvvhelper_6)\log\left(\frac{1-\msvvhelper_5}{1-\msvvhelper_6}\right) \right)$, as desired. Taking the minimum over all $\msvvhelper_4 \in [0, \extfrac]$ completes the proof of the claim. \halmos

Both claims establish an upper bound on the competitive ratio of $\MSVV$. In Figure \ref{fig:CR_weightcap} of Section \ref{subsec:results:policy}, we illustrate the piecewise-defined upper bound on \MSVV\ that results from taking the minimum of these claims for any particular \fracextname\ $\extfrac$. 

We note that while these instances are unweighted with large capacities (i.e., where $\weightcap = 1$ and $\invbidtobudget \rightarrow \infty$), the upper bound on the competitive ratio of $\MSVV$ holds in weighted, capacitated scenarios as well. To see this, consider identical instances except with a single additional \opp\ $\largeopps+1$ and a single additional \vol\ $T+1$, where $\mu_{\largeopps+1,T+1} \in (0,1)$ and $c_{\largeopps+1}=c_{min}$, corresponding to an arbitrarily large maximum heterogeneity across \vol's preferences $\weightcap$ and an arbitrarily small minimum capacity $\invbidtobudget$. In the limit of large $\largeopps$ and large $\largecapacity$,  the \fracextname\ $\extfrac$ is (asymptotically) unaffected. Similarly, the value of $\adaptivecap$ and $\OPT$ are (asymptotically) unchanged, which maintains the parameterized upper bound on the competitive ratio. \halmos
\end{proof}

\fi

\section{Omitted Details from Section \ref{sec:case_study}}
\label{app:simulations}
\subsection{Data Availability and Set of Opportunities}
\label{app:data_availability}

Through our partnership with VolunteerMatch, {we have access to the following sources of detailed data on their platform:}
\begin{enumerate}
    \item Volunteer Match's back-end database that provides opportunity-level data on characteristics such as their posting dates, locations (in-person or virtual), timings (specific dates/times or a flexible schedule), capacities (i.e., the number of volunteers needed), and the cause(s) the organization supports (out of a list of 29 including LGBTQ, seniors, hunger, etc.). To ensure consistent data quality and accuracy, we limit our analysis to the virtual \opps\ active between August 2020 and March 2021 for which we have precise data on capacity (i.e., those that request a number of \vols\ between 1 and 20). We focus on virtual \opps, as these \opps\ do not have compatibility that depends on the proximity of a \vol\ to an \opp.
    \item Google Analytics (GA) data that details user behavior on the site. 
    GA provides session-level information for all devices accessing the website, allowing us to understand the different ways users access the site. We {have access to} data for activity between August 2020 and March 2021 for devices from New York City, Miami, Austin, Alaska, Maine, Montana, Vermont, and West Virginia. Opportunities appearing in our GA dataset are those that were viewed at least once by one or more of these devices. 
\end{enumerate}

While the GA data is quite rich, it is also only a sample representing around 20\% of the overall traffic on VM and thus incomplete in the sense that we cannot perfectly connect the sign-up data to the views that generated them nor reconstruct the ranking that any volunteer might have seen. For the window between August 2020 and March 2021, we use the GA data to approximate the arrival order of internal and external traffic. In Section \ref{sec:case_study}, we focus on a simple random sample of 100 \opps\ from the 10,737 virtual \opps\ that appear in our GA dataset between August 2020 and March 2021 for which we have precise data on capacity. 

\revcolor{
\subsection{Formal Definition of $\overline{\OPT}$}
\label{app:OPT_UB}

Here we present our definition of $\overline{\OPT}$, which is the optimal solution to the following LP, denoted by D-LP:
\begin{equation*}
\label{eq:deterministic_LP}
\begin{array}{ll@{}ll}
\underset{\displaystyle \vec{x}}{\textrm{max}} & \displaystyle\sum\limits_{i \in [n]}\displaystyle\sum\limits_{t \in \inttimes} \mu_{i,t}x_{i,t} + \displaystyle\sum\limits_{i \in [n]}\displaystyle\sum\limits_{t \in \exttimes} \mathbbm{1}_{i = i_{t}^{*}} x_{i,t}&\ \ \ \ \ &\qquad \qquad \quad \qquad \text{(D-LP)} \\
&&&\\
\text{subject to}& \displaystyle\sum\limits_{t \in \inttimes}\mu_{i,t}x_{i,t} + \displaystyle\sum\limits_{t \in \exttimes}\mathbbm{1}_{i = i_{t}^{*}} x_{i,t} \leq c_{i}  & \forall i \in [n] & \ \ (\text{D-LP}-1)\\ 

       & \displaystyle\sum\limits_{i \in [n]} x_{i,t} \leq 1 \ \ \ \ \ \forall t\in [T] 
       \ \ \ \ \  & & \ \ (\text{D-LP}-2)\\
\end{array}
\end{equation*}

This program uses the set of variables $\mathbf{x} \in \mathbbm{R}^{n \times T}$.
This linear program is a deterministic fractional matching. As formalized in the proposition below, the optimal value of $\overline{\OPT}$ is thus an upper bound on the expected value of $\OPT$.

\begin{prop}
\label{prop:lp_upper_bound}
The optimal value of $\overline{\OPT}$ is an upper bound on the expected value of $\OPT$. 
\end{prop}

\begin{proof}{Proof of Proposition \ref{prop:lp_upper_bound}:}
To prove this, we will show that there exists a feasible solution to D-LP that achieves the expected value of $\OPT$. Let $\hat{x}_{i,t}$ be the ex-ante probability that \opp\ $i$ is recommended to \vol\ $t \in [T]$ under $\OPT$. To see that $\mathbf{\hat{x}}$ is a feasible solution in D-LP, note that (i) by convention, $\OPT$ is an optimal solution that only fills each \opp\ to capacity (see Definition \ref{def:opt}), and (ii) each volunteer $t \in [T]$ can receive at most one recommendation. By linearity of expectation, the expected value of \OPT\ is the same as the objective value of D-LP when we plug in $\mathbf{\hat{x}}$. Since any feasible solution to D-LP must be less than or equal to the optimal value of D-LP, we see that the expected value of $\OPT$ must be less than or equal to $\overline{\OPT}$. \halmos
\end{proof}
}

\section{Model Extensions}
\label{app:rank:fix}
\label{sec:ranking}
In many practical settings, platforms can provide more than one recommendation to \inttraf, often in the form of a ranking. Here, we discuss the ways in which our model and results can generalize to such settings, which we henceforth refer to as the \emph{ranking setting.} 

We begin this section by describing how we augment the model of Section \ref{sec:model}. Upon the arrival of an \inttraf\ \vol, we now allow the platform to present a ranking of \opps\ $\vec{\opprecommendbase} \in \oppset^{\rankmath}$, instead of a single recommendation.\footnote{We allow the domain of possible rankings $\oppset^{\rankmath}$ to consist of arbitrary ranked subsets of \opps. We only require that it includes the singleton $\{0\}$, which deterministically results in no sign-up from that \vol.}
The \vol\ \emph{views} (at most) one \opp\ from this ranked subset.\footnote{For \exttraf, we continue to follow the convention that any algorithm must recommend the single targeted \opp\ $\extrecommend$, which is then directly viewed by the \vol.}
As before, the \vol\ will \emph{sign up} for the viewed \opp\ with their pair-specific conversion probability. 
We use $\rankchoice_{i,t}(\vec{\opprecommendbase})$ to denote the probability that \vol\ $t$ signs up for \opp\ $i$ when presented with the ranking $\vec{\opprecommendbase}$. (We augment each \vol's type to include any parameters necessary to fully specify these probabilities for every possible ranking.) We use the random variable $\choicefunc_t(\vec{\opprecommendbase})$ to denote the \vol's \signup\ decision when presented with the ranking $\vec{\opprecommendbase}$, which is either $0$ or the \opp\ viewed by the \vol.



Our benchmark \OPT\ (see Definition \ref{def:opt}) generalizes to the ranking setting by simply recommending the optimal ranked subset of \opps\ to arriving \inttraf, which can again be found by solving a dynamic program of exponential size.\footnote{For any algorithm with an optimality criteria (such as $\AC$ and $\OPT$), in the presence of multiple optimal solutions, we follow the convention of choosing the optimal solution that presents the ranked subset of the smallest size, breaking ties in favor of the solution that lexicographically minimizes the indices of the ranked subset. \label{footnote:tiebreakingranking}} Likewise, the \AC\ algorithm naturally generalizes to an algorithm that we denote by $\ACrank$. The $\ACrank$ algorithm follows exactly the same steps as the $\AC$ algorithm (see Algorithm \ref{alg:acpolicy}), except instead of recommending the single \opp\ $i$ that maximizes $\convprob_{i,t}\balancefunc(\fillrate_{i,t-1})$, the $\ACrank$ algorithm recommends the ranking $\vecopprecommend{\ACrank}$ that satisfies
\begin{align}
\vecopprecommend{\ACrank} \in \text{argmax}_{\vec{\opprecommendbase} \in \oppset^{\rankmath}}  \sum_{i \in [\numopps]}\rankchoice_{i,t}(\vec{\opprecommendbase}) \cdot \balancefunc(\fillrate_{i,t-1}). \label{eq:subsetmaximization}
\end{align}
Henceforth, we assume that the platform can efficiently solve \eqref{eq:subsetmaximization}, which is a common assumption in the literature \citep{golrezaei2014real, gong2019online}. 
Given this assumption, we are able to establish results that are similar to Theorem \ref{thm:AClower}, as formalized in the following proposition.

\begin{prop}[Lower Bound on the Competitive Ratio of $\ACrank$]
\label{prop:rankinggeneralization}
Let the smallest capacity be given by $\invbidtobudget$. Then, for any \MakeLowercase{\fracextnamefull} $\extfrac$, the competitive ratio of $\ACrank$ algorithm is at least $\max\{\extfrac, e^{-1/\invbidtobudget}(1-1/e)\}$.
\end{prop}


This lower bound on the competitive ratio of the $\ACrank$ algorithm is numerically equivalent to the lower bound established in Theorem \ref{thm:AClower} when the \MCPR\ $\weightcap$ exceeds $e-1$ (beyond which the lower bound is constant in $\weightcap$). 
 The intuition developed in Section \ref{subsubsec:thm2:discussion} applies in this setting, too: we cannot guarantee that the $\ACrank$ algorithm fills \emph{any} capacity with \exttraf\ unless the \fracextname\ is sufficiently large. In fact, the instance of Example \ref{ex:weightedAC} is a special case of the ranking setting (where the platform recommends one \opp\ which the \vol\ deterministically views). Thus, the lower bound of Proposition \ref{prop:rankinggeneralization} cannot be improved, at least for that set of parameters ($\extfrac = 1-1/e,\  \invbidtobudget \rightarrow \infty, \  \weightcap \rightarrow \infty$). Furthermore, in the ranking setting, we cannot necessarily improve our result even when the \MCPR\ is bounded. \revcolor{In our base model, an \MCPR\ of $\weightcap$ is a sufficient condition to ensure that the ``relative value'' of two different (non-empty) recommendations (i.e., the ratio of their expected number of sign-ups) is bounded by $\weightcap$.
However, in the ranking setting, the ``relative value'' of two different recommendations can be quite large, regardless of the \MCPR.}
 We defer the proof of Proposition \ref{prop:rankinggeneralization} to Appendix \ref{proof:prop:rankinggeneralization}.

Though the result of Proposition \ref{prop:rankinggeneralization} holds for arbitrary choice functions, our proof technique is flexible enough to (potentially) provide stronger results when tailored to a particular choice function. For example, consider a special case of the \emph{cascade} (or sequential search) model for \vol\ choice, which has been used to model search on online platforms (see, e.g., \citealt{aggarwal2008sponsored} and \citealt{kempe2008cascade}).

\begin{defn}[\cascademodelname]
\label{def:cascade}
The \MakeLowercase{\cascademodelname} is parameterized by a \vol-specific view probability $\cascadeclick_t > 0$ and a \vol-specific exit probability $\cascadequit_{t} \geq 0$. Given a ranked subset of \opps\ (of length at most $K$), the \vol\ sequentially ``examines'' the \opps\ starting from the top (i.e., position $1$). The \vol\ views the top-ranked \opp\ independently with probability $\cascadeclick_t$. Conditional on not viewing the \opp, the \vol\ exits the platform independently with probability $\cascadequit_t$. If the \vol\ does not exit, they repeat the same process for the second-ranked \opp, and so on.
If the \vol\ reaches the end of the ranked list without viewing an \opp, they exit the platform. 
\end{defn}



The \MakeLowercase{\cascademodelname} is a special case of the cascade model in which the view probabilities depend only on the ranked position of an \opp, and are ``agnostic'' to the identity of the \opp\ itself. This property 
leads to the following observation: under the \MakeLowercase{\cascademodelname}, ranking \opps\ in descending order of $\convprob_{i,t} \cdot \balancefunc(\fillrate_{i,t-1})$ satisfies $\ACrank$'s optimality condition (as given in \eqref{eq:subsetmaximization}). Using this critical observation (formalized in Claim \ref{clm:cascadeconditioning} of Appendix \ref{proof:prop:cascademodel}), we are able to strengthen Proposition \ref{prop:rankinggeneralization} under this choice model. 



\begin{prop}[$\ACrank$ Under the {\cascademodelname}]
\label{prop:cascademodel}
Let the smallest capacity be given by $\invbidtobudget$, let the maximum conversion probability ratio (given by Definition \ref{def:MCPR}) be at most $\weightcap$, and suppose each \vol\ choice follows the \MakeLowercase{\cascademodelname} (specified in Definition \ref{def:cascade}). Then, for any \MakeLowercase{\fracextnamefull} $\extfrac$, the competitive ratio of the $\ACrank$ algorithm is at least $\compratiofunc(\beta, \invbidtobudget, \weightcap)$, as defined in the statement of Theorem \ref{thm:AClower}.
\end{prop}

The proof of Proposition \ref{prop:cascademodel} (deferred to Appendix \ref{proof:prop:cascademodel}) crucially relies on the fact that the probability of viewing an \opp\ depends only on its position in the ranking. Therefore, different rankings can only have different ``relative values'' if either (a) there are differences in conversion probabilities \emph{conditional} on a view, or (b) the rankings are of different length. The former influences our bound via the \MCPR\ $\weightcap$, while we account for the latter by leveraging the observation that the $\ACrank$ algorithm ranks \opps\ in descending order of $\convprob_{i,t} \cdot \balancefunc(\fillrate_{i,t-1})$.

\subsection{Proof of Proposition \ref{prop:rankinggeneralization}}
\label{proof:prop:rankinggeneralization} 

The proof of Proposition \ref{prop:rankinggeneralization} follows an identical approach to the proof of Theorem \ref{thm:AClower}. However, it does not require the machinery of Step 3 in the proof of Lemma \ref{lem:aclowerboundpart2}, as we do not intend to break the barrier of $1-1/e$ except in trivial cases where the \fracextname\ exceeds $1-1/e$. Up to that point (i.e., Step 3), this proof follows the exact steps of the proof of Theorem \ref{thm:AClower}. From that point, we complete the proof of Proposition \ref{prop:rankinggeneralization} by placing a further lower bound on the value of the $\ACrank$ algorithm that no longer depends on the amount of capacity filled by \exttraf\ (see Lemma \ref{lem:rankextrabound}).

To begin, we note that even in this ranking setting, if the \fracextname\ is $\extfrac$, then the $\ACrank$ algorithm will fill at least a $\extfrac$ fraction of capacity. 
\begin{lem}
\label{lem:acranklowerboundpart1}
Let the smallest capacity be given by $\invbidtobudget$. Then, for any \MakeLowercase{\fracextnamefull} $\extfrac$, the competitive ratio of the $\ACrank$ algorithm is at least~$\extfrac$.
\end{lem}
\begin{proof}{Proof of Lemma \ref{lem:acranklowerboundpart1}:}
The proof of Lemma \ref{lem:acranklowerboundpart1} is immediate and is identical to the proof of Lemma \ref{lem:aclowerboundpart1}. We simply note that the $\ACrank$ algorithm always recommends the targeted \opp\ to \exttraf. Applying the definition of the \fracextname\ (see Definition \ref{def:beta}), this feature of the \ACrank\ algorithm ensures that at least a $\extfrac$ fraction of capacity is filled in expectation. \halmos
\end{proof}

Next, we establish a lower bound of $e^{-1/\invbidtobudget}(1-1/e)$ on the competitive ratio of the $\ACrank$ algorithm, which requires more intricate analysis.
\begin{lem}
\label{lem:acranklowerboundpart2}
Let the smallest capacity be given by $\invbidtobudget$. Then, for any \MakeLowercase{\fracextnamefull} $\extfrac$, the competitive ratio of the $\ACrank$ algorithm is at least $e^{-1/\invbidtobudget}(1-1/e)$.
\end{lem}
\begin{proof}{Proof of Lemma \ref{lem:acranklowerboundpart2}:}
Fixing an instance $\instance$, we aim to lower-bound the expected amount of capacity filled under the \ACrank\ algorithm, where the expectation is taken over \emph{sample paths}. In the ranking setting, we extend our definition of a sample path such that $\samplepath = \{\randomdraw_1, \dots, \randomdraw_\horizon\}$ represents the realizations of random variables that govern \vol\ \signup\ decisions.
More specifically, we define $\randomdraw_t$ as a vector of length $|\oppset^{\rankmath}|$ (i.e., $\randomdraw_t$ has one component for every possible ranked set of recommendations). The component of $\randomdraw_t$ corresponding to the ranking $\vec{\opprecommendbase} \in \oppset^{\rankmath}$ indicates the \opp\ $i \in \oppset \cup \{0\}$ that \vol\ $t$ signs up for, conditional on the platform recommending the ranked subset $\vec{\opprecommendbase}$.\footnote{Fixing a sample path $\samplepath$, the output of $\OPT$ and $\ACrank$ are deterministic.}

For a fixed instance $\instance$ and a fixed sample path $\samplepath$, we use $\ACrank$ to denote the amount of capacity filled under the $\ACrank$ algorithm.\footnote{Even though $\ACrank$ depends on the instance and the sample path, we hereafter suppress this dependence to ease exposition (for $\ACrank$ as well as for all other quantities that depend on the instance and the sample path).} 
To provide a lower bound on $\mathbb{E}_{\samplepath}[\ACrank]$, we leverage the LP-free approach developed in \citet{goyal2019online} and \citet{goyal2020asymptotically}, which involves the creation of path-based pseudo-rewards. (For a more complete discussion of the intuition behind this approach, we kindly refer to the proof sketch of Theorem \ref{thm:AClower} in Section \ref{subsubsec:proof}.)

Before defining our pseudo-rewards in this setting, recall our convention that any algorithm (including $\OPT$ and the $\ACrank$ algorithm) always recommends the targeted \opp\ to \exttraf. As before, to ensure that we do not count \signups\ that exceed the capacity of an \opp, we define $\vecvolchoicetsuccess{\ACrank}$ as the \opp\ that \vol\ $t$ \emph{fills capacity of} under $\ACrank$. 

Furthermore, recall that for a fixed instance $\instance$ and along a fixed sample path $\samplepath$, we denote by $\bonustimes$ the subset of \inttraf\ for which $\OPT$ recommends the dummy ranking $\{0\}$; i.e., \OPT\ does not recommend any \opp. 
(Recall that \OPT\ knows \emph{a priori} how much capacity will be filled by \exttraf\ as it knows the realizations of those \vols' \signup\ decisions. This capacity is effectively reserved for \exttraf, and \inttraf\ will be used only if it can fill the remaining capacity. See Definition \ref{def:opt} and its following discussion.)

For the fixed instance $\instance$ and the fixed sample path $\samplepath$, we define the pseudo-rewards $\ballvarranking_t$ for all $t \in [\horizon]$ and $\binvarranking_i$ for all $i \in [\numopps]$ according to the following:
\begin{align}
    \ballvarranking_t(\samplepath) &= \begin{cases}
    \sum_{i \in [\numopps]} \balancefunc(\fillrate_{i,t-1})\mathbbm{1}[\vecvolchoicetsuccess{\ACrank} = i], & t \in \exttimes \cup \bonustimes \\
    \sum_{i \in [\numopps]} \balancefunc(\fillrate_{i,t-1})\mathbbm{1}[\vecvolchoicet{\OPT} = i], & t \in \inttimes \setminus \bonustimes\\
    \end{cases} \label{eq:ballvarranking}
    \\
    \binvarranking_i(\samplepath) &=  \sum_{t \in [\horizon]} \left(1-\balancefunc(\fillrate_{i,t-1})\right)\mathbbm{1}[\vecvolchoicetsuccess{\ACrank} = i] \label{eq:binvarranking}
\end{align}
We now prove that the expected sum of these pseudo-rewards serves as a lower bound on the expected value of $\ACrank$. 

\begin{lem}
\label{lem:acboundranking}
For any instance $\instance$,
\begin{equation}
      \mathbbm{E}_{\samplepath}\big[\ACrank\big] \quad \geq \quad \mathbbm{E}_{\samplepath}\left[\sum_{t \in [\horizon]} \ballvarranking_t + \sum_{i \in [\numopps]} \binvarranking_i\right],
\end{equation}
where $\ballvarranking_t$ and $\binvarranking_i$ are defined in \eqref{eq:ballvarranking} and \eqref{eq:binvarranking}, respectively.
\end{lem}
\begin{proof}{Proof of Lemma \ref{lem:acboundranking}:}
The proof follows from the definition of $\ballvarranking_t$ and $\binvarranking_i$, as well as the design of the \ACrank\ algorithm:
\begin{align}
    \mathbbm{E}_{\samplepath}[\ACrank] &= \mathbb{E}_{\samplepath} \left[ \sum_{t\in \inttimes \setminus \bonustimes}\sum_{i \in [\numopps]} \mathbbm{1}[\vecvolchoicetsuccess{\ACrank} = i] + \sum_{t\in \exttimes \cup \bonustimes}\sum_{i \in [\numopps]}\mathbbm{1}[\vecvolchoicetsuccess{\ACrank} = i]\right] \\
    &=\mathbb{E}_{\samplepath} \left[ \sum_{i \in [\numopps]} \left(\sum_{t\in \inttimes \setminus \bonustimes}\balancefunc(\fillrate_{i,t-1})\mathbbm{1}[\vecvolchoicetsuccess{\ACrank} = i] +  \sum_{t\in \inttimes \setminus \bonustimes}(1-\balancefunc(\fillrate_{i,t-1}))\mathbbm{1}[\vecvolchoicetsuccess{\ACrank} = i]\right.\right. \nonumber \\
    &\qquad \qquad \left.\left.+  \sum_{t\in \exttimes \cup \bonustimes}\balancefunc(\fillrate_{i,t-1})\mathbbm{1}[\vecvolchoicetsuccess{\ACrank} = i] +   \sum_{t\in \exttimes \cup \bonustimes}(1-\balancefunc(\fillrate_{i,t-1}))\mathbbm{1}[\vecvolchoicetsuccess{\ACrank} = i]\right) \right] \\
    & = \mathbb{E}_{\samplepath} \left[ \sum_{i \in [\numopps]} \sum_{t\in \inttimes \setminus \bonustimes}\balancefunc(\fillrate_{i,t-1})\mathbbm{1}[\vecvolchoicetsuccess{\ACrank} = i]\right] +\mathbb{E}_{\samplepath} \left[\sum_{t \in \exttimes \cup \bonustimes} \ballvarranking_t + \sum_{i \in [\numopps]}\binvarranking_i\right] \\
    & = \mathbb{E}_{\samplepath} \left[ \sum_{i \in [\numopps]} \sum_{t\in \inttimes \setminus \bonustimes}\balancefunc(\fillrate_{i,t-1})\mathbbm{1}[\vecvolchoicet{\ACrank} = i]\right] +\mathbb{E}_{\samplepath} \left[\sum_{t \in \exttimes \cup \bonustimes} \ballvarranking_t + \sum_{i \in [\numopps]}\binvarranking_i\right] \label{eq:successsignupsranking} \\
    &\geq \mathbb{E}_{\samplepath} \left[ \sum_{i \in [\numopps]} \sum_{t\in \inttimes \setminus \bonustimes}\balancefunc(\fillrate_{i,t-1})\mathbbm{1}[\vecvolchoicet{\OPT} = i]\right] +\mathbb{E}_{\samplepath} \left[\sum_{t \in \exttimes \cup \bonustimes} \ballvarranking_t + \sum_{i \in [\numopps]}\binvarranking_i\right] \label{eq:acoptconditionrankinglem1} \\
    &= \mathbbm{E}_{\samplepath}\left[\sum_{t \in [\horizon]} \ballvarranking_t + \sum_{i \in [\numopps]} \binvarranking_i\right]
\end{align}
All steps are algebraic except for \eqref{eq:successsignupsranking} and Line \eqref{eq:acoptconditionrankinglem1}. To establish the former, we will show that $\sum_{i \in [\numopps]}\balancefunc(\fillrate_{i,t-1})\mathbbm{1}[\vecvolchoicet{\ACrank} = i] = \sum_{i \in [\numopps]}\balancefunc(\fillrate_{i,t-1})\mathbbm{1}[\vecvolchoicetsuccess{\ACrank} = i]$ for $t \in \inttimes \cup \bonustimes$. We consider two cases. First, if $\fillrate_{\vecvolchoicet{\ACrank},t-1} < 1$, then $\vecvolchoicet{\ACrank} = \vecvolchoicetsuccess{\ACrank}$ and the equality holds. Alternatively, if $\fillrate_{\vecvolchoicet{\ACrank},t-1} = 1$, then $\vecvolchoicetsuccess{\ACrank} = 0$ and $\balancefunc(\fillrate_{\vecvolchoicet{\ACrank},t-1}) = 0$. Thus, both summations equal $0$, and the equality holds. 

{Inequality \eqref{eq:acoptconditionrankinglem1} follows from the $\ACrank$ algorithm's optimality condition (see Equation \ref{eq:subsetmaximization}), which ensures that it recommends the ranking that maximizes the weighted probability of generating a \signup\ (where the weight for \opp\ $i$ at time $t$ is given by $\balancefunc(\fillrate_{i,t-1})$). \revcolor{As \OPT\ does not have foreknowledge of the realization of the sign-up decisions of \inttraf, the recommendation provided by \OPT\ to any \vol\ must be independent of their \signup\ realization.} Hence, the inequality holds.  Applying the definition of the pseudo-rewards $\ballvarranking_t$ for $t \in \inttimes \setminus \bonustimes$ completes the proof of Lemma \ref{lem:acboundranking}.
} \halmos
\end{proof}

Next, we place a lower bound on the expected sum of the pseudo-rewards, which depends on the amount of capacity filled under $\OPT$ along a fixed sample path. As part of this lower bound, we define $\ACrank_{i,t}^{\inttrafmath}$ (as well as $\ACrank_{i,t}^{\exttrafmath}$ and $\OPT_{i,t}^{\inttrafmath}$) in exactly the same way as its counterpart in our base model, i.e., as the amount of \opp\ $i$'s capacity filled at time $t$ by \inttraf\ under $\ACrank$.
\begin{lem}
\label{lem:optboundranking}
For any instance $\instance$, 
\begin{align}
    \mathbbm{E}_{\samplepath}\left[\sum_{t \in [\horizon]} \ballvarranking_t + \sum_{i \in [\numopps]} \binvarranking_i\right]  \quad \geq \quad  e^{-1/\invbidtobudget}\mathbbm{E}_{\samplepath}\left[\sum_{i \in [\numopps]}\right.& \ACrank_{i,\horizon}^\exttrafmath + \ACrankbonus_{i,\horizon} +  \OPT_{i,\horizon}^{\inttrafmath} \cdot \balancefunc \left(\frac{\ACrank_{i, \horizon}^{\inttrafmath}}{\capa_i - \ACrank_{i, \horizon}^{\exttrafmath}}\right)  \nonumber \\ &\left.  +  \capa_i \left(1-\balancefunc \left(\frac{\ACrank_{i, \horizon}^{\inttrafmath} - \ACrankbonus_{i,\horizon}}{\capa_i}\right) - 1/e\right)\right], \label{eq:proprankinglemma2}
    \end{align}
    where $\ballvarranking_t$ and $\binvarranking_i$ are defined in \eqref{eq:ballvarranking} and \eqref{eq:binvarranking}, respectively.
\end{lem}

\begin{proof}{Proof of Lemma \ref{lem:optboundranking}:}
 We proceed by separately deriving lower bounds on the $\ballvarranking_t$ pseudo-rewards and the $\binvarranking_i$ pseudo-rewards. For the former, 
\begin{align}
    \sum_{t \in [\horizon]} \ballvarranking_t \quad &= \quad \sum_{t \in \exttimes \cup \bonustimes}\ballvarranking_t + \sum_{t \in \inttimes \setminus \bonustimes} \ballvarranking_t \\
    &= \quad \sum_{t \in \exttimes \cup \bonustimes}\ballvarranking_t + \sum_{t \in \inttimes \setminus \bonustimes} \sum_{i\in[\numopps]} \balancefunc(\fillrate_{i,t-1})\mathbbm{1}[\vecvolchoicet{\OPT} = i] \label{eq:bonustimesextraranking} \\
    &\geq \quad \sum_{t \in \exttimes \cup \bonustimes}\ballvarranking_t + \sum_{t \in \inttimes \setminus \bonustimes} \sum_{i\in[\numopps]} \balancefunc(\fillrate_{i,\horizon})\mathbbm{1}[\vecvolchoicet{\OPT} = i]  \label{eq:ballvarrankinglowerbound} \\
     &= \quad \sum_{t \in \exttimes \cup \bonustimes}\ballvarranking_t +\sum_{i \in [\numopps]} \balancefunc \left(\frac{\ACrank_{i, \horizon}^{\inttrafmath}}{\capa_i - \ACrank_{i, \horizon}^{\exttrafmath}}\right) \OPT_{i,\horizon}^{\inttrafmath}  \label{eq:ballvarrankingfinalbound}\end{align}
Equality in \eqref{eq:bonustimesextraranking} follows from the definition of $\ballvarranking_t$. Inequality in \eqref{eq:ballvarrankinglowerbound} holds because $\balancefunc$ is a decreasing function in its argument and $\fillrate_{i,\horizon} \geq \fillrate_{i,t-1}$ for all $t \in [\horizon]$. Equality in \eqref{eq:ballvarrankingfinalbound} comes from applying the definition of the fill rate as well as the fact that $\sum_{t \in \inttimes \setminus \bonustimes} \mathbbm{1}[\vecvolchoicet{\OPT} = i] = \OPT_{i,\horizon}^{\inttrafmath}$.

We next turn our attention to the $\binvarranking_i$ pseudo-rewards, which we further separate into two summations: 
\begin{align}
    \sum_{i \in [\numopps]}\binvarranking_i \quad =& \quad \sum_{i \in [\numopps]}\sum_{t \in \exttimes \cup \bonustimes} \left(1-\balancefunc(\fillrate_{i,t-1})\right)\mathbbm{1}[\vecvolchoicetsuccess{\ACrank} = i] + \sum_{i \in [\numopps]}\sum_{t \in \inttimes \setminus \bonustimes} \left(1-\balancefunc(\fillrate_{i,t-1})\right)\mathbbm{1}[\vecvolchoicetsuccess{\ACrank} = i]  
    \end{align}
We note that the first summation has a nice relationship with the first term in \eqref{eq:ballvarrankingfinalbound}. To see this, let us define $\ACrankbonus_{i,\horizon} = \sum_{t \in \bonustimes}\mathbbm{1}[\vecvolchoicetsuccess{\ACrank} = i]$ as the sum of \signups\ under $\ACrank$ by \vols\ who did not receive a ranking under $\OPT$. Then,
\begin{align}
   \sum_{i \in [\numopps]} \sum_{t \in \exttimes \cup \bonustimes} \left(1-\balancefunc(\fillrate_{i,t-1})\right)\mathbbm{1}[\vecvolchoicetsuccess{\ACrank} = i] \quad & = \sum_{i \in [\numopps]}\left(\sum_{t \in \exttimes \cup \bonustimes}\mathbbm{1}[\vecvolchoicetsuccess{\ACrank} = i] - \balancefunc(\fillrate_{i,t-1})\mathbbm{1}[\vecvolchoicetsuccess{\ACrank} = i]\right) \\
    &=\quad \sum_{i \in [\numopps]} \ACrank_{i,\horizon}^\exttrafmath + \ACrankbonus_{i,\horizon} - \sum_{t \in \exttimes \cup \bonustimes} \ballvarranking_t \label{eq:binvarrankingintermediate}
\end{align}
Now focusing on the second summation, which deals with \inttraf\ for which \OPT\ provides a ranking:
\begin{align}    \sum_{i \in [\numopps]}\sum_{t \in \inttimes \setminus \bonustimes} \left(1-\balancefunc(\fillrate_{i,t-1})\right)\mathbbm{1}[\vecvolchoicetsuccess{\ACrank} = i]
    \quad \geq& \quad \sum_{i \in [\numopps]}\sum_{t \in \inttimes \setminus \bonustimes} \left(1-\balancefunc\left(\frac{\ACrank_{i, t-1}^{\inttrafmath}}{\capa_i}\right)\right) \mathbbm{1}[\vecvolchoicetsuccess{\ACrank} = i] \label{eq:binvarrankinglowerbound} \\
    \quad \geq& \quad
    \sum_{i \in [\numopps]}\sum_{\counter \in [\ACrank_{i, \horizon}^{\inttrafmath} - \ACrankbonus_{i,\horizon}]} \left(1-\balancefunc\left(\frac{\counter-1}{\capa_i}\right)\right)
    \label{eq:reimannsum1ranking} \\
    \quad \geq& \quad \sum_{i \in [\numopps]}e^{-1/\capa_i}\sum_{\counter \in [\ACrank_{i, \horizon}^{\inttrafmath}- \ACrankbonus_{i,\horizon}]} \left(1-\balancefunc\left(\frac{\counter}{\capa_i}\right)\right)  \label{eq:reimannsum2ranking} \\
    \quad \geq& \quad  e^{-1/\invbidtobudget}\sum_{i \in [\numopps]}\int_{0}^{\ACrank_{i, \horizon}^{\inttrafmath}- \ACrankbonus_{i,\horizon}} 1 - \balancefunc(x/\capa_i) \ \partial x \label{eq:reimannboundranking} \\
    \quad =& \quad  e^{-1/\invbidtobudget}\sum_{i \in [\numopps]}\capa_i\left(1 - \balancefunc \left(\frac{\ACrank_{i, \horizon}^{\inttrafmath}- \ACrankbonus_{i,\horizon}}{\capa_i}\right) - 1/e\right) \label{eq:binvarrankingfinalbound}
\end{align}
In \eqref{eq:binvarrankinglowerbound}, we use the fact that $\balancefunc$ is decreasing and $\frac{\ACrank_{i, t-1}^{\inttrafmath}}{\capa_i} \leq \frac{\ACrank_{i, t-1}^{\inttrafmath}}{\capa_i - \ACrank_{i, t-1}^{\exttrafmath}} = \fillrate_{i,t-1}$. We then further reduce the argument in $\balancefunc$ in \eqref{eq:reimannsum1ranking} by noting that the lowest possible values of $\ACrank_{i, t}^{\inttrafmath}$ are $\{1, \dots,\ACrank_{i, \horizon}^{\inttrafmath}- \ACrankbonus_{i,\horizon}\}$, since $\ACrank_{i, t}^{\inttrafmath}$ increases by $1$ for any $t \in \inttimes$ where $\vecvolchoicetsuccess{\ACrank} = i$. 

The summation in \eqref{eq:reimannsum1ranking} represents a left Riemann sum of an increasing function. In \eqref{eq:reimannsum2ranking}, we utilize the fact that for any $\counter$, $1-\balancefunc((\counter-1)/\capa_i) \geq e^{1/\invbidtobudget}(1-\balancefunc(\counter/\capa_i))$. As the summation in \eqref{eq:reimannsum2ranking} is now a right Riemann sum of an increasing function, we bound the sum with an appropriate integral in \eqref{eq:reimannboundranking}. We evaluate the integral to arrive at \eqref{eq:binvarrankingfinalbound}.

Combining \eqref{eq:ballvarrankingfinalbound}, \eqref{eq:binvarrankingintermediate}, and \eqref{eq:binvarrankingfinalbound} along with the observation that $e^{-1/\invbidtobudget} < 1$, we see that for any sample path $\samplepath$,
\begin{align*} \sum_{t \in [\horizon]} \ballvarranking_t + \sum_{i \in [\numopps]} \binvarranking_i  \quad \geq \quad e^{-1/\invbidtobudget}\sum_{i \in [\numopps]}&\left( \ACrank_{i,\horizon}^\exttrafmath + \ACrankbonus_{i,\horizon} +  \OPT_{i,\horizon}^{\inttrafmath} \cdot \balancefunc \left(\frac{\ACrank_{i, \horizon}^{\inttrafmath}}{\capa_i - \ACrank_{i, \horizon}^{\exttrafmath}}\right)\right. \nonumber \\ &\left.  +  \capa_i \left(1-\balancefunc \left(\frac{\ACrank_{i, \horizon}^{\inttrafmath} - \ACrankbonus_{i,\horizon}}{\capa_i}\right) - 1/e\right)\right)
\end{align*}
Taking expectations over all sample paths completes the proof of Lemma \ref{lem:optboundranking}.
\halmos
\end{proof}

We now depart from the steps of Theorem \ref{thm:AClower} and derive a lower bound on the right hand side of \eqref{eq:proprankinglemma2} (and thus a lower bound on the sum of the pseudo-rewards) that no longer depends on $\ACrank_{i,\horizon}^\exttrafmath$ and $\ACrankbonus_{i,t}$.

\begin{lem}
\label{lem:rankextrabound}
For any instance $\instance$, any sample path $\samplepath$, and any \opp\ $i$,
\begin{align}
\ACrank_{i,\horizon}^\exttrafmath + \ACrankbonus_{i,\horizon} +  \OPT_{i,\horizon}^{\inttrafmath} \cdot \balancefunc \left(\frac{\ACrank_{i, \horizon}^{\inttrafmath}}{\capa_i - \ACrank_{i, \horizon}^{\exttrafmath}}\right)  +  \capa_i \left(1-\balancefunc \left(\frac{\ACrank_{i, \horizon}^{\inttrafmath} - \ACrankbonus_{i,\horizon}}{\capa_i}\right) - 1/e\right) \geq (1-1/e)\OPT_{i,\horizon},
\label{eq:rankextrabound}
\end{align}
where $\OPT_{i,\horizon} = \OPT_{i,\horizon}^\exttrafmath + \OPT_{i,\horizon}^\inttrafmath$.
\end{lem}
\begin{proof}{Proof of Lemma \ref{lem:rankextrabound}:}
We first note that the left hand side (LHS) of \eqref{eq:rankextrabound} is increasing in $\ACrankbonus_{i,\horizon}$. 
\begin{align}
    \frac{\partial\  \text{LHS}}{\partial \ \ACrankbonus_{i,\horizon}} &= 1 + \balancefunc'\left(\frac{\ACrank_{i, \horizon}^{\inttrafmath} - \ACrankbonus_{i,\horizon}}{\capa_i}\right) \\
    &= 1 - \text{exp}\left(\frac{\ACrank_{i, \horizon}^{\inttrafmath} - \ACrankbonus_{i,\horizon}}{\capa_i} - 1\right) \\
    & \geq 0
\end{align}
The final inequality comes from noting that $\ACrank_{i, \horizon}^{\inttrafmath} - \ACrankbonus_{i,\horizon}$ cannot exceed the capacity $\capa_i$. Therefore, we can lower-bound the LHS by plugging in $\ACrankbonus_{i,\horizon} = 0$ to yield
\begin{equation}
    \text{LHS} \quad \geq \quad \ACrank_{i,\horizon}^\exttrafmath +  \OPT_{i,\horizon}^{\inttrafmath} \cdot \balancefunc \left(\frac{\ACrank_{i, \horizon}^{\inttrafmath}}{\capa_i - \ACrank_{i, \horizon}^{\exttrafmath}}\right)  +  \capa_i \left(1-\balancefunc \left(\frac{\ACrank_{i, \horizon}^{\inttrafmath}}{\capa_i}\right) - 1/e\right)
\end{equation}
There are now two cases to consider: (i) either $\ACrank$ uses the same amount of \exttraf\ as $\OPT$ for \opp\ $i$, or (ii) \opp\ $i$ reaches capacity under $\ACrank$.\footnote{Based on Definition \ref{def:opt}, $\OPT$ will never use \inttraf\ to fill capacity that would otherwise be filled by \exttraf. As a consequence, \OPT\ uses all \exttraf\ for $i$ (or fills \opp\ $i$ with \exttraf) along each sample path. By our convention for \exttraf, $\AC$ will always \emph{recommend} the \vol's targeted \opp\ $\extrecommend$. However, if this \opp\ has already reached capacity, the \signup\ does not \emph{fill} any capacity.} 

In Case (i), we have 
\begin{align}
    \text{LHS}  &\geq  \OPT_{i,\horizon}^\exttrafmath +  \OPT_{i,\horizon}^{\inttrafmath} \cdot \balancefunc \left(\frac{\ACrank_{i, \horizon}^{\inttrafmath}}{\capa_i - \OPT_{i, \horizon}^{\exttrafmath}}\right)  +  \capa_i \left(1-\balancefunc \left(\frac{\ACrank_{i, \horizon}^{\inttrafmath}}{\capa_i}\right) - 1/e\right) \\
    &\geq  \OPT_{i,\horizon}^\exttrafmath +  \OPT_{i,\horizon}^{\inttrafmath} \cdot \balancefunc \left(\frac{\ACrank_{i, \horizon}^{\inttrafmath}}{\capa_i - \OPT_{i, \horizon}^{\exttrafmath}}\right)  +  \OPT_{i, \horizon} \left(1-\balancefunc \left(\frac{\ACrank_{i, \horizon}^{\inttrafmath}}{\capa_i}\right) - 1/e\right) \\
     &=  \OPT_{i,\horizon}^\exttrafmath +  \OPT_{i,\horizon}^{\inttrafmath} \cdot \balancefunc \left(\frac{\ACrank_{i, \horizon}^{\inttrafmath}}{\capa_i - \OPT_{i, \horizon}^{\exttrafmath}}\right)  - (\OPT_{i, \horizon}^{\exttrafmath} + \OPT_{i, \horizon}^{\inttrafmath}) \cdot \balancefunc \left(\frac{\ACrank_{i, \horizon}^{\inttrafmath}}{\capa_i}\right) +  \OPT_{i, \horizon} \left(1 - 1/e\right) \\
     &=  (\OPT_{i, \horizon}^{\exttrafmath} + \OPT_{i, \horizon}^{\inttrafmath})\cdot \exp\left(\frac{\ACrank_{i, \horizon}^{\inttrafmath}}{\capa_i} - 1\right) - \OPT_{i,\horizon}^{\inttrafmath} \cdot \text{exp} \left(\frac{\ACrank_{i, \horizon}^{\inttrafmath}}{\capa_i - \OPT_{i, \horizon}^{\exttrafmath}} -1\right)   +  \OPT_{i, \horizon} \left(1 - 1/e\right) \label{eq:case1lemranking} \\
     &=  \text{exp} \left(\frac{\ACrank_{i, \horizon}^{\inttrafmath}}{\capa_i - \OPT_{i, \horizon}^{\exttrafmath}} -1\right)\left((\OPT_{i, \horizon}^{\exttrafmath} + \OPT_{i, \horizon}^{\inttrafmath})\cdot \exp\left(\frac{-\ACrank_{i, \horizon}^{\inttrafmath} \cdot \OPT_{i, \horizon}^{\exttrafmath}}{\capa_i(\capa_i - \OPT_{i, \horizon}^{\exttrafmath})}\right) - \OPT_{i,\horizon}^{\inttrafmath}\right)   +  \OPT_{i, \horizon} \left(1 - 1/e\right) \\
     &\geq  \text{exp} \left(\frac{\ACrank_{i, \horizon}^{\inttrafmath}}{\capa_i - \OPT_{i, \horizon}^{\exttrafmath}} -1\right)\left((\OPT_{i, \horizon}^{\exttrafmath} + \OPT_{i, \horizon}^{\inttrafmath}) \left(1 -\frac{\ACrank_{i, \horizon}^{\inttrafmath}\cdot \OPT_{i, \horizon}^{\exttrafmath}}{\capa_i(\capa_i - \OPT_{i, \horizon}^{\exttrafmath})}\right) - \OPT_{i,\horizon}^{\inttrafmath}\right)   +  \OPT_{i, \horizon} \left(1 - 1/e\right) \\
    &= \text{exp} \left(\frac{\ACrank_{i, \horizon}^{\inttrafmath}}{\capa_i - \OPT_{i, \horizon}^{\exttrafmath}} -1\right)\left(\OPT_{i, \horizon}^{\exttrafmath} - (\OPT_{i, \horizon}^{\exttrafmath} + \OPT_{i, \horizon}^{\inttrafmath})\left(\frac{\ACrank_{i, \horizon}^{\inttrafmath}\cdot \OPT_{i, \horizon}^{\exttrafmath}}{\capa_i(\capa_i - \OPT_{i, \horizon}^{\exttrafmath})}\right)\right)   +  \OPT_{i, \horizon} \left(1 - 1/e\right) \\
     &\geq  \text{exp} \left(\frac{\ACrank_{i, \horizon}^{\inttrafmath}}{\capa_i - \OPT_{i, \horizon}^{\exttrafmath}} -1\right)\OPT_{i, \horizon}^{\exttrafmath} \left(1 - \frac{\ACrank_{i, \horizon}^{\inttrafmath}}{\capa_i - \OPT_{i, \horizon}^{\exttrafmath}}\right)   +  \OPT_{i, \horizon} \left(1 - 1/e\right) \label{eq:case2lemranking} \\
     &\geq  \OPT_{i, \horizon} \left(1 - 1/e\right) \label{eq:case3lemranking}
\end{align}
Equality in \eqref{eq:case1lemranking} comes from applying the definition of the function $\balancefunc$. Inequality \eqref{eq:case2lemranking} comes from noting that $\OPT_{i, \horizon}^{\exttrafmath} + \OPT_{i, \horizon}^{\inttrafmath} \leq c_i$ and \eqref{eq:case3lemranking} comes from noting that in Case (i), where the amount of \opp\ $i$'s capacity filled by \exttraf\ is the same under $\ACrank$ and $\OPT$, $\ACrank_{i, \horizon}^{\inttrafmath} + \OPT_{i, \horizon}^{\exttrafmath} = \ACrank_{i, \horizon}^{\inttrafmath} + \ACrank_{i, \horizon}^{\exttrafmath} \leq \capa_i$. This implies that $1 - \frac{\ACrank_{i, \horizon}^{\inttrafmath}}{\capa_i - \OPT_{i, \horizon}^{\exttrafmath}} \geq 0$.

In Case (ii), where \opp\ $i$ reaches capacity under $\ACrank$, we have
\begin{align}
    \text{LHS} \quad &\geq \quad \ACrank_{i,\horizon}^\exttrafmath +  \OPT_{i,\horizon}^{\inttrafmath} \cdot \balancefunc \left(\frac{\capa_i - \ACrank_{i, \horizon}^{\exttrafmath}}{\capa_i - \ACrank_{i, \horizon}^{\exttrafmath}}\right)  +  \capa_i \left(1-\balancefunc \left(\frac{\capa_i - \ACrank_{i, \horizon}^{\exttrafmath}}{\capa_i}\right) - 1/e\right) \\
    &= \quad \ACrank_{i,\horizon}^\exttrafmath +  \capa_i \left(1-\balancefunc \left(\frac{\capa_i - \ACrank_{i, \horizon}^{\exttrafmath}}{\capa_i}\right) - 1/e\right) \\
    &= \quad \ACrank_{i,\horizon}^\exttrafmath -\capa_i\cdot \balancefunc \left(1 - \frac{\ACrank_{i, \horizon}^{\exttrafmath}}{\capa_i}\right) +  \capa_i \left(1 - 1/e\right) \label{eq:case4lemranking} \\
    & \geq \quad \capa_i \left(1 - 1/e\right) \label{eq:case5lemranking} \\
    & \geq \quad \OPT_{i,T} \left(1 - 1/e\right) 
\end{align}
To establish \eqref{eq:case5lemranking}, we note that the expression in \eqref{eq:case4lemranking} is non-decreasing in $\ACrank_{i,\horizon}^\exttrafmath$, as its derivative is given by $1 - \exp(-\ACrank_{i, \horizon}^{\exttrafmath}/\capa_i) \geq 0$. Plugging in the smallest possible value for $\ACrank_{i,\horizon}^\exttrafmath$ (which is $0$) yields \eqref{eq:case5lemranking}.

This establishes \eqref{eq:rankextrabound} and completes the proof of Lemma \ref{lem:rankextrabound}. \halmos
\end{proof}

By sequentially applying Lemmas \ref{lem:acboundranking}, \ref{lem:optboundranking}, and \ref{lem:rankextrabound}, we see that we can bound the expected amount of capacity filled under $\ACrank$ via the following inequalities:
\begin{align}
      \mathbbm{E}_{\samplepath}\big[\ACrank] \quad &\geq \quad \mathbbm{E}_{\samplepath}\left[\sum_{t \in [\horizon]} \ballvarranking_t + \sum_{i \in [\numopps]} \binvarranking_i\right]  \\ 
      \quad &\geq \quad  e^{-1/\invbidtobudget}\mathbbm{E}_{\samplepath}\left[\sum_{i \in [\numopps]}\right. \ACrank_{i,\horizon}^\exttrafmath + \ACrankbonus_{i,\horizon} +  \OPT_{i,\horizon}^{\inttrafmath} \cdot \balancefunc \left(\frac{\ACrank_{i, \horizon}^{\inttrafmath}}{\capa_i - \ACrank_{i, \horizon}^{\exttrafmath}}\right)  \nonumber \\ &\left. \qquad \qquad \qquad  +  \capa_i \left(1-\balancefunc \left(\frac{\ACrank_{i, \horizon}^{\inttrafmath} - \ACrankbonus_{i,\horizon}}{\capa_i}\right) - 1/e\right)\right] \\
      & \geq \quad e^{-1/\invbidtobudget}\mathbbm{E}_{\samplepath}\left[\sum_{i \in [\numopps]} (1-1/e)\OPT_{i,T} \right] \\
      & = \quad e^{-1/\invbidtobudget} (1-1/e) \mathbbm{E}_{\samplepath}\left[\OPT\right]
\end{align}
This establishes a lower bound of $e^{-1/\invbidtobudget} (1-1/e)$ on the competitive ratio of $\ACrank$, thereby completing the proof of Lemma \ref{lem:acranklowerboundpart2}. \halmos

Together with Lemma \ref{lem:acranklowerboundpart1}, this completes the proof of Proposition \ref{prop:rankinggeneralization}.
\halmos
\end{proof}

\subsection{Proof of Proposition \ref{prop:cascademodel}}
\label{proof:prop:cascademodel}
To prove Proposition \ref{prop:cascademodel}, we use the approach of the proof of Theorem \ref{thm:AClower}. In the following, we go through the main steps of that proof (as described in Section \ref{subsubsec:proof}), and we provide detailed discussion of any adjustments needed to show that the result of Theorem \ref{thm:AClower} extends to this setting, which we henceforth refer to as the \emph{cascade setting}. We emphasize that the cascade setting is a special case of the ranking setting where we can tailor our analysis to improve the bound (which, for the ranking setting, is given by Proposition \ref{prop:rankinggeneralization}).

To begin, we note that in the cascade setting, if the \fracextname\ is $\extfrac$, then the $\ACrank$ algorithm will fill at least a $\extfrac$ fraction of capacity, as established in Lemma \ref{lem:acranklowerboundpart1}.
We next prove the following additional lower bound on the competitive ratio of the $\ACrank$ algorithm in the cascade setting.

\begin{lem}
\label{lem:accascadepart2}
Let the smallest capacity be given by $\invbidtobudget$ and let the \MCPR\ (given in Definition \ref{def:MCPR}) be at most $\weightcap$. Then, for any \MakeLowercase{\fracextnamefull} $\extfrac$, the competitive ratio of the \ACrank\ algorithm in the cascade setting is at least $z^*$ (as defined in \eqref{eq:defzstar}).
\end{lem}

This lemma is the analog (in the cascade setting) of Lemma \ref{lem:aclowerboundpart2}, and to prove this result we follow the same three steps in the proof of Lemma \ref{lem:aclowerboundpart2}, extended to this setting. 


\medskip

\noindent \textbf{Step 1: Defining Pseudo-Rewards in the Cascade Setting} 

In the cascade setting, our notion of pseudo-rewards remains dependent on both the instance and the sample path. We extend our definition of a sample path such that $\samplepath = \{\samplepathcomponent_1^v, \randomdraw_1^s \dots, \samplepathcomponent_\horizon^v, \randomdraw_\horizon^s\}$ represents the realizations of random variables that govern both \vol\ choices: the choice of which \opp\ to \emph{view} and the choice of which \opp\ to \emph{sign up for}, conditional on viewing.
As \vols' view decisions in this cascade setting are agnostic to the \opp\ in each ranked position, we define $\samplepathcomponent_t^v$ as an integer between $1$ and $K+1$, such that \vol\ $t$ views the \opp\ that is ranked in position $\samplepathcomponent_t^v$. We remind that the ranked subsets are of length at most $K$; hence, we use $\samplepathcomponent_t^v = K+1$ to indicate that the \vol\ exits the platform (at any position) without viewing an \opp. In general, a \vol\ makes two random decisions for each considered position: whether to view and whether to exit if not viewing. However, $\samplepathcomponent_t^v \in [K+1]$ is sufficient information to fully specify the outcome of $\ACrank$ and $\OPT$.

As in the base setting, we define $\randomdraw_t^s$ as a binary vector of length $n$, where the $i$\textsuperscript{th} component of $\randomdraw_t^s$ indicates whether \vol\ $t$ signs up for \opp\ $i$, conditional on viewing \opp\ $i$. We remark that, like all previous settings, given any fixed instance $\instance$ and any fixed sample path $\samplepath$, the output of $\ACrank$ and $\OPT$ are deterministic. We also remark that having $\samplepathcomponent_t^v \leq K$ is not a sufficient condition to ensure that the \vol\ views an \opp\ in that position, as it could be the case that the ranking presented to the \vol\ was shorter than $\samplepathcomponent_t^v$. In that case, again the \vol\ does not view (or sign up for) any \opp.

We further define the set $\cascadebonus$ as the set of \inttraf\ $t \in \inttimes$ for which $t$ does not \emph{view} an \opp\ under $\OPT$, along the given sample path $\samplepath$. This expands on our definition of $\bonustimes$ in the base setting: as before, $t$ is in $\cascadebonus$ if $\OPT$ does not recommend any \opps. Now, we additionally have $t$ in $\cascadebonus$ if the \vol\ would view the \opp\ ranked in position $k$ (i.e., $\samplepathcomponent_t^v = k$) but $\OPT$ provides a ranking of length less than $k$. For instance, based on our assumption that the ranking provided is at most length $K$, \vol\ $t$ will be in $\cascadebonus$ if $\samplepathcomponent_t^v = K+1$. (We emphasize that the realization of $\samplepathcomponent_t^v$ is independent from the ranking provided by \OPT\ for \vol~$t$.)

With this in mind, for the fixed instance $\instance$ and the fixed sample path $\samplepath$, we define the pseudo-rewards $\ballvarcascade_t$ for all $t \in [\horizon]$ and $\binvarcascade_i$ for all $i \in [\numopps]$ according to the following:
\begin{align}
    \ballvarcascade_t &= \begin{cases}
    \sum_{i \in [\numopps]} \balancefunc(\fillrate_{i,t-1})\mathbbm{1}[\vecvolchoicetsuccess{\ACrank} = i], & t \in \exttimes \cup \cascadebonus \\
    \sum_{i \in [\numopps]} \balancefunc(\fillrate_{i,t-1})\mathbbm{1}[\vecvolchoicet{\OPT} = i], & t \in \inttimes \setminus \cascadebonus\\
    \end{cases} \label{eq:ballvarcascade}
    \\
    \binvarcascade_i &=  \sum_{t \in [\horizon]} \left(1-\balancefunc(\fillrate_{i,t-1})\right)\mathbbm{1}[\vecvolchoicetsuccess{\ACrank} = i] \label{eq:binvarcascade}
\end{align}

\noindent \textbf{Step 2: Bounding the Value of \ACrank\ in the Cascade Setting}
 
This step of the proof involves two lemmas, which together establish a lower bound on the expected value of $\ACrank$ that depends (in part) on the expected value of $\OPT$.

\begin{lem}
\label{lem:acboundcascade}
In the cascade setting, for any instance $\instance$,
\begin{equation}
      \mathbbm{E}_{\samplepath}\big[\ACrank\big] \quad \geq \quad \mathbbm{E}_{\samplepath}\left[\sum_{t \in [\horizon]} \ballvarcascade_t + \sum_{i \in [\numopps]} \binvarcascade_i\right],
\end{equation}
where $\ballvarcascade_t$ and $\binvarcascade_i$ are defined in \eqref{eq:ballvarcascade} and \eqref{eq:binvarcascade}, respectively.
\end{lem}
\begin{proof}{Proof of Lemma \ref{lem:acboundcascade}:}
This lemma is the analog (in the cascade setting) of Lemma \ref{lem:lowerboundalg} (proven in Appendix \ref{proof:lem:lowerboundalg}). We follow the same algebraic steps, replicated below. As we later elaborate on, one particular inequality requires additional justification in the cascade setting. 

\begin{align}
    \mathbbm{E}_{\samplepath}[\ACrank] &= \mathbb{E}_{\samplepath} \left[ \sum_{t\in \inttimes \setminus \cascadebonus}\sum_{i \in [\numopps]} \mathbbm{1}\Big[\vecvolchoicetsuccess{\ACrank} = i\Big] + \sum_{t\in \exttimes \cup \cascadebonus}\sum_{i \in [\numopps]}\mathbbm{1}\Big[\vecvolchoicetsuccess{\ACrank} = i|\big]\right] \\
    &=\mathbb{E}_{\samplepath} \left[ \sum_{i \in [\numopps]} \left(\sum_{t\in \inttimes \setminus \cascadebonus}\balancefunc(\fillrate_{i,t-1})\mathbbm{1}\Big[\vecvolchoicetsuccess{\ACrank} = i\Big] +  \sum_{t\in \inttimes \setminus \cascadebonus}(1-\balancefunc(\fillrate_{i,t-1}))\mathbbm{1}\Big[\vecvolchoicetsuccess{\ACrank} = i\Big]\right.\right. \nonumber \\
    &\qquad \qquad \left.\left.+  \sum_{t\in \exttimes \cup \cascadebonus}\balancefunc(\fillrate_{i,t-1})\mathbbm{1}\Big[\vecvolchoicetsuccess{\ACrank} = i\Big] +   \sum_{t\in \exttimes \cup \cascadebonus}(1-\balancefunc(\fillrate_{i,t-1}))\mathbbm{1}\Big[\vecvolchoicetsuccess{\ACrank} = i\Big]\right) \right] \\
    & = \mathbb{E}_{\samplepath} \left[ \sum_{i \in [\numopps]} \sum_{t\in \inttimes \setminus \cascadebonus}\balancefunc(\fillrate_{i,t-1})\mathbbm{1}\Big[\vecvolchoicetsuccess{\ACrank} = i\Big]\right] +\mathbb{E}_{\samplepath} \left[\sum_{t \in \exttimes \cup \cascadebonus} \ballvarcascade_t + \sum_{i \in [\numopps]}\binvarcascade_i\right] \\
    & = \mathbb{E}_{\samplepath} \left[ \sum_{i \in [\numopps]} \sum_{t\in \inttimes \setminus \cascadebonus}\balancefunc(\fillrate_{i,t-1})\mathbbm{1}\Big[\vecvolchoicet{\ACrank} = i\Big]\right] +\mathbb{E}_{\samplepath} \left[\sum_{t \in \exttimes \cup \cascadebonus} \ballvarcascade_t + \sum_{i \in [\numopps]}\binvarcascade_i\right] \label{eq:successsignupscascade} \\
    &\geq \mathbb{E}_{\samplepath} \left[ \sum_{i \in [\numopps]} \sum_{t\in \inttimes \setminus \cascadebonus}\balancefunc(\fillrate_{i,t-1})\mathbbm{1}\Big[\vecvolchoicet{\OPT} = i\Big]\right] +\mathbb{E}_{\samplepath} \left[\sum_{t \in \exttimes \cup \cascadebonus} \ballvarcascade_t + \sum_{i \in [\numopps]}\binvarcascade_i\right] \label{eq:acoptconditioncascade} \\
    &= \mathbbm{E}_{\samplepath}\left[\sum_{t \in [\horizon]} \ballvarcascade_t + \sum_{i \in [\numopps]} \binvarcascade_i\right]
\end{align}
All steps are algebraic except for \eqref{eq:successsignupscascade} and Line \eqref{eq:acoptconditioncascade}. To establish the former, we will show that $\sum_{i \in [\numopps]}\balancefunc(\fillrate_{i,t-1})\mathbbm{1}[\vecvolchoicet{\ACrank} = i] = \sum_{i \in [\numopps]}\balancefunc(\fillrate_{i,t-1})\mathbbm{1}[\vecvolchoicetsuccess{\ACrank} = i]$. We consider two cases. First, if $\fillrate_{\vecvolchoicet{\ACrank},t-1} < 1$, then $\vecvolchoicet{\ACrank} = \vecvolchoicetsuccess{\ACrank}$ and the equality holds. Alternatively, if $\fillrate_{\vecvolchoicet{\ACrank},t-1} = 1$, then $\vecvolchoicetsuccess{\ACrank} = 0$ and $\balancefunc(\fillrate_{\vecvolchoicet{\ACrank},t-1}) = 0$. Thus, both summations equal $0$, and the equality holds. 

Establishing \eqref{eq:acoptconditioncascade} in the cascade setting requires more care, as the set $\inttimes \setminus \cascadebonus$ only includes \vols\ that \emph{viewed} an \opp\ under $\OPT$, and whether or not a \vol\ views an \opp\ under \OPT\ depends on the ranking provided by $\OPT$. 
To that end, it is sufficient to show the following inequality holds for all $t \in \inttimes$, where we define $\samplepath_{-t}$ as a sample path excluding the realizations governing the decisions of \vol\ $t$ (i.e., $\samplepathcomponent_t^v$ and $\randomdraw_t^s$).
\begin{align}
\mathbb{E}_{\samplepathcomponent_t^v, \randomdraw_t^s} \Bigg[\sum_{i \in [\numopps]}&\balancefunc(\fillrate_{i,t-1})\mathbbm{1}\Big[\vecvolchoicet{\ACrank} = i\Big] \mathbbm{1}\Big[t \notin \cascadebonus\Big] \ \big\vert \ \samplepath_{-t} \Bigg] \nonumber \\ 
& \qquad \geq \mathbb{E}_{\samplepathcomponent_t^v, \randomdraw_t^s} \left[\sum_{i \in [\numopps]} \balancefunc(\fillrate_{i,t-1})\mathbbm{1}\Big[\vecvolchoicet{\OPT} = i\Big] \mathbbm{1}\Big[t \notin \cascadebonus\Big] \ \big\vert \ \samplepath_{-t} \right] \label{eq:cascadesampleminust}
\end{align}
Applying the tower property of expectations would then establish the validity of \eqref{eq:acoptconditioncascade}.

To show that \eqref{eq:cascadesampleminust} holds, we first take advantage of the fact that, in the cascade setting, the probability of viewing an \opp\ under any algorithm (including \OPT) depends only on the \emph{length} of the ranking provided by that algorithm, and not on the identity and ordering of the \opps\ in the ranking.

To be precise, we make the following claim:
\begin{clm}
\label{clm:OPTcascade}
In the cascade setting, for a fixed instance $\instance$ and a fixed sample path $\samplepath$, for any \vol\ $t \in \inttimes$ there is a position $k^*_t(\samplepath_{-t})$ such that $t \in \inttimes \setminus \cascadebonus$ if and only if $\samplepathcomponent_t^v \leq k^*_t(\samplepath_{-t})$.
\end{clm}
\begin{proof}{Proof of Claim \ref{clm:OPTcascade}:}
Recall our convention that $\OPT$ recommends the optimal ranking which is of shortest length (breaking ties in favor of the lexicographically smallest such subset with respect to its indices). This convention -- in combination with the fact that the probability of a \vol\ viewing an \opp\ is decreasing in the \opp's rank in the cascade setting -- ensures an important property of $\OPT$: \emph{if $\OPT$ ranks a (non-dummy) \opp\ in position $k$, it will also rank (non-dummy) $\opps$ in positions $1$ through $k-1$.} To see why, suppose that this is not the case. Moving the last-ranked \opp\ up to an open position (i.e., a position occupied by a dummy \opp) shortens the ranking, and doing so weakly increases the amount of filled capacity under $\OPT$. Thus, $\OPT$ should have recommended this ranking, which establishes a contradiction. 

As a consequence, let $k^*_t$ denote the length of the ranking provided to \vol\ $t \in \inttimes$ by $\OPT$ along sample path $\samplepath$. If $t$ views an \opp\ under \OPT, then $\samplepathcomponent_t^v \leq k^*_t$. The converse also holds.

We remark that the length of this ranking (i.e., $k^*_t$) depends only on the number of \opps\ with remaining capacity for \inttraf\ at time $t-1$ (for which \vol\ $t$ has positive conversion probability). This set of \opps\ is not a function of the realizations $\samplepathcomponent_t^v$ and $\randomdraw_t^s$. \halmos
\end{proof}

In light of Claim \ref{clm:OPTcascade}, we can rewrite \eqref{eq:cascadesampleminust} as follows, using $\mathbb{P}^{\cascademath}$ to denote the probability distribution associated with the realizations $\samplepathcomponent_t^v$, which depends only on the parameters of the  \MakeLowercase{\cascademodelname}. Furthermore, we use $\vecopprecommend{\ACrank}(k)$ (resp., $\vecopprecommend{\OPT}(k)$) to denote the \opp\ ranked in position $k$ under $\ACrank$ (resp. $\OPT$).
\begin{align}
\mathbb{E}_{\samplepathcomponent_t^v, \randomdraw_t^s} \Bigg[\sum_{i \in [\numopps]} &\balancefunc(\fillrate_{i,t-1})\mathbbm{1}\Big[\vecvolchoicet{\ACrank} = i\Big] \mathbbm{1}\Big[\samplepathcomponent_t^v \leq k^*_t(\samplepath_{-t})\Big] \ \big\vert \ \samplepath_{-t} \Bigg] \nonumber \\ 
&= \ \sum_{k \in [k^*_t(\samplepath_{-t})]}\mathbb{P}^{\cascademath}\Big[\samplepathcomponent_t^v = k \ \big\vert \ \samplepathcomponent_t^v \leq k^*_t(\samplepath_{-t}) \Big]\mathbb{P}^{\cascademath}\Big[\samplepathcomponent_t^v \leq k^*_t(\samplepath_{-t})\Big]\convprob_{\vecopprecommend{\ACrank}(k),t} \balancefunc(\fillrate_{\vecopprecommend{\ACrank}(k),t-1})  \\ 
&= \ \sum_{k \in [k^*_t(\samplepath_{-t})]}\mathbb{P}^{\cascademath}\Big[\samplepathcomponent_t^v = k\Big]\convprob_{\vecopprecommend{\ACrank}(k),t} \balancefunc(\fillrate_{\vecopprecommend{\ACrank}(k),t-1})  \label{eq:cascadebayesrule1} \\ 
&\geq \ \sum_{k \in [k^*_t(\samplepath_{-t})]}\mathbb{P}^{\cascademath}\Big[\samplepathcomponent_t^v = k \Big]\convprob_{\vecopprecommend{\OPT}(k),t} \balancefunc(\fillrate_{\vecopprecommend{\OPT}(k),t-1}) \label{eq:cascadeOPTcondition} \\
&= \ \sum_{k \in [k^*_t(\samplepath_{-t})]}\mathbb{P}^{\cascademath}\Big[\samplepathcomponent_t^v = k \ \big\vert \ \samplepathcomponent_t^v \leq k^*_t(\samplepath_{-t}) \Big]\mathbb{P}^{\cascademath}\Big[\samplepathcomponent_t^v \leq k^*_t(\samplepath_{-t})\Big]\convprob_{\vecopprecommend{\OPT}(k),t} \balancefunc(\fillrate_{\vecopprecommend{\OPT}(k),t-1})  \label{eq:cascadebayesrule2} \\ 
&= \
\mathbb{E}_{\samplepathcomponent_t^v, \randomdraw_t^s} \left[\sum_{i \in [\numopps]} \balancefunc(\fillrate_{i,t-1})\mathbbm{1}\Big[\vecvolchoicet{\OPT} = i\Big] \mathbbm{1}\Big[\samplepathcomponent_t^v \leq k^*_t(\samplepath_{-t})\Big] \ \big\vert \ \samplepath_{-t} \right] \label{eq:cascadesampleminust2}
\end{align}
We note that equality in \eqref{eq:cascadebayesrule1} and \eqref{eq:cascadebayesrule2} follow from the rules of conditional probability, as for any $k \leq k^*_t(\samplepath_{-t})$, we have $\mathbb{P}^{\cascademath}\Big[\samplepathcomponent_t^v = k \ \big\vert \ \samplepathcomponent_t^v \leq k^*_t(\samplepath_{-t}) \Big] = \frac{\mathbb{P}^{\cascademath}[\samplepathcomponent_t^v  = k]}{\mathbb{P}^{\cascademath}[\samplepathcomponent_t^v \leq k^*_t(\samplepath_{-t})]}$. 
All that remains is to prove that \eqref{eq:cascadeOPTcondition} holds, which we do via the following claim:
\begin{clm}
\label{clm:cascadeconditioning}
Let $\vecopprecommend{\ACrank}$ be the ranking presented by $\ACrank$ to \vol\ $t \in \inttimes$, as given by \eqref{eq:subsetmaximization}. Then, in the cascade setting, $\vecopprecommend{\ACrank}$ also satisfies the following condition for any $k' \leq K$:
$$\vecopprecommend{\ACrank} \in \text{\emph{argmax}}_{\vec{\opprecommendbase}}\sum_{k \in [k']}\mathbb{P}^{\cascademath}\Big[\samplepathcomponent_t^v = k\Big]\convprob_{\vec{\opprecommendbase}(k),t} \balancefunc(\fillrate_{\vec{\opprecommendbase}(k),t-1}).$$
\end{clm}

\begin{proof}{Proof of Claim \ref{clm:cascadeconditioning}}
Applying the optimality condition of the $\ACrank$ algorithm (see \eqref{eq:subsetmaximization}) to the cascade setting, we see that
$$\vecopprecommend{\ACrank} \in \text{{argmax}}_{\vec{\opprecommendbase}}\sum_{k \in [K]}\mathbb{P}^{\cascademath}\Big[\samplepathcomponent_t^v = k\Big]\convprob_{\vec{\opprecommendbase}(k),t} \balancefunc(\fillrate_{\vec{\opprecommendbase}(k),t-1}).$$
To prove Claim \ref{clm:cascadeconditioning}, we need to show that that $\ACrank$ continues to satisfy this optimality condition when considering the sum over the first $k'$ terms, for any $k' \leq K$. 
In the cascade setting, the view probability depends only on an \opp's position, and $\mathbb{P}\left[\samplepathcomponent_t^v = k\right] = \cascadeclick_t\big((1-\cascadeclick_t)(1-\cascadequit_t)\big)^{k-1}$ is decreasing in the position $k$.  Therefore, the $\ACrank$ algorithm will rank \opps\ in descending order of $\convprob_{i,t}\balancefunc(\fillrate_{i,t-1})$ (breaking ties in favor of the lowest-indexed \opp), until it exhausts the maximum list size $K$. To see why, suppose $\convprob_{i,t}\balancefunc(\fillrate_{i,t-1}) > \convprob_{j,t}\balancefunc(\fillrate_{j,t-1})$, but \opp\ $i$ is ranked after \opp\ $j$ in $\vecopprecommend{\ACrank}$. In that case, switching \opp\ $i$ and \opp\ $j$ in $\vecopprecommend{\ACrank}$ would strictly increase the objective that $\ACrank$ is optimizing for, which represents a contradiction.

By an identical argument, ranking \opps\ in descending order of $\convprob_{i,t}\balancefunc(\fillrate_{i,t-1})$ also maximizes 
$$\sum_{k \in [k']}\mathbb{P}^{\cascademath}\Big[\samplepathcomponent_t^v = k\Big]\convprob_{\vec{\opprecommendbase}(k),t} \balancefunc(\fillrate_{\vec{\opprecommendbase}(k),t-1}).$$
Therefore, $\ACrank$ also satisfies this optimality condition for any $k'$, which completes the proof of Claim \ref{clm:cascadeconditioning}.
\halmos
\end{proof}

Together, Claims \ref{clm:OPTcascade} and \ref{clm:cascadeconditioning} prove that \eqref{eq:acoptconditioncascade} holds. This completes the proof of Lemma \ref{lem:acboundcascade}. \halmos
\end{proof}

\begin{lem}
\label{lem:optboundcascade}
In the cascade setting, for any instance $\instance$, 
\begin{align}
    \mathbbm{E}_{\samplepath}\left[\sum_{t \in [\horizon]} \ballvarcascade_t + \sum_{i \in [\numopps]} \binvarcascade_i\right]  \quad \geq \quad  e^{-1/\invbidtobudget}\mathbbm{E}_{\samplepath}\left[\sum_{i \in [\numopps]}\right.& \ACrank_{i,\horizon}^\exttrafmath + \ACcascadebonus_{i,\horizon} +  \OPT_{i,\horizon}^{\inttrafmath} \cdot \balancefunc \left(\frac{\ACrank_{i, \horizon}^{\inttrafmath}}{\capa_i - \ACrank_{i, \horizon}^{\exttrafmath}}\right)  \nonumber \\ &\left.  +  \capa_i \left(1-\balancefunc \left(\frac{\ACrank_{i, \horizon}^{\inttrafmath} - \ACcascadebonus_{i,\horizon}}{\capa_i}\right) - 1/e\right)\right],
    \end{align}
    where $\ballvarcascade_t$ and $\binvarcascade_i$ are defined in \eqref{eq:ballvarcascade} and \eqref{eq:binvarcascade}, respectively.
\end{lem}

Lemma \ref{lem:optboundcascade} is the analog (in the cascade setting) of Lemma \ref{lem:optboundranking}. The proof of this lemma immediately follows by taking identical steps as in the proof of Lemma \ref{lem:optboundranking}. (As the proof is algebraic and holds along each sample path, the distinction between $\bonustimes$ and $\cascadebonus$ does not impact the result.) We omit these details for the sake of brevity.

\medskip

\noindent \textbf{Step 3: Bounding the Competitive Ratio of \ACrank\ in the Cascade Setting} 

The final step of the proof involves the use of the instance-specific mathematical program $\MP$ (see Table \ref{table:MP}), which helps establish a lower bound on the competitive ratio of $\ACrank$ in the cascade setting.

 \begin{lem}
 \label{lem:MPcascade}
 In the cascade setting, for any instance $\instance$, the ratio between the expected value of $\ACrank$ (i.e., $\mathbb{E}_{\samplepath}[\ACrank]$) and the expected value of $\OPT$ (i.e., $\mathbb{E}_{\samplepath}[\OPT]$) on instance $\instance$ is at least the value of $\MP$. 
 \end{lem}

\begin{proof}{Proof of Lemma \ref{lem:MPcascade}:}
 Lemma \ref{lem:MPcascade} is the analog (in the cascade setting) of Lemma \ref{lem:MP}, and our proof follows a similar approach. To prove Lemma \ref{lem:MPcascade}, we consider the following candidate solution:

\begin{align*}
    &x_{1,i,\samplepath} = \ACrank_{i,\horizon}^{\exttrafmath}, \qquad x_{2,i,\samplepath} = \ACrank_{i,\horizon}^{\inttrafmath}, \qquad x_{3,i,\samplepath} = \ACcascadebonus_{i,\horizon}, \\
    &y_{1,i,\samplepath} = \OPT_{i,\horizon}^{\exttrafmath}, \qquad y_{2,i,\samplepath} = \OPT_{i,\horizon}^{\inttrafmath}, \qquad
    z = \frac{\mathbb{E}_{\samplepath}[\ACrank]}{\mathbb{E}_{\samplepath}[\OPT]}
\end{align*}

Such a solution has an objective value equal to the ratio $\mathbb{E}_{\samplepath}[\ACrank]/\mathbb{E}_{\samplepath}[\OPT]$ in $\MP$, and by construction it satisfies all constraints. The first five constraints hold by exactly the same rationale described in the proof of Lemma \ref{lem:MP} (see Appendix \ref{proof:lem:MP}). 

To see that the sixth constraint is satisfied, let us fix a sample path $\samplepath$ and an \opp\ $i$. The total amount of \opp\ $i$'s capacity filled by $\ACrank$ in periods $t \in \inttimes \setminus \cascadebonus$ is given by $x_{2, i, \samplepath} - x_{3, i, \samplepath}$, while the total amount of \opp\ $i$'s capacity filled by $\OPT$ in periods $t \in \inttimes \setminus \cascadebonus$ is given by $y_{2, i, \samplepath}$. Furthermore, for all $t \in \inttimes \setminus \cascadebonus$, \vol\ $t$ views an \opp\ under $\OPT$, which means it fills a unit of capacity with probability at least $\min_{i \in \oppset_t}\convprob_{i,t}$. 
For the same \vol\ $t$,  $\ACrank$ will fill a unit of capacity with probability at most $\max_{i \in \oppset_t}\convprob_{i,t}$. (We remind that $\oppset_t$ represents the subset of \opps\ $i$ for which $\convprob_{i,t} > 0$.)  As a consequence, $x_{2, i, \samplepath} - x_{3, i, \samplepath} \leq \weightcap y_{2, i, \samplepath} $, or equivalently, $x_{2, i, \samplepath} \leq \weightcap y_{2, i, \samplepath} + x_{3, i, \samplepath}$.

Based on the constructed values of $\vec{x}, \vec{y},$ and $z$, as well as the upper bound on $x_{2, i, \samplepath}$ identified above,
\begin{align}\mathbb{E}_{\samplepath}\left[\sum_{i \in [\numopps]}x_{1,i,\samplepath}\right] & = z\cdot \mathbb{E}_{\samplepath}\left[\sum_{i \in [\numopps]}y_{1,i,\samplepath} + y_{2,i,\samplepath}\right] - \mathbb{E}_{\samplepath}\left[\sum_{i \in [\numopps]}x_{2,i,\samplepath}\right]  \\
&\geq z\cdot \mathbb{E}_{\samplepath}\left[\sum_{i \in [\numopps]}y_{1,i,\samplepath} + y_{2,i,\samplepath}\right] - \mathbb{E}_{\samplepath}\left[\sum_{i \in [\numopps]}\weightcap \cdot y_{2, i, \samplepath} + x_{3, i, \samplepath}\right] \\
&= \mathbb{E}_{\samplepath}\left[\sum_{i \in [\numopps]}y_{1,i,\samplepath} \right] - \mathbb{E}_{\samplepath}\left[\sum_{i \in [\numopps]} (1-z)\cdot y_{1,i,\samplepath} + (\weightcap - z)\cdot y_{2,i,\samplepath} \right] - \mathbb{E}_{\samplepath}\left[\sum_{i \in [\numopps]}x_{3, i, \samplepath}\right] \\
&\geq \mathbb{E}_{\samplepath}\left[\sum_{i \in [\numopps]}y_{1,i,\samplepath} \right] - (\weightcap - z) \cdot \mathbb{E}_{\samplepath}\left[\sum_{i \in [\numopps]} y_{1,i,\samplepath} + y_{2,i,\samplepath} \right] - \mathbb{E}_{\samplepath}\left[\sum_{i \in [\numopps]}x_{3, i, \samplepath}\right] \label{eq:weightcapbiggerthan1cascade} \\
&\geq \beta \sum_{i \in [\numopps]} c_i - (\weightcap-z)\sum_{i \in [\numopps]} c_i  - \mathbb{E}_{\samplepath}\left[\sum_{i \in [\numopps]}x_{3, i, \samplepath}\right]
.\end{align}
Inequality \eqref{eq:weightcapbiggerthan1cascade} uses the fact that $\weightcap \geq 1$. The final inequality uses the fact that $\mathbb{E}_{\samplepath}\left[\sum_{i \in [\numopps]}y_{1,i,\samplepath} \right] = \beta \sum_{i \in [\numopps]} c_i$ based on the definitions of the optimal clairvoyant algorithm $\OPT$ and the \fracextname\ $\extfrac$ (see Definitions \ref{def:opt} and \ref{def:beta}). This final inequality establishes that our candidate solution respects constraint (vi).

The fact that the candidate solution satisfies the seventh (and final constraint) follows by applying Lemmas \ref{lem:acboundcascade} and \ref{lem:optboundcascade} from Step 2. This completes the proof of Lemma \ref{lem:MPcascade}. \halmos
\end{proof}

As established in Lemma \ref{lem:MPlower} (proven in Appendix \ref{proof:lem:MPlower}), the optimal value of $\MP$ is at least $z^*$ (as defined in \eqref{eq:defzstar}). Therefore, we have shown a lower bound on the ratio $\mathbb{E}_{\samplepath}[\ACrank]/\mathbb{E}_{\samplepath}[\OPT]$ in the cascade setting for any instance $\instance \in \instancedomain_\extfrac$, where the bound depends on only the \fracextname\ $\extfrac$, the minimum capacity $\invbidtobudget$, and the \MCPR\ $\weightcap$.

Taken as a whole, these three steps prove Lemma \ref{lem:accascadepart2}, namely, that $z^*$ is a lower bound on the competitive ratio of the \ACrank\ algorithm in the cascade setting. Thus, in combination with our observation that the competitive ratio is lower-bounded by $\extfrac$, we have shown that the competitive ratio of the $\ACrank$ algorithm is at least $\compratiofunc(\beta, \invbidtobudget, \weightcap)$, as defined in the statement of Theorem~\ref{thm:AClower}. \halmos

\if false

Based on Claim \ref{clm:OPTcascade}, to prove \eqref{eq:acoptconditioncascade}, it suffices to that the following holds for any $k' \in [K]$ and for any sample path excluding the realizations governing the decisions of \vol\ $t$ (denoted $\samplepath_{-t}$). To aid in notation, we define $\vecopprecommend{\ACrank}(k)$ (resp., $\vecopprecommend{\OPT}(k)$) to denote the \opp\ ranked in position $k$ under $\ACrank$ (resp. $\OPT$).
\begin{align}
\mathbb{E}_{\samplepathcomponent_t^v, \randomdraw_t^s} \left[\sum_{i \in [\numopps]} \balancefunc(\fillrate_{i,t-1})\mathbbm{1}[\vecvolchoicet{\ACrank} = i] \ \big\vert \ \samplepath_{-t}, \samplepathcomponent_t^v \leq k' \right] \ &= \ \sum_{k \in [k']}\mathbb{P}\Big[\samplepathcomponent_t^v = k \ \big\vert \ \samplepathcomponent_t^v \leq k' \Big]\convprob_{\vecopprecommend{\ACrank}(k),t} \balancefunc(\fillrate_{\vecopprecommend{\ACrank}(k),t-1})  \\ &\geq \ \sum_{k \in [k']}\mathbb{P}\Big[\samplepathcomponent_t^v = k \ \big\vert \ \samplepathcomponent_t^v \leq k' \Big]\convprob_{\vecopprecommend{\OPT}(k),t} \balancefunc(\fillrate_{\vecopprecommend{\OPT}(k),t-1}) \label{eq:cascadeOPTcondition} \\ &= \
     \mathbb{E}_{\samplepathcomponent_t^v, \randomdraw_t^s}\left[\sum_{i \in [\numopps]} \balancefunc(\fillrate_{i,t-1})\mathbbm{1}[\vecvolchoicet{\OPT} = i] \ \big\vert \ \samplepath_{-t}, \samplepathcomponent_t^v \leq k' \right]
\end{align}
Applying the tower property of expectations would then establish the validity of \eqref{eq:acoptconditioncascade}. 

\fi

\end{APPENDICES}



\end{document}

%% file: images/rev_upper_bounds_warmup.pgf
\begingroup%
\makeatletter%
\begin{pgfpicture}%
\pgfpathrectangle{\pgfpointorigin}{\pgfqpoint{7.500000in}{6.000000in}}%
\pgfusepath{use as bounding box, clip}%
\begin{pgfscope}%
\pgfsetbuttcap%
\pgfsetmiterjoin%
\definecolor{currentfill}{rgb}{1.000000,1.000000,1.000000}%
\pgfsetfillcolor{currentfill}%
\pgfsetlinewidth{0.000000pt}%
\definecolor{currentstroke}{rgb}{1.000000,1.000000,1.000000}%
\pgfsetstrokecolor{currentstroke}%
\pgfsetdash{}{0pt}%
\pgfpathmoveto{\pgfqpoint{0.000000in}{0.000000in}}%
\pgfpathlineto{\pgfqpoint{7.500000in}{0.000000in}}%
\pgfpathlineto{\pgfqpoint{7.500000in}{6.000000in}}%
\pgfpathlineto{\pgfqpoint{0.000000in}{6.000000in}}%
\pgfpathclose%
\pgfusepath{fill}%
\end{pgfscope}%
\begin{pgfscope}%
\pgfsetbuttcap%
\pgfsetmiterjoin%
\definecolor{currentfill}{rgb}{1.000000,1.000000,1.000000}%
\pgfsetfillcolor{currentfill}%
\pgfsetlinewidth{0.000000pt}%
\definecolor{currentstroke}{rgb}{0.000000,0.000000,0.000000}%
\pgfsetstrokecolor{currentstroke}%
\pgfsetstrokeopacity{0.000000}%
\pgfsetdash{}{0pt}%
\pgfpathmoveto{\pgfqpoint{0.957105in}{0.774972in}}%
\pgfpathlineto{\pgfqpoint{7.347222in}{0.774972in}}%
\pgfpathlineto{\pgfqpoint{7.347222in}{5.847222in}}%
\pgfpathlineto{\pgfqpoint{0.957105in}{5.847222in}}%
\pgfpathclose%
\pgfusepath{fill}%
\end{pgfscope}%
\begin{pgfscope}%
\pgfsetbuttcap%
\pgfsetroundjoin%
\definecolor{currentfill}{rgb}{0.000000,0.000000,0.000000}%
\pgfsetfillcolor{currentfill}%
\pgfsetlinewidth{0.803000pt}%
\definecolor{currentstroke}{rgb}{0.000000,0.000000,0.000000}%
\pgfsetstrokecolor{currentstroke}%
\pgfsetdash{}{0pt}%
\pgfsys@defobject{currentmarker}{\pgfqpoint{0.000000in}{-0.048611in}}{\pgfqpoint{0.000000in}{0.000000in}}{%
\pgfpathmoveto{\pgfqpoint{0.000000in}{0.000000in}}%
\pgfpathlineto{\pgfqpoint{0.000000in}{-0.048611in}}%
\pgfusepath{stroke,fill}%
}%
\begin{pgfscope}%
\pgfsys@transformshift{1.247565in}{0.774972in}%
\pgfsys@useobject{currentmarker}{}%
\end{pgfscope}%
\end{pgfscope}%
\begin{pgfscope}%
\definecolor{textcolor}{rgb}{0.000000,0.000000,0.000000}%
\pgfsetstrokecolor{textcolor}%
\pgfsetfillcolor{textcolor}%
\pgftext[x=1.247565in,y=0.677750in,,top]{\color{textcolor}\rmfamily\fontsize{18.000000}{21.600000}\selectfont \(\displaystyle {0.0}\)}%
\end{pgfscope}%
\begin{pgfscope}%
\pgfsetbuttcap%
\pgfsetroundjoin%
\definecolor{currentfill}{rgb}{0.000000,0.000000,0.000000}%
\pgfsetfillcolor{currentfill}%
\pgfsetlinewidth{0.803000pt}%
\definecolor{currentstroke}{rgb}{0.000000,0.000000,0.000000}%
\pgfsetstrokecolor{currentstroke}%
\pgfsetdash{}{0pt}%
\pgfsys@defobject{currentmarker}{\pgfqpoint{0.000000in}{-0.048611in}}{\pgfqpoint{0.000000in}{0.000000in}}{%
\pgfpathmoveto{\pgfqpoint{0.000000in}{0.000000in}}%
\pgfpathlineto{\pgfqpoint{0.000000in}{-0.048611in}}%
\pgfusepath{stroke,fill}%
}%
\begin{pgfscope}%
\pgfsys@transformshift{2.433115in}{0.774972in}%
\pgfsys@useobject{currentmarker}{}%
\end{pgfscope}%
\end{pgfscope}%
\begin{pgfscope}%
\definecolor{textcolor}{rgb}{0.000000,0.000000,0.000000}%
\pgfsetstrokecolor{textcolor}%
\pgfsetfillcolor{textcolor}%
\pgftext[x=2.433115in,y=0.677750in,,top]{\color{textcolor}\rmfamily\fontsize{18.000000}{21.600000}\selectfont \(\displaystyle {0.2}\)}%
\end{pgfscope}%
\begin{pgfscope}%
\pgfsetbuttcap%
\pgfsetroundjoin%
\definecolor{currentfill}{rgb}{0.000000,0.000000,0.000000}%
\pgfsetfillcolor{currentfill}%
\pgfsetlinewidth{0.803000pt}%
\definecolor{currentstroke}{rgb}{0.000000,0.000000,0.000000}%
\pgfsetstrokecolor{currentstroke}%
\pgfsetdash{}{0pt}%
\pgfsys@defobject{currentmarker}{\pgfqpoint{0.000000in}{-0.048611in}}{\pgfqpoint{0.000000in}{0.000000in}}{%
\pgfpathmoveto{\pgfqpoint{0.000000in}{0.000000in}}%
\pgfpathlineto{\pgfqpoint{0.000000in}{-0.048611in}}%
\pgfusepath{stroke,fill}%
}%
\begin{pgfscope}%
\pgfsys@transformshift{3.618666in}{0.774972in}%
\pgfsys@useobject{currentmarker}{}%
\end{pgfscope}%
\end{pgfscope}%
\begin{pgfscope}%
\definecolor{textcolor}{rgb}{0.000000,0.000000,0.000000}%
\pgfsetstrokecolor{textcolor}%
\pgfsetfillcolor{textcolor}%
\pgftext[x=3.618666in,y=0.677750in,,top]{\color{textcolor}\rmfamily\fontsize{18.000000}{21.600000}\selectfont \(\displaystyle {0.4}\)}%
\end{pgfscope}%
\begin{pgfscope}%
\pgfsetbuttcap%
\pgfsetroundjoin%
\definecolor{currentfill}{rgb}{0.000000,0.000000,0.000000}%
\pgfsetfillcolor{currentfill}%
\pgfsetlinewidth{0.803000pt}%
\definecolor{currentstroke}{rgb}{0.000000,0.000000,0.000000}%
\pgfsetstrokecolor{currentstroke}%
\pgfsetdash{}{0pt}%
\pgfsys@defobject{currentmarker}{\pgfqpoint{0.000000in}{-0.048611in}}{\pgfqpoint{0.000000in}{0.000000in}}{%
\pgfpathmoveto{\pgfqpoint{0.000000in}{0.000000in}}%
\pgfpathlineto{\pgfqpoint{0.000000in}{-0.048611in}}%
\pgfusepath{stroke,fill}%
}%
\begin{pgfscope}%
\pgfsys@transformshift{4.804216in}{0.774972in}%
\pgfsys@useobject{currentmarker}{}%
\end{pgfscope}%
\end{pgfscope}%
\begin{pgfscope}%
\definecolor{textcolor}{rgb}{0.000000,0.000000,0.000000}%
\pgfsetstrokecolor{textcolor}%
\pgfsetfillcolor{textcolor}%
\pgftext[x=4.804216in,y=0.677750in,,top]{\color{textcolor}\rmfamily\fontsize{18.000000}{21.600000}\selectfont \(\displaystyle {0.6}\)}%
\end{pgfscope}%
\begin{pgfscope}%
\pgfsetbuttcap%
\pgfsetroundjoin%
\definecolor{currentfill}{rgb}{0.000000,0.000000,0.000000}%
\pgfsetfillcolor{currentfill}%
\pgfsetlinewidth{0.803000pt}%
\definecolor{currentstroke}{rgb}{0.000000,0.000000,0.000000}%
\pgfsetstrokecolor{currentstroke}%
\pgfsetdash{}{0pt}%
\pgfsys@defobject{currentmarker}{\pgfqpoint{0.000000in}{-0.048611in}}{\pgfqpoint{0.000000in}{0.000000in}}{%
\pgfpathmoveto{\pgfqpoint{0.000000in}{0.000000in}}%
\pgfpathlineto{\pgfqpoint{0.000000in}{-0.048611in}}%
\pgfusepath{stroke,fill}%
}%
\begin{pgfscope}%
\pgfsys@transformshift{5.989767in}{0.774972in}%
\pgfsys@useobject{currentmarker}{}%
\end{pgfscope}%
\end{pgfscope}%
\begin{pgfscope}%
\definecolor{textcolor}{rgb}{0.000000,0.000000,0.000000}%
\pgfsetstrokecolor{textcolor}%
\pgfsetfillcolor{textcolor}%
\pgftext[x=5.989767in,y=0.677750in,,top]{\color{textcolor}\rmfamily\fontsize{18.000000}{21.600000}\selectfont \(\displaystyle {0.8}\)}%
\end{pgfscope}%
\begin{pgfscope}%
\pgfsetbuttcap%
\pgfsetroundjoin%
\definecolor{currentfill}{rgb}{0.000000,0.000000,0.000000}%
\pgfsetfillcolor{currentfill}%
\pgfsetlinewidth{0.803000pt}%
\definecolor{currentstroke}{rgb}{0.000000,0.000000,0.000000}%
\pgfsetstrokecolor{currentstroke}%
\pgfsetdash{}{0pt}%
\pgfsys@defobject{currentmarker}{\pgfqpoint{0.000000in}{-0.048611in}}{\pgfqpoint{0.000000in}{0.000000in}}{%
\pgfpathmoveto{\pgfqpoint{0.000000in}{0.000000in}}%
\pgfpathlineto{\pgfqpoint{0.000000in}{-0.048611in}}%
\pgfusepath{stroke,fill}%
}%
\begin{pgfscope}%
\pgfsys@transformshift{7.175317in}{0.774972in}%
\pgfsys@useobject{currentmarker}{}%
\end{pgfscope}%
\end{pgfscope}%
\begin{pgfscope}%
\definecolor{textcolor}{rgb}{0.000000,0.000000,0.000000}%
\pgfsetstrokecolor{textcolor}%
\pgfsetfillcolor{textcolor}%
\pgftext[x=7.175317in,y=0.677750in,,top]{\color{textcolor}\rmfamily\fontsize{18.000000}{21.600000}\selectfont \(\displaystyle {1.0}\)}%
\end{pgfscope}%
\begin{pgfscope}%
\definecolor{textcolor}{rgb}{0.000000,0.000000,0.000000}%
\pgfsetstrokecolor{textcolor}%
\pgfsetfillcolor{textcolor}%
\pgftext[x=4.152163in,y=0.408845in,,top]{\color{textcolor}\rmfamily\fontsize{22.000000}{26.400000}\selectfont Effective Fraction of External Traffic}%
\end{pgfscope}%
\begin{pgfscope}%
\pgfsetbuttcap%
\pgfsetroundjoin%
\definecolor{currentfill}{rgb}{0.000000,0.000000,0.000000}%
\pgfsetfillcolor{currentfill}%
\pgfsetlinewidth{0.803000pt}%
\definecolor{currentstroke}{rgb}{0.000000,0.000000,0.000000}%
\pgfsetstrokecolor{currentstroke}%
\pgfsetdash{}{0pt}%
\pgfsys@defobject{currentmarker}{\pgfqpoint{-0.048611in}{0.000000in}}{\pgfqpoint{-0.000000in}{0.000000in}}{%
\pgfpathmoveto{\pgfqpoint{-0.000000in}{0.000000in}}%
\pgfpathlineto{\pgfqpoint{-0.048611in}{0.000000in}}%
\pgfusepath{stroke,fill}%
}%
\begin{pgfscope}%
\pgfsys@transformshift{0.957105in}{1.234210in}%
\pgfsys@useobject{currentmarker}{}%
\end{pgfscope}%
\end{pgfscope}%
\begin{pgfscope}%
\definecolor{textcolor}{rgb}{0.000000,0.000000,0.000000}%
\pgfsetstrokecolor{textcolor}%
\pgfsetfillcolor{textcolor}%
\pgftext[x=0.464401in, y=1.150876in, left, base]{\color{textcolor}\rmfamily\fontsize{18.000000}{21.600000}\selectfont \(\displaystyle {0.65}\)}%
\end{pgfscope}%
\begin{pgfscope}%
\pgfsetbuttcap%
\pgfsetroundjoin%
\definecolor{currentfill}{rgb}{0.000000,0.000000,0.000000}%
\pgfsetfillcolor{currentfill}%
\pgfsetlinewidth{0.803000pt}%
\definecolor{currentstroke}{rgb}{0.000000,0.000000,0.000000}%
\pgfsetstrokecolor{currentstroke}%
\pgfsetdash{}{0pt}%
\pgfsys@defobject{currentmarker}{\pgfqpoint{-0.048611in}{0.000000in}}{\pgfqpoint{-0.000000in}{0.000000in}}{%
\pgfpathmoveto{\pgfqpoint{-0.000000in}{0.000000in}}%
\pgfpathlineto{\pgfqpoint{-0.048611in}{0.000000in}}%
\pgfusepath{stroke,fill}%
}%
\begin{pgfscope}%
\pgfsys@transformshift{0.957105in}{1.873718in}%
\pgfsys@useobject{currentmarker}{}%
\end{pgfscope}%
\end{pgfscope}%
\begin{pgfscope}%
\definecolor{textcolor}{rgb}{0.000000,0.000000,0.000000}%
\pgfsetstrokecolor{textcolor}%
\pgfsetfillcolor{textcolor}%
\pgftext[x=0.464401in, y=1.790385in, left, base]{\color{textcolor}\rmfamily\fontsize{18.000000}{21.600000}\selectfont \(\displaystyle {0.70}\)}%
\end{pgfscope}%
\begin{pgfscope}%
\pgfsetbuttcap%
\pgfsetroundjoin%
\definecolor{currentfill}{rgb}{0.000000,0.000000,0.000000}%
\pgfsetfillcolor{currentfill}%
\pgfsetlinewidth{0.803000pt}%
\definecolor{currentstroke}{rgb}{0.000000,0.000000,0.000000}%
\pgfsetstrokecolor{currentstroke}%
\pgfsetdash{}{0pt}%
\pgfsys@defobject{currentmarker}{\pgfqpoint{-0.048611in}{0.000000in}}{\pgfqpoint{-0.000000in}{0.000000in}}{%
\pgfpathmoveto{\pgfqpoint{-0.000000in}{0.000000in}}%
\pgfpathlineto{\pgfqpoint{-0.048611in}{0.000000in}}%
\pgfusepath{stroke,fill}%
}%
\begin{pgfscope}%
\pgfsys@transformshift{0.957105in}{2.513227in}%
\pgfsys@useobject{currentmarker}{}%
\end{pgfscope}%
\end{pgfscope}%
\begin{pgfscope}%
\definecolor{textcolor}{rgb}{0.000000,0.000000,0.000000}%
\pgfsetstrokecolor{textcolor}%
\pgfsetfillcolor{textcolor}%
\pgftext[x=0.464401in, y=2.429894in, left, base]{\color{textcolor}\rmfamily\fontsize{18.000000}{21.600000}\selectfont \(\displaystyle {0.75}\)}%
\end{pgfscope}%
\begin{pgfscope}%
\pgfsetbuttcap%
\pgfsetroundjoin%
\definecolor{currentfill}{rgb}{0.000000,0.000000,0.000000}%
\pgfsetfillcolor{currentfill}%
\pgfsetlinewidth{0.803000pt}%
\definecolor{currentstroke}{rgb}{0.000000,0.000000,0.000000}%
\pgfsetstrokecolor{currentstroke}%
\pgfsetdash{}{0pt}%
\pgfsys@defobject{currentmarker}{\pgfqpoint{-0.048611in}{0.000000in}}{\pgfqpoint{-0.000000in}{0.000000in}}{%
\pgfpathmoveto{\pgfqpoint{-0.000000in}{0.000000in}}%
\pgfpathlineto{\pgfqpoint{-0.048611in}{0.000000in}}%
\pgfusepath{stroke,fill}%
}%
\begin{pgfscope}%
\pgfsys@transformshift{0.957105in}{3.152736in}%
\pgfsys@useobject{currentmarker}{}%
\end{pgfscope}%
\end{pgfscope}%
\begin{pgfscope}%
\definecolor{textcolor}{rgb}{0.000000,0.000000,0.000000}%
\pgfsetstrokecolor{textcolor}%
\pgfsetfillcolor{textcolor}%
\pgftext[x=0.464401in, y=3.069402in, left, base]{\color{textcolor}\rmfamily\fontsize{18.000000}{21.600000}\selectfont \(\displaystyle {0.80}\)}%
\end{pgfscope}%
\begin{pgfscope}%
\pgfsetbuttcap%
\pgfsetroundjoin%
\definecolor{currentfill}{rgb}{0.000000,0.000000,0.000000}%
\pgfsetfillcolor{currentfill}%
\pgfsetlinewidth{0.803000pt}%
\definecolor{currentstroke}{rgb}{0.000000,0.000000,0.000000}%
\pgfsetstrokecolor{currentstroke}%
\pgfsetdash{}{0pt}%
\pgfsys@defobject{currentmarker}{\pgfqpoint{-0.048611in}{0.000000in}}{\pgfqpoint{-0.000000in}{0.000000in}}{%
\pgfpathmoveto{\pgfqpoint{-0.000000in}{0.000000in}}%
\pgfpathlineto{\pgfqpoint{-0.048611in}{0.000000in}}%
\pgfusepath{stroke,fill}%
}%
\begin{pgfscope}%
\pgfsys@transformshift{0.957105in}{3.792244in}%
\pgfsys@useobject{currentmarker}{}%
\end{pgfscope}%
\end{pgfscope}%
\begin{pgfscope}%
\definecolor{textcolor}{rgb}{0.000000,0.000000,0.000000}%
\pgfsetstrokecolor{textcolor}%
\pgfsetfillcolor{textcolor}%
\pgftext[x=0.464401in, y=3.708911in, left, base]{\color{textcolor}\rmfamily\fontsize{18.000000}{21.600000}\selectfont \(\displaystyle {0.85}\)}%
\end{pgfscope}%
\begin{pgfscope}%
\pgfsetbuttcap%
\pgfsetroundjoin%
\definecolor{currentfill}{rgb}{0.000000,0.000000,0.000000}%
\pgfsetfillcolor{currentfill}%
\pgfsetlinewidth{0.803000pt}%
\definecolor{currentstroke}{rgb}{0.000000,0.000000,0.000000}%
\pgfsetstrokecolor{currentstroke}%
\pgfsetdash{}{0pt}%
\pgfsys@defobject{currentmarker}{\pgfqpoint{-0.048611in}{0.000000in}}{\pgfqpoint{-0.000000in}{0.000000in}}{%
\pgfpathmoveto{\pgfqpoint{-0.000000in}{0.000000in}}%
\pgfpathlineto{\pgfqpoint{-0.048611in}{0.000000in}}%
\pgfusepath{stroke,fill}%
}%
\begin{pgfscope}%
\pgfsys@transformshift{0.957105in}{4.431753in}%
\pgfsys@useobject{currentmarker}{}%
\end{pgfscope}%
\end{pgfscope}%
\begin{pgfscope}%
\definecolor{textcolor}{rgb}{0.000000,0.000000,0.000000}%
\pgfsetstrokecolor{textcolor}%
\pgfsetfillcolor{textcolor}%
\pgftext[x=0.464401in, y=4.348420in, left, base]{\color{textcolor}\rmfamily\fontsize{18.000000}{21.600000}\selectfont \(\displaystyle {0.90}\)}%
\end{pgfscope}%
\begin{pgfscope}%
\pgfsetbuttcap%
\pgfsetroundjoin%
\definecolor{currentfill}{rgb}{0.000000,0.000000,0.000000}%
\pgfsetfillcolor{currentfill}%
\pgfsetlinewidth{0.803000pt}%
\definecolor{currentstroke}{rgb}{0.000000,0.000000,0.000000}%
\pgfsetstrokecolor{currentstroke}%
\pgfsetdash{}{0pt}%
\pgfsys@defobject{currentmarker}{\pgfqpoint{-0.048611in}{0.000000in}}{\pgfqpoint{-0.000000in}{0.000000in}}{%
\pgfpathmoveto{\pgfqpoint{-0.000000in}{0.000000in}}%
\pgfpathlineto{\pgfqpoint{-0.048611in}{0.000000in}}%
\pgfusepath{stroke,fill}%
}%
\begin{pgfscope}%
\pgfsys@transformshift{0.957105in}{5.071262in}%
\pgfsys@useobject{currentmarker}{}%
\end{pgfscope}%
\end{pgfscope}%
\begin{pgfscope}%
\definecolor{textcolor}{rgb}{0.000000,0.000000,0.000000}%
\pgfsetstrokecolor{textcolor}%
\pgfsetfillcolor{textcolor}%
\pgftext[x=0.464401in, y=4.987928in, left, base]{\color{textcolor}\rmfamily\fontsize{18.000000}{21.600000}\selectfont \(\displaystyle {0.95}\)}%
\end{pgfscope}%
\begin{pgfscope}%
\pgfsetbuttcap%
\pgfsetroundjoin%
\definecolor{currentfill}{rgb}{0.000000,0.000000,0.000000}%
\pgfsetfillcolor{currentfill}%
\pgfsetlinewidth{0.803000pt}%
\definecolor{currentstroke}{rgb}{0.000000,0.000000,0.000000}%
\pgfsetstrokecolor{currentstroke}%
\pgfsetdash{}{0pt}%
\pgfsys@defobject{currentmarker}{\pgfqpoint{-0.048611in}{0.000000in}}{\pgfqpoint{-0.000000in}{0.000000in}}{%
\pgfpathmoveto{\pgfqpoint{-0.000000in}{0.000000in}}%
\pgfpathlineto{\pgfqpoint{-0.048611in}{0.000000in}}%
\pgfusepath{stroke,fill}%
}%
\begin{pgfscope}%
\pgfsys@transformshift{0.957105in}{5.710770in}%
\pgfsys@useobject{currentmarker}{}%
\end{pgfscope}%
\end{pgfscope}%
\begin{pgfscope}%
\definecolor{textcolor}{rgb}{0.000000,0.000000,0.000000}%
\pgfsetstrokecolor{textcolor}%
\pgfsetfillcolor{textcolor}%
\pgftext[x=0.464401in, y=5.627437in, left, base]{\color{textcolor}\rmfamily\fontsize{18.000000}{21.600000}\selectfont \(\displaystyle {1.00}\)}%
\end{pgfscope}%
\begin{pgfscope}%
\definecolor{textcolor}{rgb}{0.000000,0.000000,0.000000}%
\pgfsetstrokecolor{textcolor}%
\pgfsetfillcolor{textcolor}%
\pgftext[x=0.408845in,y=3.311097in,,bottom,rotate=90.000000]{\color{textcolor}\rmfamily\fontsize{22.000000}{26.400000}\selectfont Comp. Ratio}%
\end{pgfscope}%
\begin{pgfscope}%
\pgfpathrectangle{\pgfqpoint{0.957105in}{0.774972in}}{\pgfqpoint{6.390117in}{5.072250in}}%
\pgfusepath{clip}%
\pgfsetrectcap%
\pgfsetroundjoin%
\pgfsetlinewidth{3.011250pt}%
\definecolor{currentstroke}{rgb}{0.000000,0.000000,0.501961}%
\pgfsetstrokecolor{currentstroke}%
\pgfsetdash{}{0pt}%
\pgfpathmoveto{\pgfqpoint{1.247565in}{1.005529in}}%
\pgfpathlineto{\pgfqpoint{1.247863in}{1.005765in}}%
\pgfpathlineto{\pgfqpoint{1.248766in}{1.006482in}}%
\pgfpathlineto{\pgfqpoint{1.250286in}{1.007689in}}%
\pgfpathlineto{\pgfqpoint{1.252437in}{1.009396in}}%
\pgfpathlineto{\pgfqpoint{1.255229in}{1.011613in}}%
\pgfpathlineto{\pgfqpoint{1.258678in}{1.014350in}}%
\pgfpathlineto{\pgfqpoint{1.262796in}{1.017619in}}%
\pgfpathlineto{\pgfqpoint{1.267598in}{1.021430in}}%
\pgfpathlineto{\pgfqpoint{1.273098in}{1.025796in}}%
\pgfpathlineto{\pgfqpoint{1.279310in}{1.030727in}}%
\pgfpathlineto{\pgfqpoint{1.286250in}{1.036236in}}%
\pgfpathlineto{\pgfqpoint{1.293933in}{1.042334in}}%
\pgfpathlineto{\pgfqpoint{1.302375in}{1.049035in}}%
\pgfpathlineto{\pgfqpoint{1.311592in}{1.056352in}}%
\pgfpathlineto{\pgfqpoint{1.321602in}{1.064297in}}%
\pgfpathlineto{\pgfqpoint{1.332420in}{1.072884in}}%
\pgfpathlineto{\pgfqpoint{1.344065in}{1.082128in}}%
\pgfpathlineto{\pgfqpoint{1.356555in}{1.092042in}}%
\pgfpathlineto{\pgfqpoint{1.369908in}{1.102641in}}%
\pgfpathlineto{\pgfqpoint{1.384144in}{1.113940in}}%
\pgfpathlineto{\pgfqpoint{1.399281in}{1.125956in}}%
\pgfpathlineto{\pgfqpoint{1.415340in}{1.138703in}}%
\pgfpathlineto{\pgfqpoint{1.432341in}{1.152197in}}%
\pgfpathlineto{\pgfqpoint{1.450305in}{1.166457in}}%
\pgfpathlineto{\pgfqpoint{1.469254in}{1.181498in}}%
\pgfpathlineto{\pgfqpoint{1.489209in}{1.197338in}}%
\pgfpathlineto{\pgfqpoint{1.510195in}{1.213995in}}%
\pgfpathlineto{\pgfqpoint{1.532232in}{1.231488in}}%
\pgfpathlineto{\pgfqpoint{1.555347in}{1.249835in}}%
\pgfpathlineto{\pgfqpoint{1.579562in}{1.269057in}}%
\pgfpathlineto{\pgfqpoint{1.604904in}{1.289172in}}%
\pgfpathlineto{\pgfqpoint{1.631396in}{1.310201in}}%
\pgfpathlineto{\pgfqpoint{1.659066in}{1.332164in}}%
\pgfpathlineto{\pgfqpoint{1.687941in}{1.355084in}}%
\pgfpathlineto{\pgfqpoint{1.718048in}{1.378982in}}%
\pgfpathlineto{\pgfqpoint{1.749414in}{1.403879in}}%
\pgfpathlineto{\pgfqpoint{1.782069in}{1.429799in}}%
\pgfpathlineto{\pgfqpoint{1.816042in}{1.456766in}}%
\pgfpathlineto{\pgfqpoint{1.851362in}{1.484802in}}%
\pgfpathlineto{\pgfqpoint{1.888060in}{1.513931in}}%
\pgfpathlineto{\pgfqpoint{1.926167in}{1.544179in}}%
\pgfpathlineto{\pgfqpoint{1.965715in}{1.575571in}}%
\pgfpathlineto{\pgfqpoint{2.006735in}{1.608132in}}%
\pgfpathlineto{\pgfqpoint{2.049261in}{1.641887in}}%
\pgfpathlineto{\pgfqpoint{2.093326in}{1.676864in}}%
\pgfpathlineto{\pgfqpoint{2.138963in}{1.713089in}}%
\pgfpathlineto{\pgfqpoint{2.186205in}{1.750589in}}%
\pgfpathlineto{\pgfqpoint{2.235088in}{1.789390in}}%
\pgfpathlineto{\pgfqpoint{2.285646in}{1.829521in}}%
\pgfpathlineto{\pgfqpoint{2.337914in}{1.871009in}}%
\pgfpathlineto{\pgfqpoint{2.391926in}{1.913882in}}%
\pgfpathlineto{\pgfqpoint{2.447718in}{1.958168in}}%
\pgfpathlineto{\pgfqpoint{2.505323in}{2.003893in}}%
\pgfpathlineto{\pgfqpoint{2.564778in}{2.051086in}}%
\pgfpathlineto{\pgfqpoint{2.626115in}{2.099773in}}%
\pgfpathlineto{\pgfqpoint{2.689368in}{2.149982in}}%
\pgfpathlineto{\pgfqpoint{2.754570in}{2.201737in}}%
\pgfpathlineto{\pgfqpoint{2.821753in}{2.255064in}}%
\pgfpathlineto{\pgfqpoint{2.890946in}{2.309987in}}%
\pgfpathlineto{\pgfqpoint{2.962179in}{2.366529in}}%
\pgfpathlineto{\pgfqpoint{3.035479in}{2.424712in}}%
\pgfpathlineto{\pgfqpoint{3.110869in}{2.484554in}}%
\pgfpathlineto{\pgfqpoint{3.188373in}{2.546074in}}%
\pgfpathlineto{\pgfqpoint{3.268008in}{2.609286in}}%
\pgfpathlineto{\pgfqpoint{3.349791in}{2.674202in}}%
\pgfpathlineto{\pgfqpoint{3.433731in}{2.740831in}}%
\pgfpathlineto{\pgfqpoint{3.519836in}{2.809178in}}%
\pgfpathlineto{\pgfqpoint{3.608106in}{2.879243in}}%
\pgfpathlineto{\pgfqpoint{3.698536in}{2.951023in}}%
\pgfpathlineto{\pgfqpoint{3.791113in}{3.024508in}}%
\pgfpathlineto{\pgfqpoint{3.885820in}{3.099683in}}%
\pgfpathlineto{\pgfqpoint{3.982626in}{3.176524in}}%
\pgfpathlineto{\pgfqpoint{4.081495in}{3.255003in}}%
\pgfpathlineto{\pgfqpoint{4.182380in}{3.335082in}}%
\pgfpathlineto{\pgfqpoint{4.285222in}{3.416714in}}%
\pgfpathlineto{\pgfqpoint{4.389952in}{3.499846in}}%
\pgfpathlineto{\pgfqpoint{4.496489in}{3.584410in}}%
\pgfpathlineto{\pgfqpoint{4.604737in}{3.670334in}}%
\pgfpathlineto{\pgfqpoint{4.714590in}{3.757532in}}%
\pgfpathlineto{\pgfqpoint{4.825930in}{3.845909in}}%
\pgfpathlineto{\pgfqpoint{4.938627in}{3.935364in}}%
\pgfpathlineto{\pgfqpoint{5.052539in}{4.025783in}}%
\pgfpathlineto{\pgfqpoint{5.167519in}{4.117051in}}%
\pgfpathlineto{\pgfqpoint{5.283413in}{4.209043in}}%
\pgfpathlineto{\pgfqpoint{5.400068in}{4.301639in}}%
\pgfpathlineto{\pgfqpoint{5.517330in}{4.394718in}}%
\pgfpathlineto{\pgfqpoint{5.635058in}{4.488166in}}%
\pgfpathlineto{\pgfqpoint{5.753122in}{4.581881in}}%
\pgfpathlineto{\pgfqpoint{5.871412in}{4.675776in}}%
\pgfpathlineto{\pgfqpoint{5.989840in}{4.769780in}}%
\pgfpathlineto{\pgfqpoint{6.108344in}{4.863844in}}%
\pgfpathlineto{\pgfqpoint{6.226882in}{4.957935in}}%
\pgfpathlineto{\pgfqpoint{6.345433in}{5.052037in}}%
\pgfpathlineto{\pgfqpoint{6.463987in}{5.146141in}}%
\pgfpathlineto{\pgfqpoint{6.582542in}{5.240246in}}%
\pgfpathlineto{\pgfqpoint{6.701097in}{5.334351in}}%
\pgfpathlineto{\pgfqpoint{6.819652in}{5.428456in}}%
\pgfpathlineto{\pgfqpoint{6.938207in}{5.522561in}}%
\pgfpathlineto{\pgfqpoint{7.056762in}{5.616665in}}%
\pgfusepath{stroke}%
\end{pgfscope}%
\begin{pgfscope}%
\pgfpathrectangle{\pgfqpoint{0.957105in}{0.774972in}}{\pgfqpoint{6.390117in}{5.072250in}}%
\pgfusepath{clip}%
\pgfsetbuttcap%
\pgfsetroundjoin%
\pgfsetlinewidth{3.011250pt}%
\definecolor{currentstroke}{rgb}{1.000000,0.000000,0.000000}%
\pgfsetstrokecolor{currentstroke}%
\pgfsetdash{{11.100000pt}{4.800000pt}}{0.000000pt}%
\pgfpathmoveto{\pgfqpoint{1.247565in}{1.005529in}}%
\pgfpathlineto{\pgfqpoint{1.247863in}{1.005529in}}%
\pgfpathlineto{\pgfqpoint{1.248766in}{1.005535in}}%
\pgfpathlineto{\pgfqpoint{1.250286in}{1.005551in}}%
\pgfpathlineto{\pgfqpoint{1.252437in}{1.005583in}}%
\pgfpathlineto{\pgfqpoint{1.255229in}{1.005636in}}%
\pgfpathlineto{\pgfqpoint{1.258678in}{1.005717in}}%
\pgfpathlineto{\pgfqpoint{1.262796in}{1.005834in}}%
\pgfpathlineto{\pgfqpoint{1.267598in}{1.005992in}}%
\pgfpathlineto{\pgfqpoint{1.273098in}{1.006201in}}%
\pgfpathlineto{\pgfqpoint{1.279310in}{1.006469in}}%
\pgfpathlineto{\pgfqpoint{1.286250in}{1.006804in}}%
\pgfpathlineto{\pgfqpoint{1.293933in}{1.007216in}}%
\pgfpathlineto{\pgfqpoint{1.302375in}{1.007715in}}%
\pgfpathlineto{\pgfqpoint{1.311592in}{1.008313in}}%
\pgfpathlineto{\pgfqpoint{1.321602in}{1.009020in}}%
\pgfpathlineto{\pgfqpoint{1.332420in}{1.009850in}}%
\pgfpathlineto{\pgfqpoint{1.344065in}{1.010815in}}%
\pgfpathlineto{\pgfqpoint{1.356555in}{1.011930in}}%
\pgfpathlineto{\pgfqpoint{1.369908in}{1.013209in}}%
\pgfpathlineto{\pgfqpoint{1.384144in}{1.014668in}}%
\pgfpathlineto{\pgfqpoint{1.399281in}{1.016326in}}%
\pgfpathlineto{\pgfqpoint{1.415340in}{1.018199in}}%
\pgfpathlineto{\pgfqpoint{1.432341in}{1.020307in}}%
\pgfpathlineto{\pgfqpoint{1.450305in}{1.022671in}}%
\pgfpathlineto{\pgfqpoint{1.469254in}{1.025312in}}%
\pgfpathlineto{\pgfqpoint{1.489209in}{1.028254in}}%
\pgfpathlineto{\pgfqpoint{1.510195in}{1.031522in}}%
\pgfpathlineto{\pgfqpoint{1.532232in}{1.035142in}}%
\pgfpathlineto{\pgfqpoint{1.555347in}{1.039142in}}%
\pgfpathlineto{\pgfqpoint{1.579562in}{1.043553in}}%
\pgfpathlineto{\pgfqpoint{1.604904in}{1.048404in}}%
\pgfpathlineto{\pgfqpoint{1.631396in}{1.053731in}}%
\pgfpathlineto{\pgfqpoint{1.659066in}{1.059569in}}%
\pgfpathlineto{\pgfqpoint{1.687941in}{1.065955in}}%
\pgfpathlineto{\pgfqpoint{1.718048in}{1.072930in}}%
\pgfpathlineto{\pgfqpoint{1.749414in}{1.080537in}}%
\pgfpathlineto{\pgfqpoint{1.782069in}{1.088821in}}%
\pgfpathlineto{\pgfqpoint{1.816042in}{1.097829in}}%
\pgfpathlineto{\pgfqpoint{1.851362in}{1.107613in}}%
\pgfpathlineto{\pgfqpoint{1.888060in}{1.118225in}}%
\pgfpathlineto{\pgfqpoint{1.926167in}{1.129723in}}%
\pgfpathlineto{\pgfqpoint{1.965715in}{1.142166in}}%
\pgfpathlineto{\pgfqpoint{2.006735in}{1.155618in}}%
\pgfpathlineto{\pgfqpoint{2.049261in}{1.170146in}}%
\pgfpathlineto{\pgfqpoint{2.093326in}{1.185818in}}%
\pgfpathlineto{\pgfqpoint{2.138963in}{1.202711in}}%
\pgfpathlineto{\pgfqpoint{2.186205in}{1.220900in}}%
\pgfpathlineto{\pgfqpoint{2.235088in}{1.240468in}}%
\pgfpathlineto{\pgfqpoint{2.285646in}{1.261501in}}%
\pgfpathlineto{\pgfqpoint{2.337914in}{1.284087in}}%
\pgfpathlineto{\pgfqpoint{2.391926in}{1.308321in}}%
\pgfpathlineto{\pgfqpoint{2.447718in}{1.334300in}}%
\pgfpathlineto{\pgfqpoint{2.505323in}{1.362125in}}%
\pgfpathlineto{\pgfqpoint{2.564778in}{1.391902in}}%
\pgfpathlineto{\pgfqpoint{2.626115in}{1.423738in}}%
\pgfpathlineto{\pgfqpoint{2.689368in}{1.457748in}}%
\pgfpathlineto{\pgfqpoint{2.754570in}{1.494046in}}%
\pgfpathlineto{\pgfqpoint{2.821753in}{1.532750in}}%
\pgfpathlineto{\pgfqpoint{2.890946in}{1.573981in}}%
\pgfpathlineto{\pgfqpoint{2.962179in}{1.617860in}}%
\pgfpathlineto{\pgfqpoint{3.035479in}{1.664511in}}%
\pgfpathlineto{\pgfqpoint{3.110869in}{1.714056in}}%
\pgfpathlineto{\pgfqpoint{3.188373in}{1.766617in}}%
\pgfpathlineto{\pgfqpoint{3.268008in}{1.822313in}}%
\pgfpathlineto{\pgfqpoint{3.349791in}{1.881260in}}%
\pgfpathlineto{\pgfqpoint{3.433731in}{1.943569in}}%
\pgfpathlineto{\pgfqpoint{3.519836in}{2.009342in}}%
\pgfpathlineto{\pgfqpoint{3.608106in}{2.078674in}}%
\pgfpathlineto{\pgfqpoint{3.698536in}{2.151648in}}%
\pgfpathlineto{\pgfqpoint{3.791113in}{2.228333in}}%
\pgfpathlineto{\pgfqpoint{3.885820in}{2.308781in}}%
\pgfpathlineto{\pgfqpoint{3.982626in}{2.393026in}}%
\pgfpathlineto{\pgfqpoint{4.081495in}{2.481078in}}%
\pgfpathlineto{\pgfqpoint{4.182380in}{2.572925in}}%
\pgfpathlineto{\pgfqpoint{4.285222in}{2.668525in}}%
\pgfpathlineto{\pgfqpoint{4.389952in}{2.767807in}}%
\pgfpathlineto{\pgfqpoint{4.496489in}{2.870667in}}%
\pgfpathlineto{\pgfqpoint{4.604737in}{2.976967in}}%
\pgfpathlineto{\pgfqpoint{4.714590in}{3.086534in}}%
\pgfpathlineto{\pgfqpoint{4.825930in}{3.199161in}}%
\pgfpathlineto{\pgfqpoint{4.938627in}{3.314609in}}%
\pgfpathlineto{\pgfqpoint{5.052539in}{3.432606in}}%
\pgfpathlineto{\pgfqpoint{5.167519in}{3.552858in}}%
\pgfpathlineto{\pgfqpoint{5.283413in}{3.675053in}}%
\pgfpathlineto{\pgfqpoint{5.400068in}{3.798870in}}%
\pgfpathlineto{\pgfqpoint{5.517330in}{3.923990in}}%
\pgfpathlineto{\pgfqpoint{5.635058in}{4.050109in}}%
\pgfpathlineto{\pgfqpoint{5.753122in}{4.176952in}}%
\pgfpathlineto{\pgfqpoint{5.871412in}{4.304282in}}%
\pgfpathlineto{\pgfqpoint{5.989840in}{4.431911in}}%
\pgfpathlineto{\pgfqpoint{6.108344in}{4.559701in}}%
\pgfpathlineto{\pgfqpoint{6.226882in}{4.687567in}}%
\pgfpathlineto{\pgfqpoint{6.345433in}{4.815460in}}%
\pgfpathlineto{\pgfqpoint{6.463987in}{4.943360in}}%
\pgfpathlineto{\pgfqpoint{6.582542in}{5.071262in}}%
\pgfpathlineto{\pgfqpoint{6.701097in}{5.199163in}}%
\pgfpathlineto{\pgfqpoint{6.819652in}{5.327065in}}%
\pgfpathlineto{\pgfqpoint{6.938207in}{5.454967in}}%
\pgfpathlineto{\pgfqpoint{7.056762in}{5.582868in}}%
\pgfusepath{stroke}%
\end{pgfscope}%
\begin{pgfscope}%
\pgfsetrectcap%
\pgfsetmiterjoin%
\pgfsetlinewidth{0.803000pt}%
\definecolor{currentstroke}{rgb}{0.000000,0.000000,0.000000}%
\pgfsetstrokecolor{currentstroke}%
\pgfsetdash{}{0pt}%
\pgfpathmoveto{\pgfqpoint{0.957105in}{0.774972in}}%
\pgfpathlineto{\pgfqpoint{0.957105in}{5.847222in}}%
\pgfusepath{stroke}%
\end{pgfscope}%
\begin{pgfscope}%
\pgfsetrectcap%
\pgfsetmiterjoin%
\pgfsetlinewidth{0.803000pt}%
\definecolor{currentstroke}{rgb}{0.000000,0.000000,0.000000}%
\pgfsetstrokecolor{currentstroke}%
\pgfsetdash{}{0pt}%
\pgfpathmoveto{\pgfqpoint{7.347222in}{0.774972in}}%
\pgfpathlineto{\pgfqpoint{7.347222in}{5.847222in}}%
\pgfusepath{stroke}%
\end{pgfscope}%
\begin{pgfscope}%
\pgfsetrectcap%
\pgfsetmiterjoin%
\pgfsetlinewidth{0.803000pt}%
\definecolor{currentstroke}{rgb}{0.000000,0.000000,0.000000}%
\pgfsetstrokecolor{currentstroke}%
\pgfsetdash{}{0pt}%
\pgfpathmoveto{\pgfqpoint{0.957105in}{0.774972in}}%
\pgfpathlineto{\pgfqpoint{7.347222in}{0.774972in}}%
\pgfusepath{stroke}%
\end{pgfscope}%
\begin{pgfscope}%
\pgfsetrectcap%
\pgfsetmiterjoin%
\pgfsetlinewidth{0.803000pt}%
\definecolor{currentstroke}{rgb}{0.000000,0.000000,0.000000}%
\pgfsetstrokecolor{currentstroke}%
\pgfsetdash{}{0pt}%
\pgfpathmoveto{\pgfqpoint{0.957105in}{5.847222in}}%
\pgfpathlineto{\pgfqpoint{7.347222in}{5.847222in}}%
\pgfusepath{stroke}%
\end{pgfscope}%
\begin{pgfscope}%
\pgfsetbuttcap%
\pgfsetmiterjoin%
\definecolor{currentfill}{rgb}{1.000000,1.000000,1.000000}%
\pgfsetfillcolor{currentfill}%
\pgfsetfillopacity{0.800000}%
\pgfsetlinewidth{1.003750pt}%
\definecolor{currentstroke}{rgb}{0.800000,0.800000,0.800000}%
\pgfsetstrokecolor{currentstroke}%
\pgfsetstrokeopacity{0.800000}%
\pgfsetdash{}{0pt}%
\pgfpathmoveto{\pgfqpoint{1.161271in}{4.218217in}}%
\pgfpathlineto{\pgfqpoint{4.425441in}{4.218217in}}%
\pgfpathquadraticcurveto{\pgfqpoint{4.483774in}{4.218217in}}{\pgfqpoint{4.483774in}{4.276550in}}%
\pgfpathlineto{\pgfqpoint{4.483774in}{5.643056in}}%
\pgfpathquadraticcurveto{\pgfqpoint{4.483774in}{5.701389in}}{\pgfqpoint{4.425441in}{5.701389in}}%
\pgfpathlineto{\pgfqpoint{1.161271in}{5.701389in}}%
\pgfpathquadraticcurveto{\pgfqpoint{1.102938in}{5.701389in}}{\pgfqpoint{1.102938in}{5.643056in}}%
\pgfpathlineto{\pgfqpoint{1.102938in}{4.276550in}}%
\pgfpathquadraticcurveto{\pgfqpoint{1.102938in}{4.218217in}}{\pgfqpoint{1.161271in}{4.218217in}}%
\pgfpathclose%
\pgfusepath{stroke,fill}%
\end{pgfscope}%
\begin{pgfscope}%
\pgfsetrectcap%
\pgfsetroundjoin%
\pgfsetlinewidth{3.011250pt}%
\definecolor{currentstroke}{rgb}{0.000000,0.000000,0.501961}%
\pgfsetstrokecolor{currentstroke}%
\pgfsetdash{}{0pt}%
\pgfpathmoveto{\pgfqpoint{1.365438in}{5.164881in}}%
\pgfpathlineto{\pgfqpoint{1.598771in}{5.164881in}}%
\pgfusepath{stroke}%
\end{pgfscope}%
\begin{pgfscope}%
\definecolor{textcolor}{rgb}{0.000000,0.000000,0.000000}%
\pgfsetstrokecolor{textcolor}%
\pgfsetfillcolor{textcolor}%
\pgftext[x=1.686271in, y=5.238850in, left, base]{\color{textcolor}\rmfamily\fontsize{21.000000}{25.200000}\selectfont Upper Bound, }%
\end{pgfscope}%
\begin{pgfscope}%
\definecolor{textcolor}{rgb}{0.000000,0.000000,0.000000}%
\pgfsetstrokecolor{textcolor}%
\pgfsetfillcolor{textcolor}%
\pgftext[x=1.686271in, y=4.942775in, left, base]{\color{textcolor}\rmfamily\fontsize{21.000000}{25.200000}\selectfont  AC Lower Bound}%
\end{pgfscope}%
\begin{pgfscope}%
\pgfsetbuttcap%
\pgfsetroundjoin%
\pgfsetlinewidth{3.011250pt}%
\definecolor{currentstroke}{rgb}{1.000000,0.000000,0.000000}%
\pgfsetstrokecolor{currentstroke}%
\pgfsetdash{{11.100000pt}{4.800000pt}}{0.000000pt}%
\pgfpathmoveto{\pgfqpoint{1.365438in}{4.638829in}}%
\pgfpathlineto{\pgfqpoint{1.598771in}{4.638829in}}%
\pgfusepath{stroke}%
\end{pgfscope}%
\begin{pgfscope}%
\definecolor{textcolor}{rgb}{0.000000,0.000000,0.000000}%
\pgfsetstrokecolor{textcolor}%
\pgfsetfillcolor{textcolor}%
\pgftext[x=1.686271in,y=4.536746in,left,base]{\color{textcolor}\rmfamily\fontsize{21.000000}{25.200000}\selectfont MSVV Upper Bound       }%
\end{pgfscope}%
\end{pgfpicture}%
\makeatother%
\endgroup%

%% file: images/rev_general_upper_bounds.pgf
\begingroup%
\makeatletter%
\begin{pgfpicture}%
\pgfpathrectangle{\pgfpointorigin}{\pgfqpoint{7.500000in}{6.000000in}}%
\pgfusepath{use as bounding box, clip}%
\begin{pgfscope}%
\pgfsetbuttcap%
\pgfsetmiterjoin%
\definecolor{currentfill}{rgb}{1.000000,1.000000,1.000000}%
\pgfsetfillcolor{currentfill}%
\pgfsetlinewidth{0.000000pt}%
\definecolor{currentstroke}{rgb}{1.000000,1.000000,1.000000}%
\pgfsetstrokecolor{currentstroke}%
\pgfsetdash{}{0pt}%
\pgfpathmoveto{\pgfqpoint{0.000000in}{0.000000in}}%
\pgfpathlineto{\pgfqpoint{7.500000in}{0.000000in}}%
\pgfpathlineto{\pgfqpoint{7.500000in}{6.000000in}}%
\pgfpathlineto{\pgfqpoint{0.000000in}{6.000000in}}%
\pgfpathclose%
\pgfusepath{fill}%
\end{pgfscope}%
\begin{pgfscope}%
\pgfsetbuttcap%
\pgfsetmiterjoin%
\definecolor{currentfill}{rgb}{1.000000,1.000000,1.000000}%
\pgfsetfillcolor{currentfill}%
\pgfsetlinewidth{0.000000pt}%
\definecolor{currentstroke}{rgb}{0.000000,0.000000,0.000000}%
\pgfsetstrokecolor{currentstroke}%
\pgfsetstrokeopacity{0.000000}%
\pgfsetdash{}{0pt}%
\pgfpathmoveto{\pgfqpoint{0.957105in}{0.774972in}}%
\pgfpathlineto{\pgfqpoint{7.347222in}{0.774972in}}%
\pgfpathlineto{\pgfqpoint{7.347222in}{5.847222in}}%
\pgfpathlineto{\pgfqpoint{0.957105in}{5.847222in}}%
\pgfpathclose%
\pgfusepath{fill}%
\end{pgfscope}%
\begin{pgfscope}%
\pgfsetbuttcap%
\pgfsetroundjoin%
\definecolor{currentfill}{rgb}{0.000000,0.000000,0.000000}%
\pgfsetfillcolor{currentfill}%
\pgfsetlinewidth{0.803000pt}%
\definecolor{currentstroke}{rgb}{0.000000,0.000000,0.000000}%
\pgfsetstrokecolor{currentstroke}%
\pgfsetdash{}{0pt}%
\pgfsys@defobject{currentmarker}{\pgfqpoint{0.000000in}{-0.048611in}}{\pgfqpoint{0.000000in}{0.000000in}}{%
\pgfpathmoveto{\pgfqpoint{0.000000in}{0.000000in}}%
\pgfpathlineto{\pgfqpoint{0.000000in}{-0.048611in}}%
\pgfusepath{stroke,fill}%
}%
\begin{pgfscope}%
\pgfsys@transformshift{1.247565in}{0.774972in}%
\pgfsys@useobject{currentmarker}{}%
\end{pgfscope}%
\end{pgfscope}%
\begin{pgfscope}%
\definecolor{textcolor}{rgb}{0.000000,0.000000,0.000000}%
\pgfsetstrokecolor{textcolor}%
\pgfsetfillcolor{textcolor}%
\pgftext[x=1.247565in,y=0.677750in,,top]{\color{textcolor}\rmfamily\fontsize{18.000000}{21.600000}\selectfont \(\displaystyle {0.0}\)}%
\end{pgfscope}%
\begin{pgfscope}%
\pgfsetbuttcap%
\pgfsetroundjoin%
\definecolor{currentfill}{rgb}{0.000000,0.000000,0.000000}%
\pgfsetfillcolor{currentfill}%
\pgfsetlinewidth{0.803000pt}%
\definecolor{currentstroke}{rgb}{0.000000,0.000000,0.000000}%
\pgfsetstrokecolor{currentstroke}%
\pgfsetdash{}{0pt}%
\pgfsys@defobject{currentmarker}{\pgfqpoint{0.000000in}{-0.048611in}}{\pgfqpoint{0.000000in}{0.000000in}}{%
\pgfpathmoveto{\pgfqpoint{0.000000in}{0.000000in}}%
\pgfpathlineto{\pgfqpoint{0.000000in}{-0.048611in}}%
\pgfusepath{stroke,fill}%
}%
\begin{pgfscope}%
\pgfsys@transformshift{2.421140in}{0.774972in}%
\pgfsys@useobject{currentmarker}{}%
\end{pgfscope}%
\end{pgfscope}%
\begin{pgfscope}%
\definecolor{textcolor}{rgb}{0.000000,0.000000,0.000000}%
\pgfsetstrokecolor{textcolor}%
\pgfsetfillcolor{textcolor}%
\pgftext[x=2.421140in,y=0.677750in,,top]{\color{textcolor}\rmfamily\fontsize{18.000000}{21.600000}\selectfont \(\displaystyle {0.2}\)}%
\end{pgfscope}%
\begin{pgfscope}%
\pgfsetbuttcap%
\pgfsetroundjoin%
\definecolor{currentfill}{rgb}{0.000000,0.000000,0.000000}%
\pgfsetfillcolor{currentfill}%
\pgfsetlinewidth{0.803000pt}%
\definecolor{currentstroke}{rgb}{0.000000,0.000000,0.000000}%
\pgfsetstrokecolor{currentstroke}%
\pgfsetdash{}{0pt}%
\pgfsys@defobject{currentmarker}{\pgfqpoint{0.000000in}{-0.048611in}}{\pgfqpoint{0.000000in}{0.000000in}}{%
\pgfpathmoveto{\pgfqpoint{0.000000in}{0.000000in}}%
\pgfpathlineto{\pgfqpoint{0.000000in}{-0.048611in}}%
\pgfusepath{stroke,fill}%
}%
\begin{pgfscope}%
\pgfsys@transformshift{3.594715in}{0.774972in}%
\pgfsys@useobject{currentmarker}{}%
\end{pgfscope}%
\end{pgfscope}%
\begin{pgfscope}%
\definecolor{textcolor}{rgb}{0.000000,0.000000,0.000000}%
\pgfsetstrokecolor{textcolor}%
\pgfsetfillcolor{textcolor}%
\pgftext[x=3.594715in,y=0.677750in,,top]{\color{textcolor}\rmfamily\fontsize{18.000000}{21.600000}\selectfont \(\displaystyle {0.4}\)}%
\end{pgfscope}%
\begin{pgfscope}%
\pgfsetbuttcap%
\pgfsetroundjoin%
\definecolor{currentfill}{rgb}{0.000000,0.000000,0.000000}%
\pgfsetfillcolor{currentfill}%
\pgfsetlinewidth{0.803000pt}%
\definecolor{currentstroke}{rgb}{0.000000,0.000000,0.000000}%
\pgfsetstrokecolor{currentstroke}%
\pgfsetdash{}{0pt}%
\pgfsys@defobject{currentmarker}{\pgfqpoint{0.000000in}{-0.048611in}}{\pgfqpoint{0.000000in}{0.000000in}}{%
\pgfpathmoveto{\pgfqpoint{0.000000in}{0.000000in}}%
\pgfpathlineto{\pgfqpoint{0.000000in}{-0.048611in}}%
\pgfusepath{stroke,fill}%
}%
\begin{pgfscope}%
\pgfsys@transformshift{4.768291in}{0.774972in}%
\pgfsys@useobject{currentmarker}{}%
\end{pgfscope}%
\end{pgfscope}%
\begin{pgfscope}%
\definecolor{textcolor}{rgb}{0.000000,0.000000,0.000000}%
\pgfsetstrokecolor{textcolor}%
\pgfsetfillcolor{textcolor}%
\pgftext[x=4.768291in,y=0.677750in,,top]{\color{textcolor}\rmfamily\fontsize{18.000000}{21.600000}\selectfont \(\displaystyle {0.6}\)}%
\end{pgfscope}%
\begin{pgfscope}%
\pgfsetbuttcap%
\pgfsetroundjoin%
\definecolor{currentfill}{rgb}{0.000000,0.000000,0.000000}%
\pgfsetfillcolor{currentfill}%
\pgfsetlinewidth{0.803000pt}%
\definecolor{currentstroke}{rgb}{0.000000,0.000000,0.000000}%
\pgfsetstrokecolor{currentstroke}%
\pgfsetdash{}{0pt}%
\pgfsys@defobject{currentmarker}{\pgfqpoint{0.000000in}{-0.048611in}}{\pgfqpoint{0.000000in}{0.000000in}}{%
\pgfpathmoveto{\pgfqpoint{0.000000in}{0.000000in}}%
\pgfpathlineto{\pgfqpoint{0.000000in}{-0.048611in}}%
\pgfusepath{stroke,fill}%
}%
\begin{pgfscope}%
\pgfsys@transformshift{5.941866in}{0.774972in}%
\pgfsys@useobject{currentmarker}{}%
\end{pgfscope}%
\end{pgfscope}%
\begin{pgfscope}%
\definecolor{textcolor}{rgb}{0.000000,0.000000,0.000000}%
\pgfsetstrokecolor{textcolor}%
\pgfsetfillcolor{textcolor}%
\pgftext[x=5.941866in,y=0.677750in,,top]{\color{textcolor}\rmfamily\fontsize{18.000000}{21.600000}\selectfont \(\displaystyle {0.8}\)}%
\end{pgfscope}%
\begin{pgfscope}%
\pgfsetbuttcap%
\pgfsetroundjoin%
\definecolor{currentfill}{rgb}{0.000000,0.000000,0.000000}%
\pgfsetfillcolor{currentfill}%
\pgfsetlinewidth{0.803000pt}%
\definecolor{currentstroke}{rgb}{0.000000,0.000000,0.000000}%
\pgfsetstrokecolor{currentstroke}%
\pgfsetdash{}{0pt}%
\pgfsys@defobject{currentmarker}{\pgfqpoint{0.000000in}{-0.048611in}}{\pgfqpoint{0.000000in}{0.000000in}}{%
\pgfpathmoveto{\pgfqpoint{0.000000in}{0.000000in}}%
\pgfpathlineto{\pgfqpoint{0.000000in}{-0.048611in}}%
\pgfusepath{stroke,fill}%
}%
\begin{pgfscope}%
\pgfsys@transformshift{7.115441in}{0.774972in}%
\pgfsys@useobject{currentmarker}{}%
\end{pgfscope}%
\end{pgfscope}%
\begin{pgfscope}%
\definecolor{textcolor}{rgb}{0.000000,0.000000,0.000000}%
\pgfsetstrokecolor{textcolor}%
\pgfsetfillcolor{textcolor}%
\pgftext[x=7.115441in,y=0.677750in,,top]{\color{textcolor}\rmfamily\fontsize{18.000000}{21.600000}\selectfont \(\displaystyle {1.0}\)}%
\end{pgfscope}%
\begin{pgfscope}%
\definecolor{textcolor}{rgb}{0.000000,0.000000,0.000000}%
\pgfsetstrokecolor{textcolor}%
\pgfsetfillcolor{textcolor}%
\pgftext[x=4.152163in,y=0.408845in,,top]{\color{textcolor}\rmfamily\fontsize{22.000000}{26.400000}\selectfont Effective Fraction of External Traffic}%
\end{pgfscope}%
\begin{pgfscope}%
\pgfsetbuttcap%
\pgfsetroundjoin%
\definecolor{currentfill}{rgb}{0.000000,0.000000,0.000000}%
\pgfsetfillcolor{currentfill}%
\pgfsetlinewidth{0.803000pt}%
\definecolor{currentstroke}{rgb}{0.000000,0.000000,0.000000}%
\pgfsetstrokecolor{currentstroke}%
\pgfsetdash{}{0pt}%
\pgfsys@defobject{currentmarker}{\pgfqpoint{-0.048611in}{0.000000in}}{\pgfqpoint{-0.000000in}{0.000000in}}{%
\pgfpathmoveto{\pgfqpoint{-0.000000in}{0.000000in}}%
\pgfpathlineto{\pgfqpoint{-0.048611in}{0.000000in}}%
\pgfusepath{stroke,fill}%
}%
\begin{pgfscope}%
\pgfsys@transformshift{0.957105in}{1.235866in}%
\pgfsys@useobject{currentmarker}{}%
\end{pgfscope}%
\end{pgfscope}%
\begin{pgfscope}%
\definecolor{textcolor}{rgb}{0.000000,0.000000,0.000000}%
\pgfsetstrokecolor{textcolor}%
\pgfsetfillcolor{textcolor}%
\pgftext[x=0.464401in, y=1.152533in, left, base]{\color{textcolor}\rmfamily\fontsize{18.000000}{21.600000}\selectfont \(\displaystyle {0.65}\)}%
\end{pgfscope}%
\begin{pgfscope}%
\pgfsetbuttcap%
\pgfsetroundjoin%
\definecolor{currentfill}{rgb}{0.000000,0.000000,0.000000}%
\pgfsetfillcolor{currentfill}%
\pgfsetlinewidth{0.803000pt}%
\definecolor{currentstroke}{rgb}{0.000000,0.000000,0.000000}%
\pgfsetstrokecolor{currentstroke}%
\pgfsetdash{}{0pt}%
\pgfsys@defobject{currentmarker}{\pgfqpoint{-0.048611in}{0.000000in}}{\pgfqpoint{-0.000000in}{0.000000in}}{%
\pgfpathmoveto{\pgfqpoint{-0.000000in}{0.000000in}}%
\pgfpathlineto{\pgfqpoint{-0.048611in}{0.000000in}}%
\pgfusepath{stroke,fill}%
}%
\begin{pgfscope}%
\pgfsys@transformshift{0.957105in}{1.880006in}%
\pgfsys@useobject{currentmarker}{}%
\end{pgfscope}%
\end{pgfscope}%
\begin{pgfscope}%
\definecolor{textcolor}{rgb}{0.000000,0.000000,0.000000}%
\pgfsetstrokecolor{textcolor}%
\pgfsetfillcolor{textcolor}%
\pgftext[x=0.464401in, y=1.796673in, left, base]{\color{textcolor}\rmfamily\fontsize{18.000000}{21.600000}\selectfont \(\displaystyle {0.70}\)}%
\end{pgfscope}%
\begin{pgfscope}%
\pgfsetbuttcap%
\pgfsetroundjoin%
\definecolor{currentfill}{rgb}{0.000000,0.000000,0.000000}%
\pgfsetfillcolor{currentfill}%
\pgfsetlinewidth{0.803000pt}%
\definecolor{currentstroke}{rgb}{0.000000,0.000000,0.000000}%
\pgfsetstrokecolor{currentstroke}%
\pgfsetdash{}{0pt}%
\pgfsys@defobject{currentmarker}{\pgfqpoint{-0.048611in}{0.000000in}}{\pgfqpoint{-0.000000in}{0.000000in}}{%
\pgfpathmoveto{\pgfqpoint{-0.000000in}{0.000000in}}%
\pgfpathlineto{\pgfqpoint{-0.048611in}{0.000000in}}%
\pgfusepath{stroke,fill}%
}%
\begin{pgfscope}%
\pgfsys@transformshift{0.957105in}{2.524146in}%
\pgfsys@useobject{currentmarker}{}%
\end{pgfscope}%
\end{pgfscope}%
\begin{pgfscope}%
\definecolor{textcolor}{rgb}{0.000000,0.000000,0.000000}%
\pgfsetstrokecolor{textcolor}%
\pgfsetfillcolor{textcolor}%
\pgftext[x=0.464401in, y=2.440813in, left, base]{\color{textcolor}\rmfamily\fontsize{18.000000}{21.600000}\selectfont \(\displaystyle {0.75}\)}%
\end{pgfscope}%
\begin{pgfscope}%
\pgfsetbuttcap%
\pgfsetroundjoin%
\definecolor{currentfill}{rgb}{0.000000,0.000000,0.000000}%
\pgfsetfillcolor{currentfill}%
\pgfsetlinewidth{0.803000pt}%
\definecolor{currentstroke}{rgb}{0.000000,0.000000,0.000000}%
\pgfsetstrokecolor{currentstroke}%
\pgfsetdash{}{0pt}%
\pgfsys@defobject{currentmarker}{\pgfqpoint{-0.048611in}{0.000000in}}{\pgfqpoint{-0.000000in}{0.000000in}}{%
\pgfpathmoveto{\pgfqpoint{-0.000000in}{0.000000in}}%
\pgfpathlineto{\pgfqpoint{-0.048611in}{0.000000in}}%
\pgfusepath{stroke,fill}%
}%
\begin{pgfscope}%
\pgfsys@transformshift{0.957105in}{3.168286in}%
\pgfsys@useobject{currentmarker}{}%
\end{pgfscope}%
\end{pgfscope}%
\begin{pgfscope}%
\definecolor{textcolor}{rgb}{0.000000,0.000000,0.000000}%
\pgfsetstrokecolor{textcolor}%
\pgfsetfillcolor{textcolor}%
\pgftext[x=0.464401in, y=3.084953in, left, base]{\color{textcolor}\rmfamily\fontsize{18.000000}{21.600000}\selectfont \(\displaystyle {0.80}\)}%
\end{pgfscope}%
\begin{pgfscope}%
\pgfsetbuttcap%
\pgfsetroundjoin%
\definecolor{currentfill}{rgb}{0.000000,0.000000,0.000000}%
\pgfsetfillcolor{currentfill}%
\pgfsetlinewidth{0.803000pt}%
\definecolor{currentstroke}{rgb}{0.000000,0.000000,0.000000}%
\pgfsetstrokecolor{currentstroke}%
\pgfsetdash{}{0pt}%
\pgfsys@defobject{currentmarker}{\pgfqpoint{-0.048611in}{0.000000in}}{\pgfqpoint{-0.000000in}{0.000000in}}{%
\pgfpathmoveto{\pgfqpoint{-0.000000in}{0.000000in}}%
\pgfpathlineto{\pgfqpoint{-0.048611in}{0.000000in}}%
\pgfusepath{stroke,fill}%
}%
\begin{pgfscope}%
\pgfsys@transformshift{0.957105in}{3.812427in}%
\pgfsys@useobject{currentmarker}{}%
\end{pgfscope}%
\end{pgfscope}%
\begin{pgfscope}%
\definecolor{textcolor}{rgb}{0.000000,0.000000,0.000000}%
\pgfsetstrokecolor{textcolor}%
\pgfsetfillcolor{textcolor}%
\pgftext[x=0.464401in, y=3.729093in, left, base]{\color{textcolor}\rmfamily\fontsize{18.000000}{21.600000}\selectfont \(\displaystyle {0.85}\)}%
\end{pgfscope}%
\begin{pgfscope}%
\pgfsetbuttcap%
\pgfsetroundjoin%
\definecolor{currentfill}{rgb}{0.000000,0.000000,0.000000}%
\pgfsetfillcolor{currentfill}%
\pgfsetlinewidth{0.803000pt}%
\definecolor{currentstroke}{rgb}{0.000000,0.000000,0.000000}%
\pgfsetstrokecolor{currentstroke}%
\pgfsetdash{}{0pt}%
\pgfsys@defobject{currentmarker}{\pgfqpoint{-0.048611in}{0.000000in}}{\pgfqpoint{-0.000000in}{0.000000in}}{%
\pgfpathmoveto{\pgfqpoint{-0.000000in}{0.000000in}}%
\pgfpathlineto{\pgfqpoint{-0.048611in}{0.000000in}}%
\pgfusepath{stroke,fill}%
}%
\begin{pgfscope}%
\pgfsys@transformshift{0.957105in}{4.456567in}%
\pgfsys@useobject{currentmarker}{}%
\end{pgfscope}%
\end{pgfscope}%
\begin{pgfscope}%
\definecolor{textcolor}{rgb}{0.000000,0.000000,0.000000}%
\pgfsetstrokecolor{textcolor}%
\pgfsetfillcolor{textcolor}%
\pgftext[x=0.464401in, y=4.373233in, left, base]{\color{textcolor}\rmfamily\fontsize{18.000000}{21.600000}\selectfont \(\displaystyle {0.90}\)}%
\end{pgfscope}%
\begin{pgfscope}%
\pgfsetbuttcap%
\pgfsetroundjoin%
\definecolor{currentfill}{rgb}{0.000000,0.000000,0.000000}%
\pgfsetfillcolor{currentfill}%
\pgfsetlinewidth{0.803000pt}%
\definecolor{currentstroke}{rgb}{0.000000,0.000000,0.000000}%
\pgfsetstrokecolor{currentstroke}%
\pgfsetdash{}{0pt}%
\pgfsys@defobject{currentmarker}{\pgfqpoint{-0.048611in}{0.000000in}}{\pgfqpoint{-0.000000in}{0.000000in}}{%
\pgfpathmoveto{\pgfqpoint{-0.000000in}{0.000000in}}%
\pgfpathlineto{\pgfqpoint{-0.048611in}{0.000000in}}%
\pgfusepath{stroke,fill}%
}%
\begin{pgfscope}%
\pgfsys@transformshift{0.957105in}{5.100707in}%
\pgfsys@useobject{currentmarker}{}%
\end{pgfscope}%
\end{pgfscope}%
\begin{pgfscope}%
\definecolor{textcolor}{rgb}{0.000000,0.000000,0.000000}%
\pgfsetstrokecolor{textcolor}%
\pgfsetfillcolor{textcolor}%
\pgftext[x=0.464401in, y=5.017374in, left, base]{\color{textcolor}\rmfamily\fontsize{18.000000}{21.600000}\selectfont \(\displaystyle {0.95}\)}%
\end{pgfscope}%
\begin{pgfscope}%
\pgfsetbuttcap%
\pgfsetroundjoin%
\definecolor{currentfill}{rgb}{0.000000,0.000000,0.000000}%
\pgfsetfillcolor{currentfill}%
\pgfsetlinewidth{0.803000pt}%
\definecolor{currentstroke}{rgb}{0.000000,0.000000,0.000000}%
\pgfsetstrokecolor{currentstroke}%
\pgfsetdash{}{0pt}%
\pgfsys@defobject{currentmarker}{\pgfqpoint{-0.048611in}{0.000000in}}{\pgfqpoint{-0.000000in}{0.000000in}}{%
\pgfpathmoveto{\pgfqpoint{-0.000000in}{0.000000in}}%
\pgfpathlineto{\pgfqpoint{-0.048611in}{0.000000in}}%
\pgfusepath{stroke,fill}%
}%
\begin{pgfscope}%
\pgfsys@transformshift{0.957105in}{5.744847in}%
\pgfsys@useobject{currentmarker}{}%
\end{pgfscope}%
\end{pgfscope}%
\begin{pgfscope}%
\definecolor{textcolor}{rgb}{0.000000,0.000000,0.000000}%
\pgfsetstrokecolor{textcolor}%
\pgfsetfillcolor{textcolor}%
\pgftext[x=0.464401in, y=5.661514in, left, base]{\color{textcolor}\rmfamily\fontsize{18.000000}{21.600000}\selectfont \(\displaystyle {1.00}\)}%
\end{pgfscope}%
\begin{pgfscope}%
\definecolor{textcolor}{rgb}{0.000000,0.000000,0.000000}%
\pgfsetstrokecolor{textcolor}%
\pgfsetfillcolor{textcolor}%
\pgftext[x=0.408845in,y=3.311097in,,bottom,rotate=90.000000]{\color{textcolor}\rmfamily\fontsize{22.000000}{26.400000}\selectfont Comp. Ratio}%
\end{pgfscope}%
\begin{pgfscope}%
\pgfpathrectangle{\pgfqpoint{0.957105in}{0.774972in}}{\pgfqpoint{6.390117in}{5.072250in}}%
\pgfusepath{clip}%
\pgfsetrectcap%
\pgfsetroundjoin%
\pgfsetlinewidth{2.509375pt}%
\definecolor{currentstroke}{rgb}{0.000000,0.000000,0.501961}%
\pgfsetstrokecolor{currentstroke}%
\pgfsetdash{}{0pt}%
\pgfpathmoveto{\pgfqpoint{1.247565in}{1.005529in}}%
\pgfpathlineto{\pgfqpoint{3.452930in}{1.006629in}}%
\pgfpathlineto{\pgfqpoint{3.500487in}{1.009982in}}%
\pgfpathlineto{\pgfqpoint{3.548557in}{1.015610in}}%
\pgfpathlineto{\pgfqpoint{3.597356in}{1.023578in}}%
\pgfpathlineto{\pgfqpoint{3.646473in}{1.033846in}}%
\pgfpathlineto{\pgfqpoint{3.695488in}{1.046293in}}%
\pgfpathlineto{\pgfqpoint{3.747332in}{1.061803in}}%
\pgfpathlineto{\pgfqpoint{3.798936in}{1.079585in}}%
\pgfpathlineto{\pgfqpoint{3.851953in}{1.100241in}}%
\pgfpathlineto{\pgfqpoint{3.905875in}{1.123679in}}%
\pgfpathlineto{\pgfqpoint{3.960201in}{1.149722in}}%
\pgfpathlineto{\pgfqpoint{4.014437in}{1.178104in}}%
\pgfpathlineto{\pgfqpoint{4.071538in}{1.210508in}}%
\pgfpathlineto{\pgfqpoint{4.127910in}{1.244984in}}%
\pgfpathlineto{\pgfqpoint{4.186834in}{1.283610in}}%
\pgfpathlineto{\pgfqpoint{4.248307in}{1.326670in}}%
\pgfpathlineto{\pgfqpoint{4.308238in}{1.371313in}}%
\pgfpathlineto{\pgfqpoint{4.370377in}{1.420320in}}%
\pgfpathlineto{\pgfqpoint{4.434694in}{1.473904in}}%
\pgfpathlineto{\pgfqpoint{4.501151in}{1.532263in}}%
\pgfpathlineto{\pgfqpoint{4.569700in}{1.595581in}}%
\pgfpathlineto{\pgfqpoint{4.640282in}{1.664022in}}%
\pgfpathlineto{\pgfqpoint{4.707932in}{1.732646in}}%
\pgfpathlineto{\pgfqpoint{4.777226in}{1.805948in}}%
\pgfpathlineto{\pgfqpoint{4.848089in}{1.884000in}}%
\pgfpathlineto{\pgfqpoint{4.920439in}{1.966846in}}%
\pgfpathlineto{\pgfqpoint{4.994183in}{2.054510in}}%
\pgfpathlineto{\pgfqpoint{5.069221in}{2.146983in}}%
\pgfpathlineto{\pgfqpoint{5.150935in}{2.251359in}}%
\pgfpathlineto{\pgfqpoint{5.233873in}{2.361140in}}%
\pgfpathlineto{\pgfqpoint{5.317892in}{2.476215in}}%
\pgfpathlineto{\pgfqpoint{5.402846in}{2.596440in}}%
\pgfpathlineto{\pgfqpoint{5.488587in}{2.721647in}}%
\pgfpathlineto{\pgfqpoint{5.574973in}{2.851646in}}%
\pgfpathlineto{\pgfqpoint{5.667680in}{2.995365in}}%
\pgfpathlineto{\pgfqpoint{5.760821in}{3.144056in}}%
\pgfpathlineto{\pgfqpoint{5.854269in}{3.297479in}}%
\pgfpathlineto{\pgfqpoint{5.947919in}{3.455408in}}%
\pgfpathlineto{\pgfqpoint{6.047554in}{3.627927in}}%
\pgfpathlineto{\pgfqpoint{6.147261in}{3.805116in}}%
\pgfpathlineto{\pgfqpoint{6.247000in}{3.986819in}}%
\pgfpathlineto{\pgfqpoint{6.352618in}{4.183994in}}%
\pgfpathlineto{\pgfqpoint{6.458239in}{4.385972in}}%
\pgfpathlineto{\pgfqpoint{6.563861in}{4.592652in}}%
\pgfpathlineto{\pgfqpoint{6.675350in}{4.815811in}}%
\pgfpathlineto{\pgfqpoint{6.786840in}{5.043998in}}%
\pgfpathlineto{\pgfqpoint{6.898330in}{5.277112in}}%
\pgfpathlineto{\pgfqpoint{7.015687in}{5.527712in}}%
\pgfpathlineto{\pgfqpoint{7.056762in}{5.616665in}}%
\pgfpathlineto{\pgfqpoint{7.056762in}{5.616665in}}%
\pgfusepath{stroke}%
\end{pgfscope}%
\begin{pgfscope}%
\pgfpathrectangle{\pgfqpoint{0.957105in}{0.774972in}}{\pgfqpoint{6.390117in}{5.072250in}}%
\pgfusepath{clip}%
\pgfsetbuttcap%
\pgfsetroundjoin%
\pgfsetlinewidth{2.509375pt}%
\definecolor{currentstroke}{rgb}{1.000000,0.000000,0.000000}%
\pgfsetstrokecolor{currentstroke}%
\pgfsetstrokeopacity{0.700000}%
\pgfsetdash{{9.250000pt}{4.000000pt}}{0.000000pt}%
\pgfpathmoveto{\pgfqpoint{1.247565in}{1.005529in}}%
\pgfpathlineto{\pgfqpoint{3.452930in}{1.006629in}}%
\pgfpathlineto{\pgfqpoint{3.500487in}{1.009982in}}%
\pgfpathlineto{\pgfqpoint{3.548557in}{1.015610in}}%
\pgfpathlineto{\pgfqpoint{3.650363in}{1.031025in}}%
\pgfpathlineto{\pgfqpoint{3.717435in}{1.043828in}}%
\pgfpathlineto{\pgfqpoint{3.783771in}{1.058741in}}%
\pgfpathlineto{\pgfqpoint{3.851953in}{1.076404in}}%
\pgfpathlineto{\pgfqpoint{3.917673in}{1.095696in}}%
\pgfpathlineto{\pgfqpoint{3.985368in}{1.117933in}}%
\pgfpathlineto{\pgfqpoint{4.051057in}{1.141858in}}%
\pgfpathlineto{\pgfqpoint{4.117146in}{1.168321in}}%
\pgfpathlineto{\pgfqpoint{4.183077in}{1.197180in}}%
\pgfpathlineto{\pgfqpoint{4.248307in}{1.228223in}}%
\pgfpathlineto{\pgfqpoint{4.312313in}{1.261167in}}%
\pgfpathlineto{\pgfqpoint{4.374598in}{1.295664in}}%
\pgfpathlineto{\pgfqpoint{4.439058in}{1.333982in}}%
\pgfpathlineto{\pgfqpoint{4.501151in}{1.373494in}}%
\pgfpathlineto{\pgfqpoint{4.565066in}{1.416922in}}%
\pgfpathlineto{\pgfqpoint{4.626006in}{1.461023in}}%
\pgfpathlineto{\pgfqpoint{4.688433in}{1.509019in}}%
\pgfpathlineto{\pgfqpoint{4.737432in}{1.546985in}}%
\pgfpathlineto{\pgfqpoint{4.817532in}{1.597389in}}%
\pgfpathlineto{\pgfqpoint{4.889256in}{1.645107in}}%
\pgfpathlineto{\pgfqpoint{4.957143in}{1.692922in}}%
\pgfpathlineto{\pgfqpoint{5.020839in}{1.740525in}}%
\pgfpathlineto{\pgfqpoint{5.080040in}{1.787539in}}%
\pgfpathlineto{\pgfqpoint{5.134489in}{1.833511in}}%
\pgfpathlineto{\pgfqpoint{5.189496in}{1.883028in}}%
\pgfpathlineto{\pgfqpoint{5.239442in}{1.931081in}}%
\pgfpathlineto{\pgfqpoint{5.284164in}{1.976979in}}%
\pgfpathlineto{\pgfqpoint{5.329169in}{2.026305in}}%
\pgfpathlineto{\pgfqpoint{5.368762in}{2.072668in}}%
\pgfpathlineto{\pgfqpoint{5.408539in}{2.122439in}}%
\pgfpathlineto{\pgfqpoint{5.448485in}{2.176098in}}%
\pgfpathlineto{\pgfqpoint{5.482849in}{2.225618in}}%
\pgfpathlineto{\pgfqpoint{5.517317in}{2.278849in}}%
\pgfpathlineto{\pgfqpoint{5.551882in}{2.336308in}}%
\pgfpathlineto{\pgfqpoint{5.586533in}{2.398617in}}%
\pgfpathlineto{\pgfqpoint{5.615470in}{2.454789in}}%
\pgfpathlineto{\pgfqpoint{5.644457in}{2.515380in}}%
\pgfpathlineto{\pgfqpoint{5.971353in}{3.233025in}}%
\pgfpathlineto{\pgfqpoint{7.056762in}{5.616019in}}%
\pgfpathlineto{\pgfqpoint{7.056762in}{5.616019in}}%
\pgfusepath{stroke}%
\end{pgfscope}%
\begin{pgfscope}%
\pgfpathrectangle{\pgfqpoint{0.957105in}{0.774972in}}{\pgfqpoint{6.390117in}{5.072250in}}%
\pgfusepath{clip}%
\pgfsetrectcap%
\pgfsetroundjoin%
\pgfsetlinewidth{2.509375pt}%
\definecolor{currentstroke}{rgb}{0.000000,0.000000,0.501961}%
\pgfsetstrokecolor{currentstroke}%
\pgfsetdash{}{0pt}%
\pgfpathmoveto{\pgfqpoint{1.247565in}{1.005529in}}%
\pgfpathlineto{\pgfqpoint{3.452930in}{1.006629in}}%
\pgfpathlineto{\pgfqpoint{3.500487in}{1.009982in}}%
\pgfpathlineto{\pgfqpoint{3.548557in}{1.015610in}}%
\pgfpathlineto{\pgfqpoint{3.597356in}{1.023578in}}%
\pgfpathlineto{\pgfqpoint{3.646473in}{1.033846in}}%
\pgfpathlineto{\pgfqpoint{3.695488in}{1.046293in}}%
\pgfpathlineto{\pgfqpoint{3.747332in}{1.061803in}}%
\pgfpathlineto{\pgfqpoint{3.798936in}{1.079585in}}%
\pgfpathlineto{\pgfqpoint{3.851953in}{1.100241in}}%
\pgfpathlineto{\pgfqpoint{3.905875in}{1.123679in}}%
\pgfpathlineto{\pgfqpoint{3.960201in}{1.149722in}}%
\pgfpathlineto{\pgfqpoint{4.014437in}{1.178104in}}%
\pgfpathlineto{\pgfqpoint{4.071538in}{1.210508in}}%
\pgfpathlineto{\pgfqpoint{4.127910in}{1.244984in}}%
\pgfpathlineto{\pgfqpoint{4.186834in}{1.283610in}}%
\pgfpathlineto{\pgfqpoint{4.248307in}{1.326670in}}%
\pgfpathlineto{\pgfqpoint{4.308238in}{1.371313in}}%
\pgfpathlineto{\pgfqpoint{4.370377in}{1.420320in}}%
\pgfpathlineto{\pgfqpoint{4.434694in}{1.473904in}}%
\pgfpathlineto{\pgfqpoint{4.501151in}{1.532263in}}%
\pgfpathlineto{\pgfqpoint{4.569700in}{1.595581in}}%
\pgfpathlineto{\pgfqpoint{4.640282in}{1.664022in}}%
\pgfpathlineto{\pgfqpoint{4.707932in}{1.732646in}}%
\pgfpathlineto{\pgfqpoint{4.777226in}{1.805948in}}%
\pgfpathlineto{\pgfqpoint{4.848089in}{1.884000in}}%
\pgfpathlineto{\pgfqpoint{4.920439in}{1.966846in}}%
\pgfpathlineto{\pgfqpoint{4.994183in}{2.054510in}}%
\pgfpathlineto{\pgfqpoint{5.069221in}{2.146983in}}%
\pgfpathlineto{\pgfqpoint{5.150935in}{2.251359in}}%
\pgfpathlineto{\pgfqpoint{5.233873in}{2.361140in}}%
\pgfpathlineto{\pgfqpoint{5.317892in}{2.476215in}}%
\pgfpathlineto{\pgfqpoint{5.402846in}{2.596440in}}%
\pgfpathlineto{\pgfqpoint{5.488587in}{2.721647in}}%
\pgfpathlineto{\pgfqpoint{5.574973in}{2.851646in}}%
\pgfpathlineto{\pgfqpoint{5.667680in}{2.995365in}}%
\pgfpathlineto{\pgfqpoint{5.760821in}{3.144056in}}%
\pgfpathlineto{\pgfqpoint{5.854269in}{3.297479in}}%
\pgfpathlineto{\pgfqpoint{5.947919in}{3.455408in}}%
\pgfpathlineto{\pgfqpoint{6.047554in}{3.627927in}}%
\pgfpathlineto{\pgfqpoint{6.147261in}{3.805116in}}%
\pgfpathlineto{\pgfqpoint{6.247000in}{3.986819in}}%
\pgfpathlineto{\pgfqpoint{6.352618in}{4.183994in}}%
\pgfpathlineto{\pgfqpoint{6.458239in}{4.385972in}}%
\pgfpathlineto{\pgfqpoint{6.563861in}{4.592652in}}%
\pgfpathlineto{\pgfqpoint{6.675350in}{4.815811in}}%
\pgfpathlineto{\pgfqpoint{6.786840in}{5.043998in}}%
\pgfpathlineto{\pgfqpoint{6.898330in}{5.277112in}}%
\pgfpathlineto{\pgfqpoint{7.015687in}{5.527712in}}%
\pgfpathlineto{\pgfqpoint{7.056762in}{5.616665in}}%
\pgfpathlineto{\pgfqpoint{7.056762in}{5.616665in}}%
\pgfusepath{stroke}%
\end{pgfscope}%
\begin{pgfscope}%
\pgfsetrectcap%
\pgfsetmiterjoin%
\pgfsetlinewidth{0.803000pt}%
\definecolor{currentstroke}{rgb}{0.000000,0.000000,0.000000}%
\pgfsetstrokecolor{currentstroke}%
\pgfsetdash{}{0pt}%
\pgfpathmoveto{\pgfqpoint{0.957105in}{0.774972in}}%
\pgfpathlineto{\pgfqpoint{0.957105in}{5.847222in}}%
\pgfusepath{stroke}%
\end{pgfscope}%
\begin{pgfscope}%
\pgfsetrectcap%
\pgfsetmiterjoin%
\pgfsetlinewidth{0.803000pt}%
\definecolor{currentstroke}{rgb}{0.000000,0.000000,0.000000}%
\pgfsetstrokecolor{currentstroke}%
\pgfsetdash{}{0pt}%
\pgfpathmoveto{\pgfqpoint{7.347222in}{0.774972in}}%
\pgfpathlineto{\pgfqpoint{7.347222in}{5.847222in}}%
\pgfusepath{stroke}%
\end{pgfscope}%
\begin{pgfscope}%
\pgfsetrectcap%
\pgfsetmiterjoin%
\pgfsetlinewidth{0.803000pt}%
\definecolor{currentstroke}{rgb}{0.000000,0.000000,0.000000}%
\pgfsetstrokecolor{currentstroke}%
\pgfsetdash{}{0pt}%
\pgfpathmoveto{\pgfqpoint{0.957105in}{0.774972in}}%
\pgfpathlineto{\pgfqpoint{7.347222in}{0.774972in}}%
\pgfusepath{stroke}%
\end{pgfscope}%
\begin{pgfscope}%
\pgfsetrectcap%
\pgfsetmiterjoin%
\pgfsetlinewidth{0.803000pt}%
\definecolor{currentstroke}{rgb}{0.000000,0.000000,0.000000}%
\pgfsetstrokecolor{currentstroke}%
\pgfsetdash{}{0pt}%
\pgfpathmoveto{\pgfqpoint{0.957105in}{5.847222in}}%
\pgfpathlineto{\pgfqpoint{7.347222in}{5.847222in}}%
\pgfusepath{stroke}%
\end{pgfscope}%
\begin{pgfscope}%
\pgfsetbuttcap%
\pgfsetmiterjoin%
\definecolor{currentfill}{rgb}{1.000000,1.000000,1.000000}%
\pgfsetfillcolor{currentfill}%
\pgfsetfillopacity{0.800000}%
\pgfsetlinewidth{1.003750pt}%
\definecolor{currentstroke}{rgb}{0.800000,0.800000,0.800000}%
\pgfsetstrokecolor{currentstroke}%
\pgfsetstrokeopacity{0.800000}%
\pgfsetdash{}{0pt}%
\pgfpathmoveto{\pgfqpoint{1.161271in}{4.218217in}}%
\pgfpathlineto{\pgfqpoint{4.425441in}{4.218217in}}%
\pgfpathquadraticcurveto{\pgfqpoint{4.483774in}{4.218217in}}{\pgfqpoint{4.483774in}{4.276550in}}%
\pgfpathlineto{\pgfqpoint{4.483774in}{5.643056in}}%
\pgfpathquadraticcurveto{\pgfqpoint{4.483774in}{5.701389in}}{\pgfqpoint{4.425441in}{5.701389in}}%
\pgfpathlineto{\pgfqpoint{1.161271in}{5.701389in}}%
\pgfpathquadraticcurveto{\pgfqpoint{1.102938in}{5.701389in}}{\pgfqpoint{1.102938in}{5.643056in}}%
\pgfpathlineto{\pgfqpoint{1.102938in}{4.276550in}}%
\pgfpathquadraticcurveto{\pgfqpoint{1.102938in}{4.218217in}}{\pgfqpoint{1.161271in}{4.218217in}}%
\pgfpathclose%
\pgfusepath{stroke,fill}%
\end{pgfscope}%
\begin{pgfscope}%
\pgfsetrectcap%
\pgfsetroundjoin%
\pgfsetlinewidth{2.509375pt}%
\definecolor{currentstroke}{rgb}{0.000000,0.000000,0.501961}%
\pgfsetstrokecolor{currentstroke}%
\pgfsetdash{}{0pt}%
\pgfpathmoveto{\pgfqpoint{1.365438in}{5.164881in}}%
\pgfpathlineto{\pgfqpoint{1.598771in}{5.164881in}}%
\pgfusepath{stroke}%
\end{pgfscope}%
\begin{pgfscope}%
\definecolor{textcolor}{rgb}{0.000000,0.000000,0.000000}%
\pgfsetstrokecolor{textcolor}%
\pgfsetfillcolor{textcolor}%
\pgftext[x=1.686271in, y=5.238850in, left, base]{\color{textcolor}\rmfamily\fontsize{21.000000}{25.200000}\selectfont Upper Bound, }%
\end{pgfscope}%
\begin{pgfscope}%
\definecolor{textcolor}{rgb}{0.000000,0.000000,0.000000}%
\pgfsetstrokecolor{textcolor}%
\pgfsetfillcolor{textcolor}%
\pgftext[x=1.686271in, y=4.942775in, left, base]{\color{textcolor}\rmfamily\fontsize{21.000000}{25.200000}\selectfont  AC Lower Bound}%
\end{pgfscope}%
\begin{pgfscope}%
\pgfsetbuttcap%
\pgfsetroundjoin%
\pgfsetlinewidth{2.509375pt}%
\definecolor{currentstroke}{rgb}{1.000000,0.000000,0.000000}%
\pgfsetstrokecolor{currentstroke}%
\pgfsetstrokeopacity{0.700000}%
\pgfsetdash{{9.250000pt}{4.000000pt}}{0.000000pt}%
\pgfpathmoveto{\pgfqpoint{1.365438in}{4.638829in}}%
\pgfpathlineto{\pgfqpoint{1.598771in}{4.638829in}}%
\pgfusepath{stroke}%
\end{pgfscope}%
\begin{pgfscope}%
\definecolor{textcolor}{rgb}{0.000000,0.000000,0.000000}%
\pgfsetstrokecolor{textcolor}%
\pgfsetfillcolor{textcolor}%
\pgftext[x=1.686271in,y=4.536746in,left,base]{\color{textcolor}\rmfamily\fontsize{21.000000}{25.200000}\selectfont MSVV Upper Bound       }%
\end{pgfscope}%
\end{pgfpicture}%
\makeatother%
\endgroup%

%% file: images/rev_weight_caps.pgf
\begingroup%
\makeatletter%
\begin{pgfpicture}%
\pgfpathrectangle{\pgfpointorigin}{\pgfqpoint{7.500000in}{6.000000in}}%
\pgfusepath{use as bounding box, clip}%
\begin{pgfscope}%
\pgfsetbuttcap%
\pgfsetmiterjoin%
\definecolor{currentfill}{rgb}{1.000000,1.000000,1.000000}%
\pgfsetfillcolor{currentfill}%
\pgfsetlinewidth{0.000000pt}%
\definecolor{currentstroke}{rgb}{1.000000,1.000000,1.000000}%
\pgfsetstrokecolor{currentstroke}%
\pgfsetdash{}{0pt}%
\pgfpathmoveto{\pgfqpoint{0.000000in}{0.000000in}}%
\pgfpathlineto{\pgfqpoint{7.500000in}{0.000000in}}%
\pgfpathlineto{\pgfqpoint{7.500000in}{6.000000in}}%
\pgfpathlineto{\pgfqpoint{0.000000in}{6.000000in}}%
\pgfpathclose%
\pgfusepath{fill}%
\end{pgfscope}%
\begin{pgfscope}%
\pgfsetbuttcap%
\pgfsetmiterjoin%
\definecolor{currentfill}{rgb}{1.000000,1.000000,1.000000}%
\pgfsetfillcolor{currentfill}%
\pgfsetlinewidth{0.000000pt}%
\definecolor{currentstroke}{rgb}{0.000000,0.000000,0.000000}%
\pgfsetstrokecolor{currentstroke}%
\pgfsetstrokeopacity{0.000000}%
\pgfsetdash{}{0pt}%
\pgfpathmoveto{\pgfqpoint{0.957105in}{0.774972in}}%
\pgfpathlineto{\pgfqpoint{7.347222in}{0.774972in}}%
\pgfpathlineto{\pgfqpoint{7.347222in}{5.847222in}}%
\pgfpathlineto{\pgfqpoint{0.957105in}{5.847222in}}%
\pgfpathclose%
\pgfusepath{fill}%
\end{pgfscope}%
\begin{pgfscope}%
\pgfsetbuttcap%
\pgfsetroundjoin%
\definecolor{currentfill}{rgb}{0.000000,0.000000,0.000000}%
\pgfsetfillcolor{currentfill}%
\pgfsetlinewidth{0.803000pt}%
\definecolor{currentstroke}{rgb}{0.000000,0.000000,0.000000}%
\pgfsetstrokecolor{currentstroke}%
\pgfsetdash{}{0pt}%
\pgfsys@defobject{currentmarker}{\pgfqpoint{0.000000in}{-0.048611in}}{\pgfqpoint{0.000000in}{0.000000in}}{%
\pgfpathmoveto{\pgfqpoint{0.000000in}{0.000000in}}%
\pgfpathlineto{\pgfqpoint{0.000000in}{-0.048611in}}%
\pgfusepath{stroke,fill}%
}%
\begin{pgfscope}%
\pgfsys@transformshift{1.247565in}{0.774972in}%
\pgfsys@useobject{currentmarker}{}%
\end{pgfscope}%
\end{pgfscope}%
\begin{pgfscope}%
\definecolor{textcolor}{rgb}{0.000000,0.000000,0.000000}%
\pgfsetstrokecolor{textcolor}%
\pgfsetfillcolor{textcolor}%
\pgftext[x=1.247565in,y=0.677750in,,top]{\color{textcolor}\rmfamily\fontsize{18.000000}{21.600000}\selectfont \(\displaystyle {0.0}\)}%
\end{pgfscope}%
\begin{pgfscope}%
\pgfsetbuttcap%
\pgfsetroundjoin%
\definecolor{currentfill}{rgb}{0.000000,0.000000,0.000000}%
\pgfsetfillcolor{currentfill}%
\pgfsetlinewidth{0.803000pt}%
\definecolor{currentstroke}{rgb}{0.000000,0.000000,0.000000}%
\pgfsetstrokecolor{currentstroke}%
\pgfsetdash{}{0pt}%
\pgfsys@defobject{currentmarker}{\pgfqpoint{0.000000in}{-0.048611in}}{\pgfqpoint{0.000000in}{0.000000in}}{%
\pgfpathmoveto{\pgfqpoint{0.000000in}{0.000000in}}%
\pgfpathlineto{\pgfqpoint{0.000000in}{-0.048611in}}%
\pgfusepath{stroke,fill}%
}%
\begin{pgfscope}%
\pgfsys@transformshift{2.409544in}{0.774972in}%
\pgfsys@useobject{currentmarker}{}%
\end{pgfscope}%
\end{pgfscope}%
\begin{pgfscope}%
\definecolor{textcolor}{rgb}{0.000000,0.000000,0.000000}%
\pgfsetstrokecolor{textcolor}%
\pgfsetfillcolor{textcolor}%
\pgftext[x=2.409544in,y=0.677750in,,top]{\color{textcolor}\rmfamily\fontsize{18.000000}{21.600000}\selectfont \(\displaystyle {0.2}\)}%
\end{pgfscope}%
\begin{pgfscope}%
\pgfsetbuttcap%
\pgfsetroundjoin%
\definecolor{currentfill}{rgb}{0.000000,0.000000,0.000000}%
\pgfsetfillcolor{currentfill}%
\pgfsetlinewidth{0.803000pt}%
\definecolor{currentstroke}{rgb}{0.000000,0.000000,0.000000}%
\pgfsetstrokecolor{currentstroke}%
\pgfsetdash{}{0pt}%
\pgfsys@defobject{currentmarker}{\pgfqpoint{0.000000in}{-0.048611in}}{\pgfqpoint{0.000000in}{0.000000in}}{%
\pgfpathmoveto{\pgfqpoint{0.000000in}{0.000000in}}%
\pgfpathlineto{\pgfqpoint{0.000000in}{-0.048611in}}%
\pgfusepath{stroke,fill}%
}%
\begin{pgfscope}%
\pgfsys@transformshift{3.571524in}{0.774972in}%
\pgfsys@useobject{currentmarker}{}%
\end{pgfscope}%
\end{pgfscope}%
\begin{pgfscope}%
\definecolor{textcolor}{rgb}{0.000000,0.000000,0.000000}%
\pgfsetstrokecolor{textcolor}%
\pgfsetfillcolor{textcolor}%
\pgftext[x=3.571524in,y=0.677750in,,top]{\color{textcolor}\rmfamily\fontsize{18.000000}{21.600000}\selectfont \(\displaystyle {0.4}\)}%
\end{pgfscope}%
\begin{pgfscope}%
\pgfsetbuttcap%
\pgfsetroundjoin%
\definecolor{currentfill}{rgb}{0.000000,0.000000,0.000000}%
\pgfsetfillcolor{currentfill}%
\pgfsetlinewidth{0.803000pt}%
\definecolor{currentstroke}{rgb}{0.000000,0.000000,0.000000}%
\pgfsetstrokecolor{currentstroke}%
\pgfsetdash{}{0pt}%
\pgfsys@defobject{currentmarker}{\pgfqpoint{0.000000in}{-0.048611in}}{\pgfqpoint{0.000000in}{0.000000in}}{%
\pgfpathmoveto{\pgfqpoint{0.000000in}{0.000000in}}%
\pgfpathlineto{\pgfqpoint{0.000000in}{-0.048611in}}%
\pgfusepath{stroke,fill}%
}%
\begin{pgfscope}%
\pgfsys@transformshift{4.733504in}{0.774972in}%
\pgfsys@useobject{currentmarker}{}%
\end{pgfscope}%
\end{pgfscope}%
\begin{pgfscope}%
\definecolor{textcolor}{rgb}{0.000000,0.000000,0.000000}%
\pgfsetstrokecolor{textcolor}%
\pgfsetfillcolor{textcolor}%
\pgftext[x=4.733504in,y=0.677750in,,top]{\color{textcolor}\rmfamily\fontsize{18.000000}{21.600000}\selectfont \(\displaystyle {0.6}\)}%
\end{pgfscope}%
\begin{pgfscope}%
\pgfsetbuttcap%
\pgfsetroundjoin%
\definecolor{currentfill}{rgb}{0.000000,0.000000,0.000000}%
\pgfsetfillcolor{currentfill}%
\pgfsetlinewidth{0.803000pt}%
\definecolor{currentstroke}{rgb}{0.000000,0.000000,0.000000}%
\pgfsetstrokecolor{currentstroke}%
\pgfsetdash{}{0pt}%
\pgfsys@defobject{currentmarker}{\pgfqpoint{0.000000in}{-0.048611in}}{\pgfqpoint{0.000000in}{0.000000in}}{%
\pgfpathmoveto{\pgfqpoint{0.000000in}{0.000000in}}%
\pgfpathlineto{\pgfqpoint{0.000000in}{-0.048611in}}%
\pgfusepath{stroke,fill}%
}%
\begin{pgfscope}%
\pgfsys@transformshift{5.895483in}{0.774972in}%
\pgfsys@useobject{currentmarker}{}%
\end{pgfscope}%
\end{pgfscope}%
\begin{pgfscope}%
\definecolor{textcolor}{rgb}{0.000000,0.000000,0.000000}%
\pgfsetstrokecolor{textcolor}%
\pgfsetfillcolor{textcolor}%
\pgftext[x=5.895483in,y=0.677750in,,top]{\color{textcolor}\rmfamily\fontsize{18.000000}{21.600000}\selectfont \(\displaystyle {0.8}\)}%
\end{pgfscope}%
\begin{pgfscope}%
\pgfsetbuttcap%
\pgfsetroundjoin%
\definecolor{currentfill}{rgb}{0.000000,0.000000,0.000000}%
\pgfsetfillcolor{currentfill}%
\pgfsetlinewidth{0.803000pt}%
\definecolor{currentstroke}{rgb}{0.000000,0.000000,0.000000}%
\pgfsetstrokecolor{currentstroke}%
\pgfsetdash{}{0pt}%
\pgfsys@defobject{currentmarker}{\pgfqpoint{0.000000in}{-0.048611in}}{\pgfqpoint{0.000000in}{0.000000in}}{%
\pgfpathmoveto{\pgfqpoint{0.000000in}{0.000000in}}%
\pgfpathlineto{\pgfqpoint{0.000000in}{-0.048611in}}%
\pgfusepath{stroke,fill}%
}%
\begin{pgfscope}%
\pgfsys@transformshift{7.057463in}{0.774972in}%
\pgfsys@useobject{currentmarker}{}%
\end{pgfscope}%
\end{pgfscope}%
\begin{pgfscope}%
\definecolor{textcolor}{rgb}{0.000000,0.000000,0.000000}%
\pgfsetstrokecolor{textcolor}%
\pgfsetfillcolor{textcolor}%
\pgftext[x=7.057463in,y=0.677750in,,top]{\color{textcolor}\rmfamily\fontsize{18.000000}{21.600000}\selectfont \(\displaystyle {1.0}\)}%
\end{pgfscope}%
\begin{pgfscope}%
\definecolor{textcolor}{rgb}{0.000000,0.000000,0.000000}%
\pgfsetstrokecolor{textcolor}%
\pgfsetfillcolor{textcolor}%
\pgftext[x=4.152163in,y=0.408845in,,top]{\color{textcolor}\rmfamily\fontsize{22.000000}{26.400000}\selectfont Effective Fraction of External Traffic}%
\end{pgfscope}%
\begin{pgfscope}%
\pgfsetbuttcap%
\pgfsetroundjoin%
\definecolor{currentfill}{rgb}{0.000000,0.000000,0.000000}%
\pgfsetfillcolor{currentfill}%
\pgfsetlinewidth{0.803000pt}%
\definecolor{currentstroke}{rgb}{0.000000,0.000000,0.000000}%
\pgfsetstrokecolor{currentstroke}%
\pgfsetdash{}{0pt}%
\pgfsys@defobject{currentmarker}{\pgfqpoint{-0.048611in}{0.000000in}}{\pgfqpoint{-0.000000in}{0.000000in}}{%
\pgfpathmoveto{\pgfqpoint{-0.000000in}{0.000000in}}%
\pgfpathlineto{\pgfqpoint{-0.048611in}{0.000000in}}%
\pgfusepath{stroke,fill}%
}%
\begin{pgfscope}%
\pgfsys@transformshift{0.957105in}{1.229710in}%
\pgfsys@useobject{currentmarker}{}%
\end{pgfscope}%
\end{pgfscope}%
\begin{pgfscope}%
\definecolor{textcolor}{rgb}{0.000000,0.000000,0.000000}%
\pgfsetstrokecolor{textcolor}%
\pgfsetfillcolor{textcolor}%
\pgftext[x=0.464401in, y=1.146376in, left, base]{\color{textcolor}\rmfamily\fontsize{18.000000}{21.600000}\selectfont \(\displaystyle {0.65}\)}%
\end{pgfscope}%
\begin{pgfscope}%
\pgfsetbuttcap%
\pgfsetroundjoin%
\definecolor{currentfill}{rgb}{0.000000,0.000000,0.000000}%
\pgfsetfillcolor{currentfill}%
\pgfsetlinewidth{0.803000pt}%
\definecolor{currentstroke}{rgb}{0.000000,0.000000,0.000000}%
\pgfsetstrokecolor{currentstroke}%
\pgfsetdash{}{0pt}%
\pgfsys@defobject{currentmarker}{\pgfqpoint{-0.048611in}{0.000000in}}{\pgfqpoint{-0.000000in}{0.000000in}}{%
\pgfpathmoveto{\pgfqpoint{-0.000000in}{0.000000in}}%
\pgfpathlineto{\pgfqpoint{-0.048611in}{0.000000in}}%
\pgfusepath{stroke,fill}%
}%
\begin{pgfscope}%
\pgfsys@transformshift{0.957105in}{1.856634in}%
\pgfsys@useobject{currentmarker}{}%
\end{pgfscope}%
\end{pgfscope}%
\begin{pgfscope}%
\definecolor{textcolor}{rgb}{0.000000,0.000000,0.000000}%
\pgfsetstrokecolor{textcolor}%
\pgfsetfillcolor{textcolor}%
\pgftext[x=0.464401in, y=1.773300in, left, base]{\color{textcolor}\rmfamily\fontsize{18.000000}{21.600000}\selectfont \(\displaystyle {0.70}\)}%
\end{pgfscope}%
\begin{pgfscope}%
\pgfsetbuttcap%
\pgfsetroundjoin%
\definecolor{currentfill}{rgb}{0.000000,0.000000,0.000000}%
\pgfsetfillcolor{currentfill}%
\pgfsetlinewidth{0.803000pt}%
\definecolor{currentstroke}{rgb}{0.000000,0.000000,0.000000}%
\pgfsetstrokecolor{currentstroke}%
\pgfsetdash{}{0pt}%
\pgfsys@defobject{currentmarker}{\pgfqpoint{-0.048611in}{0.000000in}}{\pgfqpoint{-0.000000in}{0.000000in}}{%
\pgfpathmoveto{\pgfqpoint{-0.000000in}{0.000000in}}%
\pgfpathlineto{\pgfqpoint{-0.048611in}{0.000000in}}%
\pgfusepath{stroke,fill}%
}%
\begin{pgfscope}%
\pgfsys@transformshift{0.957105in}{2.483557in}%
\pgfsys@useobject{currentmarker}{}%
\end{pgfscope}%
\end{pgfscope}%
\begin{pgfscope}%
\definecolor{textcolor}{rgb}{0.000000,0.000000,0.000000}%
\pgfsetstrokecolor{textcolor}%
\pgfsetfillcolor{textcolor}%
\pgftext[x=0.464401in, y=2.400224in, left, base]{\color{textcolor}\rmfamily\fontsize{18.000000}{21.600000}\selectfont \(\displaystyle {0.75}\)}%
\end{pgfscope}%
\begin{pgfscope}%
\pgfsetbuttcap%
\pgfsetroundjoin%
\definecolor{currentfill}{rgb}{0.000000,0.000000,0.000000}%
\pgfsetfillcolor{currentfill}%
\pgfsetlinewidth{0.803000pt}%
\definecolor{currentstroke}{rgb}{0.000000,0.000000,0.000000}%
\pgfsetstrokecolor{currentstroke}%
\pgfsetdash{}{0pt}%
\pgfsys@defobject{currentmarker}{\pgfqpoint{-0.048611in}{0.000000in}}{\pgfqpoint{-0.000000in}{0.000000in}}{%
\pgfpathmoveto{\pgfqpoint{-0.000000in}{0.000000in}}%
\pgfpathlineto{\pgfqpoint{-0.048611in}{0.000000in}}%
\pgfusepath{stroke,fill}%
}%
\begin{pgfscope}%
\pgfsys@transformshift{0.957105in}{3.110481in}%
\pgfsys@useobject{currentmarker}{}%
\end{pgfscope}%
\end{pgfscope}%
\begin{pgfscope}%
\definecolor{textcolor}{rgb}{0.000000,0.000000,0.000000}%
\pgfsetstrokecolor{textcolor}%
\pgfsetfillcolor{textcolor}%
\pgftext[x=0.464401in, y=3.027148in, left, base]{\color{textcolor}\rmfamily\fontsize{18.000000}{21.600000}\selectfont \(\displaystyle {0.80}\)}%
\end{pgfscope}%
\begin{pgfscope}%
\pgfsetbuttcap%
\pgfsetroundjoin%
\definecolor{currentfill}{rgb}{0.000000,0.000000,0.000000}%
\pgfsetfillcolor{currentfill}%
\pgfsetlinewidth{0.803000pt}%
\definecolor{currentstroke}{rgb}{0.000000,0.000000,0.000000}%
\pgfsetstrokecolor{currentstroke}%
\pgfsetdash{}{0pt}%
\pgfsys@defobject{currentmarker}{\pgfqpoint{-0.048611in}{0.000000in}}{\pgfqpoint{-0.000000in}{0.000000in}}{%
\pgfpathmoveto{\pgfqpoint{-0.000000in}{0.000000in}}%
\pgfpathlineto{\pgfqpoint{-0.048611in}{0.000000in}}%
\pgfusepath{stroke,fill}%
}%
\begin{pgfscope}%
\pgfsys@transformshift{0.957105in}{3.737405in}%
\pgfsys@useobject{currentmarker}{}%
\end{pgfscope}%
\end{pgfscope}%
\begin{pgfscope}%
\definecolor{textcolor}{rgb}{0.000000,0.000000,0.000000}%
\pgfsetstrokecolor{textcolor}%
\pgfsetfillcolor{textcolor}%
\pgftext[x=0.464401in, y=3.654072in, left, base]{\color{textcolor}\rmfamily\fontsize{18.000000}{21.600000}\selectfont \(\displaystyle {0.85}\)}%
\end{pgfscope}%
\begin{pgfscope}%
\pgfsetbuttcap%
\pgfsetroundjoin%
\definecolor{currentfill}{rgb}{0.000000,0.000000,0.000000}%
\pgfsetfillcolor{currentfill}%
\pgfsetlinewidth{0.803000pt}%
\definecolor{currentstroke}{rgb}{0.000000,0.000000,0.000000}%
\pgfsetstrokecolor{currentstroke}%
\pgfsetdash{}{0pt}%
\pgfsys@defobject{currentmarker}{\pgfqpoint{-0.048611in}{0.000000in}}{\pgfqpoint{-0.000000in}{0.000000in}}{%
\pgfpathmoveto{\pgfqpoint{-0.000000in}{0.000000in}}%
\pgfpathlineto{\pgfqpoint{-0.048611in}{0.000000in}}%
\pgfusepath{stroke,fill}%
}%
\begin{pgfscope}%
\pgfsys@transformshift{0.957105in}{4.364329in}%
\pgfsys@useobject{currentmarker}{}%
\end{pgfscope}%
\end{pgfscope}%
\begin{pgfscope}%
\definecolor{textcolor}{rgb}{0.000000,0.000000,0.000000}%
\pgfsetstrokecolor{textcolor}%
\pgfsetfillcolor{textcolor}%
\pgftext[x=0.464401in, y=4.280996in, left, base]{\color{textcolor}\rmfamily\fontsize{18.000000}{21.600000}\selectfont \(\displaystyle {0.90}\)}%
\end{pgfscope}%
\begin{pgfscope}%
\pgfsetbuttcap%
\pgfsetroundjoin%
\definecolor{currentfill}{rgb}{0.000000,0.000000,0.000000}%
\pgfsetfillcolor{currentfill}%
\pgfsetlinewidth{0.803000pt}%
\definecolor{currentstroke}{rgb}{0.000000,0.000000,0.000000}%
\pgfsetstrokecolor{currentstroke}%
\pgfsetdash{}{0pt}%
\pgfsys@defobject{currentmarker}{\pgfqpoint{-0.048611in}{0.000000in}}{\pgfqpoint{-0.000000in}{0.000000in}}{%
\pgfpathmoveto{\pgfqpoint{-0.000000in}{0.000000in}}%
\pgfpathlineto{\pgfqpoint{-0.048611in}{0.000000in}}%
\pgfusepath{stroke,fill}%
}%
\begin{pgfscope}%
\pgfsys@transformshift{0.957105in}{4.991253in}%
\pgfsys@useobject{currentmarker}{}%
\end{pgfscope}%
\end{pgfscope}%
\begin{pgfscope}%
\definecolor{textcolor}{rgb}{0.000000,0.000000,0.000000}%
\pgfsetstrokecolor{textcolor}%
\pgfsetfillcolor{textcolor}%
\pgftext[x=0.464401in, y=4.907920in, left, base]{\color{textcolor}\rmfamily\fontsize{18.000000}{21.600000}\selectfont \(\displaystyle {0.95}\)}%
\end{pgfscope}%
\begin{pgfscope}%
\pgfsetbuttcap%
\pgfsetroundjoin%
\definecolor{currentfill}{rgb}{0.000000,0.000000,0.000000}%
\pgfsetfillcolor{currentfill}%
\pgfsetlinewidth{0.803000pt}%
\definecolor{currentstroke}{rgb}{0.000000,0.000000,0.000000}%
\pgfsetstrokecolor{currentstroke}%
\pgfsetdash{}{0pt}%
\pgfsys@defobject{currentmarker}{\pgfqpoint{-0.048611in}{0.000000in}}{\pgfqpoint{-0.000000in}{0.000000in}}{%
\pgfpathmoveto{\pgfqpoint{-0.000000in}{0.000000in}}%
\pgfpathlineto{\pgfqpoint{-0.048611in}{0.000000in}}%
\pgfusepath{stroke,fill}%
}%
\begin{pgfscope}%
\pgfsys@transformshift{0.957105in}{5.618177in}%
\pgfsys@useobject{currentmarker}{}%
\end{pgfscope}%
\end{pgfscope}%
\begin{pgfscope}%
\definecolor{textcolor}{rgb}{0.000000,0.000000,0.000000}%
\pgfsetstrokecolor{textcolor}%
\pgfsetfillcolor{textcolor}%
\pgftext[x=0.464401in, y=5.534844in, left, base]{\color{textcolor}\rmfamily\fontsize{18.000000}{21.600000}\selectfont \(\displaystyle {1.00}\)}%
\end{pgfscope}%
\begin{pgfscope}%
\definecolor{textcolor}{rgb}{0.000000,0.000000,0.000000}%
\pgfsetstrokecolor{textcolor}%
\pgfsetfillcolor{textcolor}%
\pgftext[x=0.408845in,y=3.311097in,,bottom,rotate=90.000000]{\color{textcolor}\rmfamily\fontsize{22.000000}{26.400000}\selectfont Comp. Ratio}%
\end{pgfscope}%
\begin{pgfscope}%
\pgfpathrectangle{\pgfqpoint{0.957105in}{0.774972in}}{\pgfqpoint{6.390117in}{5.072250in}}%
\pgfusepath{clip}%
\pgfsetrectcap%
\pgfsetroundjoin%
\pgfsetlinewidth{2.509375pt}%
\definecolor{currentstroke}{rgb}{0.000000,0.000000,0.501961}%
\pgfsetstrokecolor{currentstroke}%
\pgfsetdash{}{0pt}%
\pgfpathmoveto{\pgfqpoint{1.247565in}{1.005529in}}%
\pgfpathlineto{\pgfqpoint{3.431386in}{1.006611in}}%
\pgfpathlineto{\pgfqpoint{3.477865in}{1.009829in}}%
\pgfpathlineto{\pgfqpoint{3.524344in}{1.015138in}}%
\pgfpathlineto{\pgfqpoint{3.570823in}{1.022494in}}%
\pgfpathlineto{\pgfqpoint{3.623113in}{1.033167in}}%
\pgfpathlineto{\pgfqpoint{3.675402in}{1.046324in}}%
\pgfpathlineto{\pgfqpoint{3.727691in}{1.061911in}}%
\pgfpathlineto{\pgfqpoint{3.779980in}{1.079878in}}%
\pgfpathlineto{\pgfqpoint{3.832269in}{1.100175in}}%
\pgfpathlineto{\pgfqpoint{3.884558in}{1.122755in}}%
\pgfpathlineto{\pgfqpoint{3.936847in}{1.147573in}}%
\pgfpathlineto{\pgfqpoint{3.994946in}{1.177720in}}%
\pgfpathlineto{\pgfqpoint{4.053045in}{1.210519in}}%
\pgfpathlineto{\pgfqpoint{4.111144in}{1.245914in}}%
\pgfpathlineto{\pgfqpoint{4.169243in}{1.283854in}}%
\pgfpathlineto{\pgfqpoint{4.227342in}{1.324287in}}%
\pgfpathlineto{\pgfqpoint{4.291251in}{1.371586in}}%
\pgfpathlineto{\pgfqpoint{4.355160in}{1.421780in}}%
\pgfpathlineto{\pgfqpoint{4.419069in}{1.474812in}}%
\pgfpathlineto{\pgfqpoint{4.482977in}{1.530622in}}%
\pgfpathlineto{\pgfqpoint{4.552696in}{1.594613in}}%
\pgfpathlineto{\pgfqpoint{4.622415in}{1.661777in}}%
\pgfpathlineto{\pgfqpoint{4.692134in}{1.732049in}}%
\pgfpathlineto{\pgfqpoint{4.761853in}{1.805368in}}%
\pgfpathlineto{\pgfqpoint{4.837381in}{1.888163in}}%
\pgfpathlineto{\pgfqpoint{4.912910in}{1.974387in}}%
\pgfpathlineto{\pgfqpoint{4.988439in}{2.063971in}}%
\pgfpathlineto{\pgfqpoint{5.063967in}{2.156846in}}%
\pgfpathlineto{\pgfqpoint{5.145306in}{2.260471in}}%
\pgfpathlineto{\pgfqpoint{5.226644in}{2.367760in}}%
\pgfpathlineto{\pgfqpoint{5.307983in}{2.478637in}}%
\pgfpathlineto{\pgfqpoint{5.395131in}{2.601334in}}%
\pgfpathlineto{\pgfqpoint{5.482280in}{2.727984in}}%
\pgfpathlineto{\pgfqpoint{5.569428in}{2.858504in}}%
\pgfpathlineto{\pgfqpoint{5.662387in}{3.001905in}}%
\pgfpathlineto{\pgfqpoint{5.755345in}{3.149530in}}%
\pgfpathlineto{\pgfqpoint{5.848304in}{3.301292in}}%
\pgfpathlineto{\pgfqpoint{5.947072in}{3.466979in}}%
\pgfpathlineto{\pgfqpoint{6.045840in}{3.637147in}}%
\pgfpathlineto{\pgfqpoint{6.144608in}{3.811702in}}%
\pgfpathlineto{\pgfqpoint{6.249186in}{4.001209in}}%
\pgfpathlineto{\pgfqpoint{6.353765in}{4.195436in}}%
\pgfpathlineto{\pgfqpoint{6.458343in}{4.394286in}}%
\pgfpathlineto{\pgfqpoint{6.568731in}{4.609095in}}%
\pgfpathlineto{\pgfqpoint{6.679119in}{4.828847in}}%
\pgfpathlineto{\pgfqpoint{6.789507in}{5.053441in}}%
\pgfpathlineto{\pgfqpoint{6.905705in}{5.294980in}}%
\pgfpathlineto{\pgfqpoint{7.021903in}{5.541670in}}%
\pgfpathlineto{\pgfqpoint{7.056762in}{5.616665in}}%
\pgfpathlineto{\pgfqpoint{7.056762in}{5.616665in}}%
\pgfusepath{stroke}%
\end{pgfscope}%
\begin{pgfscope}%
\pgfpathrectangle{\pgfqpoint{0.957105in}{0.774972in}}{\pgfqpoint{6.390117in}{5.072250in}}%
\pgfusepath{clip}%
\pgfsetbuttcap%
\pgfsetroundjoin%
\pgfsetlinewidth{2.509375pt}%
\definecolor{currentstroke}{rgb}{0.254902,0.184314,0.533333}%
\pgfsetstrokecolor{currentstroke}%
\pgfsetstrokeopacity{0.700000}%
\pgfsetdash{{2.500000pt}{4.125000pt}}{0.000000pt}%
\pgfpathmoveto{\pgfqpoint{1.247565in}{1.016555in}}%
\pgfpathlineto{\pgfqpoint{3.565714in}{1.016555in}}%
\pgfpathlineto{\pgfqpoint{3.571524in}{1.029094in}}%
\pgfpathlineto{\pgfqpoint{3.658672in}{1.029094in}}%
\pgfpathlineto{\pgfqpoint{3.664482in}{1.041632in}}%
\pgfpathlineto{\pgfqpoint{3.722581in}{1.041632in}}%
\pgfpathlineto{\pgfqpoint{3.728391in}{1.054171in}}%
\pgfpathlineto{\pgfqpoint{3.774870in}{1.054171in}}%
\pgfpathlineto{\pgfqpoint{3.780680in}{1.066709in}}%
\pgfpathlineto{\pgfqpoint{3.821349in}{1.066709in}}%
\pgfpathlineto{\pgfqpoint{3.827159in}{1.079248in}}%
\pgfpathlineto{\pgfqpoint{3.862019in}{1.079248in}}%
\pgfpathlineto{\pgfqpoint{3.867829in}{1.091786in}}%
\pgfpathlineto{\pgfqpoint{3.896878in}{1.091786in}}%
\pgfpathlineto{\pgfqpoint{3.902688in}{1.104325in}}%
\pgfpathlineto{\pgfqpoint{3.931738in}{1.104325in}}%
\pgfpathlineto{\pgfqpoint{3.937547in}{1.116863in}}%
\pgfpathlineto{\pgfqpoint{3.966597in}{1.116863in}}%
\pgfpathlineto{\pgfqpoint{3.972407in}{1.129402in}}%
\pgfpathlineto{\pgfqpoint{3.995646in}{1.129402in}}%
\pgfpathlineto{\pgfqpoint{4.001456in}{1.141940in}}%
\pgfpathlineto{\pgfqpoint{4.024696in}{1.141940in}}%
\pgfpathlineto{\pgfqpoint{4.030506in}{1.154479in}}%
\pgfpathlineto{\pgfqpoint{4.053745in}{1.154479in}}%
\pgfpathlineto{\pgfqpoint{4.059555in}{1.167017in}}%
\pgfpathlineto{\pgfqpoint{4.082795in}{1.167017in}}%
\pgfpathlineto{\pgfqpoint{4.088605in}{1.179556in}}%
\pgfpathlineto{\pgfqpoint{4.106035in}{1.179556in}}%
\pgfpathlineto{\pgfqpoint{4.111844in}{1.192094in}}%
\pgfpathlineto{\pgfqpoint{4.129274in}{1.192094in}}%
\pgfpathlineto{\pgfqpoint{4.135084in}{1.204633in}}%
\pgfpathlineto{\pgfqpoint{4.152514in}{1.204633in}}%
\pgfpathlineto{\pgfqpoint{4.158324in}{1.217171in}}%
\pgfpathlineto{\pgfqpoint{4.175753in}{1.217171in}}%
\pgfpathlineto{\pgfqpoint{4.181563in}{1.229710in}}%
\pgfpathlineto{\pgfqpoint{4.198993in}{1.229710in}}%
\pgfpathlineto{\pgfqpoint{4.204803in}{1.242248in}}%
\pgfpathlineto{\pgfqpoint{4.222232in}{1.242248in}}%
\pgfpathlineto{\pgfqpoint{4.228042in}{1.254787in}}%
\pgfpathlineto{\pgfqpoint{4.239662in}{1.254787in}}%
\pgfpathlineto{\pgfqpoint{4.245472in}{1.267325in}}%
\pgfpathlineto{\pgfqpoint{4.262902in}{1.267325in}}%
\pgfpathlineto{\pgfqpoint{4.268712in}{1.279864in}}%
\pgfpathlineto{\pgfqpoint{4.280331in}{1.279864in}}%
\pgfpathlineto{\pgfqpoint{4.286141in}{1.292402in}}%
\pgfpathlineto{\pgfqpoint{4.303571in}{1.292402in}}%
\pgfpathlineto{\pgfqpoint{4.309381in}{1.304940in}}%
\pgfpathlineto{\pgfqpoint{4.321001in}{1.304940in}}%
\pgfpathlineto{\pgfqpoint{4.326811in}{1.317479in}}%
\pgfpathlineto{\pgfqpoint{4.338430in}{1.317479in}}%
\pgfpathlineto{\pgfqpoint{4.344240in}{1.330017in}}%
\pgfpathlineto{\pgfqpoint{4.355860in}{1.330017in}}%
\pgfpathlineto{\pgfqpoint{4.361670in}{1.342556in}}%
\pgfpathlineto{\pgfqpoint{4.373290in}{1.342556in}}%
\pgfpathlineto{\pgfqpoint{4.379100in}{1.355094in}}%
\pgfpathlineto{\pgfqpoint{4.390720in}{1.355094in}}%
\pgfpathlineto{\pgfqpoint{4.396529in}{1.367633in}}%
\pgfpathlineto{\pgfqpoint{4.408149in}{1.367633in}}%
\pgfpathlineto{\pgfqpoint{4.413959in}{1.380171in}}%
\pgfpathlineto{\pgfqpoint{4.425579in}{1.380171in}}%
\pgfpathlineto{\pgfqpoint{4.431389in}{1.392710in}}%
\pgfpathlineto{\pgfqpoint{4.443009in}{1.392710in}}%
\pgfpathlineto{\pgfqpoint{4.448818in}{1.405248in}}%
\pgfpathlineto{\pgfqpoint{4.454628in}{1.405248in}}%
\pgfpathlineto{\pgfqpoint{4.460438in}{1.417787in}}%
\pgfpathlineto{\pgfqpoint{4.472058in}{1.417787in}}%
\pgfpathlineto{\pgfqpoint{4.477868in}{1.430325in}}%
\pgfpathlineto{\pgfqpoint{4.489488in}{1.430325in}}%
\pgfpathlineto{\pgfqpoint{4.495298in}{1.442864in}}%
\pgfpathlineto{\pgfqpoint{4.501108in}{1.442864in}}%
\pgfpathlineto{\pgfqpoint{4.506917in}{1.455402in}}%
\pgfpathlineto{\pgfqpoint{4.518537in}{1.455402in}}%
\pgfpathlineto{\pgfqpoint{4.524347in}{1.467941in}}%
\pgfpathlineto{\pgfqpoint{4.535967in}{1.467941in}}%
\pgfpathlineto{\pgfqpoint{4.541777in}{1.480479in}}%
\pgfpathlineto{\pgfqpoint{4.547587in}{1.480479in}}%
\pgfpathlineto{\pgfqpoint{4.553397in}{1.493018in}}%
\pgfpathlineto{\pgfqpoint{4.565016in}{1.493018in}}%
\pgfpathlineto{\pgfqpoint{4.570826in}{1.505556in}}%
\pgfpathlineto{\pgfqpoint{4.576636in}{1.505556in}}%
\pgfpathlineto{\pgfqpoint{4.582446in}{1.518095in}}%
\pgfpathlineto{\pgfqpoint{4.588256in}{1.518095in}}%
\pgfpathlineto{\pgfqpoint{4.594066in}{1.530633in}}%
\pgfpathlineto{\pgfqpoint{4.605686in}{1.530633in}}%
\pgfpathlineto{\pgfqpoint{4.611496in}{1.543172in}}%
\pgfpathlineto{\pgfqpoint{4.617306in}{1.543172in}}%
\pgfpathlineto{\pgfqpoint{4.623115in}{1.555710in}}%
\pgfpathlineto{\pgfqpoint{4.634735in}{1.555710in}}%
\pgfpathlineto{\pgfqpoint{4.640545in}{1.568249in}}%
\pgfpathlineto{\pgfqpoint{4.646355in}{1.568249in}}%
\pgfpathlineto{\pgfqpoint{4.652165in}{1.580787in}}%
\pgfpathlineto{\pgfqpoint{4.657975in}{1.580787in}}%
\pgfpathlineto{\pgfqpoint{4.663785in}{1.593325in}}%
\pgfpathlineto{\pgfqpoint{4.669595in}{1.593325in}}%
\pgfpathlineto{\pgfqpoint{4.675405in}{1.605864in}}%
\pgfpathlineto{\pgfqpoint{4.687024in}{1.605864in}}%
\pgfpathlineto{\pgfqpoint{4.692834in}{1.618402in}}%
\pgfpathlineto{\pgfqpoint{4.698644in}{1.618402in}}%
\pgfpathlineto{\pgfqpoint{4.704454in}{1.630941in}}%
\pgfpathlineto{\pgfqpoint{4.710264in}{1.630941in}}%
\pgfpathlineto{\pgfqpoint{4.716074in}{1.643479in}}%
\pgfpathlineto{\pgfqpoint{4.721884in}{1.643479in}}%
\pgfpathlineto{\pgfqpoint{4.727694in}{1.656018in}}%
\pgfpathlineto{\pgfqpoint{4.733504in}{1.656018in}}%
\pgfpathlineto{\pgfqpoint{4.739313in}{1.668556in}}%
\pgfpathlineto{\pgfqpoint{4.745123in}{1.668556in}}%
\pgfpathlineto{\pgfqpoint{4.750933in}{1.681095in}}%
\pgfpathlineto{\pgfqpoint{4.762553in}{1.681095in}}%
\pgfpathlineto{\pgfqpoint{4.768363in}{1.693633in}}%
\pgfpathlineto{\pgfqpoint{4.774173in}{1.693633in}}%
\pgfpathlineto{\pgfqpoint{4.779983in}{1.706172in}}%
\pgfpathlineto{\pgfqpoint{4.785793in}{1.706172in}}%
\pgfpathlineto{\pgfqpoint{4.791602in}{1.718710in}}%
\pgfpathlineto{\pgfqpoint{4.797412in}{1.718710in}}%
\pgfpathlineto{\pgfqpoint{4.803222in}{1.731249in}}%
\pgfpathlineto{\pgfqpoint{4.809032in}{1.731249in}}%
\pgfpathlineto{\pgfqpoint{4.814842in}{1.743787in}}%
\pgfpathlineto{\pgfqpoint{4.820652in}{1.743787in}}%
\pgfpathlineto{\pgfqpoint{4.826462in}{1.756326in}}%
\pgfpathlineto{\pgfqpoint{4.832272in}{1.756326in}}%
\pgfpathlineto{\pgfqpoint{4.838082in}{1.768864in}}%
\pgfpathlineto{\pgfqpoint{4.843892in}{1.768864in}}%
\pgfpathlineto{\pgfqpoint{4.849701in}{1.781403in}}%
\pgfpathlineto{\pgfqpoint{4.855511in}{1.781403in}}%
\pgfpathlineto{\pgfqpoint{4.861321in}{1.793941in}}%
\pgfpathlineto{\pgfqpoint{4.867131in}{1.793941in}}%
\pgfpathlineto{\pgfqpoint{4.872941in}{1.806480in}}%
\pgfpathlineto{\pgfqpoint{4.878751in}{1.806480in}}%
\pgfpathlineto{\pgfqpoint{4.890371in}{1.831557in}}%
\pgfpathlineto{\pgfqpoint{4.896181in}{1.831557in}}%
\pgfpathlineto{\pgfqpoint{4.901991in}{1.844095in}}%
\pgfpathlineto{\pgfqpoint{4.907800in}{1.844095in}}%
\pgfpathlineto{\pgfqpoint{4.913610in}{1.856634in}}%
\pgfpathlineto{\pgfqpoint{4.919420in}{1.856634in}}%
\pgfpathlineto{\pgfqpoint{4.925230in}{1.869172in}}%
\pgfpathlineto{\pgfqpoint{4.931040in}{1.869172in}}%
\pgfpathlineto{\pgfqpoint{4.936850in}{1.881710in}}%
\pgfpathlineto{\pgfqpoint{4.942660in}{1.881710in}}%
\pgfpathlineto{\pgfqpoint{4.948470in}{1.894249in}}%
\pgfpathlineto{\pgfqpoint{4.954280in}{1.894249in}}%
\pgfpathlineto{\pgfqpoint{4.965899in}{1.919326in}}%
\pgfpathlineto{\pgfqpoint{4.971709in}{1.919326in}}%
\pgfpathlineto{\pgfqpoint{4.977519in}{1.931864in}}%
\pgfpathlineto{\pgfqpoint{4.983329in}{1.931864in}}%
\pgfpathlineto{\pgfqpoint{4.989139in}{1.944403in}}%
\pgfpathlineto{\pgfqpoint{4.994949in}{1.944403in}}%
\pgfpathlineto{\pgfqpoint{5.000759in}{1.956941in}}%
\pgfpathlineto{\pgfqpoint{5.006569in}{1.956941in}}%
\pgfpathlineto{\pgfqpoint{5.018189in}{1.982018in}}%
\pgfpathlineto{\pgfqpoint{5.023998in}{1.982018in}}%
\pgfpathlineto{\pgfqpoint{5.029808in}{1.994557in}}%
\pgfpathlineto{\pgfqpoint{5.035618in}{1.994557in}}%
\pgfpathlineto{\pgfqpoint{5.041428in}{2.007095in}}%
\pgfpathlineto{\pgfqpoint{5.047238in}{2.007095in}}%
\pgfpathlineto{\pgfqpoint{5.058858in}{2.032172in}}%
\pgfpathlineto{\pgfqpoint{5.064668in}{2.032172in}}%
\pgfpathlineto{\pgfqpoint{5.070478in}{2.044711in}}%
\pgfpathlineto{\pgfqpoint{5.076287in}{2.044711in}}%
\pgfpathlineto{\pgfqpoint{5.087907in}{2.069788in}}%
\pgfpathlineto{\pgfqpoint{5.093717in}{2.069788in}}%
\pgfpathlineto{\pgfqpoint{5.099527in}{2.082326in}}%
\pgfpathlineto{\pgfqpoint{5.105337in}{2.082326in}}%
\pgfpathlineto{\pgfqpoint{5.116957in}{2.107403in}}%
\pgfpathlineto{\pgfqpoint{5.122767in}{2.107403in}}%
\pgfpathlineto{\pgfqpoint{5.128577in}{2.119942in}}%
\pgfpathlineto{\pgfqpoint{5.134386in}{2.119942in}}%
\pgfpathlineto{\pgfqpoint{5.146006in}{2.145019in}}%
\pgfpathlineto{\pgfqpoint{5.151816in}{2.145019in}}%
\pgfpathlineto{\pgfqpoint{5.157626in}{2.157557in}}%
\pgfpathlineto{\pgfqpoint{5.163436in}{2.157557in}}%
\pgfpathlineto{\pgfqpoint{5.175056in}{2.182634in}}%
\pgfpathlineto{\pgfqpoint{5.180866in}{2.182634in}}%
\pgfpathlineto{\pgfqpoint{5.192485in}{2.207711in}}%
\pgfpathlineto{\pgfqpoint{5.198295in}{2.207711in}}%
\pgfpathlineto{\pgfqpoint{5.204105in}{2.220249in}}%
\pgfpathlineto{\pgfqpoint{5.209915in}{2.220249in}}%
\pgfpathlineto{\pgfqpoint{5.221535in}{2.245326in}}%
\pgfpathlineto{\pgfqpoint{5.227345in}{2.245326in}}%
\pgfpathlineto{\pgfqpoint{5.238965in}{2.270403in}}%
\pgfpathlineto{\pgfqpoint{5.244775in}{2.270403in}}%
\pgfpathlineto{\pgfqpoint{5.256394in}{2.295480in}}%
\pgfpathlineto{\pgfqpoint{5.262204in}{2.295480in}}%
\pgfpathlineto{\pgfqpoint{5.273824in}{2.320557in}}%
\pgfpathlineto{\pgfqpoint{5.279634in}{2.320557in}}%
\pgfpathlineto{\pgfqpoint{5.291254in}{2.345634in}}%
\pgfpathlineto{\pgfqpoint{5.297064in}{2.345634in}}%
\pgfpathlineto{\pgfqpoint{5.308683in}{2.370711in}}%
\pgfpathlineto{\pgfqpoint{5.314493in}{2.370711in}}%
\pgfpathlineto{\pgfqpoint{5.326113in}{2.395788in}}%
\pgfpathlineto{\pgfqpoint{5.331923in}{2.395788in}}%
\pgfpathlineto{\pgfqpoint{5.343543in}{2.420865in}}%
\pgfpathlineto{\pgfqpoint{5.349353in}{2.420865in}}%
\pgfpathlineto{\pgfqpoint{5.360973in}{2.445942in}}%
\pgfpathlineto{\pgfqpoint{5.366782in}{2.445942in}}%
\pgfpathlineto{\pgfqpoint{5.378402in}{2.471019in}}%
\pgfpathlineto{\pgfqpoint{5.384212in}{2.471019in}}%
\pgfpathlineto{\pgfqpoint{5.395832in}{2.496096in}}%
\pgfpathlineto{\pgfqpoint{5.401642in}{2.496096in}}%
\pgfpathlineto{\pgfqpoint{5.413262in}{2.521173in}}%
\pgfpathlineto{\pgfqpoint{5.419071in}{2.521173in}}%
\pgfpathlineto{\pgfqpoint{5.436501in}{2.558788in}}%
\pgfpathlineto{\pgfqpoint{5.442311in}{2.558788in}}%
\pgfpathlineto{\pgfqpoint{5.453931in}{2.583865in}}%
\pgfpathlineto{\pgfqpoint{5.459741in}{2.583865in}}%
\pgfpathlineto{\pgfqpoint{5.477170in}{2.621481in}}%
\pgfpathlineto{\pgfqpoint{5.482980in}{2.621481in}}%
\pgfpathlineto{\pgfqpoint{5.494600in}{2.646558in}}%
\pgfpathlineto{\pgfqpoint{5.500410in}{2.646558in}}%
\pgfpathlineto{\pgfqpoint{5.517840in}{2.684173in}}%
\pgfpathlineto{\pgfqpoint{5.523650in}{2.684173in}}%
\pgfpathlineto{\pgfqpoint{5.541079in}{2.721789in}}%
\pgfpathlineto{\pgfqpoint{5.546889in}{2.721789in}}%
\pgfpathlineto{\pgfqpoint{5.558509in}{2.746865in}}%
\pgfpathlineto{\pgfqpoint{5.564319in}{2.746865in}}%
\pgfpathlineto{\pgfqpoint{5.581749in}{2.784481in}}%
\pgfpathlineto{\pgfqpoint{5.587559in}{2.784481in}}%
\pgfpathlineto{\pgfqpoint{5.604988in}{2.822096in}}%
\pgfpathlineto{\pgfqpoint{5.610798in}{2.822096in}}%
\pgfpathlineto{\pgfqpoint{5.628228in}{2.859712in}}%
\pgfpathlineto{\pgfqpoint{5.634038in}{2.859712in}}%
\pgfpathlineto{\pgfqpoint{5.651467in}{2.897327in}}%
\pgfpathlineto{\pgfqpoint{5.657277in}{2.897327in}}%
\pgfpathlineto{\pgfqpoint{5.680517in}{2.947481in}}%
\pgfpathlineto{\pgfqpoint{5.686327in}{2.947481in}}%
\pgfpathlineto{\pgfqpoint{5.703756in}{2.985097in}}%
\pgfpathlineto{\pgfqpoint{5.709566in}{2.985097in}}%
\pgfpathlineto{\pgfqpoint{5.732806in}{3.035250in}}%
\pgfpathlineto{\pgfqpoint{5.738616in}{3.035250in}}%
\pgfpathlineto{\pgfqpoint{5.756046in}{3.072866in}}%
\pgfpathlineto{\pgfqpoint{5.761855in}{3.072866in}}%
\pgfpathlineto{\pgfqpoint{5.785095in}{3.123020in}}%
\pgfpathlineto{\pgfqpoint{5.790905in}{3.123020in}}%
\pgfpathlineto{\pgfqpoint{5.814145in}{3.173174in}}%
\pgfpathlineto{\pgfqpoint{5.819954in}{3.173174in}}%
\pgfpathlineto{\pgfqpoint{5.843194in}{3.223328in}}%
\pgfpathlineto{\pgfqpoint{5.849004in}{3.223328in}}%
\pgfpathlineto{\pgfqpoint{5.878053in}{3.286020in}}%
\pgfpathlineto{\pgfqpoint{5.883863in}{3.286020in}}%
\pgfpathlineto{\pgfqpoint{5.912913in}{3.348712in}}%
\pgfpathlineto{\pgfqpoint{5.918723in}{3.348712in}}%
\pgfpathlineto{\pgfqpoint{5.941962in}{3.398866in}}%
\pgfpathlineto{\pgfqpoint{5.947772in}{3.398866in}}%
\pgfpathlineto{\pgfqpoint{5.976822in}{3.461559in}}%
\pgfpathlineto{\pgfqpoint{5.982632in}{3.461559in}}%
\pgfpathlineto{\pgfqpoint{6.017491in}{3.536790in}}%
\pgfpathlineto{\pgfqpoint{6.023301in}{3.536790in}}%
\pgfpathlineto{\pgfqpoint{6.058160in}{3.612020in}}%
\pgfpathlineto{\pgfqpoint{6.063970in}{3.612020in}}%
\pgfpathlineto{\pgfqpoint{6.098830in}{3.687251in}}%
\pgfpathlineto{\pgfqpoint{6.104639in}{3.687251in}}%
\pgfpathlineto{\pgfqpoint{6.139499in}{3.762482in}}%
\pgfpathlineto{\pgfqpoint{6.145309in}{3.762482in}}%
\pgfpathlineto{\pgfqpoint{6.185978in}{3.850252in}}%
\pgfpathlineto{\pgfqpoint{6.191788in}{3.850252in}}%
\pgfpathlineto{\pgfqpoint{6.238267in}{3.950559in}}%
\pgfpathlineto{\pgfqpoint{6.244077in}{3.950559in}}%
\pgfpathlineto{\pgfqpoint{6.296366in}{4.063406in}}%
\pgfpathlineto{\pgfqpoint{6.302176in}{4.063406in}}%
\pgfpathlineto{\pgfqpoint{6.354465in}{4.176252in}}%
\pgfpathlineto{\pgfqpoint{6.360275in}{4.176252in}}%
\pgfpathlineto{\pgfqpoint{6.424184in}{4.314175in}}%
\pgfpathlineto{\pgfqpoint{6.429994in}{4.314175in}}%
\pgfpathlineto{\pgfqpoint{6.505522in}{4.477175in}}%
\pgfpathlineto{\pgfqpoint{6.511332in}{4.477175in}}%
\pgfpathlineto{\pgfqpoint{6.598481in}{4.665253in}}%
\pgfpathlineto{\pgfqpoint{6.604291in}{4.665253in}}%
\pgfpathlineto{\pgfqpoint{6.732108in}{4.941099in}}%
\pgfpathlineto{\pgfqpoint{6.737918in}{4.941099in}}%
\pgfpathlineto{\pgfqpoint{6.993554in}{5.492792in}}%
\pgfpathlineto{\pgfqpoint{6.993554in}{5.492792in}}%
\pgfusepath{stroke}%
\end{pgfscope}%
\begin{pgfscope}%
\pgfpathrectangle{\pgfqpoint{0.957105in}{0.774972in}}{\pgfqpoint{6.390117in}{5.072250in}}%
\pgfusepath{clip}%
\pgfsetbuttcap%
\pgfsetroundjoin%
\pgfsetlinewidth{2.509375pt}%
\definecolor{currentstroke}{rgb}{0.447059,0.168627,0.415686}%
\pgfsetstrokecolor{currentstroke}%
\pgfsetstrokeopacity{0.700000}%
\pgfsetdash{{9.250000pt}{4.000000pt}}{0.000000pt}%
\pgfpathmoveto{\pgfqpoint{1.247565in}{1.016555in}}%
\pgfpathlineto{\pgfqpoint{4.146704in}{1.016555in}}%
\pgfpathlineto{\pgfqpoint{4.152514in}{1.029094in}}%
\pgfpathlineto{\pgfqpoint{4.222232in}{1.029094in}}%
\pgfpathlineto{\pgfqpoint{4.228042in}{1.041632in}}%
\pgfpathlineto{\pgfqpoint{4.274522in}{1.041632in}}%
\pgfpathlineto{\pgfqpoint{4.280331in}{1.054171in}}%
\pgfpathlineto{\pgfqpoint{4.321001in}{1.054171in}}%
\pgfpathlineto{\pgfqpoint{4.326811in}{1.066709in}}%
\pgfpathlineto{\pgfqpoint{4.361670in}{1.066709in}}%
\pgfpathlineto{\pgfqpoint{4.367480in}{1.079248in}}%
\pgfpathlineto{\pgfqpoint{4.396529in}{1.079248in}}%
\pgfpathlineto{\pgfqpoint{4.402339in}{1.091786in}}%
\pgfpathlineto{\pgfqpoint{4.425579in}{1.091786in}}%
\pgfpathlineto{\pgfqpoint{4.431389in}{1.104325in}}%
\pgfpathlineto{\pgfqpoint{4.454628in}{1.104325in}}%
\pgfpathlineto{\pgfqpoint{4.460438in}{1.116863in}}%
\pgfpathlineto{\pgfqpoint{4.483678in}{1.116863in}}%
\pgfpathlineto{\pgfqpoint{4.489488in}{1.129402in}}%
\pgfpathlineto{\pgfqpoint{4.506917in}{1.129402in}}%
\pgfpathlineto{\pgfqpoint{4.512727in}{1.141940in}}%
\pgfpathlineto{\pgfqpoint{4.530157in}{1.141940in}}%
\pgfpathlineto{\pgfqpoint{4.535967in}{1.154479in}}%
\pgfpathlineto{\pgfqpoint{4.553397in}{1.154479in}}%
\pgfpathlineto{\pgfqpoint{4.559207in}{1.167017in}}%
\pgfpathlineto{\pgfqpoint{4.576636in}{1.167017in}}%
\pgfpathlineto{\pgfqpoint{4.582446in}{1.179556in}}%
\pgfpathlineto{\pgfqpoint{4.599876in}{1.179556in}}%
\pgfpathlineto{\pgfqpoint{4.605686in}{1.192094in}}%
\pgfpathlineto{\pgfqpoint{4.617306in}{1.192094in}}%
\pgfpathlineto{\pgfqpoint{4.623115in}{1.204633in}}%
\pgfpathlineto{\pgfqpoint{4.640545in}{1.204633in}}%
\pgfpathlineto{\pgfqpoint{4.646355in}{1.217171in}}%
\pgfpathlineto{\pgfqpoint{4.657975in}{1.217171in}}%
\pgfpathlineto{\pgfqpoint{4.663785in}{1.229710in}}%
\pgfpathlineto{\pgfqpoint{4.675405in}{1.229710in}}%
\pgfpathlineto{\pgfqpoint{4.681214in}{1.242248in}}%
\pgfpathlineto{\pgfqpoint{4.692834in}{1.242248in}}%
\pgfpathlineto{\pgfqpoint{4.698644in}{1.254787in}}%
\pgfpathlineto{\pgfqpoint{4.710264in}{1.254787in}}%
\pgfpathlineto{\pgfqpoint{4.716074in}{1.267325in}}%
\pgfpathlineto{\pgfqpoint{4.727694in}{1.267325in}}%
\pgfpathlineto{\pgfqpoint{4.733504in}{1.279864in}}%
\pgfpathlineto{\pgfqpoint{4.745123in}{1.279864in}}%
\pgfpathlineto{\pgfqpoint{4.750933in}{1.292402in}}%
\pgfpathlineto{\pgfqpoint{4.762553in}{1.292402in}}%
\pgfpathlineto{\pgfqpoint{4.768363in}{1.304940in}}%
\pgfpathlineto{\pgfqpoint{4.774173in}{1.304940in}}%
\pgfpathlineto{\pgfqpoint{4.779983in}{1.317479in}}%
\pgfpathlineto{\pgfqpoint{4.791602in}{1.317479in}}%
\pgfpathlineto{\pgfqpoint{4.797412in}{1.330017in}}%
\pgfpathlineto{\pgfqpoint{4.803222in}{1.330017in}}%
\pgfpathlineto{\pgfqpoint{4.809032in}{1.342556in}}%
\pgfpathlineto{\pgfqpoint{4.820652in}{1.342556in}}%
\pgfpathlineto{\pgfqpoint{4.826462in}{1.355094in}}%
\pgfpathlineto{\pgfqpoint{4.832272in}{1.355094in}}%
\pgfpathlineto{\pgfqpoint{4.838082in}{1.367633in}}%
\pgfpathlineto{\pgfqpoint{4.849701in}{1.367633in}}%
\pgfpathlineto{\pgfqpoint{4.855511in}{1.380171in}}%
\pgfpathlineto{\pgfqpoint{4.861321in}{1.380171in}}%
\pgfpathlineto{\pgfqpoint{4.867131in}{1.392710in}}%
\pgfpathlineto{\pgfqpoint{4.878751in}{1.392710in}}%
\pgfpathlineto{\pgfqpoint{4.884561in}{1.405248in}}%
\pgfpathlineto{\pgfqpoint{4.890371in}{1.405248in}}%
\pgfpathlineto{\pgfqpoint{4.896181in}{1.417787in}}%
\pgfpathlineto{\pgfqpoint{4.901991in}{1.417787in}}%
\pgfpathlineto{\pgfqpoint{4.907800in}{1.430325in}}%
\pgfpathlineto{\pgfqpoint{4.913610in}{1.430325in}}%
\pgfpathlineto{\pgfqpoint{4.919420in}{1.442864in}}%
\pgfpathlineto{\pgfqpoint{4.931040in}{1.442864in}}%
\pgfpathlineto{\pgfqpoint{4.936850in}{1.455402in}}%
\pgfpathlineto{\pgfqpoint{4.942660in}{1.455402in}}%
\pgfpathlineto{\pgfqpoint{4.948470in}{1.467941in}}%
\pgfpathlineto{\pgfqpoint{4.954280in}{1.467941in}}%
\pgfpathlineto{\pgfqpoint{4.960090in}{1.480479in}}%
\pgfpathlineto{\pgfqpoint{4.965899in}{1.480479in}}%
\pgfpathlineto{\pgfqpoint{4.971709in}{1.493018in}}%
\pgfpathlineto{\pgfqpoint{4.977519in}{1.493018in}}%
\pgfpathlineto{\pgfqpoint{4.983329in}{1.505556in}}%
\pgfpathlineto{\pgfqpoint{4.989139in}{1.505556in}}%
\pgfpathlineto{\pgfqpoint{4.994949in}{1.518095in}}%
\pgfpathlineto{\pgfqpoint{5.000759in}{1.518095in}}%
\pgfpathlineto{\pgfqpoint{5.006569in}{1.530633in}}%
\pgfpathlineto{\pgfqpoint{5.012379in}{1.530633in}}%
\pgfpathlineto{\pgfqpoint{5.018189in}{1.543172in}}%
\pgfpathlineto{\pgfqpoint{5.023998in}{1.543172in}}%
\pgfpathlineto{\pgfqpoint{5.029808in}{1.555710in}}%
\pgfpathlineto{\pgfqpoint{5.035618in}{1.555710in}}%
\pgfpathlineto{\pgfqpoint{5.041428in}{1.568249in}}%
\pgfpathlineto{\pgfqpoint{5.047238in}{1.568249in}}%
\pgfpathlineto{\pgfqpoint{5.053048in}{1.580787in}}%
\pgfpathlineto{\pgfqpoint{5.058858in}{1.580787in}}%
\pgfpathlineto{\pgfqpoint{5.064668in}{1.593325in}}%
\pgfpathlineto{\pgfqpoint{5.070478in}{1.593325in}}%
\pgfpathlineto{\pgfqpoint{5.076287in}{1.605864in}}%
\pgfpathlineto{\pgfqpoint{5.082097in}{1.605864in}}%
\pgfpathlineto{\pgfqpoint{5.093717in}{1.630941in}}%
\pgfpathlineto{\pgfqpoint{5.099527in}{1.630941in}}%
\pgfpathlineto{\pgfqpoint{5.105337in}{1.643479in}}%
\pgfpathlineto{\pgfqpoint{5.111147in}{1.643479in}}%
\pgfpathlineto{\pgfqpoint{5.116957in}{1.656018in}}%
\pgfpathlineto{\pgfqpoint{5.122767in}{1.656018in}}%
\pgfpathlineto{\pgfqpoint{5.128577in}{1.668556in}}%
\pgfpathlineto{\pgfqpoint{5.134386in}{1.668556in}}%
\pgfpathlineto{\pgfqpoint{5.146006in}{1.693633in}}%
\pgfpathlineto{\pgfqpoint{5.151816in}{1.693633in}}%
\pgfpathlineto{\pgfqpoint{5.157626in}{1.706172in}}%
\pgfpathlineto{\pgfqpoint{5.163436in}{1.706172in}}%
\pgfpathlineto{\pgfqpoint{5.169246in}{1.718710in}}%
\pgfpathlineto{\pgfqpoint{5.175056in}{1.718710in}}%
\pgfpathlineto{\pgfqpoint{5.186676in}{1.743787in}}%
\pgfpathlineto{\pgfqpoint{5.192485in}{1.743787in}}%
\pgfpathlineto{\pgfqpoint{5.198295in}{1.756326in}}%
\pgfpathlineto{\pgfqpoint{5.204105in}{1.756326in}}%
\pgfpathlineto{\pgfqpoint{5.215725in}{1.781403in}}%
\pgfpathlineto{\pgfqpoint{5.221535in}{1.781403in}}%
\pgfpathlineto{\pgfqpoint{5.233155in}{1.806480in}}%
\pgfpathlineto{\pgfqpoint{5.238965in}{1.806480in}}%
\pgfpathlineto{\pgfqpoint{5.244775in}{1.819018in}}%
\pgfpathlineto{\pgfqpoint{5.250584in}{1.819018in}}%
\pgfpathlineto{\pgfqpoint{5.262204in}{1.844095in}}%
\pgfpathlineto{\pgfqpoint{5.268014in}{1.844095in}}%
\pgfpathlineto{\pgfqpoint{5.279634in}{1.869172in}}%
\pgfpathlineto{\pgfqpoint{5.285444in}{1.869172in}}%
\pgfpathlineto{\pgfqpoint{5.297064in}{1.894249in}}%
\pgfpathlineto{\pgfqpoint{5.302874in}{1.894249in}}%
\pgfpathlineto{\pgfqpoint{5.314493in}{1.919326in}}%
\pgfpathlineto{\pgfqpoint{5.320303in}{1.919326in}}%
\pgfpathlineto{\pgfqpoint{5.331923in}{1.944403in}}%
\pgfpathlineto{\pgfqpoint{5.337733in}{1.944403in}}%
\pgfpathlineto{\pgfqpoint{5.349353in}{1.969480in}}%
\pgfpathlineto{\pgfqpoint{5.355163in}{1.969480in}}%
\pgfpathlineto{\pgfqpoint{5.366782in}{1.994557in}}%
\pgfpathlineto{\pgfqpoint{5.372592in}{1.994557in}}%
\pgfpathlineto{\pgfqpoint{5.384212in}{2.019634in}}%
\pgfpathlineto{\pgfqpoint{5.390022in}{2.019634in}}%
\pgfpathlineto{\pgfqpoint{6.993554in}{5.480254in}}%
\pgfpathlineto{\pgfqpoint{6.993554in}{5.480254in}}%
\pgfusepath{stroke}%
\end{pgfscope}%
\begin{pgfscope}%
\pgfpathrectangle{\pgfqpoint{0.957105in}{0.774972in}}{\pgfqpoint{6.390117in}{5.072250in}}%
\pgfusepath{clip}%
\pgfsetrectcap%
\pgfsetroundjoin%
\pgfsetlinewidth{2.509375pt}%
\definecolor{currentstroke}{rgb}{0.635294,0.149020,0.294118}%
\pgfsetstrokecolor{currentstroke}%
\pgfsetstrokeopacity{0.700000}%
\pgfsetdash{}{0pt}%
\pgfpathmoveto{\pgfqpoint{1.247565in}{1.016555in}}%
\pgfpathlineto{\pgfqpoint{4.565016in}{1.016555in}}%
\pgfpathlineto{\pgfqpoint{4.570826in}{1.029094in}}%
\pgfpathlineto{\pgfqpoint{4.628925in}{1.029094in}}%
\pgfpathlineto{\pgfqpoint{4.634735in}{1.041632in}}%
\pgfpathlineto{\pgfqpoint{4.675405in}{1.041632in}}%
\pgfpathlineto{\pgfqpoint{4.681214in}{1.054171in}}%
\pgfpathlineto{\pgfqpoint{4.710264in}{1.054171in}}%
\pgfpathlineto{\pgfqpoint{4.716074in}{1.066709in}}%
\pgfpathlineto{\pgfqpoint{4.745123in}{1.066709in}}%
\pgfpathlineto{\pgfqpoint{4.750933in}{1.079248in}}%
\pgfpathlineto{\pgfqpoint{4.774173in}{1.079248in}}%
\pgfpathlineto{\pgfqpoint{4.779983in}{1.091786in}}%
\pgfpathlineto{\pgfqpoint{4.803222in}{1.091786in}}%
\pgfpathlineto{\pgfqpoint{4.809032in}{1.104325in}}%
\pgfpathlineto{\pgfqpoint{4.826462in}{1.104325in}}%
\pgfpathlineto{\pgfqpoint{4.832272in}{1.116863in}}%
\pgfpathlineto{\pgfqpoint{4.849701in}{1.116863in}}%
\pgfpathlineto{\pgfqpoint{4.855511in}{1.129402in}}%
\pgfpathlineto{\pgfqpoint{4.872941in}{1.129402in}}%
\pgfpathlineto{\pgfqpoint{4.878751in}{1.141940in}}%
\pgfpathlineto{\pgfqpoint{4.890371in}{1.141940in}}%
\pgfpathlineto{\pgfqpoint{4.896181in}{1.154479in}}%
\pgfpathlineto{\pgfqpoint{4.913610in}{1.154479in}}%
\pgfpathlineto{\pgfqpoint{4.919420in}{1.167017in}}%
\pgfpathlineto{\pgfqpoint{4.931040in}{1.167017in}}%
\pgfpathlineto{\pgfqpoint{4.936850in}{1.179556in}}%
\pgfpathlineto{\pgfqpoint{4.948470in}{1.179556in}}%
\pgfpathlineto{\pgfqpoint{4.954280in}{1.192094in}}%
\pgfpathlineto{\pgfqpoint{4.965899in}{1.192094in}}%
\pgfpathlineto{\pgfqpoint{4.971709in}{1.204633in}}%
\pgfpathlineto{\pgfqpoint{4.983329in}{1.204633in}}%
\pgfpathlineto{\pgfqpoint{4.989139in}{1.217171in}}%
\pgfpathlineto{\pgfqpoint{5.000759in}{1.217171in}}%
\pgfpathlineto{\pgfqpoint{5.006569in}{1.229710in}}%
\pgfpathlineto{\pgfqpoint{5.012379in}{1.229710in}}%
\pgfpathlineto{\pgfqpoint{5.018189in}{1.242248in}}%
\pgfpathlineto{\pgfqpoint{5.029808in}{1.242248in}}%
\pgfpathlineto{\pgfqpoint{6.993554in}{5.480254in}}%
\pgfpathlineto{\pgfqpoint{6.993554in}{5.480254in}}%
\pgfusepath{stroke}%
\end{pgfscope}%
\begin{pgfscope}%
\pgfpathrectangle{\pgfqpoint{0.957105in}{0.774972in}}{\pgfqpoint{6.390117in}{5.072250in}}%
\pgfusepath{clip}%
\pgfsetrectcap%
\pgfsetroundjoin%
\pgfsetlinewidth{2.509375pt}%
\definecolor{currentstroke}{rgb}{0.898039,0.000000,0.000000}%
\pgfsetstrokecolor{currentstroke}%
\pgfsetstrokeopacity{0.700000}%
\pgfsetdash{}{0pt}%
\pgfpathmoveto{\pgfqpoint{1.247565in}{1.016555in}}%
\pgfpathlineto{\pgfqpoint{4.925230in}{1.016555in}}%
\pgfpathlineto{\pgfqpoint{6.993554in}{5.480254in}}%
\pgfpathlineto{\pgfqpoint{6.993554in}{5.480254in}}%
\pgfusepath{stroke}%
\end{pgfscope}%
\begin{pgfscope}%
\pgfsetrectcap%
\pgfsetmiterjoin%
\pgfsetlinewidth{0.803000pt}%
\definecolor{currentstroke}{rgb}{0.000000,0.000000,0.000000}%
\pgfsetstrokecolor{currentstroke}%
\pgfsetdash{}{0pt}%
\pgfpathmoveto{\pgfqpoint{0.957105in}{0.774972in}}%
\pgfpathlineto{\pgfqpoint{0.957105in}{5.847222in}}%
\pgfusepath{stroke}%
\end{pgfscope}%
\begin{pgfscope}%
\pgfsetrectcap%
\pgfsetmiterjoin%
\pgfsetlinewidth{0.803000pt}%
\definecolor{currentstroke}{rgb}{0.000000,0.000000,0.000000}%
\pgfsetstrokecolor{currentstroke}%
\pgfsetdash{}{0pt}%
\pgfpathmoveto{\pgfqpoint{7.347222in}{0.774972in}}%
\pgfpathlineto{\pgfqpoint{7.347222in}{5.847222in}}%
\pgfusepath{stroke}%
\end{pgfscope}%
\begin{pgfscope}%
\pgfsetrectcap%
\pgfsetmiterjoin%
\pgfsetlinewidth{0.803000pt}%
\definecolor{currentstroke}{rgb}{0.000000,0.000000,0.000000}%
\pgfsetstrokecolor{currentstroke}%
\pgfsetdash{}{0pt}%
\pgfpathmoveto{\pgfqpoint{0.957105in}{0.774972in}}%
\pgfpathlineto{\pgfqpoint{7.347222in}{0.774972in}}%
\pgfusepath{stroke}%
\end{pgfscope}%
\begin{pgfscope}%
\pgfsetrectcap%
\pgfsetmiterjoin%
\pgfsetlinewidth{0.803000pt}%
\definecolor{currentstroke}{rgb}{0.000000,0.000000,0.000000}%
\pgfsetstrokecolor{currentstroke}%
\pgfsetdash{}{0pt}%
\pgfpathmoveto{\pgfqpoint{0.957105in}{5.847222in}}%
\pgfpathlineto{\pgfqpoint{7.347222in}{5.847222in}}%
\pgfusepath{stroke}%
\end{pgfscope}%
\begin{pgfscope}%
\pgfsetbuttcap%
\pgfsetmiterjoin%
\definecolor{currentfill}{rgb}{1.000000,1.000000,1.000000}%
\pgfsetfillcolor{currentfill}%
\pgfsetfillopacity{0.800000}%
\pgfsetlinewidth{1.003750pt}%
\definecolor{currentstroke}{rgb}{0.800000,0.800000,0.800000}%
\pgfsetstrokecolor{currentstroke}%
\pgfsetstrokeopacity{0.800000}%
\pgfsetdash{}{0pt}%
\pgfpathmoveto{\pgfqpoint{1.161271in}{3.292076in}}%
\pgfpathlineto{\pgfqpoint{5.248795in}{3.292076in}}%
\pgfpathquadraticcurveto{\pgfqpoint{5.307128in}{3.292076in}}{\pgfqpoint{5.307128in}{3.350410in}}%
\pgfpathlineto{\pgfqpoint{5.307128in}{5.643056in}}%
\pgfpathquadraticcurveto{\pgfqpoint{5.307128in}{5.701389in}}{\pgfqpoint{5.248795in}{5.701389in}}%
\pgfpathlineto{\pgfqpoint{1.161271in}{5.701389in}}%
\pgfpathquadraticcurveto{\pgfqpoint{1.102938in}{5.701389in}}{\pgfqpoint{1.102938in}{5.643056in}}%
\pgfpathlineto{\pgfqpoint{1.102938in}{3.350410in}}%
\pgfpathquadraticcurveto{\pgfqpoint{1.102938in}{3.292076in}}{\pgfqpoint{1.161271in}{3.292076in}}%
\pgfpathclose%
\pgfusepath{stroke,fill}%
\end{pgfscope}%
\begin{pgfscope}%
\pgfsetrectcap%
\pgfsetroundjoin%
\pgfsetlinewidth{2.509375pt}%
\definecolor{currentstroke}{rgb}{0.000000,0.000000,0.501961}%
\pgfsetstrokecolor{currentstroke}%
\pgfsetdash{}{0pt}%
\pgfpathmoveto{\pgfqpoint{1.365438in}{5.336806in}}%
\pgfpathlineto{\pgfqpoint{1.598771in}{5.336806in}}%
\pgfusepath{stroke}%
\end{pgfscope}%
\begin{pgfscope}%
\definecolor{textcolor}{rgb}{0.000000,0.000000,0.000000}%
\pgfsetstrokecolor{textcolor}%
\pgfsetfillcolor{textcolor}%
\pgftext[x=1.686271in,y=5.234722in,left,base]{\color{textcolor}\rmfamily\fontsize{21.000000}{25.200000}\selectfont Upper Bound}%
\end{pgfscope}%
\begin{pgfscope}%
\pgfsetbuttcap%
\pgfsetroundjoin%
\pgfsetlinewidth{2.509375pt}%
\definecolor{currentstroke}{rgb}{0.254902,0.184314,0.533333}%
\pgfsetstrokecolor{currentstroke}%
\pgfsetstrokeopacity{0.700000}%
\pgfsetdash{{2.500000pt}{4.125000pt}}{0.000000pt}%
\pgfpathmoveto{\pgfqpoint{1.365438in}{4.930776in}}%
\pgfpathlineto{\pgfqpoint{1.598771in}{4.930776in}}%
\pgfusepath{stroke}%
\end{pgfscope}%
\begin{pgfscope}%
\definecolor{textcolor}{rgb}{0.000000,0.000000,0.000000}%
\pgfsetstrokecolor{textcolor}%
\pgfsetfillcolor{textcolor}%
\pgftext[x=1.686271in,y=4.828693in,left,base]{\color{textcolor}\rmfamily\fontsize{21.000000}{25.200000}\selectfont AC Lower Bound \(\displaystyle \sigma=1.0\)   }%
\end{pgfscope}%
\begin{pgfscope}%
\pgfsetbuttcap%
\pgfsetroundjoin%
\pgfsetlinewidth{2.509375pt}%
\definecolor{currentstroke}{rgb}{0.447059,0.168627,0.415686}%
\pgfsetstrokecolor{currentstroke}%
\pgfsetstrokeopacity{0.700000}%
\pgfsetdash{{9.250000pt}{4.000000pt}}{0.000000pt}%
\pgfpathmoveto{\pgfqpoint{1.365438in}{4.524747in}}%
\pgfpathlineto{\pgfqpoint{1.598771in}{4.524747in}}%
\pgfusepath{stroke}%
\end{pgfscope}%
\begin{pgfscope}%
\definecolor{textcolor}{rgb}{0.000000,0.000000,0.000000}%
\pgfsetstrokecolor{textcolor}%
\pgfsetfillcolor{textcolor}%
\pgftext[x=1.686271in,y=4.422664in,left,base]{\color{textcolor}\rmfamily\fontsize{21.000000}{25.200000}\selectfont AC Lower Bound \(\displaystyle \sigma=1.2\)  }%
\end{pgfscope}%
\begin{pgfscope}%
\pgfsetrectcap%
\pgfsetroundjoin%
\pgfsetlinewidth{2.509375pt}%
\definecolor{currentstroke}{rgb}{0.635294,0.149020,0.294118}%
\pgfsetstrokecolor{currentstroke}%
\pgfsetstrokeopacity{0.700000}%
\pgfsetdash{}{0pt}%
\pgfpathmoveto{\pgfqpoint{1.365438in}{4.118718in}}%
\pgfpathlineto{\pgfqpoint{1.598771in}{4.118718in}}%
\pgfusepath{stroke}%
\end{pgfscope}%
\begin{pgfscope}%
\definecolor{textcolor}{rgb}{0.000000,0.000000,0.000000}%
\pgfsetstrokecolor{textcolor}%
\pgfsetfillcolor{textcolor}%
\pgftext[x=1.686271in,y=4.016635in,left,base]{\color{textcolor}\rmfamily\fontsize{21.000000}{25.200000}\selectfont AC Lower Bound \(\displaystyle \sigma=1.4\)  }%
\end{pgfscope}%
\begin{pgfscope}%
\pgfsetrectcap%
\pgfsetroundjoin%
\pgfsetlinewidth{2.509375pt}%
\definecolor{currentstroke}{rgb}{0.898039,0.000000,0.000000}%
\pgfsetstrokecolor{currentstroke}%
\pgfsetstrokeopacity{0.700000}%
\pgfsetdash{}{0pt}%
\pgfpathmoveto{\pgfqpoint{1.365438in}{3.712689in}}%
\pgfpathlineto{\pgfqpoint{1.598771in}{3.712689in}}%
\pgfusepath{stroke}%
\end{pgfscope}%
\begin{pgfscope}%
\definecolor{textcolor}{rgb}{0.000000,0.000000,0.000000}%
\pgfsetstrokecolor{textcolor}%
\pgfsetfillcolor{textcolor}%
\pgftext[x=1.686271in,y=3.610606in,left,base]{\color{textcolor}\rmfamily\fontsize{21.000000}{25.200000}\selectfont AC Lower Bound \(\displaystyle \sigma \geq e-1\)  }%
\end{pgfscope}%
\end{pgfpicture}%
\makeatother%
\endgroup%